\newcommand{\DoPrePrint}{0} 
\ifnum\DoPrePrint=1
  \documentclass[aps,prd,preprint,showpacs,superscriptaddress,nofootinbib,linenumbers,floatfix]{revtex4-1}
\else
  \documentclass[aps,prd,twocolumn,showpacs,superscriptaddress,nofootinbib,floatfix]{revtex4-1}
\fi

\usepackage{graphicx}  % for figures
\usepackage{xspace}
\usepackage{units}
\usepackage{multirow}

\newcommand{\sizecheck}{0} % 0 to do nothing; 1 to check size
\newcommand{\APPsupp}{1}   % 0 to do nothing; 1 to put the appendix in a supplement
\ifnum\APPsupp=1
  \newcommand{\SuppLocation}{in the Appendix}
\else
  \newcommand{\SuppLocation}{at URL}
\fi

\newif\ifpdf
\ifx\pdfoutput\undefined
   \pdffalse
\else
   \pdfoutput=1
   \pdftrue
\fi
\usepackage{graphicx}
\ifpdf
   \usepackage{epstopdf}
\fi
\makeatletter
\def\p@subsubsection{}
\def\p@subsection{}
\def\p@section{}
\makeatother

\begin{document}

% Detectors
\newcommand{\minerva}{MINERvA\xspace}
\newcommand{\minos}{MINOS\xspace}
\newcommand{\argoneut}{ArgoNeuT\xspace}

% Particles
\newcommand{\nue}{\ensuremath{\nu_{e}}\xspace}
\newcommand{\nuebar}{\ensuremath{\overline{\nu}_{e}}\xspace}
\newcommand{\numu}{\ensuremath{\nu_{\mu}}\xspace}
\newcommand{\numubar}{\ensuremath{\overline{\nu}_{\mu}}\xspace}
\newcommand{\nutau}{\ensuremath{\nu_{\tau}}\xspace}
\newcommand{\nutaubar}{\ensuremath{\overline{\nu}_{\tau}}\xspace}
\newcommand{\muplus}{\ensuremath{\mu^{+}}\xspace}
\newcommand{\muminus}{\ensuremath{\mu^{-}}\xspace}
\newcommand{\piplus}{\ensuremath{\pi^{+}}\xspace}
\newcommand{\piminus}{\ensuremath{\pi^{-}}\xspace}
\newcommand{\picharge}{\ensuremath{\pi^{\pm}}\xspace}
\newcommand{\pizero}{\ensuremath{\pi^{0}}\xspace}
\newcommand{\kcharge}{\ensuremath{K^{\pm}}\xspace}
\newcommand{\Pom}{I$\!$P}  

% Neutrino Oscillation
\newcommand{\fstate}[1]{\ensuremath{|\nu_{#1}\rangle}\xspace}
\newcommand{\mstate}[1]{\ensuremath{|\nu_{#1}\rangle}\xspace}

% Flux
\newcommand{\pC}{\ensuremath{pC}\xspace}
\newcommand{\pT}{\ensuremath{p_{T}}\xspace} 
\newcommand{\xF}{\ensuremath{x_{F}}\xspace} 
\newcommand{\veve}{\ensuremath{\nu+e^{-}\to\nu+e^{-}}\xspace}

% Calibrations
\newcommand{\dEdx}{\ensuremath{\frac{dE}{dx}}\xspace}
\newcommand{\meu}{\ensuremath{M}\xspace}
\newcommand{\meutime}{\ensuremath{M(t)}\xspace}
\newcommand{\meutrial}{\ensuremath{M_{T}}\xspace}
\newcommand{\ly}{\ensuremath{Y}\xspace}
\newcommand{\lytrial}{\ensuremath{Y_{T}}\xspace}
\newcommand{\traw}{\ensuremath{T_{raw}}\xspace}
\newcommand{\tcal}{\ensuremath{T_{cal}}\xspace}
\newcommand{\ttrack}{\ensuremath{T_{track}}\xspace}
\newcommand{\tcor}{\ensuremath{T_{cor}}\xspace}
\newcommand{\ttof}{\ensuremath{T_{tof}}\xspace}
\newcommand{\tpath}{\ensuremath{T_{path}(c)}\xspace}
\newcommand{\tslew}{\ensuremath{T_{slew}(PE)}\xspace}
\newcommand{\tslewt}{\ensuremath{T_{slew}(t)}\xspace}
\newcommand{\tslewi}{\ensuremath{T_{slew,i}}\xspace}
\newcommand{\toffset}{\ensuremath{T_{offset}(c)}\xspace}
\newcommand{\toffsett}{\ensuremath{T_{offset}(c,t)}\xspace}
\newcommand{\toffseti}{\ensuremath{T_{offset,i}}\xspace}

% Kinematic vars
\newcommand{\enu}{\ensuremath{E_{\nu}}\xspace}
\newcommand{\emu}{\ensuremath{E_{\mu}}\xspace}
\newcommand{\epi}{\ensuremath{E_{\pi}}\xspace}
\newcommand{\thetapi}{\ensuremath{\theta_{\pi}}\xspace}
\newcommand{\thetamu}{\ensuremath{\theta_{\mu}}\xspace}
\newcommand{\qsq}{\ensuremath{Q^{2}}\xspace}
\newcommand{\pnu}{\ensuremath{{p}_{\mu}}\xspace}
\newcommand{\pmu}{\ensuremath{{p}_{\mu}}\xspace}
\newcommand{\pmuvec}{\ensuremath{\vec{p}_{\mu}}\xspace}
\newcommand{\tabs}{\ensuremath{|t|}\xspace}
\newcommand{\invm}{$W$\xspace}
\newcommand{\wgen}{\ensuremath{W_{gen}}\xspace}
\newcommand{\Tp}{\ensuremath{T_{p}}\xspace}
\newcommand{\tdiff}{\ensuremath{|t|_{diff}}\xspace}
\newcommand{\tmin}{\ensuremath{|t|_{min}}\xspace}
\newcommand{\trel}{\ensuremath{|t|-|t|_{min}}\xspace}

% Units
\newcommand{\gevpercsq}{\ensuremath{(\text{GeV/c})^{2}}\xspace}

% Event selection
\newcommand{\evtx}{\ensuremath{E_{vtx}}\xspace}
\newcommand{\evrange}{\ensuremath{2.0 < \enu < 20\ \text{GeV}}\xspace}
\newcommand{\dedx}{$dE/dx$\xspace}

% Processes
\newcommand{\cohnumu}{$\numu A\to\mu^{-}\pi^{+}A$\xspace}
\newcommand{\cohnumubar}{$\numubar A\to\mu^{+}\pi^{-}A$\xspace}
\newcommand{\diffractive}{\ensuremath{\overset{(-)}{\nu}_\mu p\to\mu^\mp\pi^\pm p}\xspace}
\newcommand{\diffractivenumu}{\ensuremath{\numu p\to\mu^{-}\pi^{+}p}\xspace}
\newcommand{\diffractivenumubar}{\ensuremath{\numubar p\to\mu^{+}\pi^{-}p}\xspace}

% Cross sections
\newcommand{\sigenu}{\ensuremath{\sigma(\enu)}\xspace}
\newcommand{\dsigdxi}{\ensuremath{\frac{d\sigma}{d\xi}}\xspace}
\newcommand{\dsigdepi}{\ensuremath{\frac{d\sigma}{d\epi}}\xspace}
\newcommand{\dsigdthetapi}{\ensuremath{\frac{d\sigma}{d\thetapi}}\xspace}
\newcommand{\dsigdqsq}{\ensuremath{\frac{d\sigma}{d\qsq}}\xspace}
\newcommand{\dsigdt}{\ensuremath{\frac{d\sigma}{d|t|}}\xspace}
\newcommand{\dsigdtrel}{\ensuremath{\frac{d\sigma}{d(|t|-|t|_{min})}}\xspace}

% Comparisons
\newcommand{\lt}{$<$\xspace}
\newcommand{\gt}{$>$\xspace}

\newcommand{\chisq}{\ensuremath{\chi^{2}}\xspace}

% Latin
\newcommand{\etal}{\textit{et al.}\xspace}
\newcommand{\ie}{\textit{i.e.}\xspace}
\newcommand{\eg}{\textit{e.g.}\xspace}
\newcommand{\insitu}{\textit{in situ}\xspace}
\newcommand{\vs}{\textit{vs.}\xspace}

\title{Measurement of Total and Differential Cross Sections of Neutrino and Antineutrino Coherent $\pi^\pm$ Production on Carbon}


%% MANUAL PARTS OF AUTHOR LIST
\author{A.~Mislivec}                       \affiliation{\Rochester}
\author{A.~Higuera}\thanks{\higueraThanks} \affiliation{\Rochester}  \affiliation{\Guanajuato}

%% AUTOMATIC LIST (EDITED AS ABOVE)
%%%
%%% This file created automagically via Glaucus using this URL:
%%% http://mnvevd1.fnal.gov/Glaucus/web/latex_paper.cgi?paper_id=52
%%% Options: do_thanks
%%%

%% List of institution addresses, in command form.
\newcommand{\Rutgers}{Rutgers, The State University of New Jersey, Piscataway, New Jersey 08854, USA}
\newcommand{\Hampton}{Hampton University, Dept. of Physics, Hampton, VA 23668, USA}
\newcommand{\Dortmund}{Institute of Physics, Dortmund University, 44221, Germany }
\newcommand{\Otterbein}{Department of Physics, Otterbein University, 1 South Grove Street, Westerville, OH, 43081 USA}
\newcommand{\JMU}{James Madison University, Harrisonburg, Virginia 22807, USA}
\newcommand{\Florida}{University of Florida, Department of Physics, Gainesville, FL 32611}
\newcommand{\UCIrvine}{Department of Physics and Astronomy, University of California, Irvine, Irvine, California 92697-4575, USA}
\newcommand{\CBPF}{Centro Brasileiro de Pesquisas F\'{i}sicas, Rua Dr. Xavier Sigaud 150, Urca, Rio de Janeiro, Rio de Janeiro, 22290-180, Brazil}
\newcommand{\PUCP}{Secci\'{o}n F\'{i}sica, Departamento de Ciencias, Pontificia Universidad Cat\'{o}lica del Per\'{u}, Apartado 1761, Lima, Per\'{u}}
\newcommand{\INRM}{Institute for Nuclear Research of the Russian Academy of Sciences, 117312 Moscow, Russia}
\newcommand{\Jlab}{Jefferson Lab, 12000 Jefferson Avenue, Newport News, VA 23606, USA}
\newcommand{\Pittsburgh}{Department of Physics and Astronomy, University of Pittsburgh, Pittsburgh, Pennsylvania 15260, USA}
\newcommand{\Guanajuato}{Campus Le\'{o}n y Campus Guanajuato, Universidad de Guanajuato, Lascurain de Retana No. 5, Colonia Centro, Guanajuato 36000, Guanajuato M\'{e}xico.}
\newcommand{\Athens}{Department of Physics, University of Athens, GR-15771 Athens, Greece}
\newcommand{\Tufts}{Physics Department, Tufts University, Medford, Massachusetts 02155, USA}
\newcommand{\WM}{Department of Physics, College of William \& Mary, Williamsburg, Virginia 23187, USA}
\newcommand{\FNAL}{Fermi National Accelerator Laboratory, Batavia, Illinois 60510, USA}
\newcommand{\Purdue}{Department of Chemistry and Physics, Purdue University Calumet, Hammond, Indiana 46323, USA}
\newcommand{\MCLA}{Massachusetts College of Liberal Arts, 375 Church Street, North Adams, MA 01247}
\newcommand{\UMD}{Department of Physics, University of Minnesota -- Duluth, Duluth, Minnesota 55812, USA}
\newcommand{\Northwestern}{Northwestern University, Evanston, Illinois 60208}
\newcommand{\UNI}{Universidad Nacional de Ingenier\'{i}a, Apartado 31139, Lima, Per\'{u}}
\newcommand{\Rochester}{University of Rochester, Rochester, New York 14627 USA}
\newcommand{\Austin}{Department of Physics, University of Texas, 1 University Station, Austin, Texas 78712, USA}
\newcommand{\USM}{Departamento de F\'{i}sica, Universidad T\'{e}cnica Federico Santa Mar\'{i}a, Avenida Espa\~{n}a 1680 Casilla 110-V, Valpara\'{i}so, Chile}
\newcommand{\Geneva}{University of Geneva, 1211 Geneva 4, Switzerland}
\newcommand{\Chicago}{Enrico Fermi Institute, University of Chicago, Chicago, IL 60637 USA}
\newcommand{\hired}{}
\newcommand{\OregonState}{Department of Physics, Oregon State University, Corvallis, Oregon 97331, USA}
\newcommand{\oxford}{}
\newcommand{\umiss}{University of Mississippi, Oxford, Mississippi 38677, USA}
\newcommand{\upenn}{209 S. 33rd St. Philadelphia, PA 19104}
\newcommand{\AMU}{AMU Campus, Aligarh, Uttar Pradesh 202001, India}
\newcommand{\wroclaw}{University of Wroclaw, plac Uniwersytecki 1, 50-137 Wrocław, Poland}
\newcommand{\Mohali}{Knowledge city, Sector 81, SAS Nagar, Manauli PO 140306}
\newcommand{\damartinThanks}{now at Illinois Institute of Technology, Chicago, IL 60616, USA}
\newcommand{\edgarchThanks}{Science Faculty, Universidad Nacional de Ingenier\'ia, Lima 25, Lima Per\'u}
\newcommand{\higueraThanks}{now at University of Houston, Houston, TX 77204, USA}
\newcommand{\chrisMarshallThanks}{now at Lawrence Berkeley National Laboratory, Berkeley, CA 94720, USA}
\newcommand{\joelmousseauThanks}{now at University of Michigan, Ann Arbor, MI 48109, USA}
\newcommand{\gzgThanks}{Deceased}

% 64 total signatories.

\author{L.~Aliaga}                        \affiliation{\WM}  \affiliation{\PUCP}
\author{L.~Bellantoni}                     \affiliation{\FNAL}
\author{A.~Bercellie}                     \affiliation{\Rochester}
\author{M.~Betancourt}                    \affiliation{\FNAL}
\author{A.~Bodek}                         \affiliation{\Rochester}
\author{A.~Bravar}                        \affiliation{\Geneva}
\author{H.~Budd}                          \affiliation{\Rochester}
\author{G.~F.~R.~Caceres~V.}              \affiliation{\CBPF}
\author{T.~Cai}                           \affiliation{\Rochester}
\author{D.A.~Martinez~Caicedo}\thanks{\damartinThanks}  \affiliation{\CBPF}  \affiliation{\FNAL}
\author{M.F.~Carneiro}                    \affiliation{\OregonState}
\author{E.~Chavarria}					  \affiliation{\UNI}
\author{H.~da~Motta}                      \affiliation{\CBPF}
\author{S.A.~Dytman}                      \affiliation{\Pittsburgh}
\author{G.A.~D\'{i}az~}                   \affiliation{\Rochester}  \affiliation{\PUCP}
\author{J.~Felix}                         \affiliation{\Guanajuato}
\author{L.~Fields}                        \affiliation{\FNAL}  \affiliation{\Northwestern}
\author{R.~Fine}                          \affiliation{\Rochester}
\author{A.M.~Gago}                        \affiliation{\PUCP}
\author{R.~Galindo}                        \affiliation{\USM}
\author{H.~Gallagher}                     \affiliation{\Tufts}
\author{A.~Ghosh}                         \affiliation{\USM}  \affiliation{\CBPF}
\author{R.~Gran}                          \affiliation{\UMD}
\author{D.A.~Harris}                      \affiliation{\FNAL}
\author{K.~Hurtado}                       \affiliation{\CBPF}  \affiliation{\UNI}
\author{D.~Jena}						  \affiliation{\FNAL}
\author{J.~Kleykamp}                      \affiliation{\Rochester}
\author{M.~Kordosky}                      \affiliation{\WM}
\author{T.~Le}                            \affiliation{\Tufts}  \affiliation{\Rutgers}
\author{E.~Maher}                         \affiliation{\MCLA}
\author{S.~Manly}                         \affiliation{\Rochester}
\author{W.A.~Mann}                        \affiliation{\Tufts}
\author{C.M.~Marshall}\thanks{\chrisMarshallThanks}  \affiliation{\Rochester}
\author{K.S.~McFarland}                    \affiliation{\Rochester}  \affiliation{\FNAL}
\author{B.~Messerly}                      \affiliation{\Pittsburgh}
\author{J.~Miller}                        \affiliation{\USM}
\author{J.G.~Morf\'{i}n}                  \affiliation{\FNAL}
\author{J.~Mousseau}\thanks{\joelmousseauThanks}  \affiliation{\Florida}
\author{D.~Naples}                        \affiliation{\Pittsburgh}
\author{J.K.~Nelson}                      \affiliation{\WM}
\author{C.~Nguyen}						  \affiliation{\Florida}
\author{A.~Norrick}                       \affiliation{\WM}
\author{Nuruzzaman}                       \affiliation{\Rutgers}  \affiliation{\USM}
\author{V.~Paolone}                       \affiliation{\Pittsburgh}
\author{G.N.~Perdue}                      \affiliation{\FNAL}  \affiliation{\Rochester}
\author{M.A.~Ram\'{i}rez}                 \affiliation{\Guanajuato}
\author{R.D.~Ransome}                     \affiliation{\Rutgers}
\author{H.~Ray}                           \affiliation{\Florida}
\author{L.~Ren}                           \affiliation{\Pittsburgh}
\author{D.~Rimal}                         \affiliation{\Florida}
\author{P.A.~Rodrigues}                   \affiliation{\umiss}  \affiliation{\Rochester}
\author{D.~Ruterbories}                   \affiliation{\Rochester}
\author{H.~Schellman}                     \affiliation{\OregonState}  \affiliation{\Northwestern}
\author{C.J.~Solano~Salinas}              \affiliation{\UNI}
\author{M.~Sultana}                       \affiliation{\Rochester}
\author{S.~S\'{a}nchez~Falero}            \affiliation{\PUCP}
\author{N.~Tagg}                          \affiliation{\Otterbein}
\author{E.~Valencia}                      \affiliation{\WM}  \affiliation{\Guanajuato}
\author{M.Wospakrik}                      \affiliation{\Florida}
\author{B.~Yaeggy}                        \affiliation{\USM}
\author{G.~Zavala}\thanks{\gzgThanks}     \affiliation{\Guanajuato}

\collaboration{The MINER$\nu$A Collaboration}\ \noaffiliation
\date{\today}

\pacs{}
\begin{abstract}
Neutrino induced coherent charged pion production on nuclei, $\overline{\nu}_\mu A\to\mu^\pm\pi^\mp A$, is a rare inelastic
interaction in which the four-momentum squared transfered to the nucleus is nearly zero, leaving
it intact.  We identify such events in the scintillator of \minerva by reconstructing \tabs from the final state
pion and muon momenta and by removing events with evidence of  energetic nuclear recoil or production of other final state
particles.  We measure the total neutrino and antineutrino cross sections as a function of neutrino energy between 2 and 20\,GeV
and measure flux integrated differential cross sections as a function of $Q^2$, $E_\pi$ and $\theta_\pi$.  The $Q^2$ dependence and equality
of the neutrino and anti-neutrino cross-sections at finite $Q^2$ provide a confirmation of Adler's PCAC hypothesis.
\end{abstract}
\ifnum\sizecheck=0  
\maketitle
\fi

\section{Introduction}
\label{ch:introduce}

Neutrino-nucleus coherent pion production is an inelastic interaction that produces a lepton and a pion in the forward
direction while leaving the nucleus in its initial state.  The charged current (CC) processes are
\begin{eqnarray}
\nu_{l} + \text{A} \to l^{-} + \pi^{+} + \text{A} \nonumber \\
\overline{\nu}_{l} + \text{A} \to l^{+} + \pi^{-} + \text{A}
\end{eqnarray}
and the neutral current (NC) processes are
\begin{eqnarray}
\nu_{l} + \text{A} \to \nu_{l} + \pi^{0} + \text{A} \nonumber \\
\overline{\nu}_{l} + \text{A} \to \overline{\nu}_{l} + \pi^{0} + \text{A}
\end{eqnarray}
where A is the nucleus.  For the interaction to preserve the initial state of the nucleus, the absolute value of the square of the four-momentum
exchanged with the nucleus, \tabs, must be small.
%, i.e., $\tabs\lesssim 1/R^2$, where $R$ is the radius of the nucleus.
In addition, the particle(s) exchanged with the nucleus can only carry vacuum quantum numbers in coherent scattering.  

%The square of the four-momentum exchanged with the nucleus is defined as
%\begin{equation}
%|t| = |(q-p_{\pi})^{2}| = |(p_{\nu} - p_{\mu} - p_{\pi})^{2}|,
%\end{equation}
%where $p_{\nu}$, $p_{\mu}$, and $p_{\pi}$ are the four-momentum of the neutrino, muon, and pion, respectively.  

%For neutrino (antineutrino) interactions on carbon nuclei at neutrino energy $\enu=$ 1.0 GeV, theoretical models (see
%Sec.~\ref{sec:models}) predict the rate of coherent pion production to be only $\sim$1\% ($\sim$3\%) of the total
%interaction rate for both CC and NC interactions.

Coherent pion production is not a common process; for neutrino (antineutrino) interactions on carbon nuclei at
$\enu \sim$\unit[3]{GeV}, theoretical models (see Sec.~\ref{sec:MC_bkg_Data}) predict the rate of coherent pion production
to be only $\sim$1\% ($\sim$3\%) of the total interaction rate for both CC and NC interactions.  Nonetheless, coherent
pion production is an important background for neutrino oscillation experiments, which typically operate in the range
\unit[1]{GeV}$\lesssim \enu \lesssim$ \unit[10]{GeV}.  NC coherent pion
production is an important background to $\numu\to\nue$ and $\numubar\to\nuebar$ oscillation measurements where the oscillation
signals are $\nue + N \to e^{-} + X$ and $\nuebar + N \to e^{+} + X$,
where $N$ is a target nucleon and $X$ is the hadronic final state.  NC coherent pion production yields only a \pizero\ to be
observed in the detector and if one of the two final state photons is not observed, the other can be mistaken for an electron
or positron.  CC coherent pion production is a background to measurements of \numu and \numubar disappearance at low neutrino
energy ($\enu\lesssim$ 1 GeV) where quasielastic scattering
\begin{eqnarray}
\numu + n \to \mu^{-} + p \nonumber \\
\numubar + p \to \mu^{+} + n
\end{eqnarray}
is the primary interaction process.  Coherent pion production can be mistaken for quasielastic scattering when the
$\pi^{\pm}$ is misidentified as a proton or is not detected.

This paper presents precise measurements of the \numu and \numubar CC coherent pion production cross sections on carbon
for 2 $<\enu<$ 20 GeV.  The cross sections are measured as a function of the pion energy, pion angle, and \qsq, which characterize the
coherent pion production kinematics.  A subset of these results based on this same data was published earlier~\cite{bib:minerva_coh}.
The results presented in this paper have a revised treatment of backgrounds, additional differential distributions, a new prediction of the neutrino and
antineutrino fluxes in the beam and the correlations between the neutrino and antineutrino measurements.  Also presented is a
study from the data of the possible contributions of diffractive scattering to this measurement of coherent pion production.
Accordingly, the results in this paper supersede those of Ref.~\cite{bib:minerva_coh}.

This paper is organized as follows.  Sections \ref{sec:models} and \ref{sec:before2015} describe two existing models and the experimental
state of the field prior to this work.  Section \ref{sec:diffractiveIntro} introduces diffractive scattering of neutrinos off hydrogen,
an important and closely related process that is studied here.  Sections \ref{sec:NuMI_N_me}-\ref{sec:event_selection} describe the experiment,
the event selection and the reconstruction of candidate events.  Sections \ref{sec:bg_tuning}-\ref{sec:systematics} explain the measurement
of the cross sections and their systematic uncertainties.  Sections \ref{sec:measured_cross_sections}-\ref{sec:conclusions} present the
results and their interpretation.

\section{Models}
\label{sec:models}

\begin{figure}[tp]
\centering
\mbox{
\includegraphics[width=0.49\columnwidth]{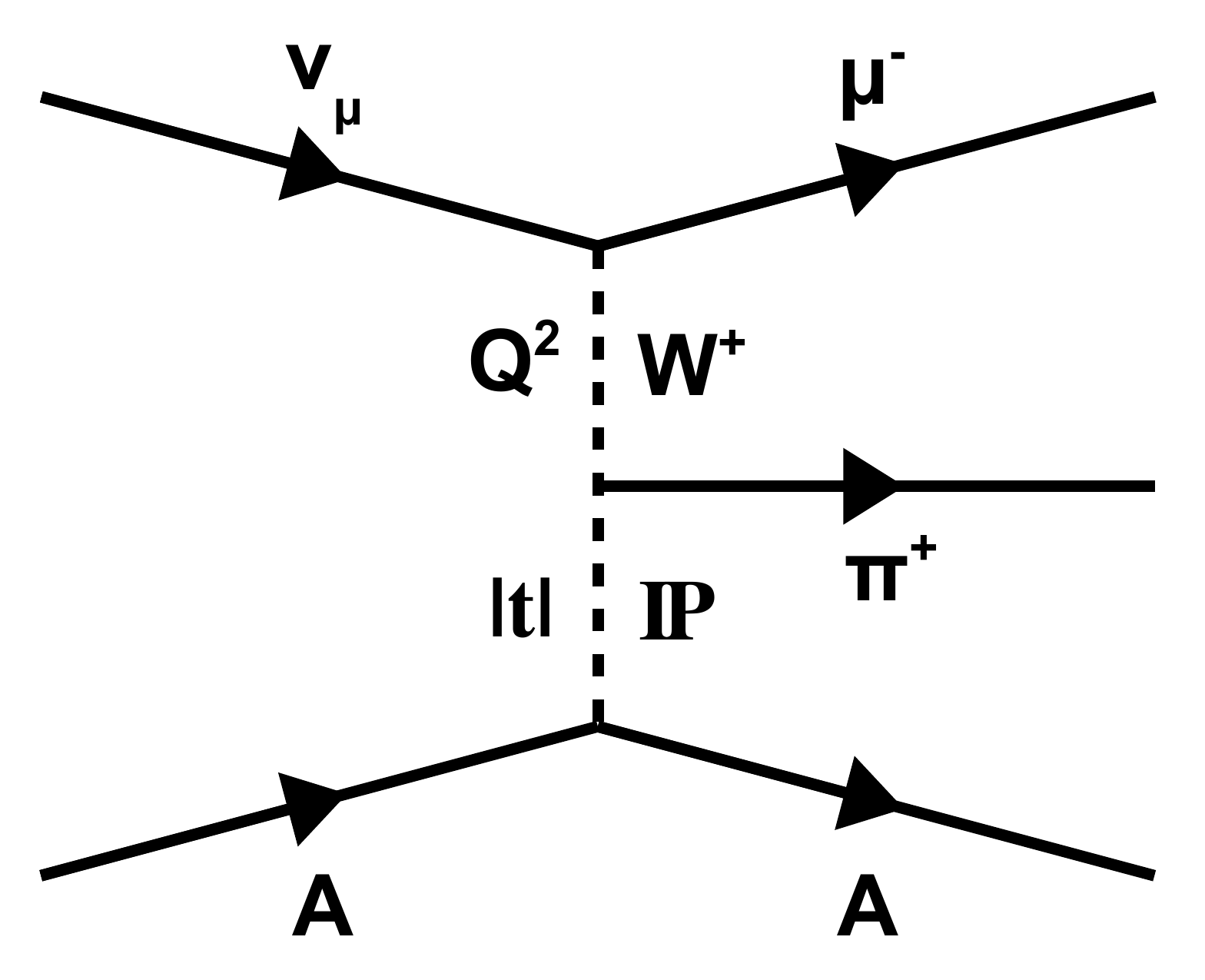}
\includegraphics[width=0.49\columnwidth]{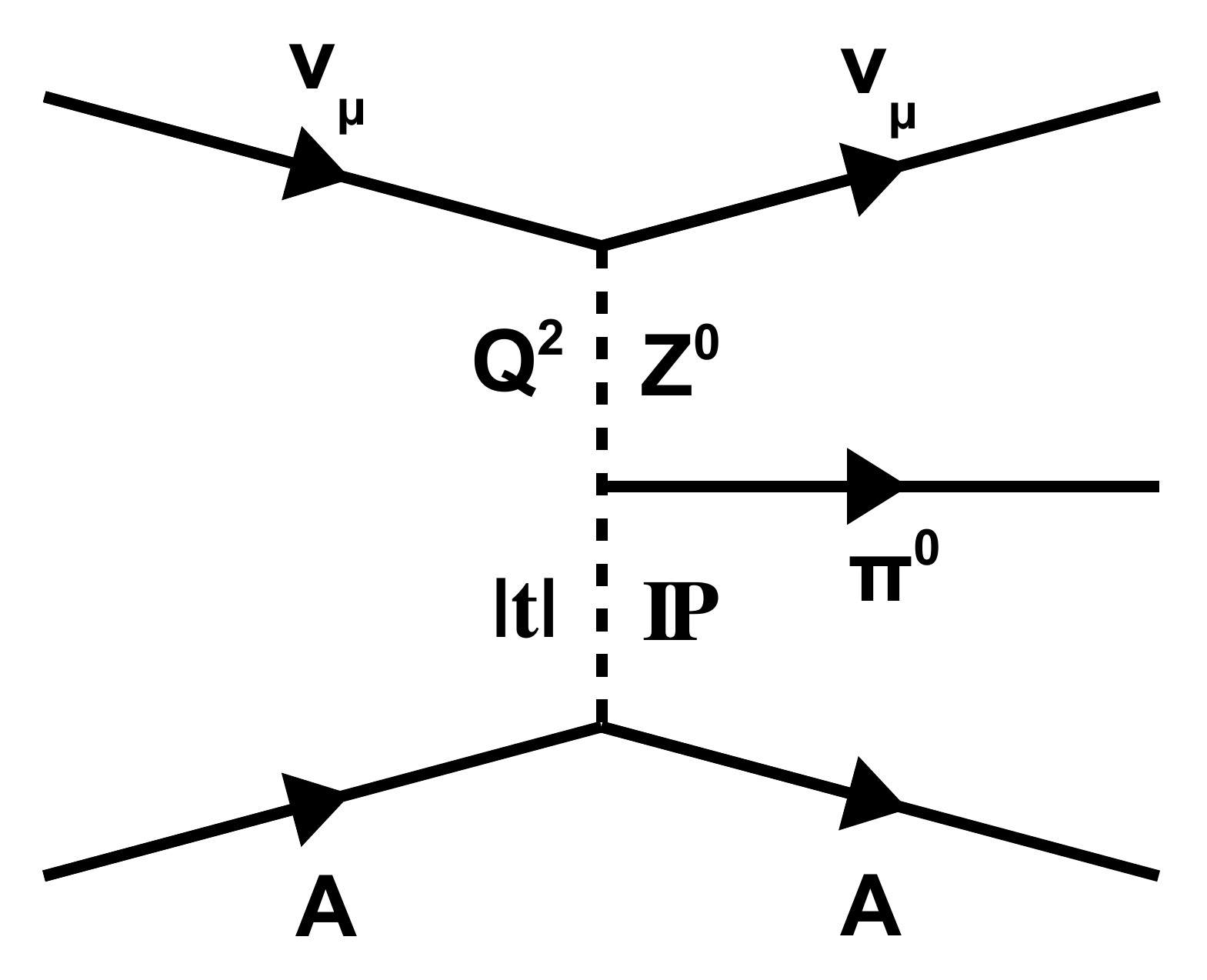}}
\caption[CC and NC neutrino-nucleus coherent pion production in the PCAC picture]{CC (left) and NC (right) neutrino-nucleus
																				  coherent pion production in the PCAC picture.}
\label{fig:feyman_pcac_coh_cc_nc}
\end{figure}

PCAC coherent
models~\cite{bib:coh_pcac3,bib:coh_pcac4,bib:coh_pcac5,bib:coh_pcac6,bib:coh_pcac7,bib:coh_pcac10}
and \cite{bib:RS,bib:RS2,bib:berger_sehgal, bib:kartavtsev, bib:manny}
are a class of coherent pion production models that are based on
Adler's PCAC theorem~\cite{bib:adler_pcac} (Partial Conservation of Axial Current, or in modern language,
spontaneous breaking of chiral symmetry in QCD) which can be used to relate coherent pion production
at $Q^2=-(p_{\nu}-p_{l})^{2}=0$ to elastic pion-nucleus scattering.  Here, $p_{\nu}$ ($p_{l}$) is the
four-momentum of the incoming neutrino (outgoing charged lepton).  In this picture of coherent pion production,
the intermediate weak boson fluctuates to a virtual pion, which scatters elastically off the nucleus, as shown in
the Pomeron (\Pom) diagram of Fig.~\ref{fig:feyman_pcac_coh_cc_nc}.

\subsection{Rein-Sehgal Model}
\label{sec:Rein-Sehgal}

The Rein-Sehgal coherent model~\cite{bib:RS,bib:RS2} is the most widely used PCAC coherent pion production model in
neutrino event generators.  The Rein-Sehgal model extrapolates the Adler coherent cross section result to $\qsq > 0$
using a multiplicative axial-vector dipole form factor
\begin{equation}
F_{A} = \frac{M_{A}^{2}}{Q^{2} + M_{A}^{2}},
\end{equation}
where $M_{A} \approx$ 1 GeV is the axial vector mass.  The Rein-Sehgal model assumes no vector current contribution in
this extrapolation, and therefore predicts equal cross sections for neutrinos and antineutrinos.

The Rein-Sehgal model calculates the $\pi^{\pm}$A elastic cross section using charged pion-nucleon ($\pi^{\pm}N$)
scattering data.

The differential CC coherent pion production cross section calculated by the Rein-Sehgal model is
\begin{eqnarray}
\label{eq:rs_cccoh_xsec}
\frac{d\sigma_{coh}^{CC}}{d\qsq dyd\tabs} &= \frac{G^{2}f_{\pi^{\pm}}^{2}}{2\pi}\frac{(1-y)}{y}\frac{M_{A}^{2}}{Q^{2} + M_{A}^{2}}A^{2} \nonumber \\
&\times \exp\left(-\frac{1}{3}R_{0}^{2}A^{2/3}\tabs-\frac{9A^{1/3}}{16\pi R_{0}^{2}}\sigma_{inel}^{\pi^{\pm}N}(\epi)\right)\nonumber\\
&\times \frac{1}{16\pi}\left(\sigma_{tot}^{\pi^{\pm}N}(\epi)\right)^{2}(1+r^{2}),
%\right]_{E_{\pi}=yE_{\nu}}.
%\frac{d\sigma_{coh}^{CC}}{d\qsq dyd\tabs} &= \frac{G^{2}f_{\pi^{\pm}}^{2}}{2\pi}\frac{(1-y)}{y}\frac{M_{A}^{2}}{Q^{2} + M_{A}^{2}} \nonumber \\
%&\times \left[A^{2}\exp\left(-\frac{1}{3}R_{0}^{2}A^{2/3}\tabs\right)\exp\left(-\frac{9A^{1/3}}{16\pi R_{0}^{2}}\sigma_{inel}^{\pi^{\pm}\text{N}}\right)\frac{1}{16\pi}\Big[\sigma_{tot}^{\pi^{\pm}\text{N}}\Big]^{2}(1+r^{2})\right]_{E_{\pi}=yE_{\nu}}. %\nonumber \\
%&\times \left[\left(1-\frac{1}{2}\frac{Q^{2}_{min}}{Q^{2}+m^{2}_{\pi}}\right)^{2}+\frac{y}{4}\frac{Q^{2}_{min}(Q^{2}-Q^{2}_{min})}{(Q^{2}+m^{2}_{\pi})^{2}}\right] \nonumber \\
%&\times \theta(\qsq - \qsq_{min})\theta(y - y_{min})\theta(y_{max} - y)
\end{eqnarray}
where the pion energy $\epi=y\enu$, $A$ is the atomic number of the target nucleus, $f_\pi$ is the pion decay constant,
$R_0$ is the nuclear length scale $\sim 1$~fm and $r = \text{Re}f(0)/\text{Im}f(0)$ is the ratio of the real and imaginary
parts of the $\pi^{\pm}N$ forward scattering amplitude.  The exponential dependence in \tabs is the consequence of a simple
gaussian model for the nuclear form factor in $\pi A$ elastic scattering. The Rein-Sehgal model calculates the NC
differential cross section from Eq.~(\ref{eq:rs_cccoh_xsec}) using $f_{\pi^{0}} = f_{\pi^{\pm}}/\sqrt{2}$ and assuming that
the $\pi^{\pm}N$ and $\pi^{0}N$ cross sections are equal.

The Rein-Sehgal model corrects the CC differential cross section (Eq.~\ref{eq:rs_cccoh_xsec}) for the mass of the final state
lepton \cite{bib:RS2,bib:lepton_mass_corr}.  The correction, proposed by Adler \cite{bib:lepton_mass_corr}, is
\begin{eqnarray}
C &= \left(1-\frac{1}{2}\frac{Q^{2}_{min}}{Q^{2}+m^{2}_{\pi}}\right)^{2}+\frac{y}{4}\frac{Q^{2}_{min}(Q^{2}-Q^{2}_{min})}{(Q^{2}+m^{2}_{\pi})^{2}} \\
&\times \theta(\qsq - \qsq_{min})\theta(y - y_{min})\theta(y_{max} - y), \nonumber
\end{eqnarray}
where $Q^{2}_{min} = m^{2}_{l}y/(1-y)$ is the kinematic minimum \qsq, $y_{min} = m_{\pi}/\enu$ and $y_{max} = 1 - m_{l}/\enu$ are the kinematic minimum and maximum $y$, and $m_{l}$ and $m_{\pi}$ are the final state lepton and pion masses.

The Rein-Sehgal model predicts that both $\sigma_{coh}^{CC}(\enu)$ and $\sigma_{coh}^{NC}(\enu)$
scale with $A$ approximately as $A^{1/3}$, as a result of the nuclear coherence condition and pion absorption effects.
%, and \tabs-dependence.   Leo asks: oh?

\subsection{Berger-Sehgal Model}
\label{sec::Berger-Sehgal}

\begin{figure}[tpb]
\centering
%\includegraphics[width=1.0\columnwidth]{figures/CoherentChapter/rs_bs_pion_carbon_elastic_xsec.pdf}
%\caption[Rein-Sehgal and Berger-Sehgal pion-carbon elastic scattering cross sections]{The plot on the left is the Rein-Sehgal (dashed line) and Berger-Sehgal (solid line) predictions of the pion-carbon elastic scattering cross section as a function of the pion momentum in the laboratory frame.  The table on the right are the parameters for the Berger-Sehgal pion-carbon elastic scattering cross section determined from fitting pion-carbon-elastic scattering data.  The plot and table are from \cite{bib:berger_sehgal}}
\includegraphics[width=1.0\columnwidth]{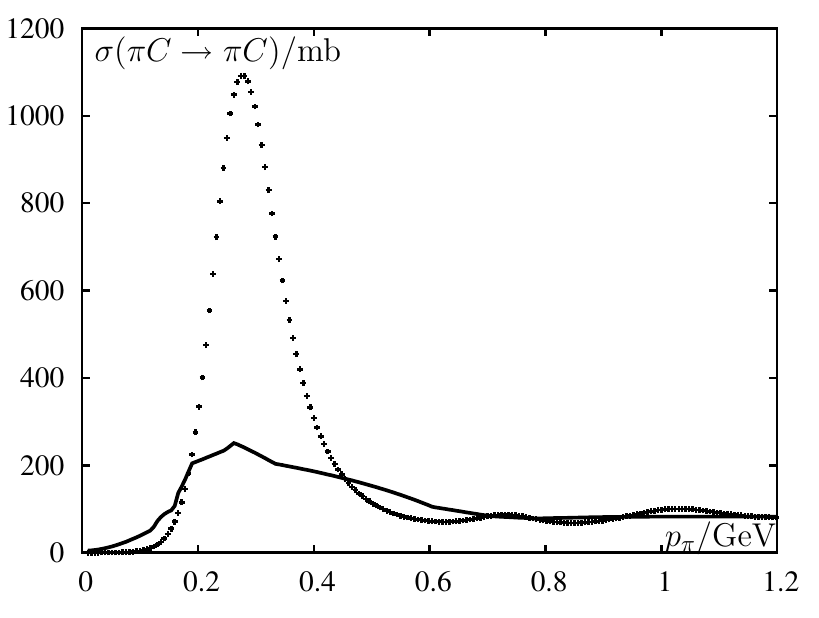}
\caption[Rein-Sehgal and Berger-Sehgal pion-carbon elastic scattering cross sections]{The Rein-Sehgal (dashed line)
		 and Berger-Sehgal (solid line) pion-carbon elastic scattering cross sections as a function of the pion momentum
		 in the laboratory frame.  From \cite{bib:berger_sehgal}.}
\label{fig:rs_bs_pion_carbon_elastic_xsec}
\end{figure}

The cross section of Eq.~(\ref{eq:rs_cccoh_xsec}) is differential in three variables but coherent scattering only occurs in
a limited region of the phase space.  Newer models~\cite{bib:berger_sehgal, bib:kartavtsev, bib:manny} use data from pion-nucleus
elastic scattering.  These show that coherent scattering predominantly occurs at \qsq$\sim\mathcal{O}(m_\pi^2), ~\nu^2 = (E_\nu - E_l)^2 \gg \qsq$
and $\tabs\lesssim 1/R^2$, where $E_\nu$ ($E_l$) is the energy of the incoming neutrino (outgoing lepton) and $R$ is the radius of the nucleus.
In particular, coherent scattering creates a sharp increase in the slope of $d\sigma/d\tabs$ as \tabs approaches
\tmin$\sim \frac{ (Q^2+m_\pi^2)^2}{2\nu}$, the minimum kinematically possible \tabs.
The Berger-Sehgal PCAC coherent model \cite{bib:berger_sehgal} is a modification of the Rein-Sehgal model where the
parameterization of the $\pi^{\pm}$A elastic scattering cross section is instead fit to charged pion-carbon ($\pi^{\pm}$C)
elastic scattering data and scaled to other nuclei.  This approach avoids some of the uncertainties from modeling nuclear effects,
\textit{e.g.} pion absorption, in the Rein-Sehgal parameterization.  In the Berger-Sehgal model, coherent cross sections scale
as $A^{2/3}$.
%The Berger-Sehgal model parameterizes the differential$\pi^{\pm}$C elastic cross section as
%\begin{equation}
%\frac{d\sigma_{el}^{\pi^{\pm}\text{C}}}{d\tabs} = A_{1}e^{-b_{1}\tabs},
%\label{eq:bs_piC_xsec}
%\end{equation}
%where the normalization $A_{1}$ and slope $b_{1}$ are functions of \epi and are fit to data (Figure~\ref{fig:rs_bs_pion_carbon_elastic_xsec}).
The comparison of Rein-Sehgal and Berger-Sehgal calculations of $\sigma_{el}^{\pi^{\pm}\text{C}}$ in
Fig.~\ref{fig:rs_bs_pion_carbon_elastic_xsec} shows that while the two calculations agree for $|\vec{p}_{\pi}|\gtrsim0.7$ GeV,
the Rein-Sehgal predicts a much larger cross section in the $\Delta$ resonance dominated region, $|\vec{p}_{\pi}|\sim0.3$ GeV.

\section{Earlier Measurements}
\label{sec:before2015}

Coherent CC interactions can be identified by requiring that the observed final state consist only of a charged lepton and
pion (the target nucleus is not observed since the energy transferred to the nucleus is small) and small \tabs.  From 
the assumption of zero energy transfer to the nucleus,
\begin{eqnarray}
\tabs &= |( p_{\nu_{l}} - p_{l} - p_{\pi})^{2}| \nonumber \\
& \approx \left(\displaystyle \sum_{i=\mu,\pi} E_i-p_{i,L}\right)^2+\left| \displaystyle \sum_{i=\mu,\pi}\vec{p}_{i,T}\right| ^2,
\label{eq:t_calc} 
\end{eqnarray}
where $p_{\nu_{l}}$, $p_{l}$, and $p_{\pi}$ are the four-momenta of the neutrino, charged lepton, and pion, respectively,
and $\vec{p}_{T}$ and $p_{L}$ are transverse and longitudinal momenta with respect to the incoming neutrino direction.
The neutrino four-momentum is determined by assuming that its direction is that of the neutrino beam and its energy is
$\enu = E_{l} + \epi$.
%where the energy transferred to the nucleus is ignored since it will be small at low \tabs.

For NC coherent interactions, the final state neutrino is not observed and the observed final state is a lone $\pi^{0}$.
Neither \enu nor \tabs can be measured.
Since \tabs is not available, experiments isolate NC coherent interactions using the condition
\begin{equation}
\epi(1-\cos\thetapi) < \frac{1}{R}, 
\end{equation}
where \thetapi is the angle between the pion and incoming neutrino directions.  This condition follows from the
$\tabs \lesssim 1/R^2$ condition for coherent scattering at \mbox{$\qsq\approx$ 0 \cite{bib:lackner}}.  However, imposing
selection requirements on the pion kinematics creates implicit selection requirements on the lepton kinematics and
\qsq (see Eq.~\ref{eq:t_calc}).
Consequently, NC coherent pion production measurements must make model-dependent assumptions about the \qsq dependence of the
coherent scattering cross section when correcting for the inefficiency of this selection requirement.  

Since \enu cannot be measured, NC coherent pion production measurements are averaged over the energy spectrum of a neutrino
beam.  These experimental limitations of NC coherent pion production measurements increase the importance of CC coherent
pion production measurements, which in the PCAC picture provide a constraint on the NC reaction.

Many measurements of 
NC~\cite{bib:coh_nc_AP,bib:coh_nc_garg,bib:coh_nc_charm,bib:coh_nc_cc_skat,bib:coh_nc_fnal,bib:coh_nc_nomad,bib:coh_nc_sb,bib:coh_nc_mb,bib:coh_nc_minos}
and
CC~\cite{bib:coh_nc_cc_skat,bib:coh_cc_bebc1,bib:coh_cc_fnal1,bib:coh_cc_bebc2,bib:coh_cc_fnal2,bib:coh_cc_fnal3,bib:coh_cc_charm,
bib:coh_cc_k2k,bib:coh_cc_sb,bib:coh_cc_argo}
%MOVE THESE TO LATER SINCE THIS IS A LIST OF "BEFORE 2015" - well, it was.  But now it isn't
coherent pion production have been made at neutrino energies of 1 $\lesssim\enu\lesssim$ 100 GeV using both \numu and \numubar
beams and a variety of scattering target materials (carbon, neon, aluminum, argon, Freon, glass, marble and iron).  Measurements
made before the discovery of neutrino oscillations were mostly made at $\enu>$ 10 GeV.
Precise measurements at 1 $\lesssim\enu\lesssim$ 10 GeV are now needed for constraining backgrounds in
oscillation measurements in high intensity \numu and \numubar beams.
% that operate at 1 $\lesssim\enu\lesssim$ 10 GeV.

\begin{figure}[tpb]
\centering
\includegraphics[width=0.8\columnwidth]{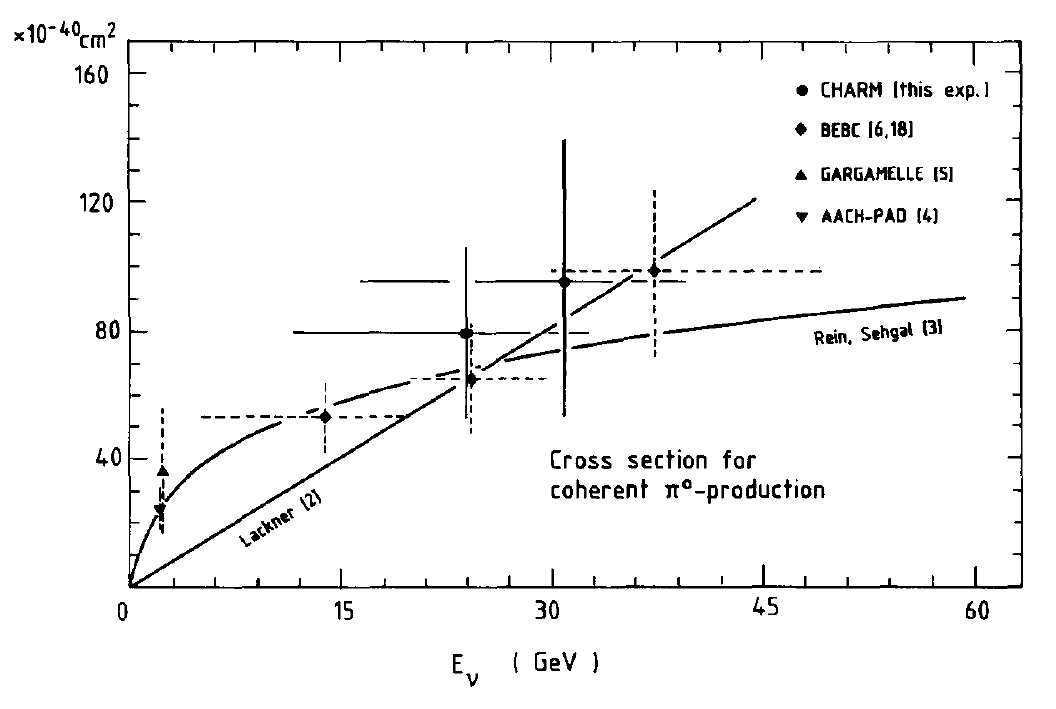}
\caption[Early measurements of NC coherent pion production]
	{Early measurements of the NC coherent pion production cross section.  The horizontal error bars represent the range
	of neutrino energies sampled by the measurement.  The figure is from \cite{bib:coh_nc_charm}.}
\label{fig:charm_coh_nc}
\end{figure}

Early measurements of the NC coherent pion production cross section as a function of
\enu~\cite{bib:coh_nc_AP,bib:coh_nc_garg,bib:coh_nc_charm} are shown in Fig.~\ref{fig:charm_coh_nc}.  They were made
using different scattering target materials and, for the purpose of comparison, are scaled in Fig.~\ref{fig:charm_coh_nc}
to a marble scattering target with an effective $A$ = 20 using the $A^{1/3}$ dependence of the cross section.
While the Rein-Sehgal model prediction agrees with the measured cross sections within their uncertainties, the uncertainties
are large ($\gtrsim$ 30\% \cite{bib:coh_nc_nomad}).

\begin{figure}[tpb]
\centering
\includegraphics[width=0.8\columnwidth]{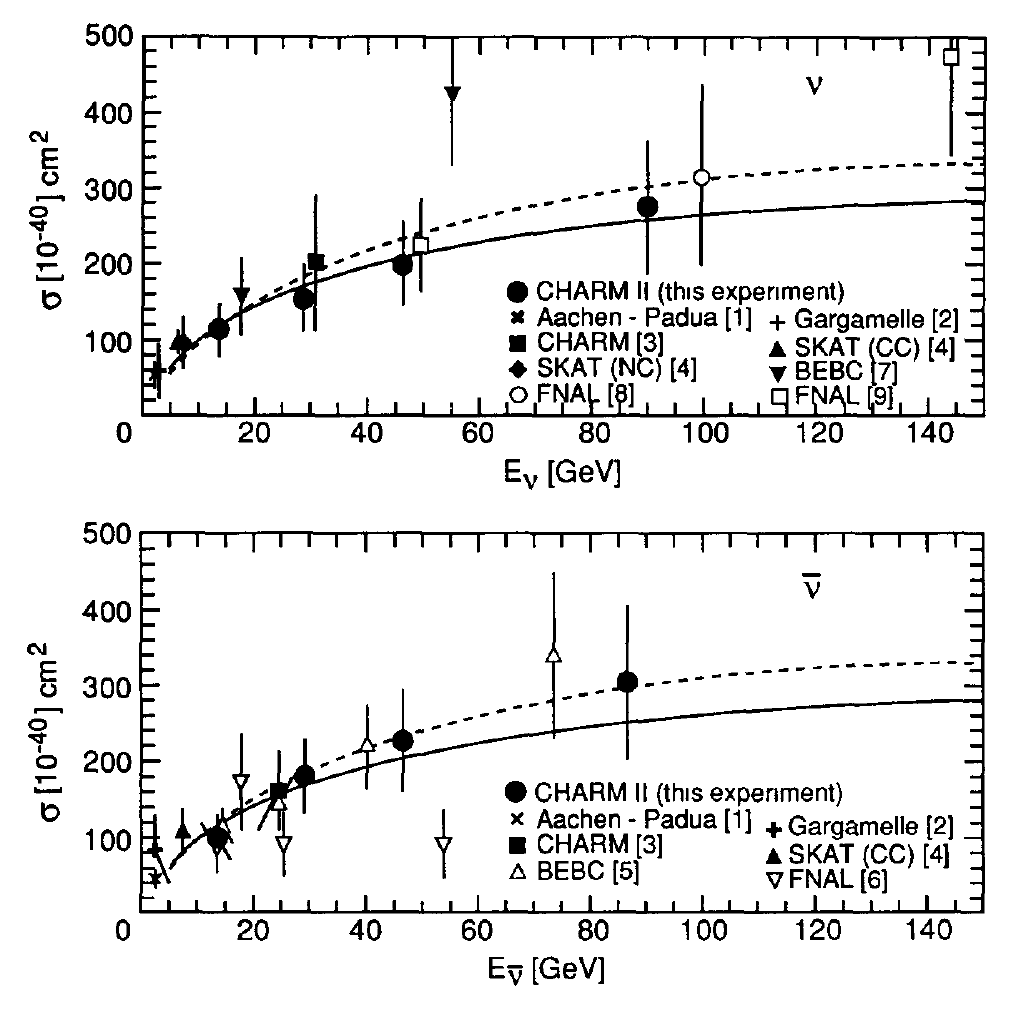}
\caption[Early measurements of CC coherent pion production]{Early measurements of the \numu (top) and \numubar (bottom) CC
	coherent pion production cross sections.  The solid line is the Rein-Sehgal model prediction.  The figure is
	from~\cite{bib:coh_cc_charm}.}
\label{fig:charm_coh_cc}
\end{figure}

Early measurements of the \numu and \numubar CC coherent pion production cross sections as a function of
\enu~\cite{bib:coh_nc_cc_skat,bib:coh_cc_bebc1,bib:coh_cc_fnal1,bib:coh_cc_bebc2,bib:coh_cc_fnal2,bib:coh_cc_fnal3,bib:coh_cc_charm},
along with the early measurements of the NC coherent pion production cross
section~\cite{bib:coh_nc_AP,bib:coh_nc_garg,bib:coh_nc_charm,bib:coh_nc_cc_skat}, are shown in Fig.~\ref{fig:charm_coh_cc}.
For comparison, the measurements in Fig.~\ref{fig:charm_coh_cc} are scaled to a glass scattering target with an effective
$A$ = 20.1, and the NC measurements are additionally scaled by a factor of 2 per the relation between the CC and NC coherent
cross sections from Adler's PCAC theorem.  These early CC measurements isolated CC coherent interactions by requiring a
forward $\mu^{\mp}$ and $\pi^{\pm}$, the absence of additional particles emerging from the interaction vertex, and small \tabs.
The Rein-Sehgal model agrees well with most of the measurements, which supports the predicted $A^{1/3}$ dependence and the
2-to-1 relationship between the CC and NC coherent cross sections.

The only measurements of NC coherent pion production at $\enu<$ 2 GeV were made recently by the MiniBooNE~\cite{bib:coh_nc_mb}
and SciBooNE \cite{bib:coh_nc_sb} experiments.  The MiniBooNE measurement was made using a mineral oil target (CH$_{2}$) at a
peak \enu of 0.7 GeV.  MiniBooNE measured the NC coherent pion production cross section to be $(19.5\pm2.7)$\% of the
total (coherent + non-coherent) NC single $\pi^{0}$ production cross section.  The SciBooNE measurement was made using a
polystyrene target (CH) at an average \enu  of 0.8 GeV.  SciBooNE measured the ratio of the NC coherent pion
production cross section to the \numu CC total (all scattering processes) cross section to be $(1.16\pm0.24)\times10^{-2}$.  
%These measurements provide strong evidence for NC coherent pion production at $\enu<$ 2 GeV.

\begin{figure}[tpb]
\centering
\includegraphics[width=0.8\columnwidth]{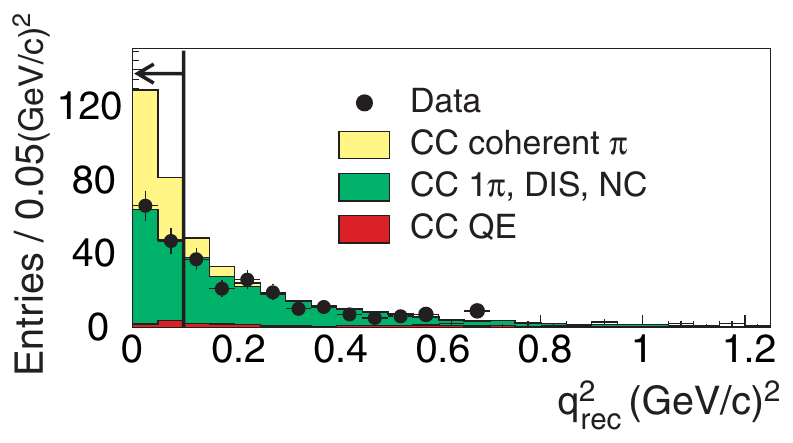}
\caption[Search for \numu CC coherent pion production at K2K]
	{The data and simulated \qsq distributions for \numu CC coherent pion production candidates at K2K.  The figure is
	from~\cite{bib:coh_cc_k2k}.}
\label{fig:k2k_coh_cc}
\end{figure}

\begin{figure}[tpb]
\centering
\includegraphics[width=0.8\columnwidth]{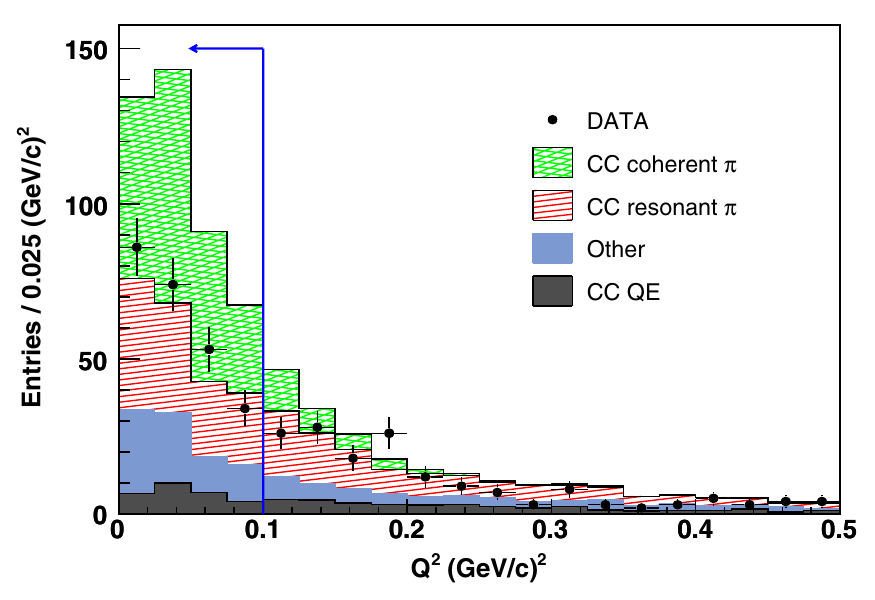}
\caption[Search for \numu CC coherent pion production at SciBooNE]{The data and simulated \qsq distributions for \numu CC
	coherent pion production candidates at SciBooNE.  The figure is from~\cite{bib:coh_cc_sb}.}
\label{fig:sb_coh_cc}
\end{figure}

The first searches for \numu CC coherent pion production at $\enu\lesssim$ 2 GeV were  performed  by the K2K~\cite{bib:coh_cc_k2k}
and SciBooNE~\cite{bib:coh_cc_sb} experiments.  These experiments used the same CH-based detector in two different neutrino beams.  
Neither could measure \tabs since the detector did not provide adequate containment of the pion to allow accurate measurement of
the pion energy, and instead searched for coherent scattering at low \qsq.
Neither experiment found evidence for CC coherent interactions after subtracting the predicted non-coherent background
(Figs.~\ref{fig:k2k_coh_cc} and~\ref{fig:sb_coh_cc}).  The experiments reported upper limits on the ratio of the \numu CC
coherent pion production cross section to the \numu CC total cross section, which are listed in Table~\ref{tab:k2k_sb_coh_cc}.

The non-observation of \numu CC coherent pion production at $\enu<$ 2 GeV is in contradiction to the the Rein-Sehgal coherent model.
It should be noted that a search for coherent pion production in \qsq is model dependent.  In addition, both the K2K and
SciBooNE measurements constrained their background prediction using interactions with activity near the interaction vertex in
addition to that from the muon and pion.  The background prediction is therefore sensitive to the modeling of nuclear effects
which are poorly understood (see Sec.~\ref{sec:bg_tuning}).

\begin{table}[h!]
%\small
\begin{center}
\begin{tabular}{ c | c | c }
%\hline
%Name & Symbol & Charge & Mass  \\
Experiment & $\sigma_{CC}^{coh} / \sigma_{CC}^{tot}$ (90\% C.L.) & $\langle\enu\rangle$ (GeV) \\
\hline
K2K & $<$ 0.6 $\times$ 10$^{-2}$ & 1.3 \\
SciBooNE & $<$ 0.67 $\times$ 10$^{-2}$ & 1.1 \\
SciBooNE & $<$ 1.36 $\times$ 10$^{-2}$ & 2.2 \\
\end{tabular}
\end{center}
\caption[Upper limits on the \numu CC coherent pion production cross section reported by K2K and SciBooNE]
	{The upper limits on the ratio of the \numu CC coherent pion production cross section to the \numu CC total cross
	section, $\sigma_{CC}^{coh} / \sigma_{CC}^{tot}$, reported by K2K~\cite{bib:coh_cc_k2k} and SciBooNE~\cite{bib:coh_cc_sb}.}
\label{tab:k2k_sb_coh_cc}
\end{table}

The non-observation of CC coherent pion production at $\enu\lesssim$ 2 GeV posed a problem for both theorists and neutrino
oscillation experiments; production models are unable to reconcile it with the observation of the NC reaction at the same \enu.
To account for the theoretical and experimental disagreement, the T2K neutrino oscillation measurement, which operates at
\enu $\approx\xspace$ 0.6 GeV, applied a 100\% uncertainty on their predicted CC coherent interaction
rate, while applying a 30\% uncertainty on their predicted NC coherent interaction rate \cite{bib:t2k_osc}.  
%Neutrino-nucleus interactions are the largest source of systematic uncertainty in the T2K $\numu\to\nue$ oscillation and $\numu$ disappearance measurements (Figure~\ref{fig:t2k_osc_unc}), and the predicted rate of coherent interactions is a significant contribution to the uncertainty \cite{bib:t2k_osc}.

%\begin{figure}[!htpb]
%\centering
%\includegraphics[width=0.7\columnwidth]{figures/CoherentChapter/t2k_osc_unc.pdf}
%\caption[T2K systematic uncertainties]{The systematic uncertainties on the predictions of \numu CC and \nue CC interactions in the Super Kamiokande (SK) far detector of the T2K neutrino oscillation experiment.  Constraints on the neutrino beam flux and neutrino interaction predictions are obtained from the ND280 near detector.  The table is from \cite{bib:t2k_osc}.}
%\label{fig:t2k_osc_unc}
%\end{figure}

Precise measurements of neutrino-nucleus coherent pion production at 1 $\lesssim\enu\lesssim$ 10 GeV are needed for
testing coherent pion production models and reducing systematic uncertainties in neutrino oscillation measurements.

\section{Diffractive Scattering}
\label{sec:diffractiveIntro}

\begin{figure}[tpb]
\begin{center}
\includegraphics[width=0.8\columnwidth]{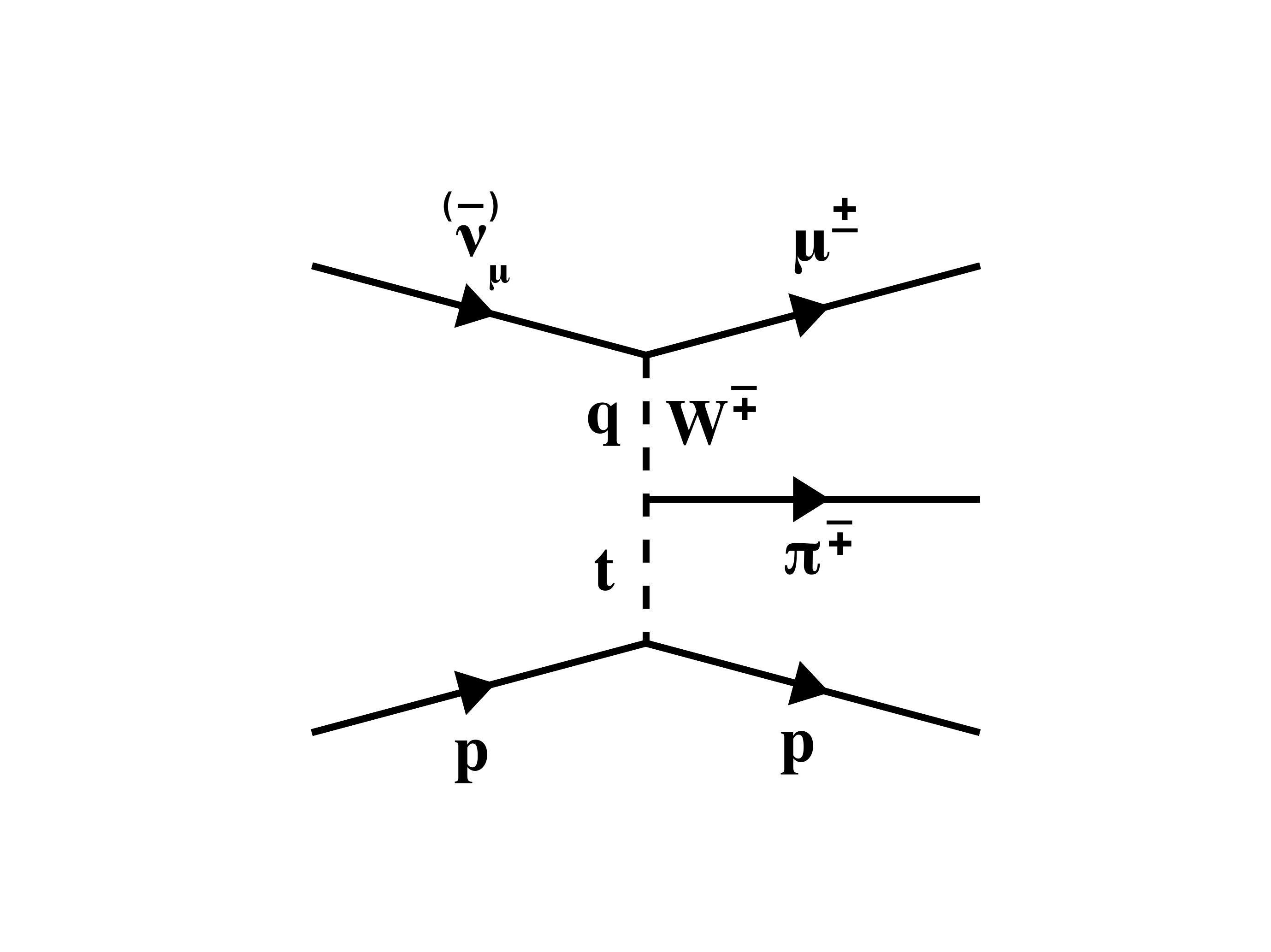}
\caption[Diffractive scattering in the PCAC picture]{Diffractive pion production on a free proton.}
\label{fig:diffractive_diagram}
\end{center}
\end{figure}

Diffractive pion production, $\nu_\mu p\to \mu^-\pi^+ p$ (Fig.~\ref{fig:diffractive_diagram})
is a process similar to coherent scattering but for single proton.  It produces a muon and charged pion in the forward direction
while leaving the nucleus, {\it i.e.} the proton, intact and with minimal recoil.  
Like coherent scattering, diffractive production occurs at low \tabs.  Unlike coherent scattering, the recoil
proton does absorb some kinetic energy, making it potentially visible in the detector.  There are also resonant and non-resonant
processes that occur in a broader range of \tabs.  At low \tabs these may interfere with the diffractive process
which complicates the prediction of the process.

Diffractive scattering is experimentally indistinguishable from coherent scattering when the recoil proton is
below detection threshold.  A \numu / \numubar CC sample in the \minerva tracker may contain diffractive
scattering interactions since the CH scintillator contains free
protons.  Diffractive scattering was not included in the simulated backgrounds to the coherent process, so
the measured coherent cross section may therefore contain a contribution from it.  It is not coherent scattering on carbon
and is therefore a potential background to this measurement.  One of the new results presented here is that in fact this
potential background has a rate consistent with zero.

An important difference between coherent and diffractive scattering is the \tabs-dependence of the cross sections.
In the PCAC picture of diffractive scattering, the intermediate weak boson fluctuates to a pion, which scatters
elastically off the target proton.  The expectation therefore is that the \tabs-dependence comes from the
pion-proton/nucleus elastic scattering cross section, which is assumed to fall exponentially with \tabs as
$\exp(-b\tabs)$ for the same reason that it is a reasonable choice in the Rein-Sehgal model of coherent scattering~\cite{bib:RS}.  The exponential slope $b$ is given by
\begin{equation}
b=\frac{1}{3}R_{0}^{2}A^{2/3},
\end{equation}
where $R_{0}\sim$1 fm is the nuclear length scale and A is the number of nucleons in the target.  The predicted
exponential slope for coherent scattering on carbon ($A=12$) is $\sim$40 (GeV/c)$^{-2}$, and the predicted exponential
slope for diffractive scattering ($A=1$) is $\sim$8 (GeV/c)$^{-2}$.  The diffractive cross section therefore falls more
slowly with \tabs than the coherent cross section.  The squared four-momentum exchanged with the target proton in
diffractive scattering \tdiff is related to the recoil proton kinetic energy \Tp by
\begin{eqnarray}
\label{eq:t_diff}
%|t|_{diff} &=|(p_{\nu}-p_{\mu}-p_{\pi})^{2}| \nonumber \\
%&= |(p_{p,f}-p_{p,i})^{2}| \nonumber \\
%%&= | m_{p}^{2} + m_{p}^{2} - 2E_{p,i}E_{p,f} - \vec{p}_{p,i} \cdot \vec{p}_{p,f}| \nonumber \\
%%&= |2m_{p}(m_{p}-E_{p,f})| \nonumber \\
%&= 2m_{p}T_{p},
|t|_{diff} & = |(p_{\nu}-p_{\mu}-p_{\pi})^{2}| \nonumber \\
           & = |(p_{p,f}-p_{p,i})^{2}| = 2m_{p}T_{p},
\end{eqnarray}
where $p_{\nu}$ is the neutrino four-momentum, $p_{\mu}$ is the muon four-momentum, $p_{\pi}$ is the pion four-momentum,
$p_{p,i}$ and $p_{p,f}$ are the target (initial state) and recoil (final state) proton four-momentum and $m_{p}$ is the
proton mass.
%, $E_{p,i}$ and $E_{p,f}$ are the target and recoiling proton energy, $\vec{p}_{p,i}$ and $\vec{p}_{p,f}$
%are the target and recoil proton three-momentum, and 
The target proton is assumed to be on-shell and at rest.
%($E_{p,i} = m_{p}, |\vec{p}_{p,i}| = 0$).
The amount of energy deposited in the detector by the recoil proton in a diffractive scattering interaction determines
whether the interaction is accepted or rejected by the vertex energy cut (Sec. ~\ref{sec:evtx_cut}) which requires that energy near the interaction vertex is consistent with the energy deposited by only a muon and a pion.  Accepted events are therefore restricted to small \Tp,
and equivalently small \tabs.  The small \tabs diffractive acceptance in conjunction with the slowly falling
\tabs-dependence of the diffractive cross section results in a small contribution of diffractive scattering to the
measured coherent cross sections.

Predictions for the diffractive pion production and the possible contribution to the signal sample are discussed in Sec.~\ref{sec:diffractiveMeasure}.

\section{Neutrino Beam and Detector}
\label{sec:NuMI_N_me}

The \minerva experiment is in the NuMI beamline at Fermilab.  Both the beamline and detector are described in detail
elsewhere~\cite{bib:numi_beam,bib:minerva_nim}; here is a short summary.

A beam of \unit[120]{GeV} protons strikes a graphite target, and the charged mesons produced there are focused by two
magnetic horns into a helium-filled decay pipe \unit[675]{m} long.  The horns focus positive (negative) mesons, resulting in
a $\numu$ ($\numubar$) enriched beam with a peak neutrino energy of \unit[3.5]{GeV}.  Muons produced in meson decays are
absorbed in the \unit[240]{m} of rock downstream of the decay pipe.  The neutrino beam is simulated in a Geant4-based~\cite{bib:geant1}
model weighted to reproduce hadron production measurements~\cite{bib:flux}.

Roughly $85\%$ of the neutrinos are produced by interactions of
protons on carbon, and data from the NA49 and Barton {\em et. al.} experiments~\cite{bib:na49,bib:barton} is used to
constrain the production of the charged pions and kaons that decay to neutrinos.
Proton interaction rates with aluminum, iron, and helium nuclei were tuned to the \pC data using an $A$-dependent scaling;
neutron interaction rates with carbon were also tuned to the \pC data using isospin symmetry.  Rates for interactions
where there are no applicable measurements were taken from the hadron interaction model used in the simulation
(FTFP-BERT)\footnote{{F}TFP shower model in Geant4 version 9.2 patch 03.}.  Predictions from the weighted model were
compared against in-situ measurements of neutrino scattering events with low hadronic recoil to validate the model.
An \insitu measurement~\cite{bib:veve} of $\nu + e^- \rightarrow e^- + \nu$ provided an additional constraint on the flux, giving
a roughly \unit[2-4]{\%} reduction in the flux prediction and an $\sim$ \unit[1]{\%} reduction in the flux uncertainty.  The
uncertainty in the prediction of the neutrino flux, discussed in more detail later, is set by (a) the precision in these hadron
production measurements, (b) uncertainties in the beam line focusing system and alignment~\cite{bib:numi_align} and (c) comparisons
between different hadron production models in regions not covered by hadron production data.
%the \insitu $\nu+e$ elastic scattering measurement\cite{bib:veve}.

\begin{figure}[tpb]
\centering
\includegraphics[width=0.8\linewidth]{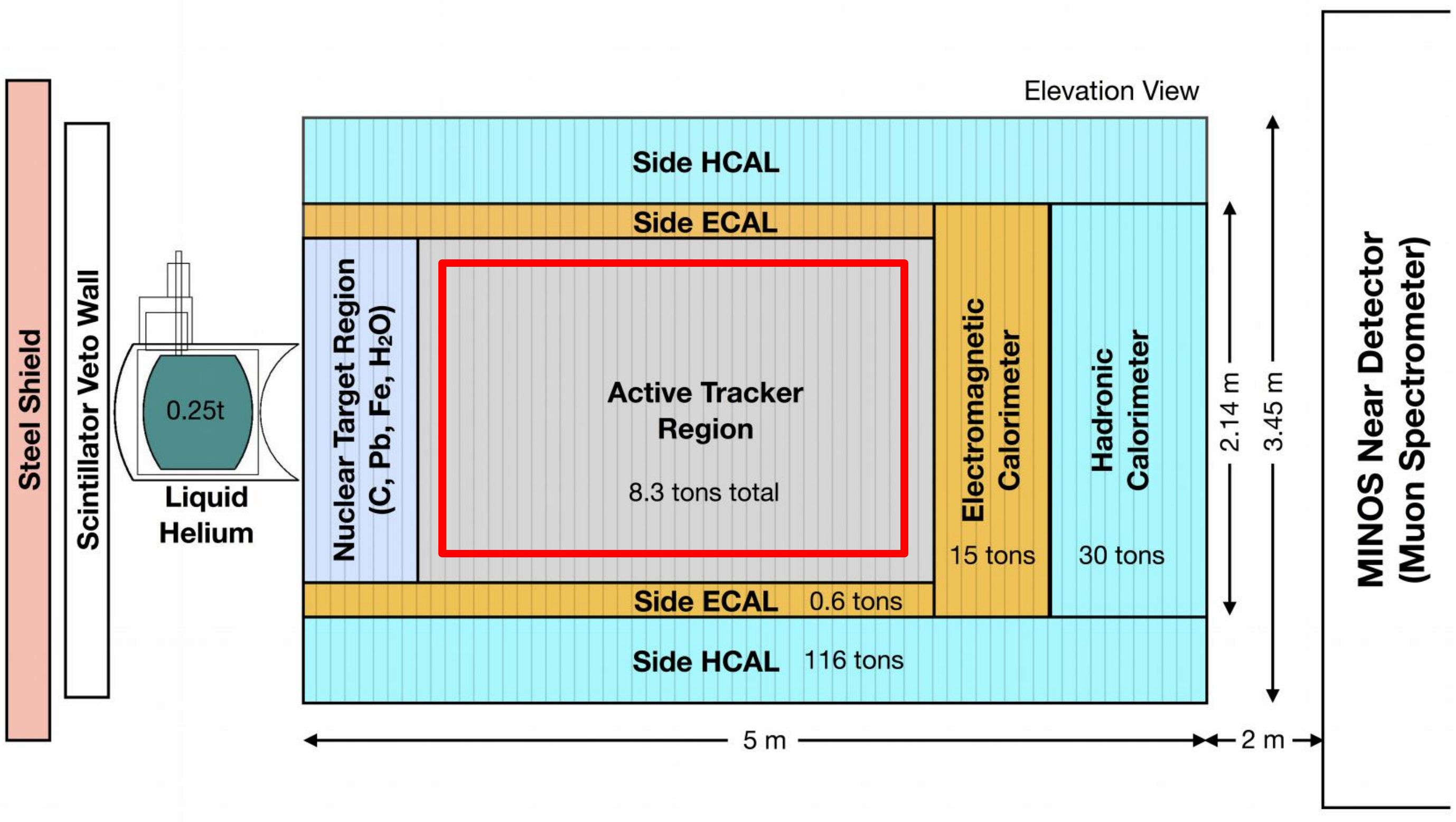}
\includegraphics[width=0.6\linewidth]{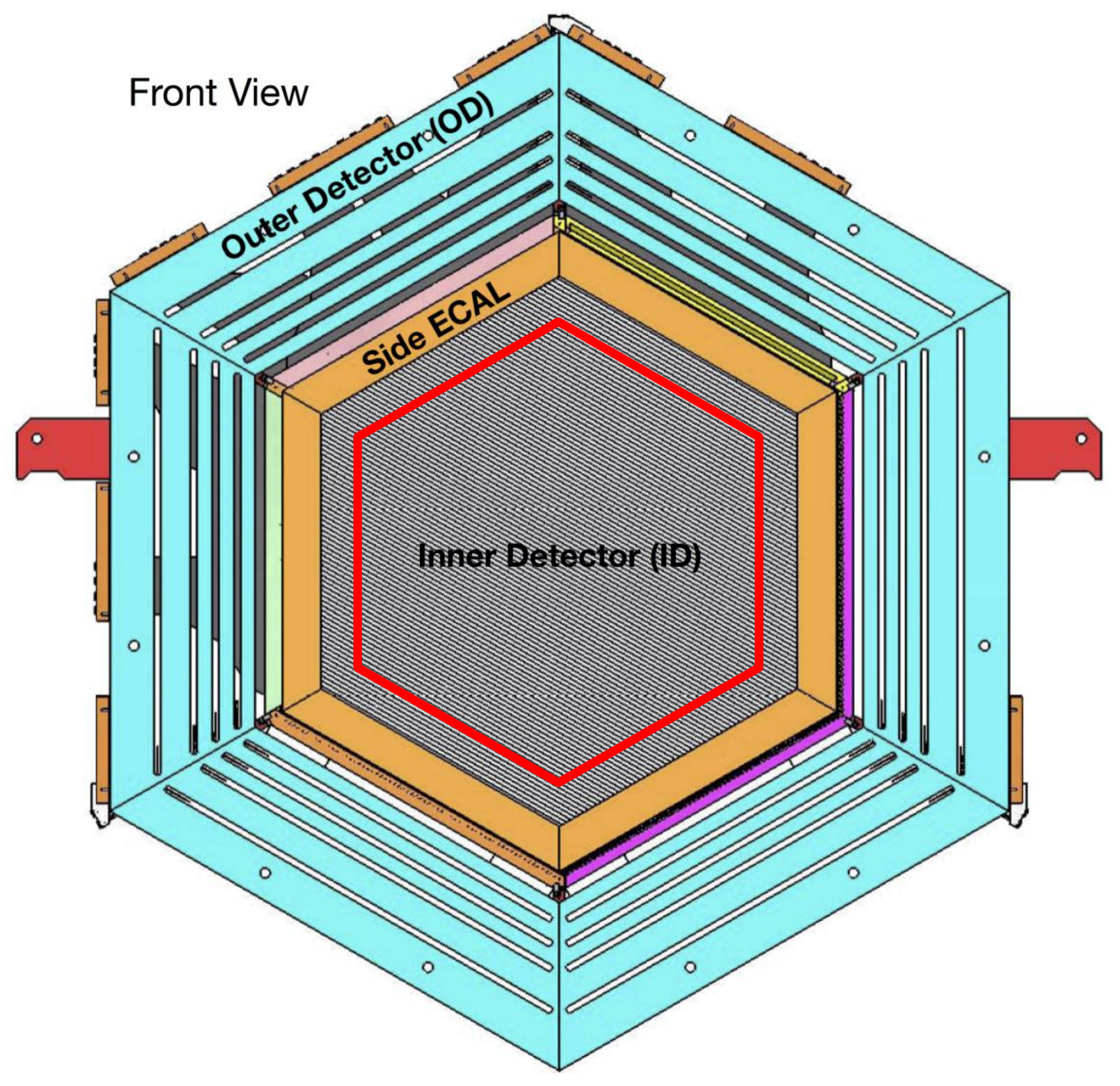}
\caption[Fiducial volume]{The \minerva detector.  The fiducial volume is illustrated by the red rectangle and hexagon (approximate scale)}
\label{fig:fiducial_volume}
\end{figure}

The \minerva detector (Fig.~\ref{fig:fiducial_volume}) consists of a central tracker composed of scintillator strips surrounded by
electromagnetic and hadronic calorimeters on the sides and downstream end of the detector.  The triangular
$3.4\times1.7$~cm$^2$ strips are perpendicular to the
z-axis\footnote{The y-axis points along the zenith and the beam is directed downward by \unit[58]{mrad} in the y-z plane.} and
are arranged in hexagonal planes.  Up to two planes are assembled into a supporting frame, and this assembly is called a module.
Three plane orientations, or views, with $0^\circ$ and $\pm 60^\circ$ rotations around the z-axis, enable reconstruction of the neutrino
interaction point and the tracks of outgoing charged particles in three dimensions. The \unit[3.0]{ns}
timing resolution per plane allows separation of multiple interactions within a single beam spill.  
\minerva is located \unit[2]{m} upstream of the MINOS near detector, a magnetized iron spectrometer~\cite{bib:minos_nim}
which is used to reconstruct the momentum and charge of $\mu^\pm$.

The fiducial volume for this analysis is contained within the tracker.  The fiducial volume boundaries are recessed from the
boundaries of the tracker to minimize contamination from interactions on non-carbon nuclei (\textit{e.g.} lead) in the
upstream nuclear targets region, side and downstream surrounding electromagnetic calorimeters (ECALs) and hadronic calorimeters (HCALs).
%that result from from mis-reconstructing the interaction vertex position.  
%The fiducial volume spans the central 108 (56) scintillator planes of the tracker in the full (partial) detector, which excludes the 8 most upstream and 8 most downstream planes of the tracker.  In the transverse (XY) plane, the fiducial volume edges are defined by a hexagon with an 85 cm apothem centered at the detector axis, which are recessed $\sim$10 cm from the inside edges of the side ECAL.  
The full detector fiducial volume mass is 5.47 metric tons.

The \minerva\ detector records the energy and time of energy deposits (hits) in each scintillator strip.  Hits are first grouped
chronologically and then clusters of energy are formed by spatially grouping the hits in each scintillator plane.  Clusters with
energy $>\unit[1]{MeV}$ are then matched among the three views to create a track.  The  $\mu^\pm$ candidate is a track that exits
the back of \minerva matching a track of the expected charge entering the front of \minos. The most upstream cluster on the muon
track in MINERvA is taken to be the interaction vertex.  The transverse position resolution of each track cluster
is \unit[2.7]{mm} and the angular resolution of the muon track is better than \unit[10]{mrad} in each view.  The reconstruction
of the muon in the MINOS spectrometer gives a typical muon momentum of $11\%$.  
Charged $\pi^\pm$ reconstruction requires a second track originating from the vertex.

The \minerva detector's response is simulated by a tuned Geant4-based
%~\cite{bib:geant1}
program.  
The energy scale of the detector is set by ensuring that both the photostatistics and the reconstructed energy deposited by
momentum-analyzed through-going muons agree in data and simulation.  The calorimetric constants used to reconstruct the energy of
$\pi^\pm$ showers and the corrections for passive material are determined from the simulation~\cite{bib:minerva_nim} and verified
by a test beam measurement~\cite{bib:minerva_Tbeam}.

\section{Signal, Backgrounds and Data}
\label{sec:MC_bkg_Data}

\subsection{Experimental Signature}
\label{sec:exp_sig}

\begin{figure}[tpb]
\centering
\mbox{\includegraphics[width=\linewidth]{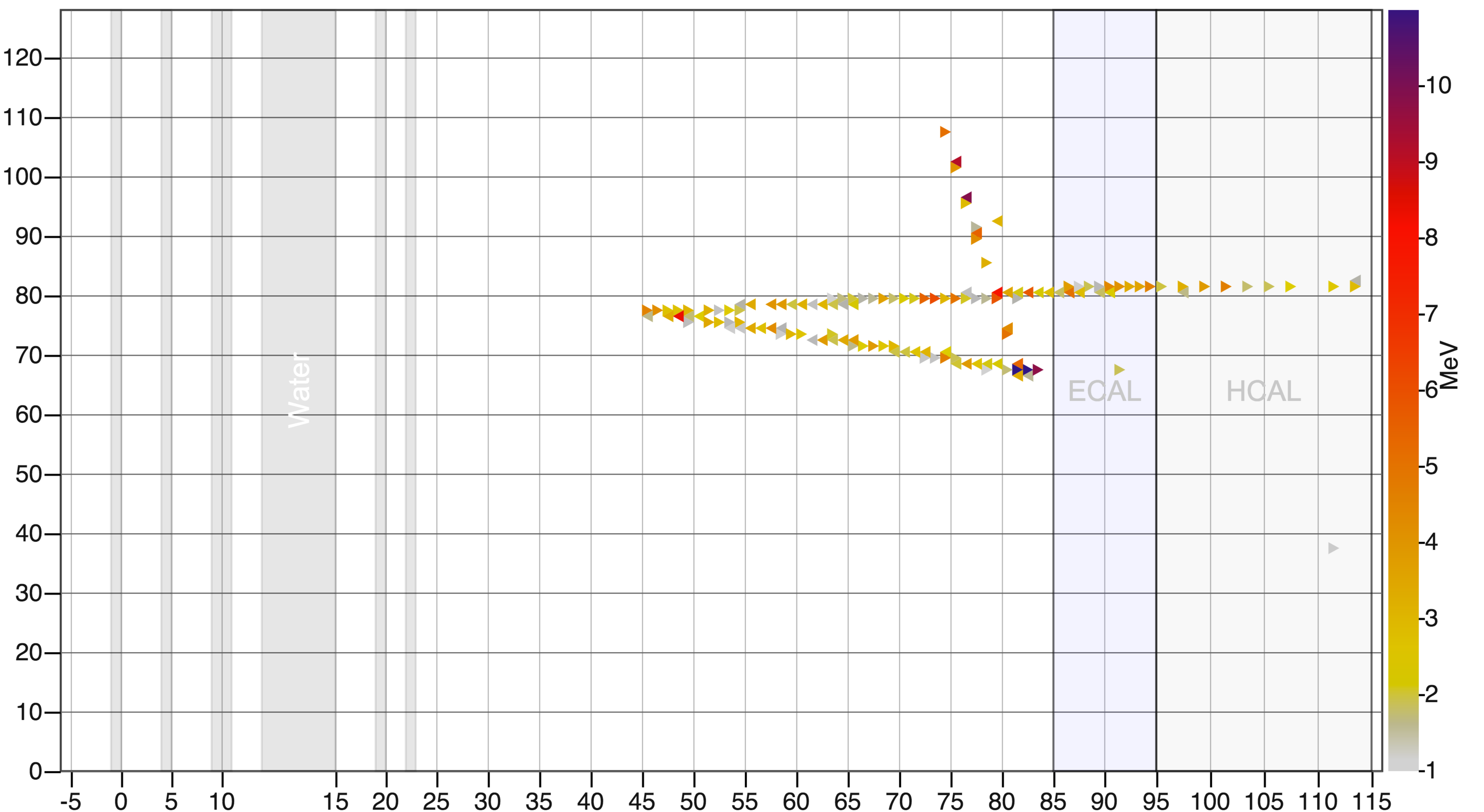}}
\caption[\numu coherent scattering candidate]{\small A data \numu coherent scattering candidate.  This visualization shows energy deposited in the vertical scintillator strips of the detector, with the numbering on the $x$-axis representing module number, the numbering on the $y$-axis representing strip number, and the color scale indicating the energy deposited in each strip.}
\label{fig:coherent_candidate}
\end{figure}

In \minerva, coherent scattering appears as two forward tracks originating from a common vertex with no additional visible
energy near the vertex (Fig.~\ref{fig:coherent_candidate}).  One, a muon, typically exits the downstream end of \minerva and
enters MINOS, producing a minimum ionizing track in both detectors.  The other, a pion, produces a minimumally ionizing track
before stopping or interacting hadronically within \minerva.  Visible energy near the vertex in addition to that from the
minimally ionizing muon and pion is indicative of nuclear breakup in an incoherent interaction.  In addition, the \minerva
detector enables reconstruction of the squared four-momentum transfered to the nucleus, \tabs.

\subsection{Signal and Background Definitions}
\label{sec:mc_signal_background}

Signal interactions are simulated using the implementation of the Rein-Sehgal model~\cite{bib:RS} in
GENIE~\cite{bib:genie_manual} Monte Carlo (MC), version 2.6.2, with the modifications described in
Sec.~\ref{sec:mc_reweight}.  Events in the simulation are categorized as either signal or background as follows:

\begin{enumerate}
\item {\it Signal (``Coherent")} - Interactions that produce a final state consisting of a muon, a charged pion, and
the initial state nucleus.  The only GENIE interactions that produce this final state are \numu(\numubar) CC
coherent pion production interactions.  Coherent interactions on non-carbon nuclei are categorized as signal rather than
background to avoid dependence of the background prediction on the signal model.  The correction for non-carbon nuclei is
described in Sec.~\ref{sec:target_norm}.
\item {\it Charged-qurrent Quasielastic (``QE")} - \numu(\numubar) CC quasielastic interactions as modeled in GENIE.
\item {\it Non-Quasielastic, $W< 1$ .4 GeV} - \numu(\numubar) CC interactions, excluding quasielastic, with true invariant mass $W<$ 1.4 GeV, where $W$ is the invariant mass of the hadronic recoil.  This category is primarily delta resonance production, but also includes non-resonant pion production.
\item {\it 1.4 $<W<$ 2.0 GeV} - \numu(\numubar) CC interactions with true invariant mass 1.4 $<W<$ 2.0 GeV.  This is the
transition region from delta resonance production to deep inelastic scattering.
\item {\it $W>$ 2.0 GeV} - \numu(\numubar) CC interactions with true invariant mass $W>$ 2.0 GeV.  This is the deep inelastic scattering (DIS) region.
\item {\it Other} - All other background interactions.  This category is primarily wrong sign interactions (\textit{i.e.} \numubar
instead of \numu interactions and vice versa), but also includes non-\numu(\numubar) and NC interactions.  Wrong
sign CC coherent pion production interactions are included in this category but are a small contribution.
\end{enumerate}

\subsection{Data and MC Samples}
\label{sec:data_mc_samples}

This analysis uses \minerva data taken between November 2009 and April 2012 in the low energy \numu and \numubar  beam
configurations.  The neutrino flux is normalized to the number of protons incident on the NuMI target (POT).  The total
POT for data taken  in the \numu (\numubar) beam configuration was $3.04\times10^{20}$ ($2.00\times10^{20}$).
In $\numubar$ mode, $45\%$ of the POT were taken with a partial detector with 52$\%$ of the fiducial mass, meaning
that $\sim$ 30\% of the antineutrino coherent scattering events should occur in this partial detector.

The MC sample was generated with appropriate neutrino beam configurations, and was overlaid with data taken with the appropriate
configuration to mimic pile-up effects in the data.  

High-statistics signal-only MC samples were generated for estimating the resolution of the reconstructed kinematics, unfolding
matrices, and selection efficiency for \numu and \numubar CC coherent scattering.

\subsubsection{Weighting Simulated Events}
\label{sec:mc_reweight}

The MC was weighted to account for improvements in the understanding of (a) the neutrino flux, (b) the rate of neutrino
interactions with single-pion final states, (c) the pion angle distribution in delta resonance decay, (d) the coherent pion production
cross section in GENIE, and (e)\minos muon tracking efficiency.

The \minerva MC was generated with a flux prediction from a GEANT4 model of the NuMI target.  The MC was
subsequently weighted to reflect the flux prediction constrained by external hadron production data as described in
Sec.~\ref{sec:NuMI_N_me}.

The GENIE prediction for single-pion final states was constrained by reanalyzed \numu-deuterium scattering data~\cite{bib:D2Data}.
The axial vector mass for resonant pion production ($M^{RES}_{A}$) and corrections to the resonant pion production and nonresonant
single pion production normalizations in GENIE were extracted from a fit of GENIE to the re-analyzed data for single-pion final
states~\cite{bib:D2Fit}.  Table~\ref{tab:deuterium_fit} lists the values extracted from the fit.  The default $M^{RES}_{A}$ in
GENIE is 1.12 $\pm$ 0.22 GeV.  Resonant interactions and nonresonant single pion production interactions in the MC were weighted
to the values from the fit.

\begin{table}[bp] \small
\begin{center}
\begin{tabular}{ l | c }
Parameter & Value \\
\hline
$M^{RES}_{A}$ (GeV) & 0.94 $\pm$ 0.05 \\
Resonant Normalization Correction & 1.15 $\pm$ 0.07 \\
Non-Resonant 1$\pi$ Normalization Correction & 0.46 $\pm$ 0.04 \\
\end{tabular}
\end{center}
\caption{Parameters for single pion production in GENIE extracted from the fit of GENIE to re-analyzed \numu-deuterium scattering data from Ref.~\cite{bib:D2Data}.}
\label{tab:deuterium_fit}
\end{table}

\newcommand{\pthreetwo}{\ensuremath{p_\frac{3}{2}}\xspace}
\newcommand{\ponetwo}{\ensuremath{p_\frac{1}{2}}\xspace}
\newcommand{\legendrepoly}{\ensuremath{P_2(\cos\theta)}\xspace}

GENIE assumes isotropic decay of baryon resonances from neutrino production, but the Rein-Seghal model
%GENIE simulates the decay of the baryon resonance in order to simulate final state interactions of the decay products within the nucleus.  For simplicity in generating events, 
%, where the decay products have a uniform angle distribution in the center of mass frame of the resonance with respect to the angular momentum quantization axis of the resonance 
predicts anisotropic baryon resonance decay.  For $\Delta\to N\pi$ delta resonance decays, the angular distribution of the pion is given by
\begin{equation}
\label{eq:delta_decay_pion_angle}
%W_{\pi}(\cos\theta) = 1 - p_\frac{3}{2}P_2(\cos\theta) + p_\frac{1}{2}P_2(\cos\theta),
W_{\pi}(\cos\theta) = 1 - \pthreetwo\legendrepoly + \ponetwo\legendrepoly,
\end{equation}
where $\theta$ is the pion angle in the $\Delta$ center of mass frame with respect to the $\Delta$ angular quantization axis,
\pthreetwo and \ponetwo are coefficients for the $m=\frac{3}{2}$ and $m=\frac{1}{2}$ states of the $\Delta$, respectively,
and \legendrepoly is the 2nd-order Legendre polynomial.  
For $\Delta\to N\pi$, GENIE simulates isotropic $\Delta$ decay with $\pthreetwo = \ponetwo = 0.5$, whereas the Rein-Sehgal resonance production model predicts non-isotropic $\Delta$ decay with $\pthreetwo = 0.75$ and $\ponetwo = 0.25$.  
%For propagating the uncertainty from simulating isotropic $\Delta^{++}\to N\pi$ decay, GENIE provides event-by-event weights that warp the isotropic angle distribution of the decay pions to the non-isotropic Rein-Sehgal prediction.  Using these weights, t
%%
%%I (Leo) can't figure out this sentence, and the citation is to a MINERvA DocDB, not a released result.
%%
%The pion angle distribution measured from \minerva data of \numu charged current $\pi^{+}$ production was shown to prefer
%the GENIE prediction for these processes with non-isotropic $\Delta^{++}\to N\pi$ decay per Rein-Sehgal \cite{bib:eberly_delta_decay}.  
%The convention within \minerva, which is applied in this analysis, is to 
We weight the isotropic pion angle distribution from $\Delta^{++}\to N\pi$ decays as generated by GENIE to half the anisotropy
predicted by the Rein-Seghal resonance production model, and consider both unweighted and fully-weighted anisotropies in setting systematic uncertainties.

We used the default INTRANUKE / hA model of final state interactions within the target nucleus (FSI) in GENIE.

The GENIE implementation of the Rein-Sehgal coherent scattering model uses measured total and inelastic cross sections
for pion-proton and pion-deuterium scattering to calculate the pion-nucleus elastic scattering cross section.
GENIE 2.6.2 contained an error in indexing the pion-proton and pion-deuterium cross section data tables.  This error was
corrected by weighting each MC signal event.  The correction is less than $5\%$ in all \epi bins in the analysis, except for
$750<\epi<1000$~MeV, where the correction is $\sim 10\%$.  
%Figure~\ref{fig:genie_coh_bug} illustrates the size of the correction as the ratio of the corrected to uncorrected generated event rate as a function of true \epi for \numu and \numubar charged current coherent interactions on carbon in the fiducial volume.

%\begin{figure}[tpb]
%\centering
%\mbox{
%\includegraphics[width=0.5\linewidth]{figures/Efficiency/h_Epi_GenieCohBugWeight_FluxConstrained_WeightedSignalModel_minerva1.pdf}
%\includegraphics[width=0.5\linewidth]{figures/Efficiency/h_Epi_GenieCohBugWeight_FluxConstrained_WeightedSignalModel_minerva5.pdf}}
%\caption{The ratio of the generated event rate corrected for the GENIE coherent cross section error to the uncorrected generated event rate as a function of true \epi for \numu (left) and \numubar (right) charged current coherent interactions on carbon in the fiducial volume.}
%\label{fig:genie_coh_bug}
%\end{figure}

The efficiency of tracking the muon in \minos differed between data and MC due to pile-up not being simulated in \minos.
The efficiency was measured in both data and MC by projecting muon tracks that exited \minerva into \minos and
measuring the muon reconstruction rate in \minos.  The differences in efficiency between the data and MC were corrected
for by weighting each MC event; the corrections were typically less than 5\%.

\section{Reconstruction}
\label{sec:kinematics_reco}

%The muon and pion directions are measured from their reconstructed tracks.
The muon is identified as the track that exits the downstream end of \minerva and matches a track in MINOS.  The reconstructed
interaction vertex position is defined as the most upstream point along the \minerva muon track.  The pion track is the second track
originating from the interaction vertex.  
Reconstruction produces, for each track, a 3D direction vector of its respective particle in each scintillator plane along the track.  
To reduce the effects of scattering, the direction of both the muon and pion are taken as the direction vector in the most upstream plane along
their respective track.  The angle between the directions of the muon (pion) and incoming neutrino is denoted by \thetamu (\thetapi).  The direction
of the incoming neutrino is assumed to be parallel to neutrino beam axis.

The muon energy \emu is reconstructed from the muon's tracks in \minerva and \minos.  The energy of the muon at its entrance point to MINOS is
reconstructed from the \minos track's range (curvature in the magnetic field) if the muon stops inside (exits) MINOS.  This energy is added to the
calculated muon energy loss by ionization along the \minerva track.  
%For the partial detector, the calculated muon energy loss by ionization in \argoneut is included in \emu.

The pion energy \epi is reconstructed by calorimetry
%.  This approach is motivated by the majority of charged pions in \minerva interacting hadronically with nuclei in the detector.  These interactions produce a variety of secondary particles, each of which produces a different detector response.  In particular, neutrons can be produced which, if they interact, only deposit a fraction of their energy in the form of scattered low energy protons and thereby constitute missing pion energy.  Calorimetry corrects the visible energy from a pion and its secondary particles to give, on average, the correct \epi.  \epi is reconstructed 
under assumption that all hadronic (\textit{i.e.} non-muon) visible energy results from interactions of the pion.  To minimize sensitivity to
mismodeled vertex activity for background events, hadronic visible energy within \unit[200]{mm} of the interaction vertex
is excluded from the \epi reconstruction and replaced by a constant \unit[60]{MeV} which is the average calorimetric
energy deposited by a minimum ionizing pion over \unit[200]{mm} in the tracker. Energy deposited in parts of the detector
with passive layers (the side and downstream ECALs and HCALs) is corrected for the energy not observed in those passive
layers.  An overall calorimetric scale factor of $1.7$ is applied for $\pi^\pm$ response, as obtained from the simulation of signal events.
Clusters of hits are included if they are between \unit[-20]{ns} and \unit[+30]{ns} of the reconstructed interaction vertex time.  This requirement
minimizes contamination from pile-up. 

The reconstructed neutrino energy \enu is $\emu + \epi$,
where \emu and \epi are the reconstructed muon and pion energies, respectively.
This calculation assumes zero energy transfer to the nucleus, which is a good approximation for coherent scattering since the nucleus remains in its ground state and the recoil of the nucleus is small due to small \tabs.

\begin{figure*}[tpb]
\centering
\mbox{
\includegraphics[width=0.49\linewidth]{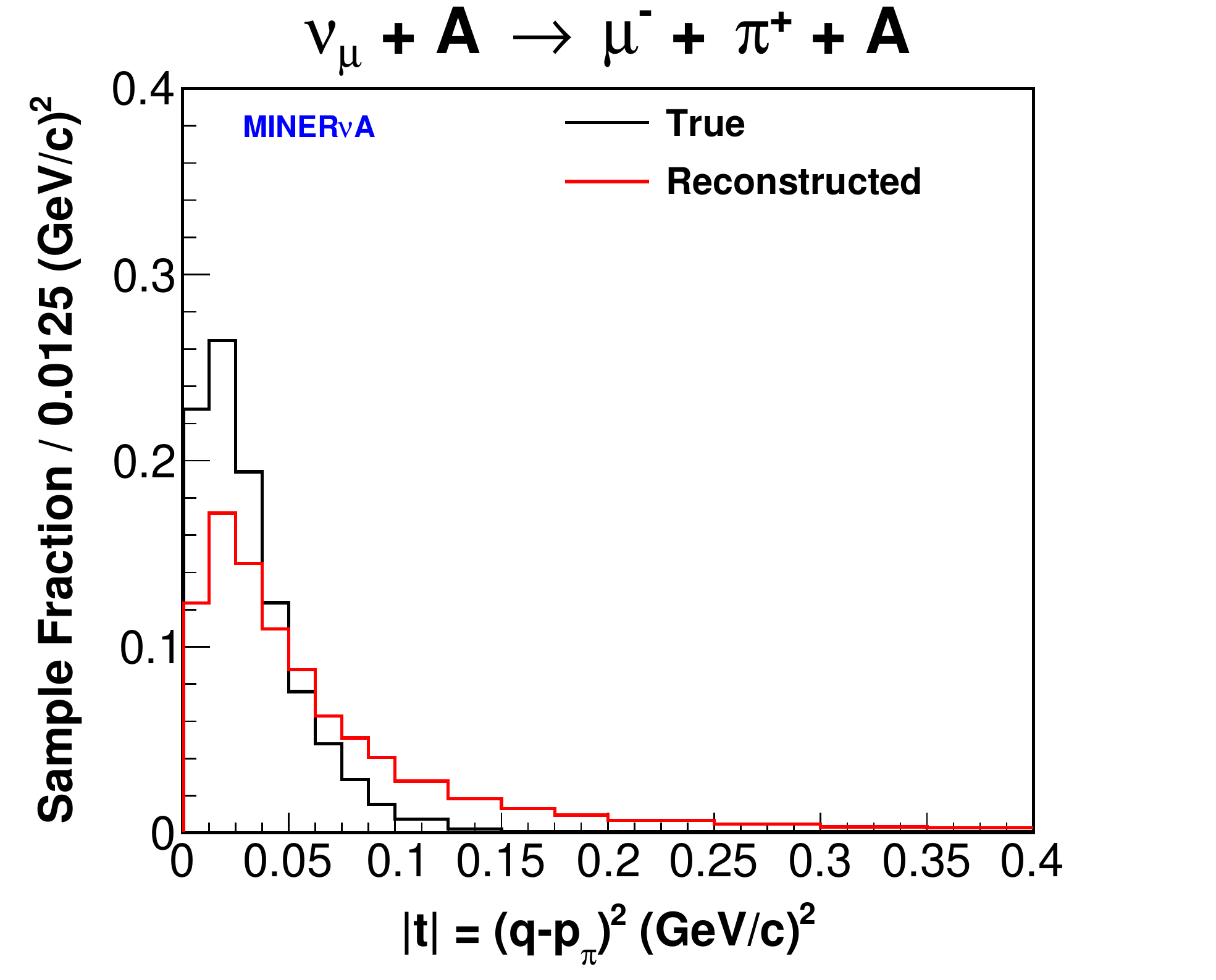}
\includegraphics[width=0.49\linewidth]{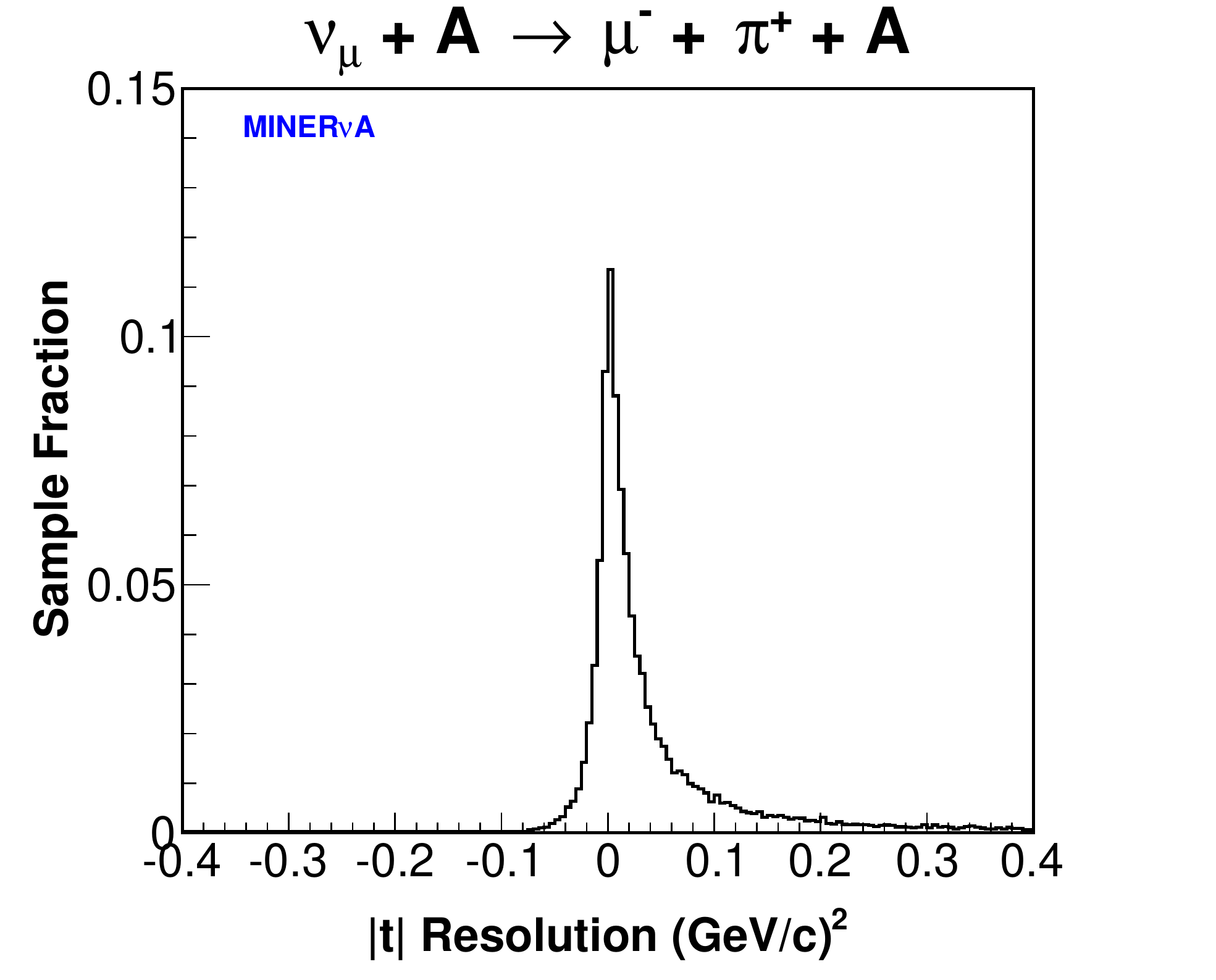}}
%\mbox{
%\includegraphics[width=0.49\linewidth]{figures/Resolution/h_t_True_Reco_FluxConstrained_WeightedSignalModel_minerva5.pdf}
%\includegraphics[width=0.49\linewidth]{figures/Resolution/h_t_Resolution_FluxConstrained_WeightedSignalModel_minerva5.pdf}}
%\mbox{
%\includegraphics[width=0.49\linewidth]{figures/Resolution/h_t_True_Reco_FluxConstrained_WeightedSignalModel_downstream.pdf}
%\includegraphics[width=0.49\linewidth]{figures/Resolution/h_t_Resolution_FluxConstrained_WeightedSignalModel_downstream.pdf}}
\caption[Reconstructed and true \tabs distributions and reconstructed \tabs resolution]{The reconstructed and true \tabs
	distributions (left) and the \tabs resolution, defined as \mbox{$|t|_{\rm\textstyle reco}-|t|_{\rm\textstyle true}$}, (right) for reconstructed \numu CC coherent MC interactions inside the fiducial volume.}
% The top plots are for \numu interactions in the full detector, and the middle (bottom) plots are for \numubar interactions in the full (partial) detector.}
\label{fig:resolution_t}
\end{figure*}

The reconstructed square of the four-momentum transferred to the nucleus \tabs is calculated as
\begin{equation}
\label{eq:tabs_reco}
|t|=|(q-p_{\pi})^{2}|=|(p_{\nu}-p_{\mu}-p_{\pi})^{2}|,
\end{equation}
where $p_{\nu}$, $p_{\mu}$, and $p_{\pi}$ are the reconstructed four-momentum vectors for the neutrino, muon, and pion, respectively.  The reconstructed four-momentum 
vector for each particle is calculated from the particle's reconstructed energy and direction.
%(as mentioned previously, the neutrino direction is assumed to be parallel to the neutrino beam axis)
Likewise, the reconstructed squared four-momentum transferred from the lepton system \qsq is calculated as
\begin{equation}
\label{eq:qsq_reco}
Q^{2}=-q^{2}=(p_{\nu}-p_{\mu})^{2}.
\end{equation}
The resolution of the reconstructed interaction vertex position is $\sim$\unit[6]{mm} in the directions transverse to the beam and
$\sim$\unit[12]{mm} parallel to the beam.  Two-dimensional angular resolutions are $\sim$\unit[1]{$^\circ$}, although approximately
40\% of the events where the $\pi^\pm$ scatters near the vertex have a significantly worse resolution, $\sim$\unit[10]{$^\circ$}.
Muon energy resolution is $\sim$\unit[8]{\%}, and $\pi^\pm$ energy resolution is $\sim$\unit[25]{\%}.  The resolution in \qsq is 
$\sim$\unit[0.03]{GeV$^2$}, with about one in three events with worse resolution, up to $\sim$\unit[0.13]{GeV$^2$}.
%The resolution on \tabs is $\sim$\unit[0.015]{(GeV/c)$^2$} for most events with about one in three events having a resolution of $\sim$\unit[0.06]{(GeV/c)$^2$}, mostly due to scattering of the final state $\pi^\pm$.
The resolution in \tabs\ is shown in Fig.~\ref{fig:resolution_t}.

%\input{tables/ResolutionTables_FluxConstrained_WeightedSignalModel_Angles.tex}
%\input{tables/ResolutionTables_FluxConstrained_WeightedSignalModel_Emu.tex}
%\input{tables/ResolutionTables_FluxConstrained_WeightedSignalModel_Epi.tex}
%\input{tables/ResolutionTables_FluxConstrained_WeightedSignalModel_Ev.tex}
%\input{tables/ResolutionTables_FluxConstrained_WeightedSignalModel_Q2.tex}
%\input{tables/ResolutionTables_FluxConstrained_WeightedSignalModel_t.tex}

%\clearpage

\section{Event Selection}
\label{sec:event_selection}

%This section describes the procedure for selecting coherent-like interactions from data and MC.  The procedure consists of a series of cuts that require each interaction match \minerva's experimental signature for coherent scattering (Section~\ref{sec:exp_sig}).

This section describes the procedure for selecting candidate coherent scattering events
% (\ie events that match the experimental signature for coherent scattering)
from the data and MC for the measurement of the coherent scattering cross sections.
%The event selection procedure consists of a series of requirements, referred to as cuts, that correspond to the features of the
%experimental signature and determine whether each event is accepted or rejected.
%The MC is used to estimate the signal and background acceptance for each cut, and the background rate in the coherent candidate samples.

\subsection{Reconstruction and Fiducial Volume Cuts}
\label{sec:reco_and_fiducial}

%The event selection first requires each event have exactly one track in \minerva that is matched to a track in \minos.  The matched tracks identify the muon since other particles tend to either stop or interact within \minerva.
%This requirement gives a sample $>$99\% pure in charged-current \numu/\numubar events.  In addition, the \minos track enables determination of the muon charge and reconstruction of the muon energy.
Requiring exactly one reconstructed muon as described in Sec.~\ref{sec:kinematics_reco} gives a sample that is more than 99\% pure \numu or \numubar CC events.

Each event is required to have exactly one additional track originating at the interaction vertex and pointing in the forward direction.  For coherent events, this track identifies the pion and measures its direction.

The interaction vertex is required to be located within the fiducial volume.
%The interaction vertex is located at the origin of the muon track.  For coherent events, the muon track origin gives the best estimate of the interaction vertex position.
Dead time in the front end electronics~\cite{bib:minerva_daq} can result in an interaction upstream of the fiducial volume faking an interaction inside of it.
This occurs when a portion of the visible energy deposited inside the fiducial volume by a muon from an upstream interaction is lost
% due to dead time,
resulting in the muon track appearing to originate inside the fiducial volume.
Many of these ``dead time" events are also rejected by the requirement for a second track originating at the event vertex.
%While most of these ``dead time" events are rejected by the requirement for a second track originating at the event vertex, a dead time cut is also applied to reject these events.  
%For the dead time cut, the muon track is projected into the four planes immediately upstream of the event vertex.  
Specifically, candidate events are rejected
%if two or more TriP-t chips reading out the strips 
if strips on two or more planes intersected by the upstream track projection
had dead time.  Dead time is modeled in the MC by overlaying the MC with data and simulating the charge integration period for the
channels in the MC containing charge from the data overlay.

\subsection{Muon Charge Cut}

\begin{figure*}[tpb]
\centering
\mbox{
\includegraphics[width=0.49\linewidth]{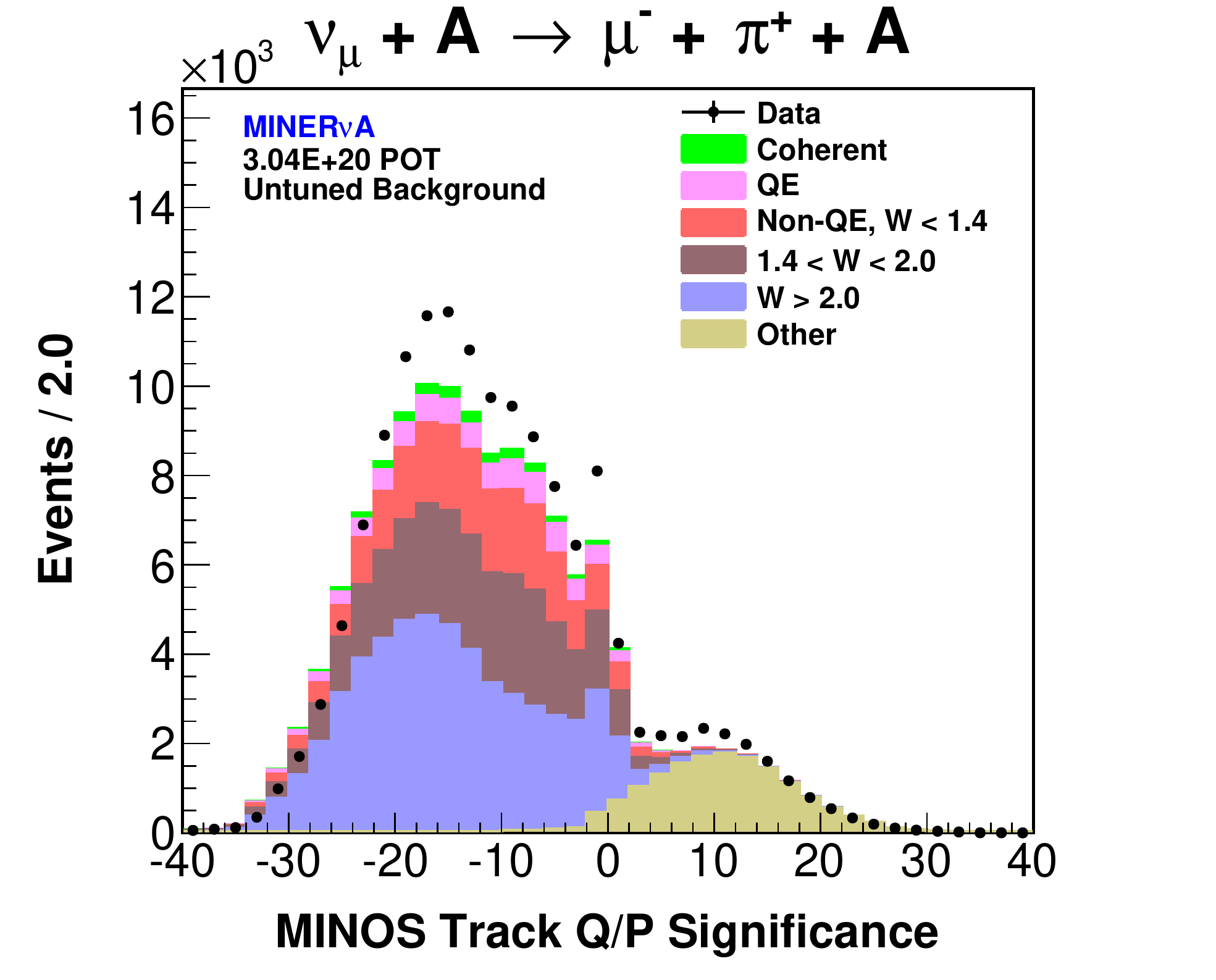}
\includegraphics[width=0.49\linewidth]{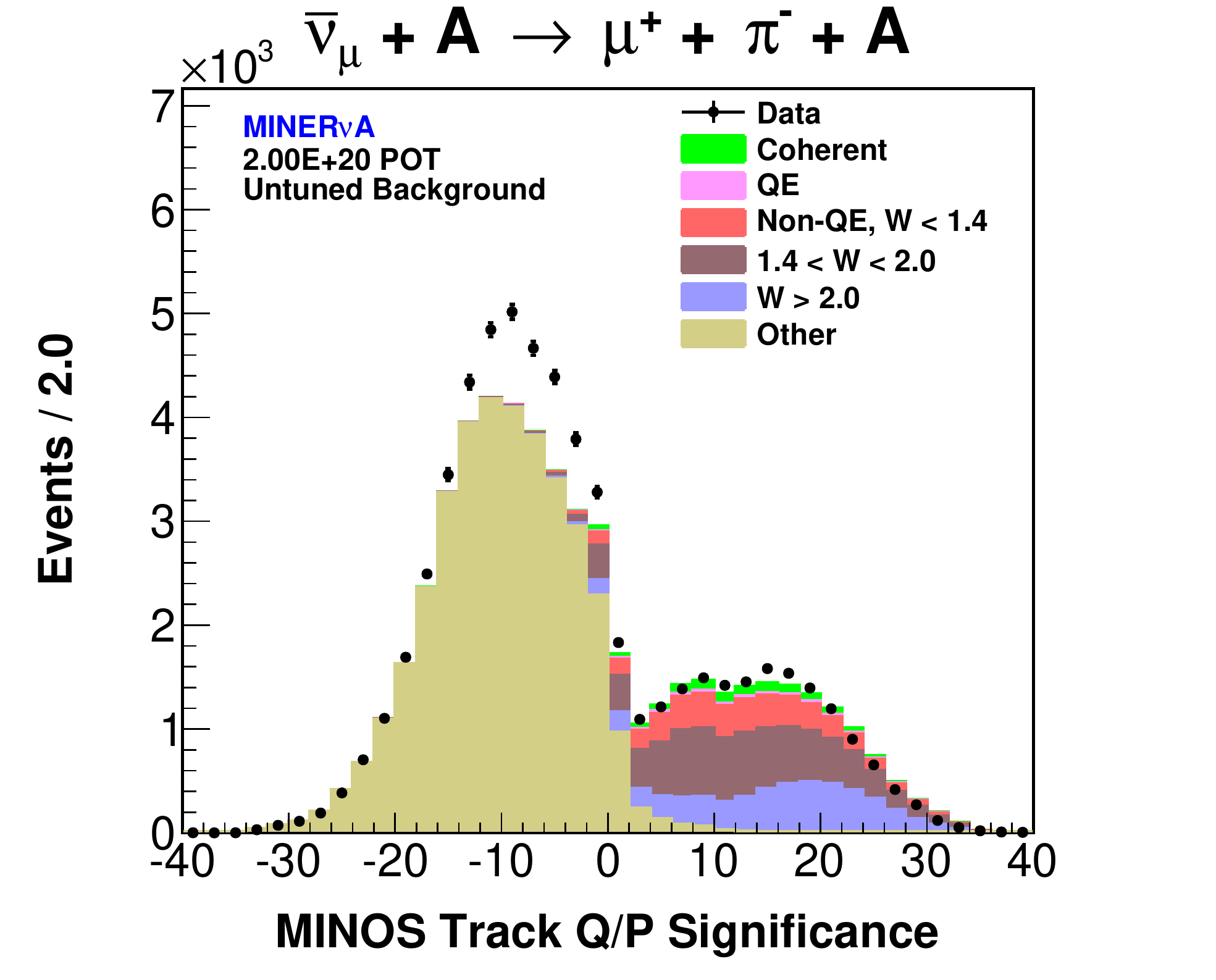}}
\caption{The \numu (left) and \numubar (right) $q/p$ divided by its uncertainty for the matched \minos track for events that pass the reconstruction and fiducial volume cuts.}
\label{fig:qp_sig}
\end{figure*}

The muon charge is used to select either \numu or \numubar CC events.  The muon charge is measured by the quantity $q/p$ extracted from the \minos track fit,
where $q$ and $p$ are the muon charge and momentum.  Selected \numu (\numubar) events have $q/p>0$ ($q/p<0$).
%$q/p$ is proportional to the reciprocal of the radius of curvature of the muon track in the magnetic field.  Selected \numu (\numubar) events have $q/p>0$ ($q/p<0$).  
Fig.~\ref{fig:qp_sig} shows $q/p$ divided by the uncertainty on $q/p$ from the fit for events in the \numu and \numubar samples that pass the
reconstruction and fiducial volume cuts of Sec.~\ref{sec:reco_and_fiducial}.
%These distributions demonstrate that the background category ``Other" is composed primarily of wrong-sign events.
Prior to the $q/p$ cut, the \numubar sample contains more wrong-sign (\numu) events than \numubar events.  This is because \numu
composes 15\% of the flux in the \numubar beam configuration and CC \numu interactions produce a tracked hadron, either
a proton or a pion, more often than CC \numubar interactions.

\subsection{Neutrino Energy Cut}

\begin{figure*}[tpb]
\centering
\mbox{\includegraphics[width=0.49\linewidth]{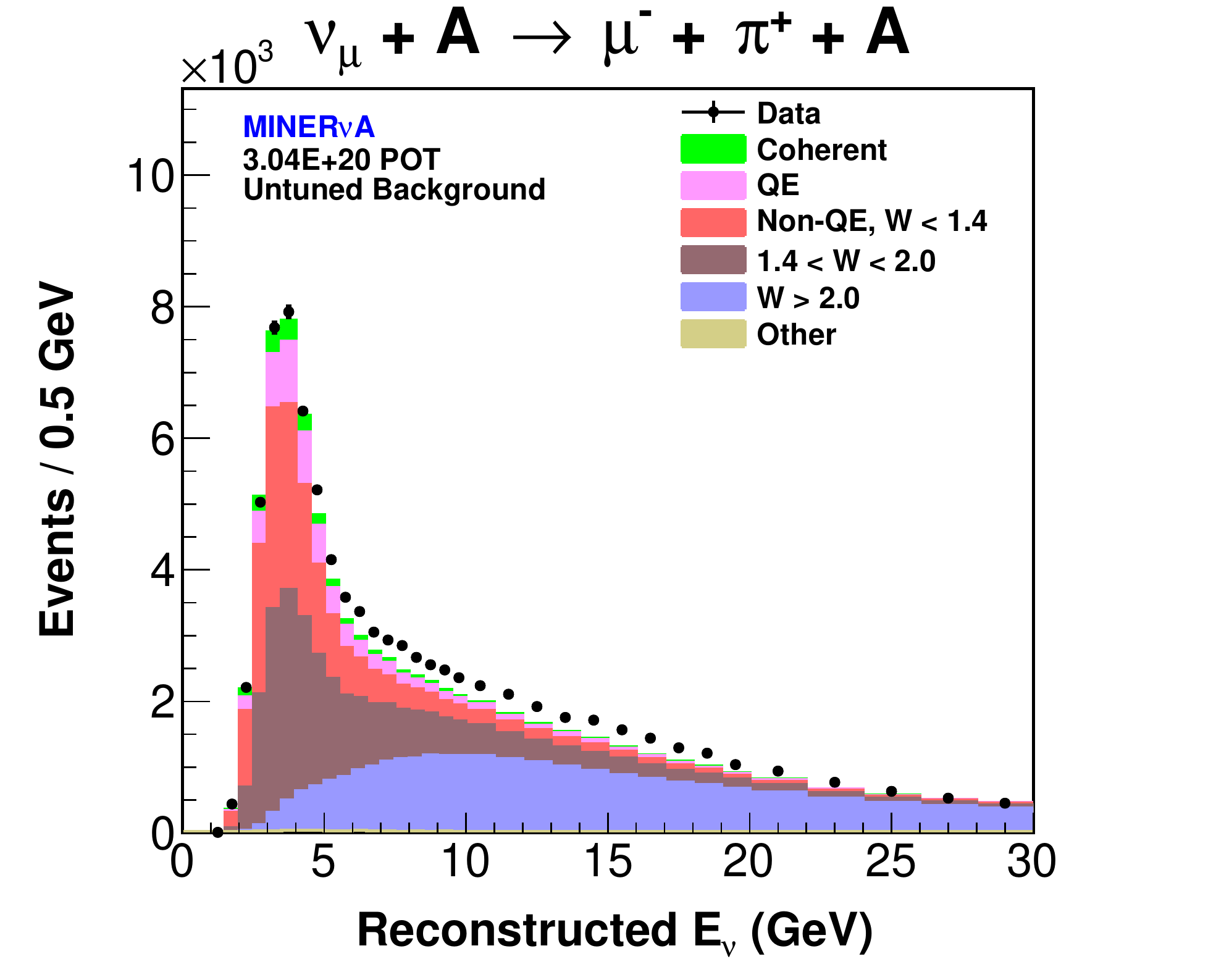}
\includegraphics[width=0.49\linewidth]{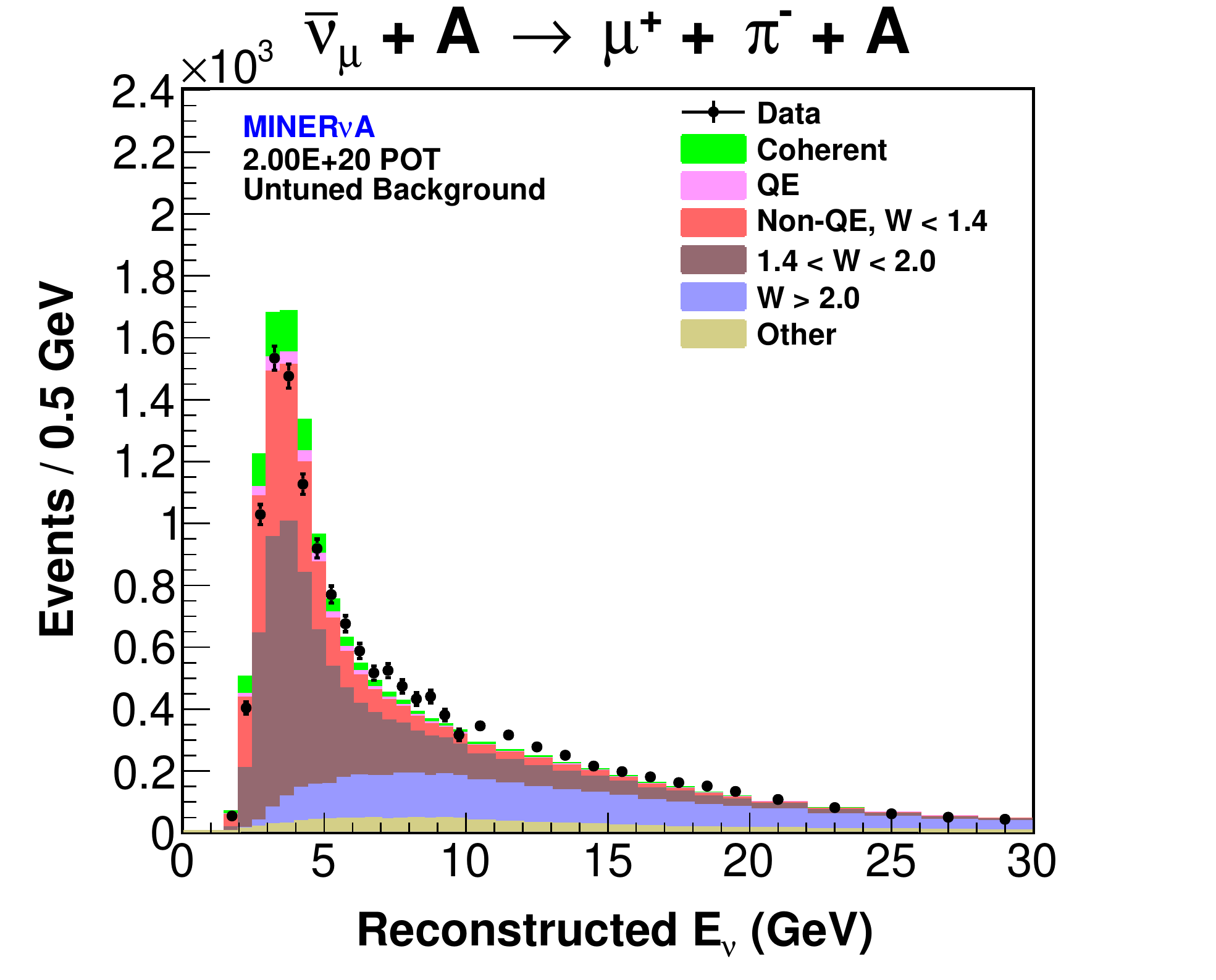}}
\caption{The \enu distribution for events in the \numu (left) and \numubar (right) sample that pass all selection cuts up through the matched \minos track $q/p$ cut.  Events with \evrange are selected.}
\label{fig:enu_cut}
\end{figure*}

The neutrino energy of the \numu and \numubar samples (Fig.~\ref{fig:enu_cut}) is restricted to \evrange.
%The \enu\gt\unit[2.0]{GeV}
%%1.5 GeV
%requirement therefore excludes mis-reconstructed muons.
The \enu\lt\unit[20]{GeV} requirement primarily excludes neutrinos resulting from kaon production at the
NuMI target which are not well constrained.  Note that muons that originate in
the tracker and are tracked in \minos have \emu\gt1.5 GeV.
%The \enu cut also provides a well-defined neutrino energy range for comparing the measured coherent crosssections to model predictions.

\subsection{Proton Identification (Score) Cut}
\label{sec:proton_score_cut}

\newcommand{\edepm}{\ensuremath{\Delta E^{meas}_{i}}\xspace}
\newcommand{\edepp}{\ensuremath{\Delta E^{pred}_{i}}\xspace}
\newcommand{\sigedep}{\ensuremath{\sigma^{comb}_{i}}\xspace}

The visible energy along the hadron track is analyzed to reject events with a final state proton.  The likelihood that the hadron track corresponds
to a proton, referred to as the proton score, is calculated by comparing the visible energy in the clusters along the hadron track to the predicted
energy deposition of a stopping proton and pion.
%In predicting the energy deposition, the proton/pion originates (stops) at the begin (end) point of the hadron track.
The predicted energy deposition in each cluster is the product of the average \dedx in the scintillator, calculated by the Bethe-Bloch equation, and
the path length of the track in the scintillator.  In calculating \dedx, the pion/proton momentum at each cluster is estimated by range along the
hadron track.  The reduced \chisq for comparing the visible energy to the predicted pion/proton energy deposition in the clusters along the hadron
track is
\begin{equation}
\label{eq:proton_score_chi2}
\chisq = \frac{1}{n}\displaystyle\sum_{i=1}^{n}\frac{\edepm-\edepp}{\sigedep},
\end{equation}
where \edepm is the measured deposited energy (\ie visible energy), \edepp is the predicted energy deposition, and \sigedep is the
combined uncertainty on \edepm and \edepp in the $i^{th}$ cluster along the hadron track with $n$ total clusters.
The combined uncertainty \sigedep consists of range fluctuations on the calculated \dedx, photo-statistical uncertainty on the
measured and predicted deposited energy, and uncertainty on the path length of the track in the scintillator~\cite{bib:walton_thesis}.
The proton score is calculated as
\begin{equation}
\label{eq:proton_score}
\text{Proton Score} = 1.0 - \frac{\chisq_{p}}{\sqrt{(\chisq_{p})^{2}+(\chisq_{\pi})^{2}}},
\end{equation}
where $\chisq_{p}$ and $\chisq_{\pi}$ are the reduced \chisq for the predicted proton and pion energy deposition, respectively.

A
%hadron track representing a
particle that interacts in the detector may have one or more secondary tracks that emerge from its endpoint, representing either the
scattered incident particle or particles produced in the interaction.  For tracks with exactly one secondary track, the proton score is
calculated from the secondary track, because proton/pion discrimination is at the Bragg peak. 
%The proton score is not calculated for hadron tracks with two or more secondary tracks, since it is ambiguous which secondary track
%corresponds to the scattered proton/pion.  The proton score is also not calculated for hadron/secondary tracks associated with
%activity in the outer detector where the proton score is unreliable due to the thick layers of steel.
Events where the proton score was not calculated are selected.  This occurs when there is more than one secondary track, or when the
track exits the tracker. In the latter case, the proton score is unreliable because of the absorber layers in the calorimeters of the
outer detector.

\begin{figure*}[tpb]
\centering
\mbox{
\includegraphics[width=0.49\linewidth]{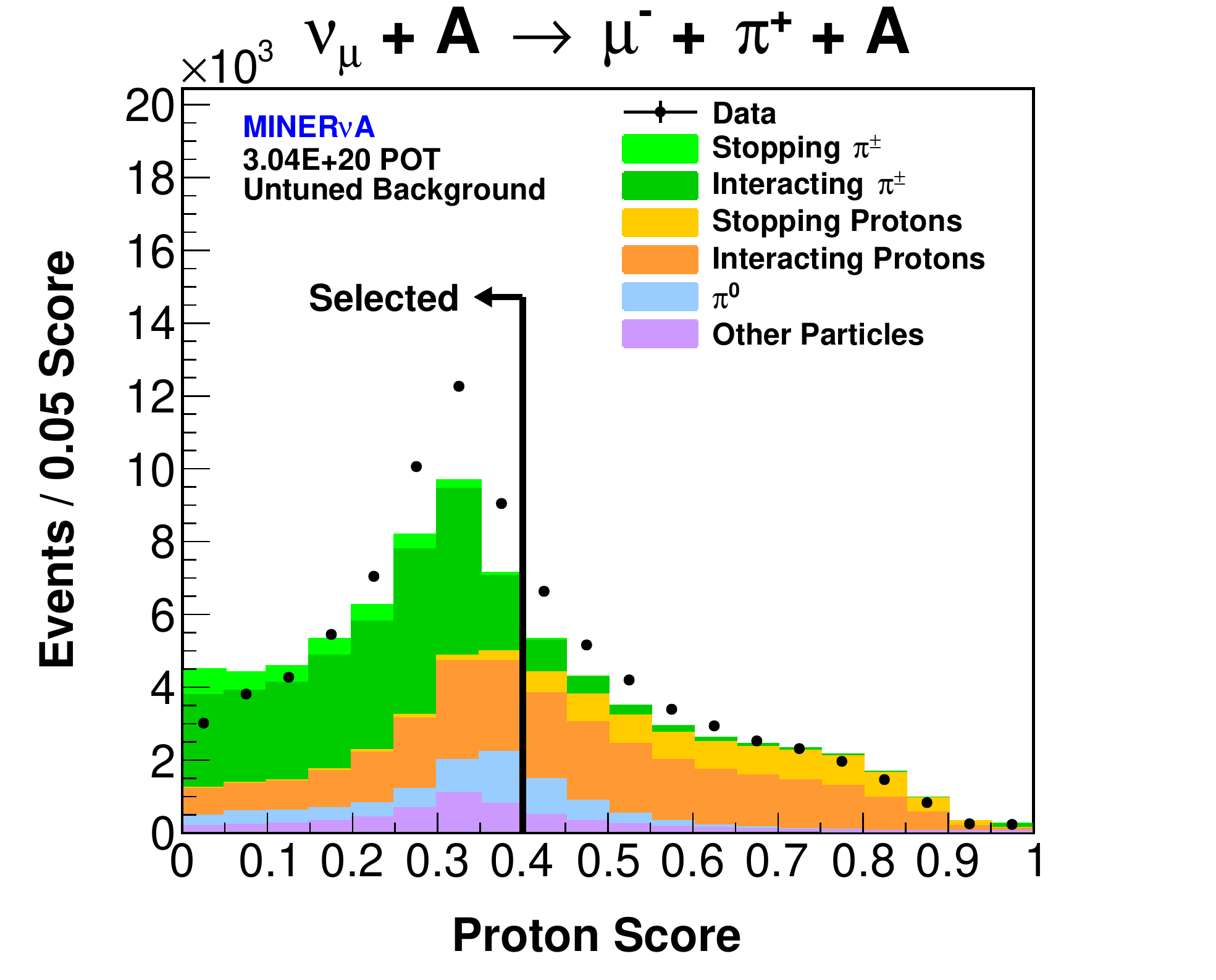}
\includegraphics[width=0.49\linewidth]{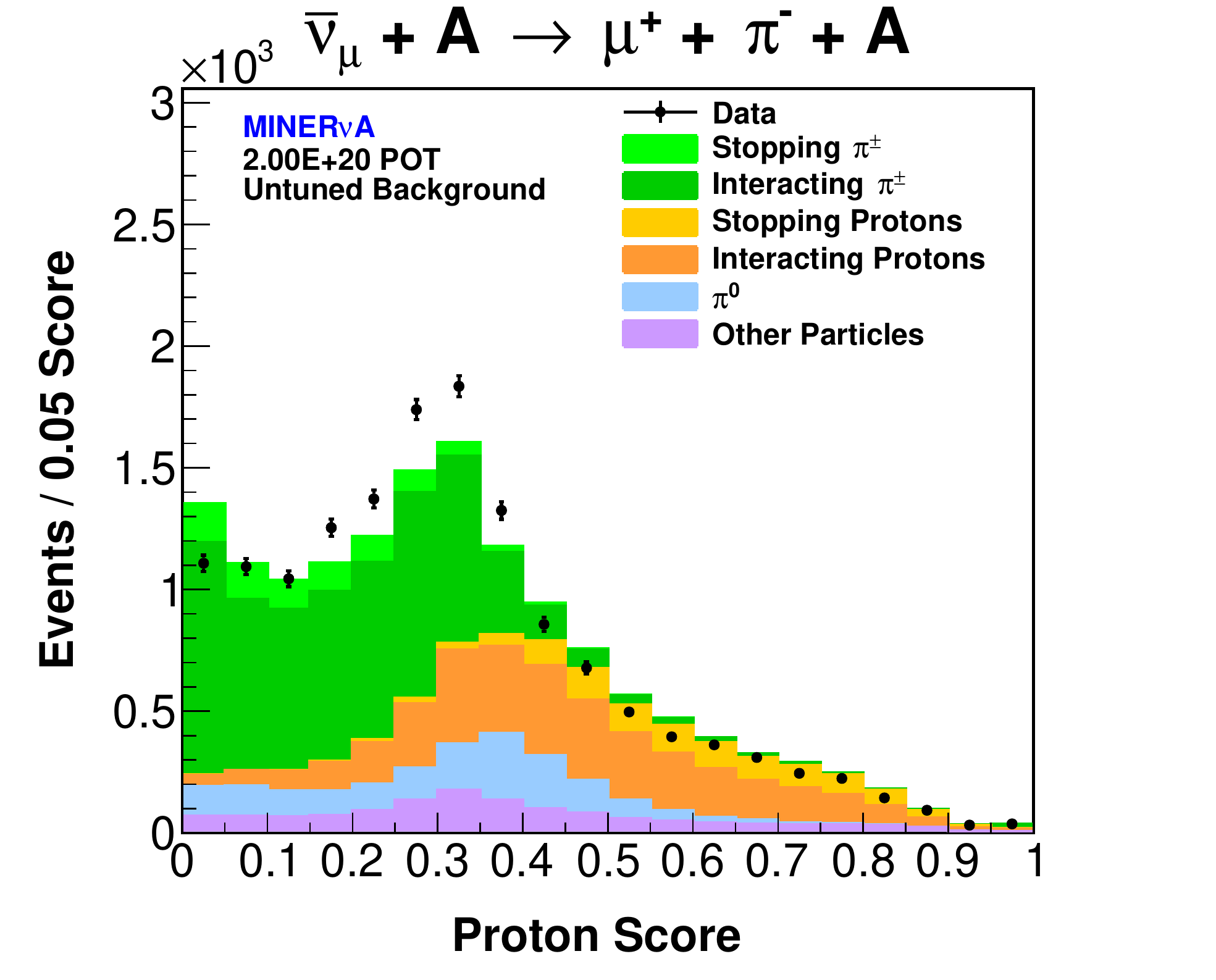}}
\mbox{
\includegraphics[width=0.49\linewidth]{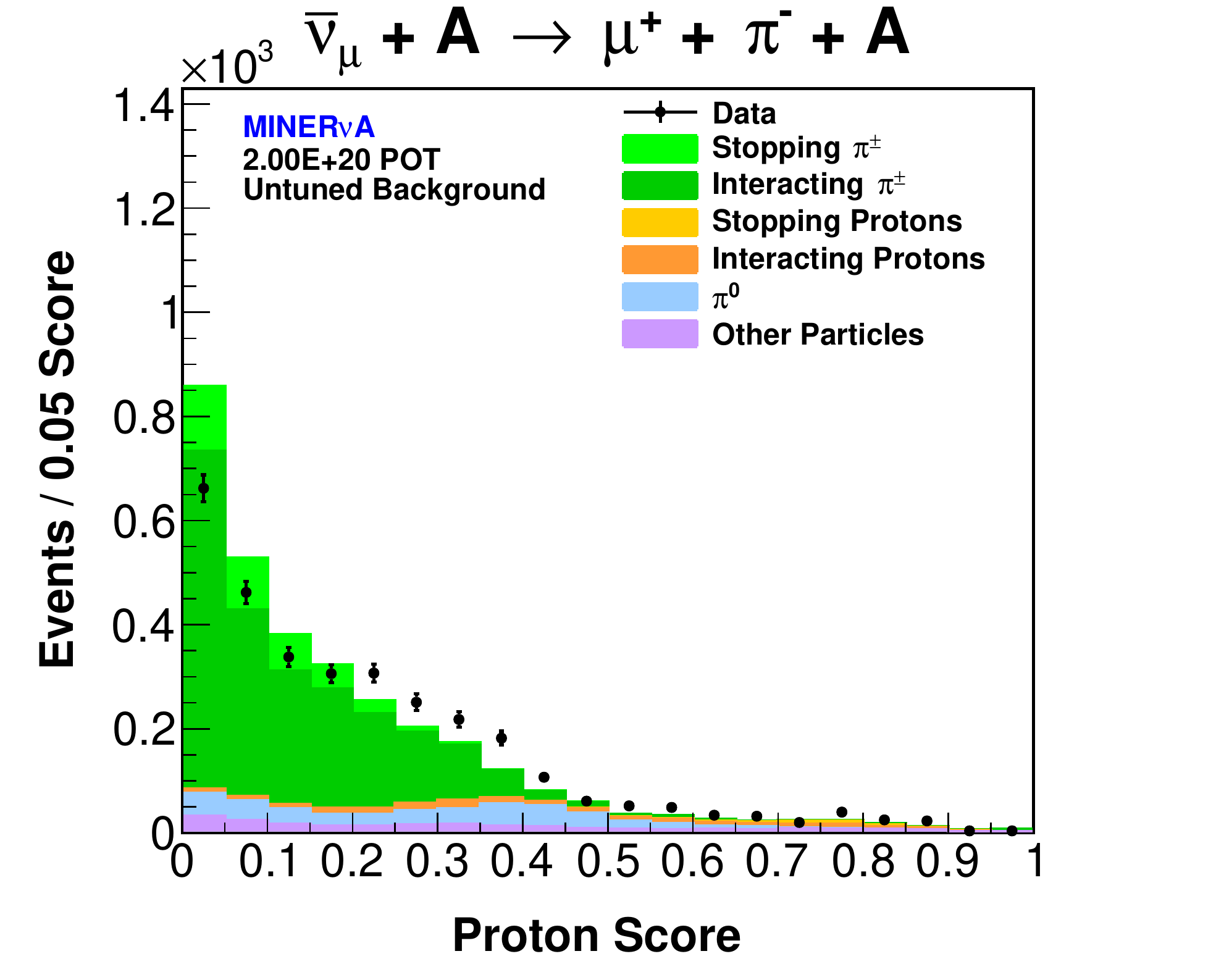}}
\caption{The top left (right) plot shows the proton score distribution for events, broken down by the true identity of the candidate hadron track, in the \numu (\numubar) sample before applying cuts on proton score, on the vertex energy and on \tabs.  The bottom plot shows the same \numubar sample, but after the vertex energy cut.  The vertex energy cut rejects most events in the \numubar sample with a tracked proton.}
\label{fig:numubar_proton_score}
\end{figure*}

Figure~\ref{fig:numubar_proton_score} shows the proton score distribution for events in the \numu and \numubar samples.
% that pass all cuts up through the \enu cut, where the MC is categorized by the final state particle represented by the hadron track.  Pions and protons are categorized by whether they stopped or interacted in the detector.  
The category ``Other Particles" consists primarily of charged kaons and neutrons that interacted near the event vertex.  The disagreement
between the data and MC in the proton score distribution is attributed to the MC prediction of neutrino pion production, which is subsequently
tuned to data (see Sec.~\ref{sec:bg_tuning}).
% (Section~\ref{sec:bg_tuning}).  
The peak at 0.3 is due to clusters along the hadron track containing energy deposition from multiple particles. 
%This is demonstrated by the peak in DIS events (W \gt 2 GeV) at 0.3 proton score (Figure~\ref{fig:proton_score}).  DIS interactions tend to produce multiple final state hadrons which lends to overlapping activity.
Events in the \numu sample with proton score \lt 0.4 are selected.  Most events
in the \numubar sample with a tracked proton are rejected by the vertex energy cut, (see Sec.~\ref{sec:evtx_cut}) since these events tend to have additional final state charged
hadrons due to charge conservation.  A proton score cut is not imposed on the \numubar sample to maximize the signal selection efficiency;
in this sample, the quasielastic process is not a background.

%\begin{figure}[tpb]
%\centering
%\mbox{
%\includegraphics[width=0.49\linewidth]{figures/EventSelection/h_ProtonScore_ByPart_cut0_CutArrows_ThetaPiCor_FluxConstrained_BkgdTuned_neutrino.pdf}
%\includegraphics[width=0.49\linewidth]{figures/EventSelection/h_ProtonScore_ByPart_cut0_ThetaPiCor_FluxConstrained_BkgdTuned_antineutrino.pdf}}
%\mbox{
%\includegraphics[width=0.49\linewidth]{figures/EventSelection/h_ProtonScore_cut0_CutArrows_ThetaPiCor_FluxConstrained_BkgdTuned_neutrino.pdf}
%\includegraphics[width=0.49\linewidth]{figures/EventSelection/h_ProtonScore_cut0_ThetaPiCor_FluxConstrained_BkgdTuned_antineutrino.pdf}}
%\caption{The proton score distribution for events in the \numu (left) and \numubar (right) sample that pass all selection cuts up through the \enu cut.  In the top plots, the MC is categorized by the final state particle represented by the hadron track.  In the bottom plots, the MC is categorized by the neutrino interaction type.  Events in the \numu sample are required to have proton score \lt 0.4.}
%\label{fig:proton_score}
%\end{figure}

\subsection{Vertex Energy Cut}
\label{sec:evtx_cut}

\begin{figure*}[tpb]
\centering
\mbox{
\includegraphics[width=0.49\linewidth]{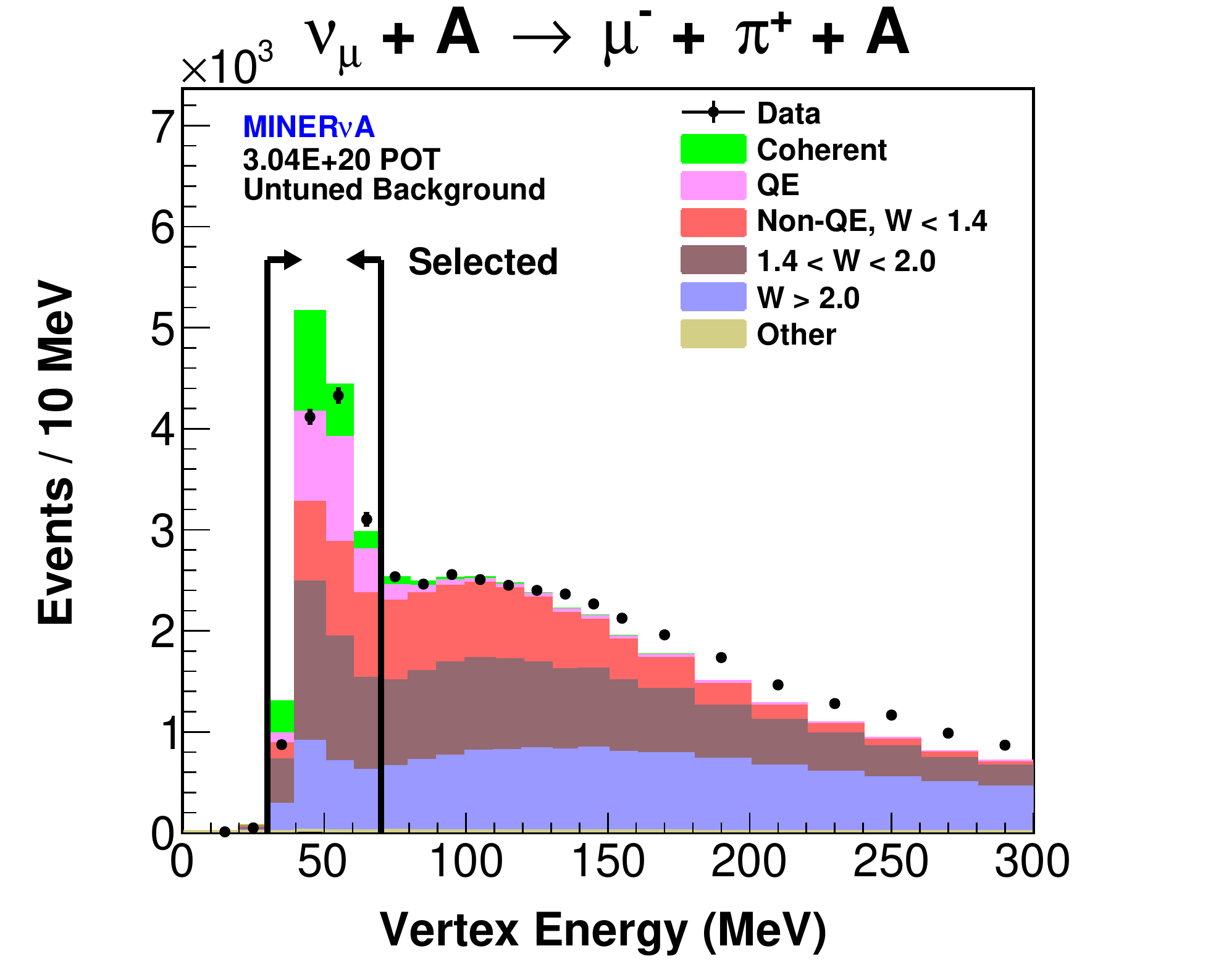}
\includegraphics[width=0.49\linewidth]{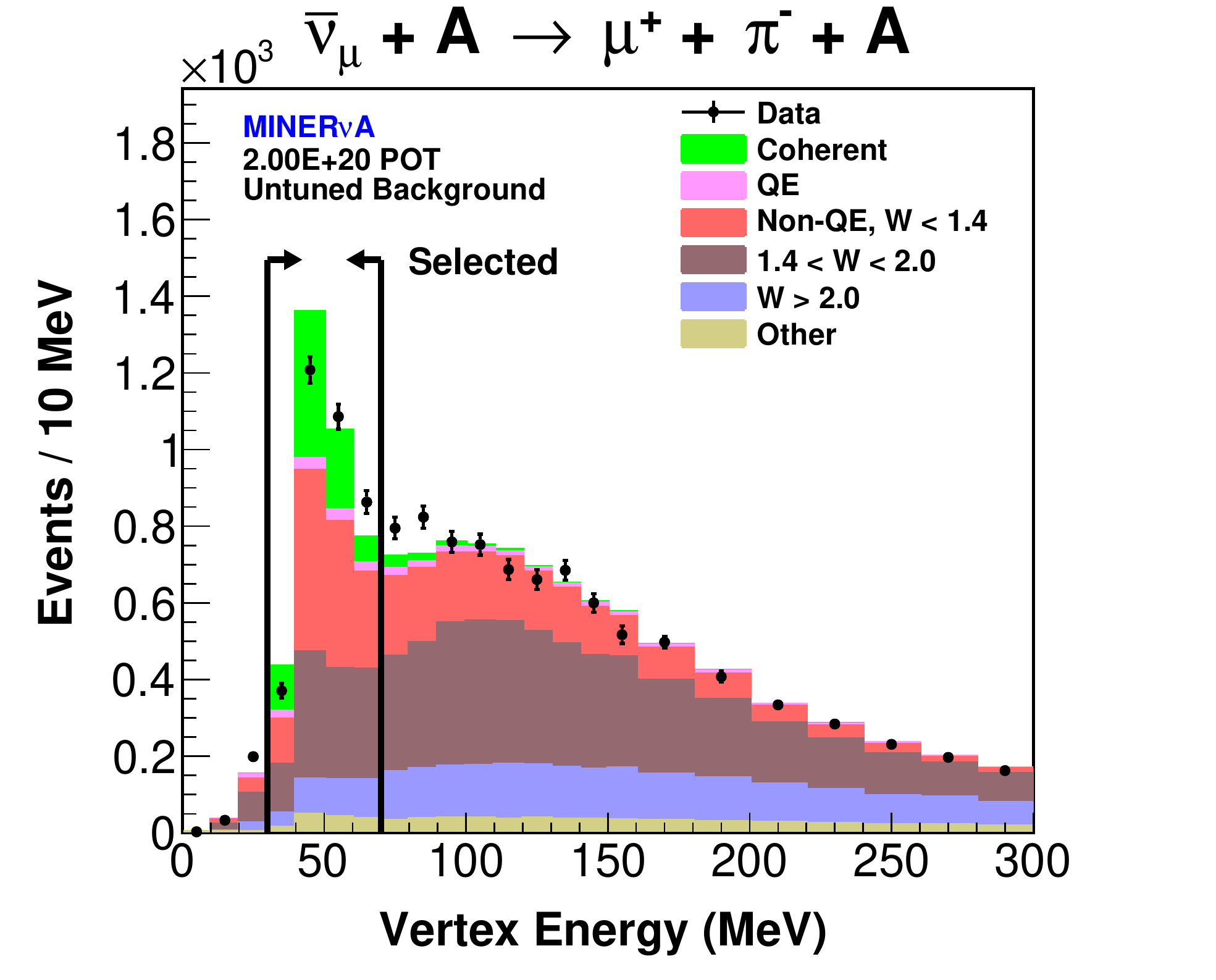}}
\caption{The left (right) plot shows the \evtx distribution for events in the \numu (\numubar) sample that pass all selection cuts up through the proton score (\enu) cut.  Events in the range 30 \lt \evtx \lt 70 MeV are selected.}
\label{fig:vertex_energy}
\end{figure*}

Coherent scattering produces a muon and charged pion in the forward direction while leaving the nucleus intact.
Energy near the interaction vertex of each event is required to be consistent with the energy deposited by only a minimum ionizing muon and pion.  
%Additional energy indicates break up of the nucleus resulting from an incoherent interaction. 
Vertex energy \evtx is defined as the sum of the energies of clusters on the two vertex tracks within $\pm$5 planes ($\sim$\unit[110]{mm}
in the longitudinal direction) from the event vertex, and clusters not on the two vertex tracks within $\pm$5 planes and \unit[200]{mm} in
the transverse direction from the event vertex.  In calculating \evtx, the cluster energies are corrected for passive material and the
on-track cluster energies are corrected for the angle between their respective track and the perpendicular to the scintillator planes.  
This correction to normal incidence minimizes the dependence of the \evtx cut on the muon and pion kinematics.  
%The $\pm$5 plane longitudinal range was optimized to sample enough planes to smooth fluctuations in single plane energy deposition by the muon and pion while minimizing the amount of sampled material traversed by the pion, which in turn minimizes the loss in signal acceptance due to pion interactions which increase \evtx.  
The $\pm$\unit[200]{mm} range in the transverse direction extends to the edge of the scintillator plane from the edge of the \unit[850]{mm}
fiducial volume apothem.  Figure~\ref{fig:vertex_energy} shows the \evtx distribution for events in the \numu (\numubar) sample that pass all selection cuts up through the proton score (\enu) cut.  Events with 30 \lt \evtx \lt 70 MeV are selected.  Requiring \evtx \gt 30 MeV rejects events with a tracked $\gamma$ from the decay of a final state $\pi^{0}$, where the energy deposited by the $\gamma$ via $\gamma\to e^{+}e^{-}$ is separated from the event vertex.

\subsection{\tabs Cut}
\label{sec:t_cut}

\begin{figure*}[tpb]
\centering
\mbox{
\includegraphics[width=0.49\linewidth]{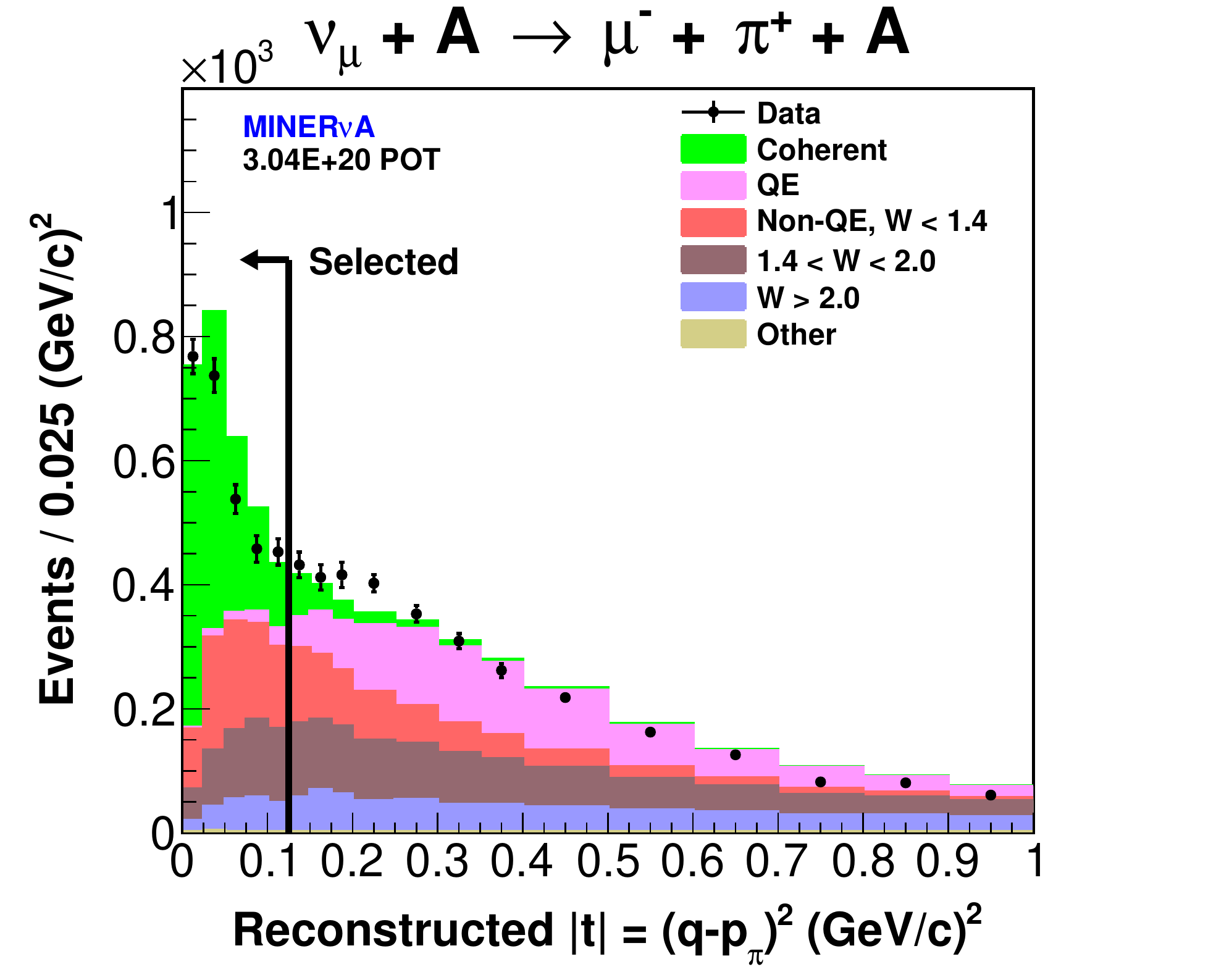}
\includegraphics[width=0.49\linewidth]{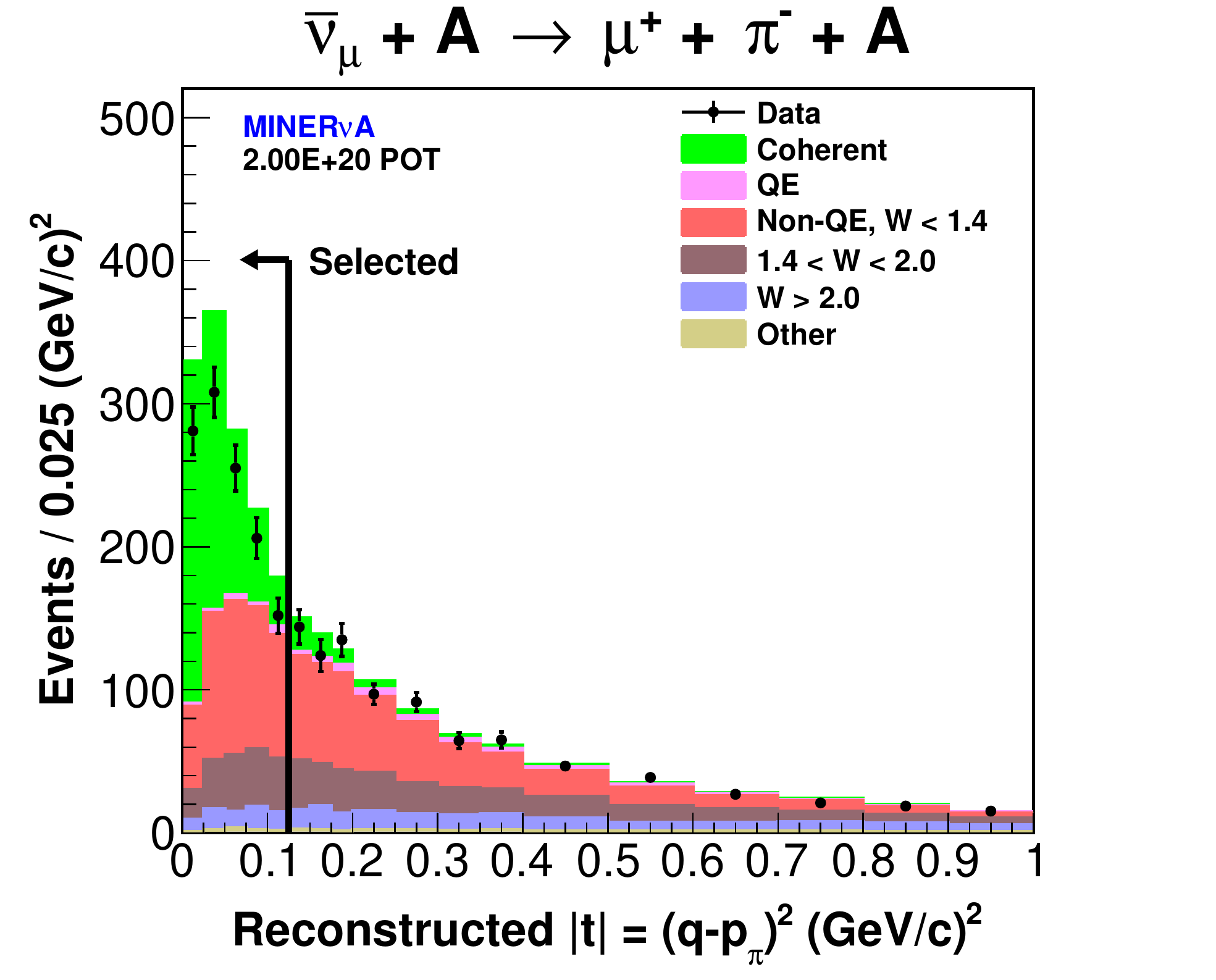}}
\caption{The \numu (left) and \numubar (right) \tabs distributions, after background tuning, for events passing all cuts up through the \evtx cut.  Events with \tabs\lt0.125 \gevpercsq are selected.}
\label{fig:t_tuned_arrow}
\end{figure*}

The final cut in the event selection is on \tabs, which is necessarily small for coherent scattering; large \tabs is
indicative of nuclear break-up in incoherent interactions.  \tabs is calculated from the reconstructed four-momenta of
the neutrino, muon and pion.  The \numu and \numubar \tabs distributions with the tuned background prediction are shown
in Fig.~\ref{fig:t_tuned_arrow} for events passing all cuts up through the \evtx cut.  Events with
\tabs\lt0.125 \gevpercsq are selected.

%The coherent selection efficiency of the \tabs cut is dependent on the \tabs dependence of the coherent cross section, the reconstructed \tabs resolution, and the selected \tabs range.  The coherent selection efficiency is an input to the measurement of the coherent cross sections and is estimated using coherent events in the MC (Section~\ref{sec:efficiency}).  
A cut on \tabs could introduce a model dependence into the measured cross sections because of the \tabs dependence of the coherent
cross section model in the MC.  In both the Rein-Sehgal and Berger-Sehgal models, the \tabs dependence arises from
the pion-nucleus elastic scattering cross section, which falls as $\exp(-b\tabs)$ where $b$ is a free parameter.
GENIE calculates the coherent scattering cross section on carbon using the Rein-Sehgal coherent model with $b\sim40$ (GeV/c)$^{-2}$.   In GENIE, 99\% of the coherent events on carbon have true \tabs\lt0.125 \gevpercsq.
The pion-carbon elastic scattering cross section in the Berger-Sehgal model was fit to pion-carbon elastic scattering data for
incident pion kinetic energies $\lesssim1$ GeV, giving $b\gtrsim60$ (GeV/c)$^{-2}$~\cite{bib:berger_sehgal}.  In addition,
an exponential slope $b\sim60$ (GeV/c)$^{-2}$ was measured from data of \piplus and \piminus elastic scattering on carbon at
$\sim2$ GeV/c incident pion momentum~\cite{bib:pion_carbon_slope}.  The pion-carbon elastic scattering data suggest the cross
section for coherent scattering on carbon in the MC should fall faster in \tabs, which would result in $>$99\% of MC coherent
events on carbon having true \tabs\lt0.125 \gevpercsq.  To select low \tabs events, this analysis requires \tabs\lt0.125 \gevpercsq.
The high value of the \tabs cut makes its efficiency independent of variations in the \tabs distribution in different models.
%and is dependent on the reconstructed \tabs resolution only.

\section{Background Tuning}
\label{sec:bg_tuning}

\subsection{Sideband Scale Factors}

\begin{figure*}[tpb]
\centering
\mbox{\includegraphics[width=0.49\linewidth]{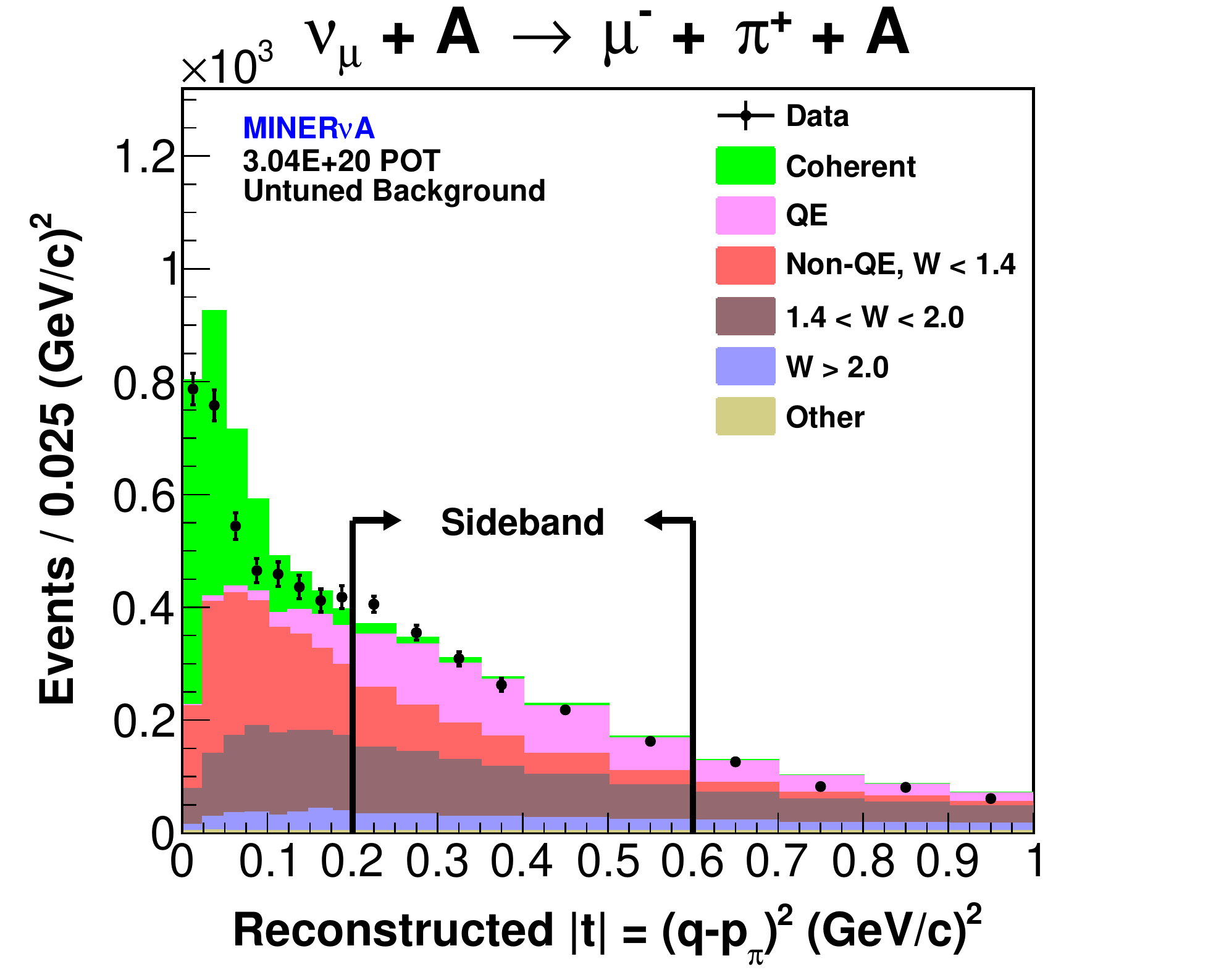}
\includegraphics[width=0.49\linewidth]{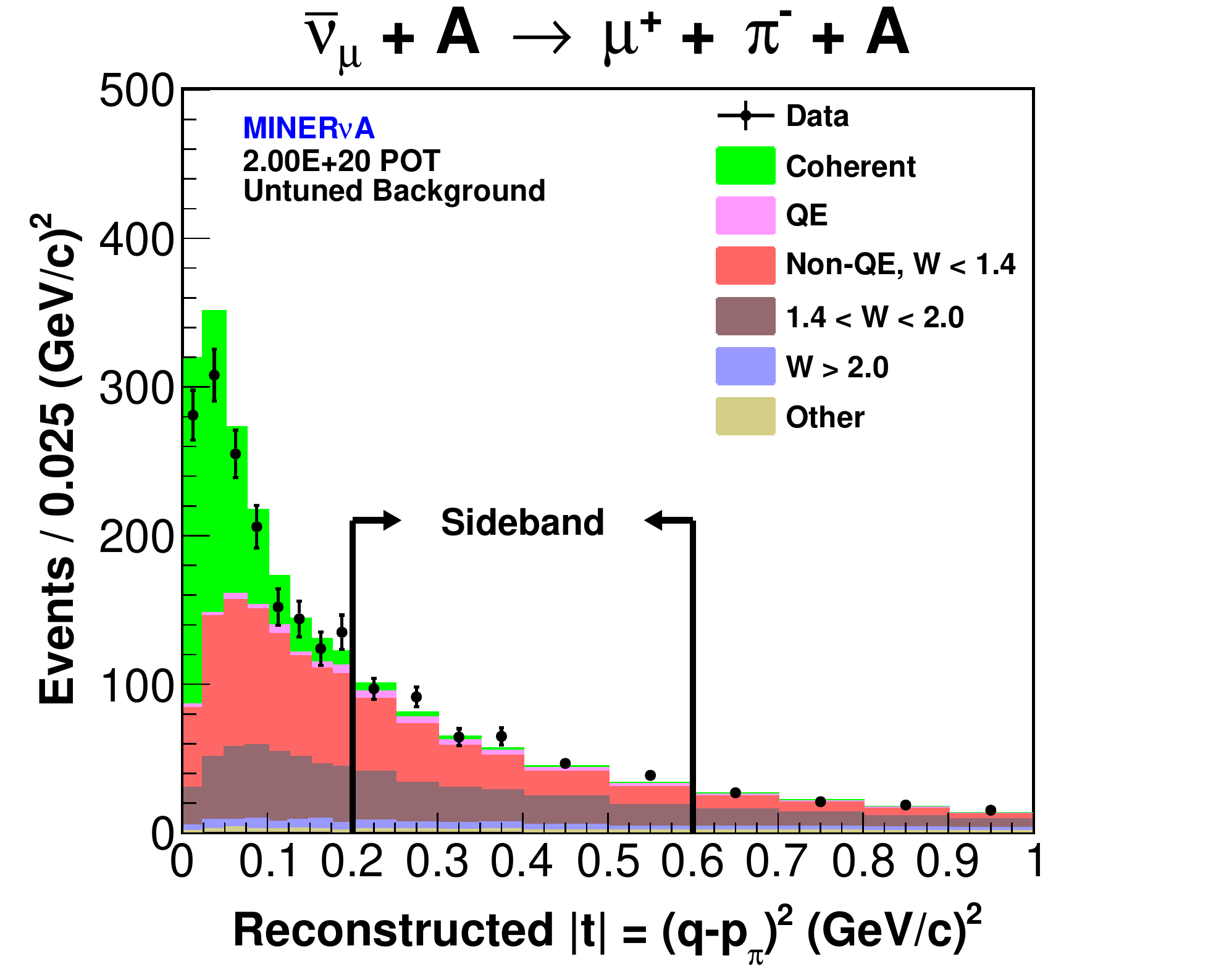}}
\caption{The \numu (left) and \numubar (right) \tabs distributions, before background tuning, for events that pass all selection cuts up through the vertex energy and proton score cuts.  The sideband used for the background tuning are events with 0.2 \lt \tabs \lt 0.6 \gevpercsq.}
\label{fig:sideband_region}
\end{figure*}

The \numu and \numubar coherent candidate samples contain significant backgrounds.
%that must be subtracted from the data in order to measure the coherent cross sections.  
%The MC is used to estimate the rate of the background coherent candidates, which can have large uncertainties due to uncertainties on the flux and underlying background interaction models. 
Uncertainties from the MC estimate are minimized by tuning the estimated background rates to data in a sideband.
%After applying all selection cuts, incoherent backgrounds remain in the selected coherent-like sample.  In order to measure the coherent cross sections, the rates of the remaining backgrounds must be estimated using the MC and subtracted from the data.  The MC prediction of the rates of the remaining backgrounds have large uncertainties due to uncertainties on the flux and underlying background interaction models.  These uncertainties are minimized by constraining the rates of the remaining backgrounds predicted by the MC to data in a sideband.
The sideband is defined as events with 0.2 \lt \tabs \lt 0.6 \gevpercsq that pass all selection cuts up through the vertex
energy cut (Fig.~\ref{fig:sideband_region}).  
The requirement that events in the sideband pass the vertex energy cut minimizes sensitivity of the background tuning to mismodeled vertex activity (Sec.~\ref{sec:sys_vertex_energy}).  This mismodeling could result in disagreement between data and the MC in the background acceptance of the vertex energy cut, and performing the background tuning after imposing the vertex energy cut will correct this disagreement.

\begin{figure*}[tpb]
\centering
\mbox{\includegraphics[width=0.49\linewidth]{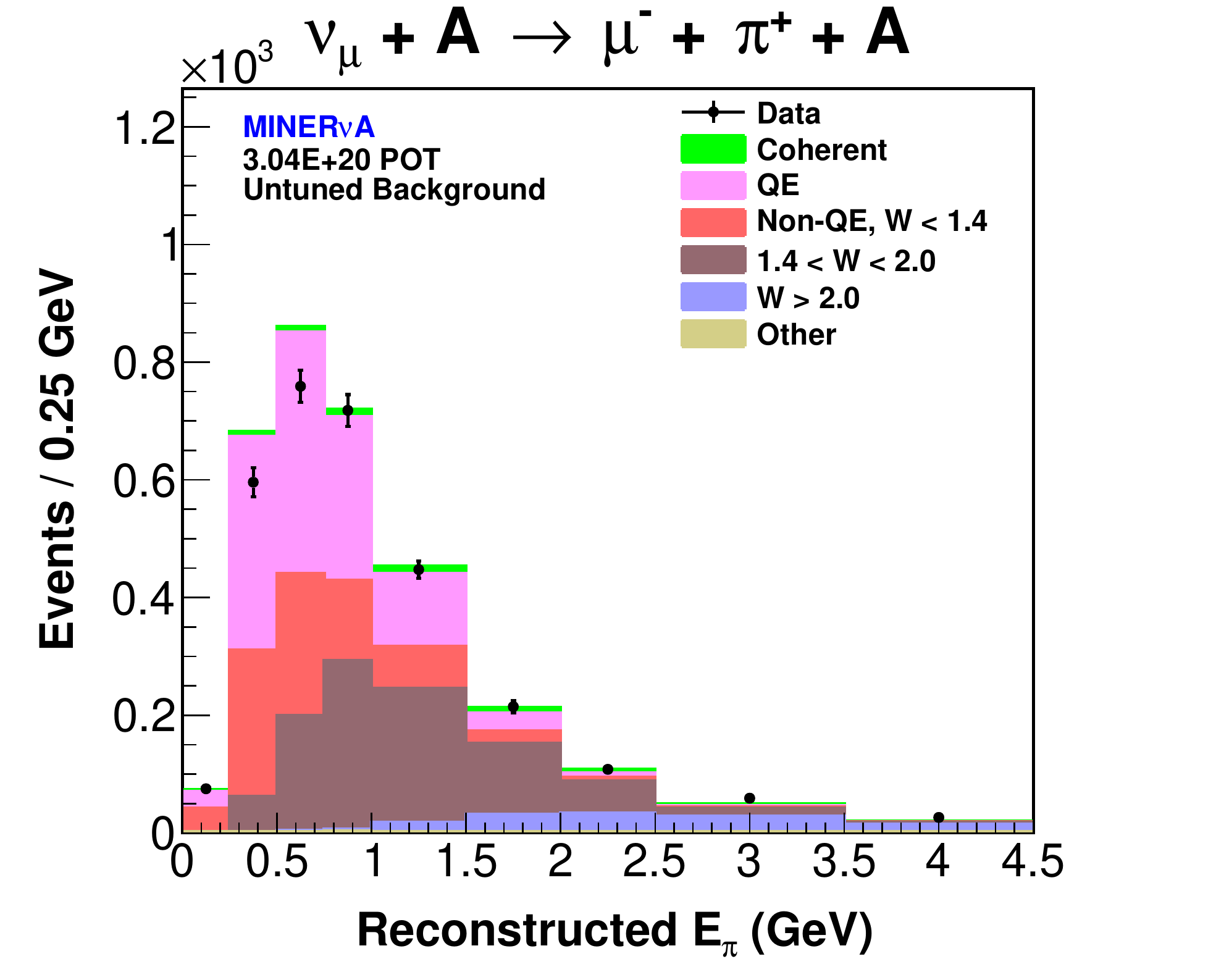}
\includegraphics[width=0.49\linewidth]{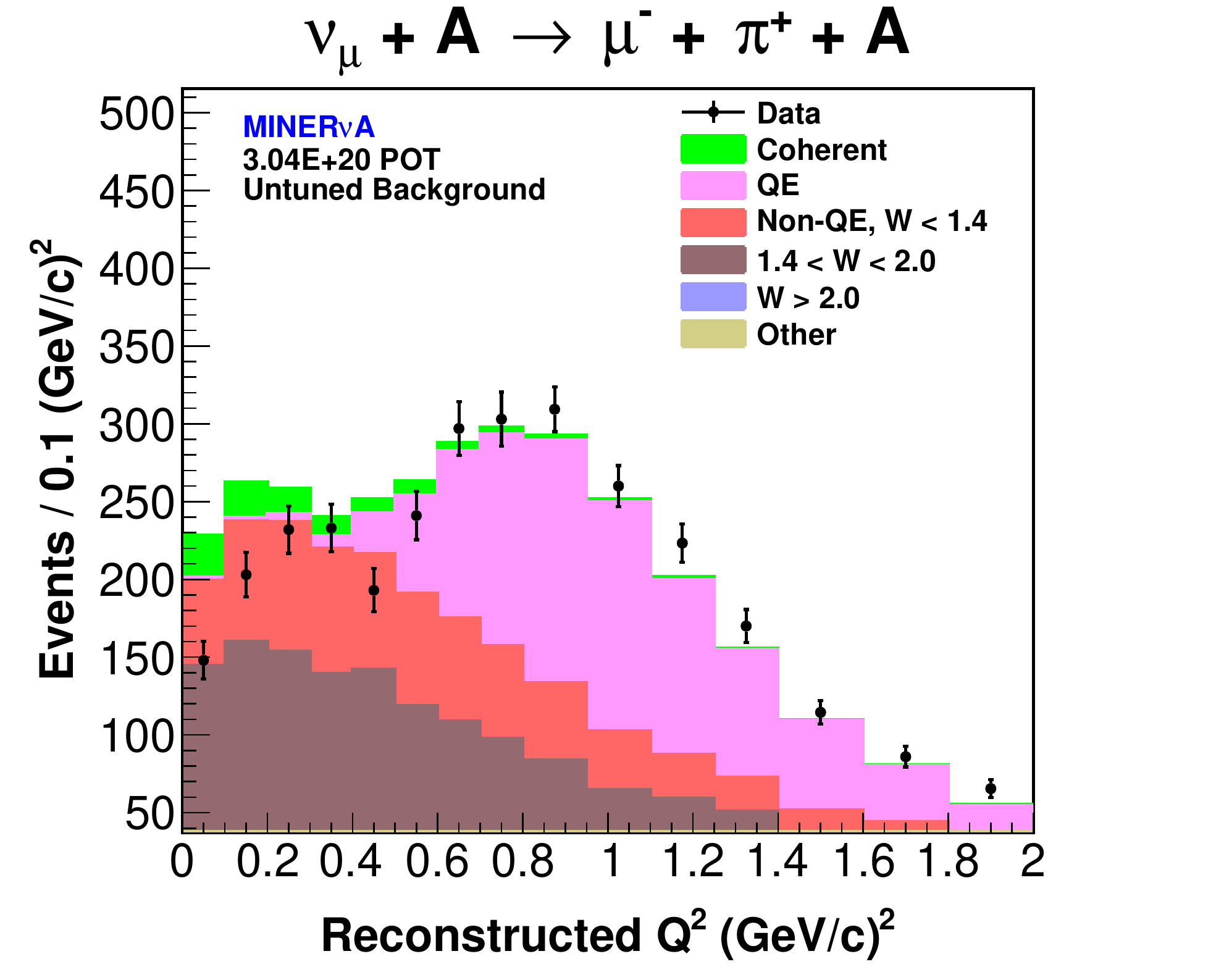}}
\mbox{\includegraphics[width=0.49\linewidth]{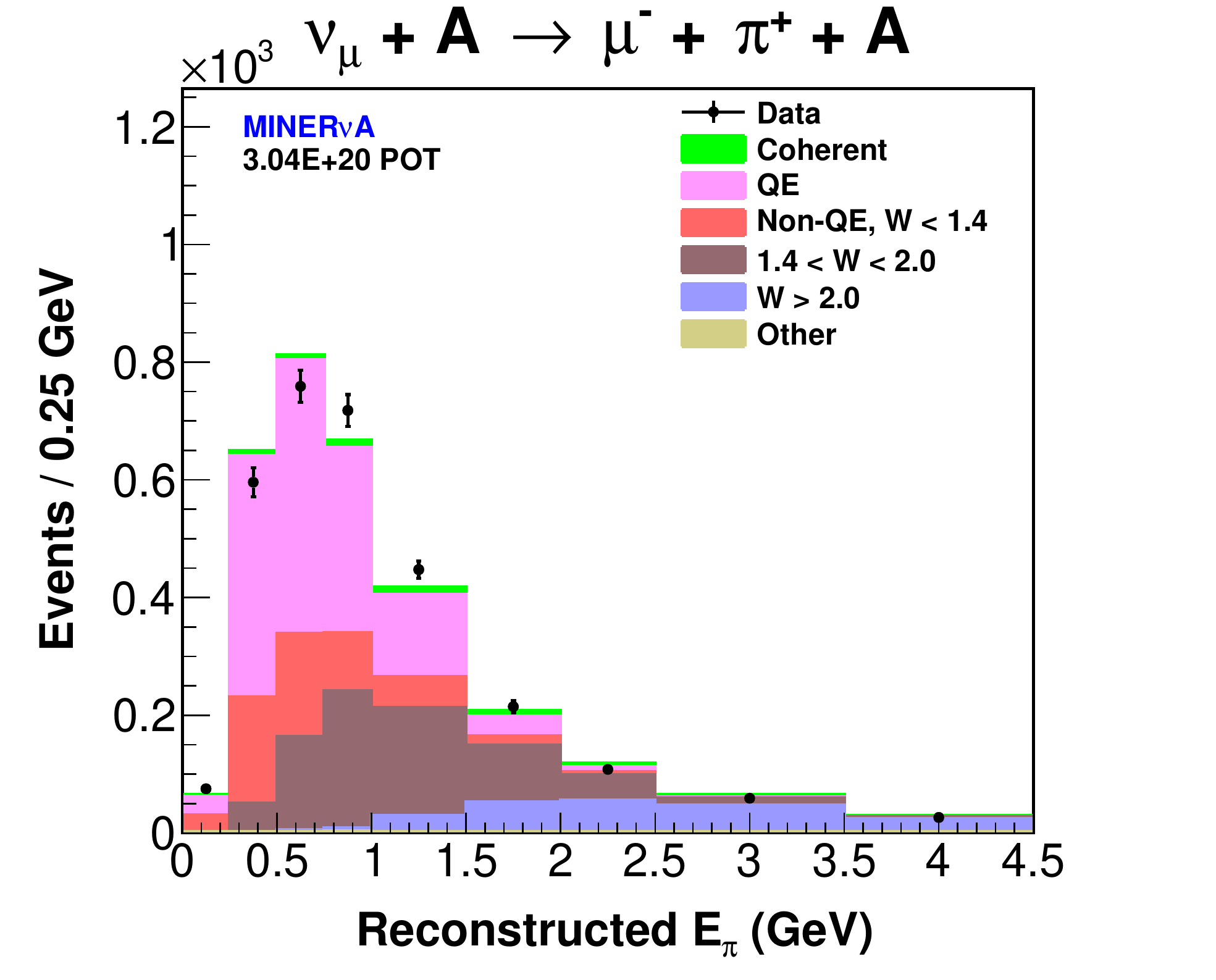}
\includegraphics[width=0.49\linewidth]{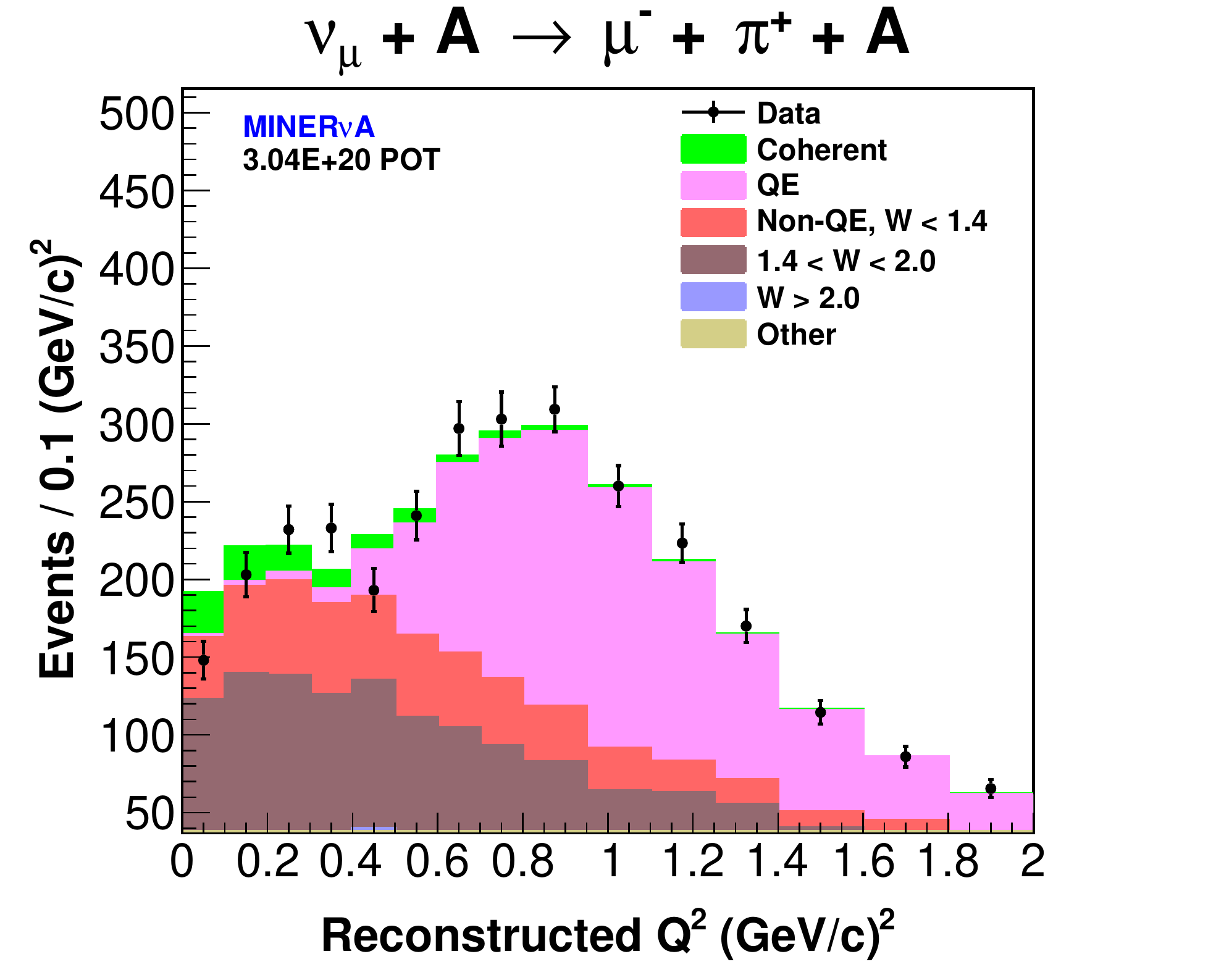}}
\caption{The \numu sideband \epi (left) and \qsq (right) distributions before (top) and after (bottom) background tuning}
\label{fig:bg_tuning_numu}
\end{figure*}

The background tuning extracts a correction to the normalization of each background.  These 
background scale factors are determined by varying the normalizations of the backgrounds in a fit of the total MC to
data in the sideband.  
%Choosing a sideband distribution for the background tuning that gives separation of the backgrounds facilitates the fit and reduces uncertainties on the background scale factors.  
The \numu sideband \epi distribution (Fig.~\ref{fig:bg_tuning_numu}) provides separation of the resonance, transition, and
DIS processes, while the \numu sideband \qsq distribution (Fig.~\ref{fig:bg_tuning_numu}) separates resonance and QE
processes.  The \numu background tuning therefore is performed with a fit to bins of \qsq and \epi.  
%The \chisq minimized in fitting the \numu background scale factors is calculated as
%\begin{equation}
%\label{eq:bg_tuning_chi2_numu}
%\chisq = \sum_{i}\sum_{j} \frac{(N^{Data}_{ij}-\sum_{k}\alpha_{k}N^{MC}_{ijk})^{2}}{\sum_{k}\alpha_{k}N^{MC}_{ijk}},
%\end{equation}
%where $i$ and $j$ are \epi and \qsq bins, respectively, $k$ is a MC event category, $\alpha_{k}$ is the scale factor for the normalization of category $k$, $N^{MC}_{ijk}$ is the number of MC events from category $k$ in \qsq vs. \epi bin $ij$, and $N^{Data}_{ij}$ is the number of data events in bin $ij$.  
Coherent scattering is a small contribution to the \numu sideband and its scale factor is fixed to 1.0 in the fit.
Other backgrounds are also a small contribution to the \numu sample and their scale factor is also fixed to 1.0.
%to minimize the number of free parameters in the fit.
The \numu background scale factors extracted from the fit are listed in Table~\ref{tab:bg_scale_factors}.
The \numu sideband \epi and \qsq distributions after applying the background scale factors to the MC are shown in Fig.~\ref{fig:bg_tuning_numu}.

\begin{figure*}[tpb]
\centering
\mbox{\includegraphics[width=0.49\linewidth]{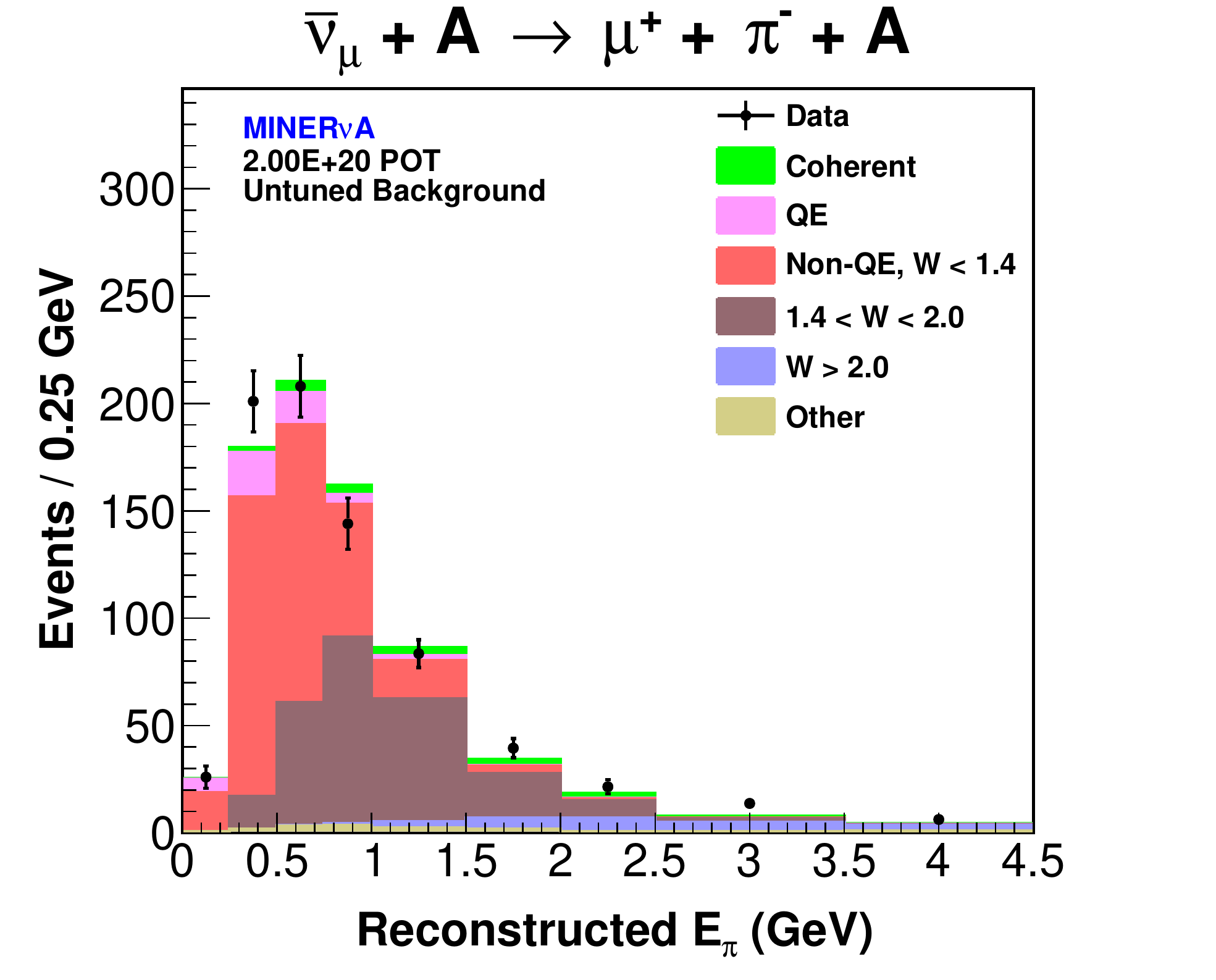}
\includegraphics[width=0.49\linewidth]{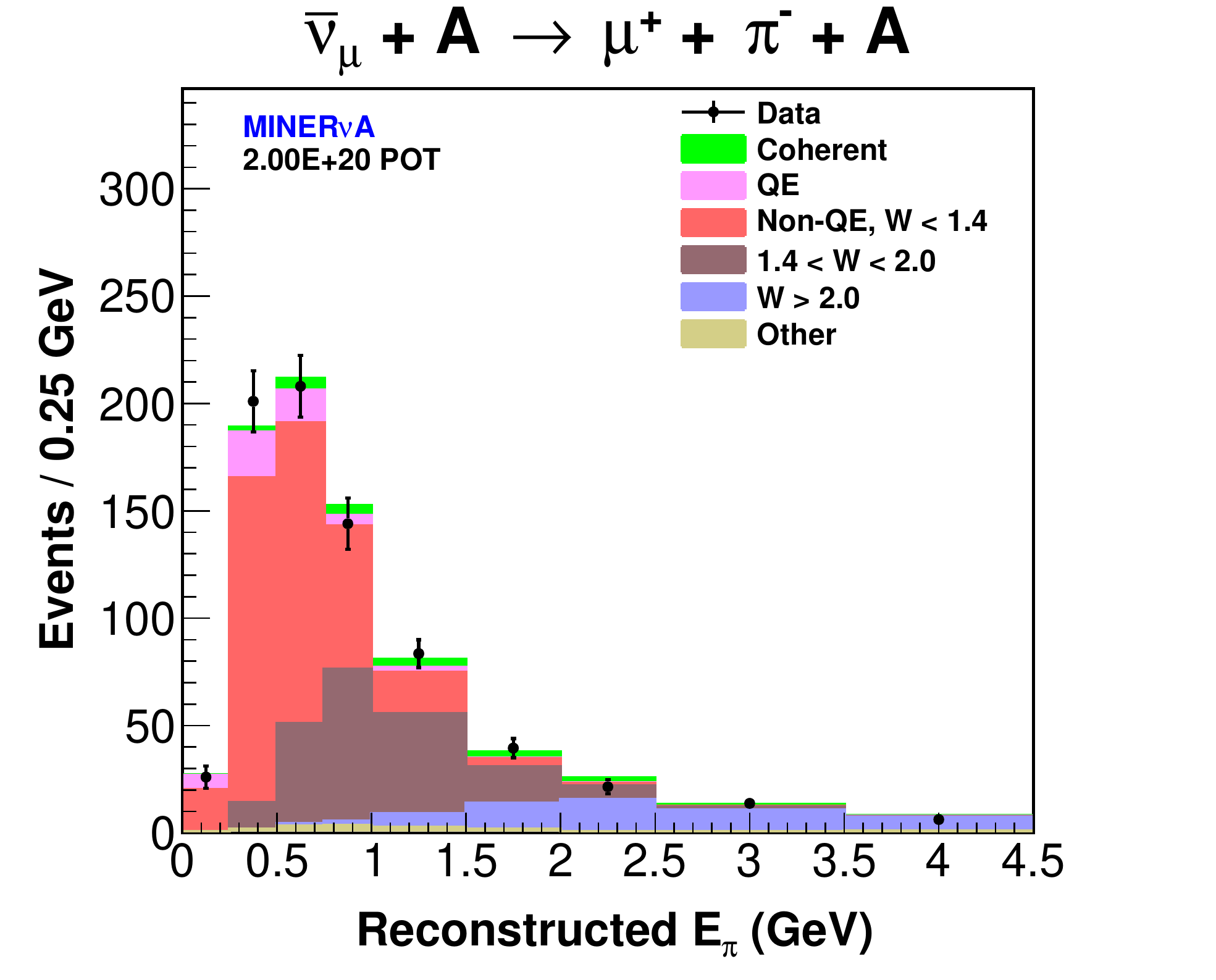}}
\caption{The \numubar sideband \epi distribution before (left) and after (right) background tuning}
\label{fig:bg_tuning_numubar}
\end{figure*}

The \numubar background tuning is performed with a fit to the \numubar sideband \epi distribution
(Fig.~\ref{fig:bg_tuning_numubar}) only.
Unlike the \numu sample, QE is a small contribution to the \numubar sample since the recoil neutron rarely produces a reconstructed track.
This scale factor is fixed at 1.0.
%The \chisq minimized in fitting the \numubar background scale factors is calculated as
%\begin{equation}
%\label{eq:bg_tuning_chi2_numubar}
%\chisq = \sum_{i} \frac{(N^{Data}_{i}-\sum_{k}\alpha_{k}N^{MC}_{ik})^{2}}{\sum_{k}\alpha_{k}N^{MC}_{ik}},
%\end{equation}
%where $N^{Data}_{i}$ is the number of data events in \epi bin $i$, $\alpha_{k}$ is the scale factor for the normalization of MC event category $k$, and $N^{MC}_{ik}$ is the number of MC events from category $k$ in bin $i$.  The scale factors for coherent scattering and the background category ``Other" are fixed to 1.0 in the fit for the reasons given for the \numu background tuning.  QE is a small contribution to the \numubar sample and its scale factor is fixed to 1.0 to minimize the number of free parameters in the fit.  
The \numubar background scale factors extracted from the fit are listed in Table~\ref{tab:bg_scale_factors}.
The \numubar sideband \epi distribution after applying the background scale factors to the MC is shown in Fig.~\ref{fig:bg_tuning_numubar}.

\begin{table}[bp] \small
\begin{center}
\begin{tabular}{ l | c | c }
Background & \numu Sample & \numubar Sample  \\
\hline
Charged-current quasielastic & 1.13$ \pm$0.04 & 1.0 (fixed) \\
Non-Quasielastic, \wgen \lt 1.4 GeV & 0.73$\pm$0.08 & 1.07$\pm$0.08 \\
1.4 \lt \wgen \lt 2.0 GeV & 0.81$\pm$0.05 & 0.79$\pm$0.09 \\
\wgen \gt 2.0 GeV & 1.7$\pm$0.2 & 2.3$\pm$0.3 \\
Other & 1.0 (fixed) & 1.0 (fixed) \\
\end{tabular}
\end{center}
\caption{\small Background scale factors}
\label{tab:bg_scale_factors}
\end{table}

Background tuning corrects the normalization and provides evidence that the \epi and \qsq distributions of the backgrounds are likely correct.  It does not correct all possible sources of uncertainty in the kinematics for each background.  Mismodeling of the
kinematics of the backgrounds is a source of systematic uncertainty as described later.
%on the background prediction and the measured cross sections.  
%The evaluation of this uncertainty is described in Section~\ref{sec:systematics}.

\subsection{Pion Angle Weighting}
\label{sec:thetapi_weighting}

\begin{figure*}[tpb]
\centering
\mbox{\includegraphics[width=0.49\linewidth]{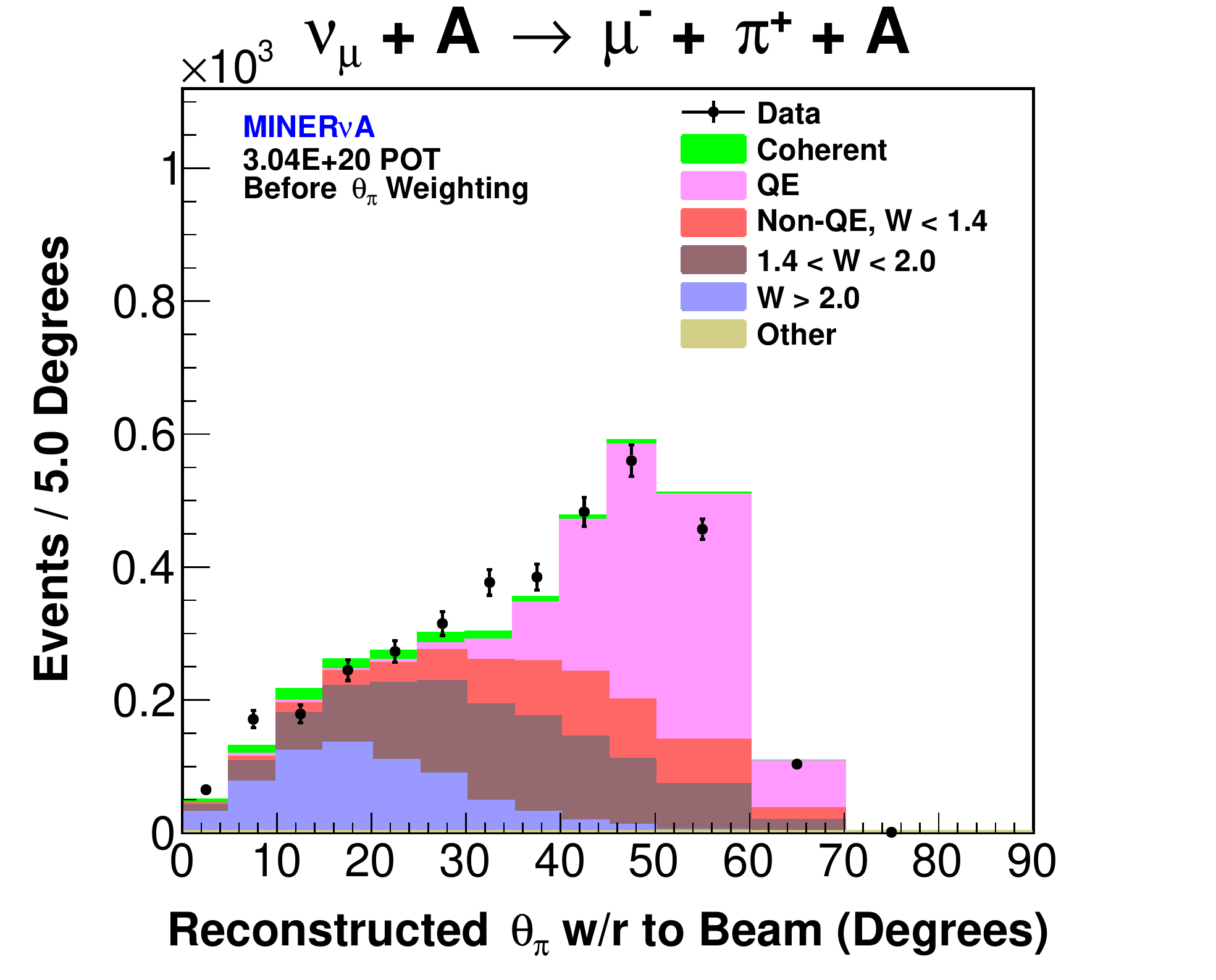}
\includegraphics[width=0.49\linewidth]{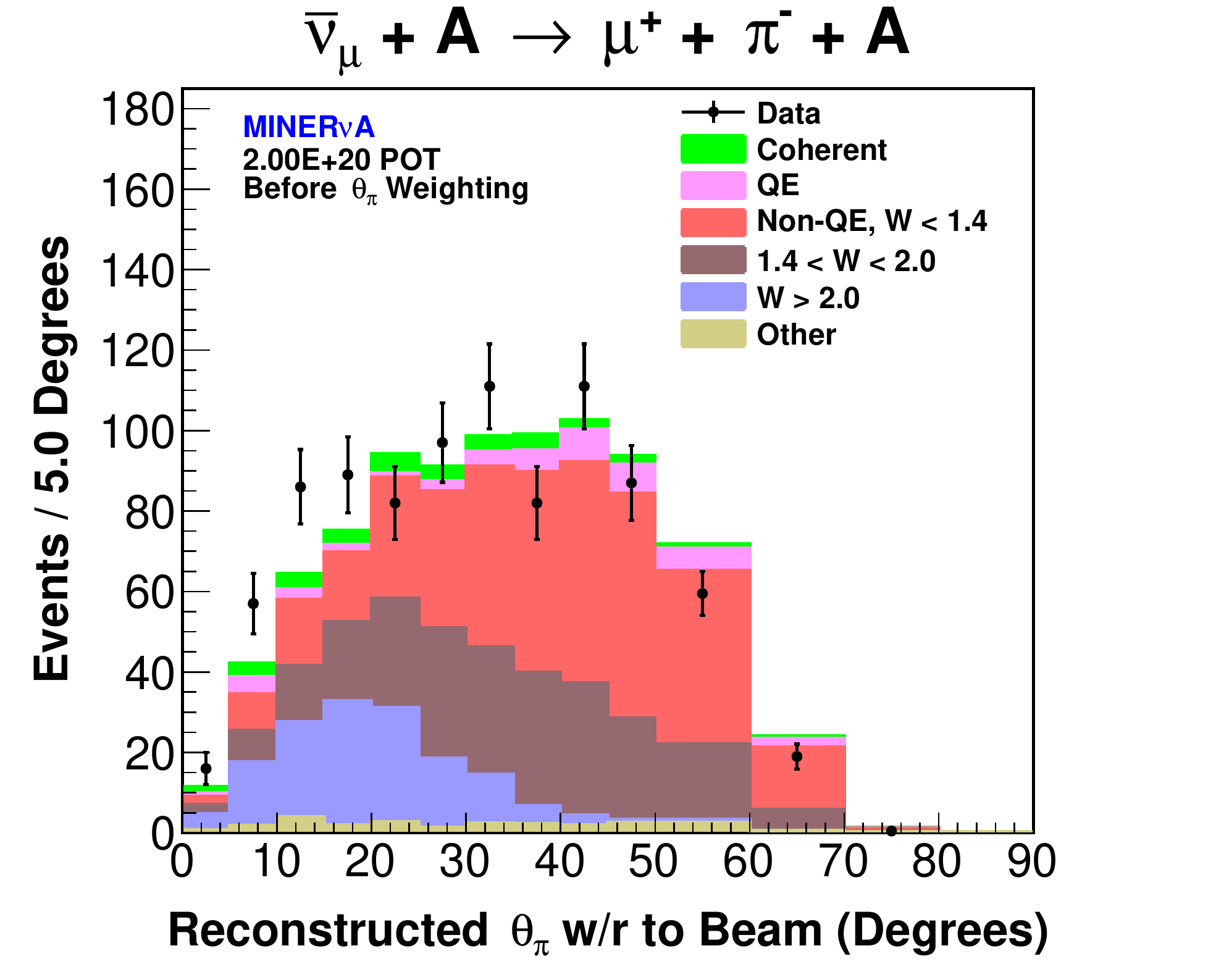}}
\mbox{\includegraphics[width=0.49\linewidth]{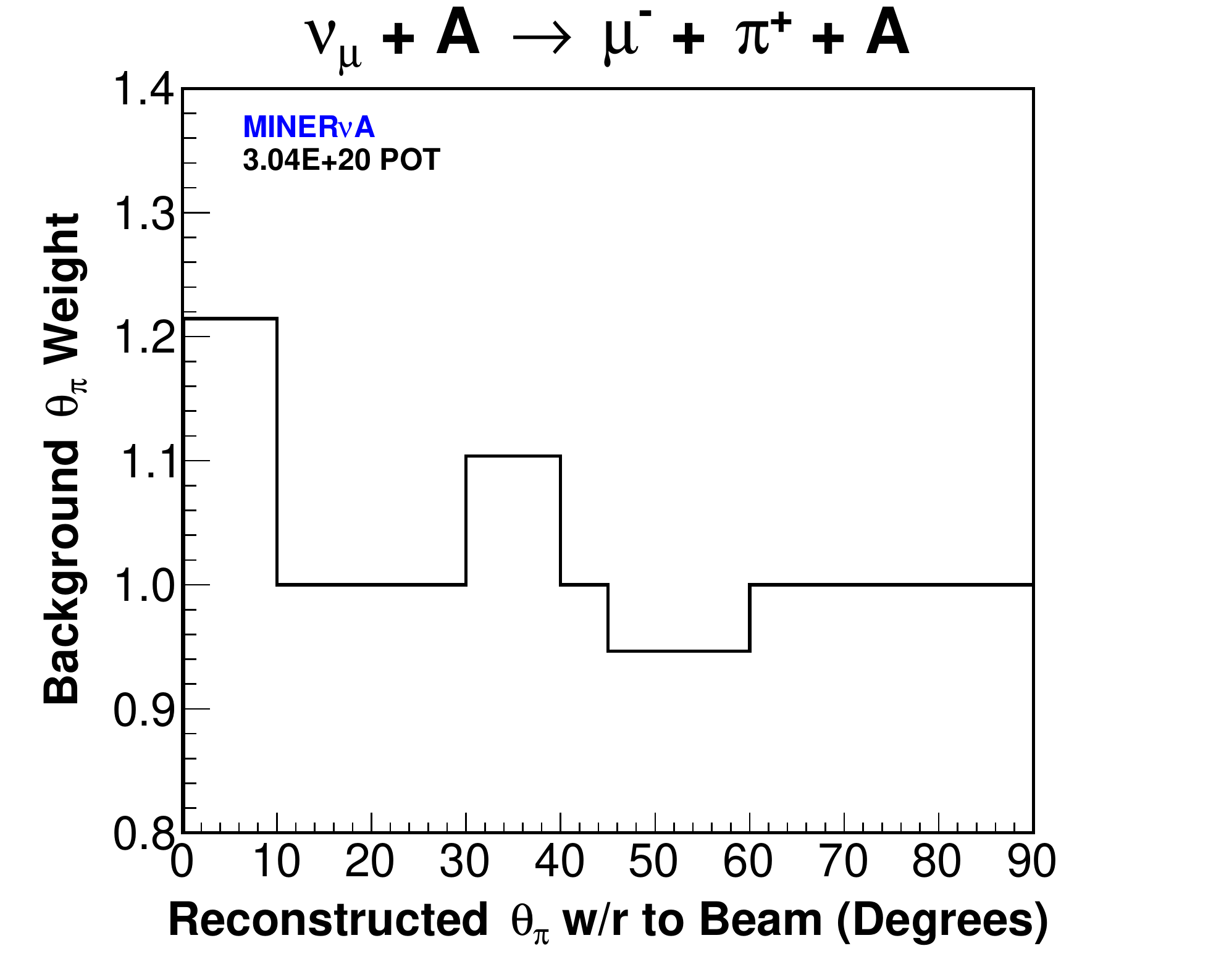}
\includegraphics[width=0.49\linewidth]{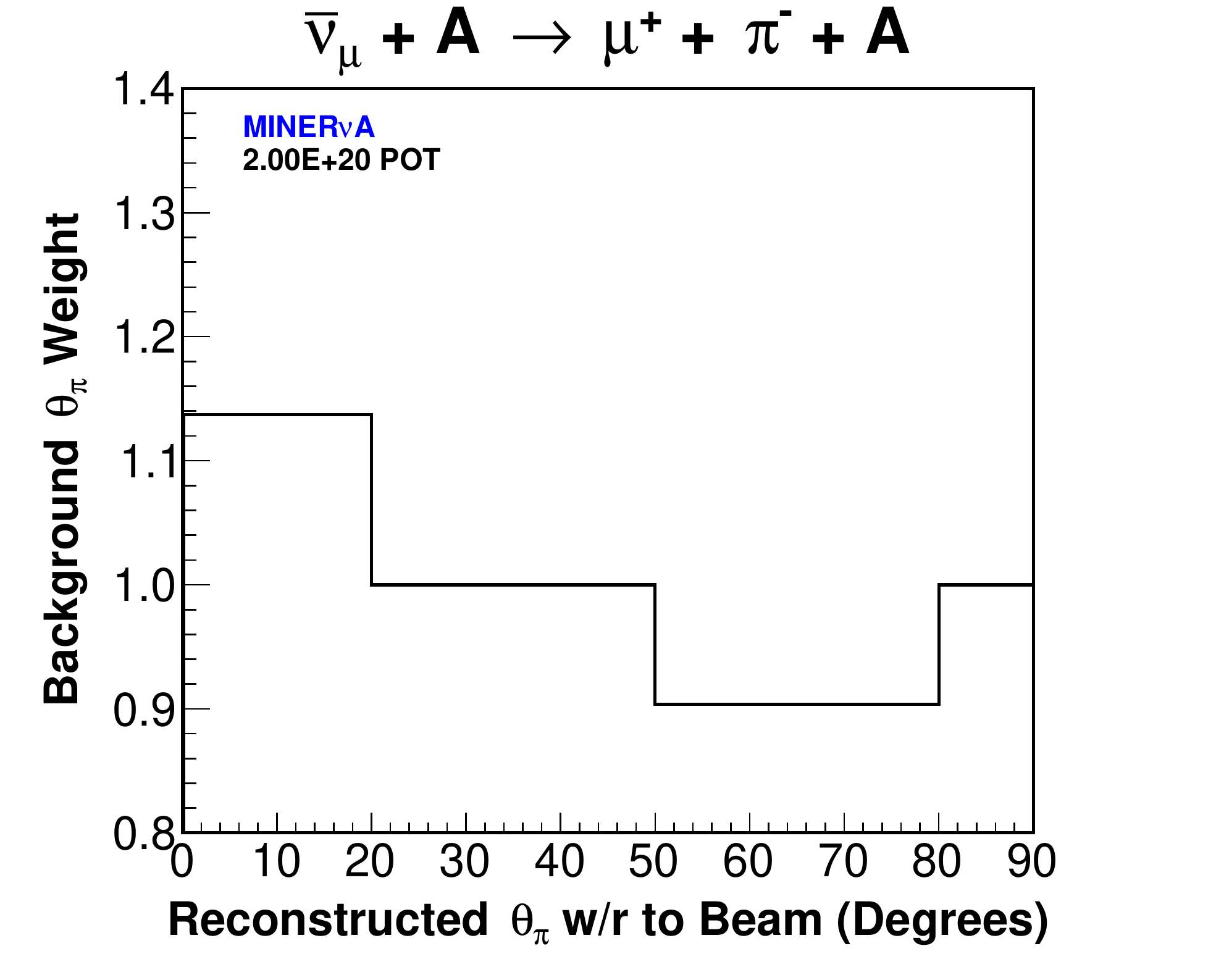}}
\mbox{\includegraphics[width=0.49\linewidth]{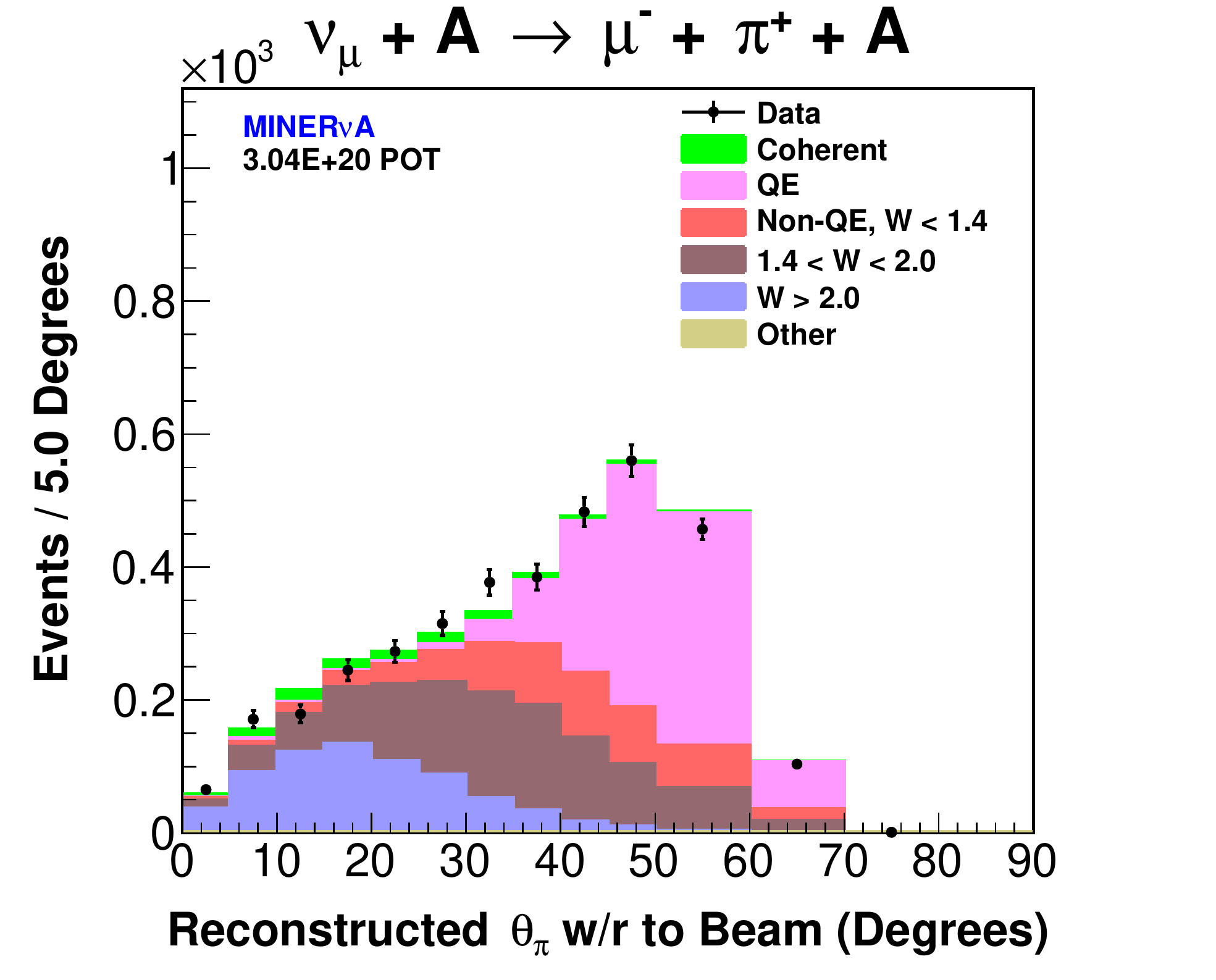}
\includegraphics[width=0.49\linewidth]{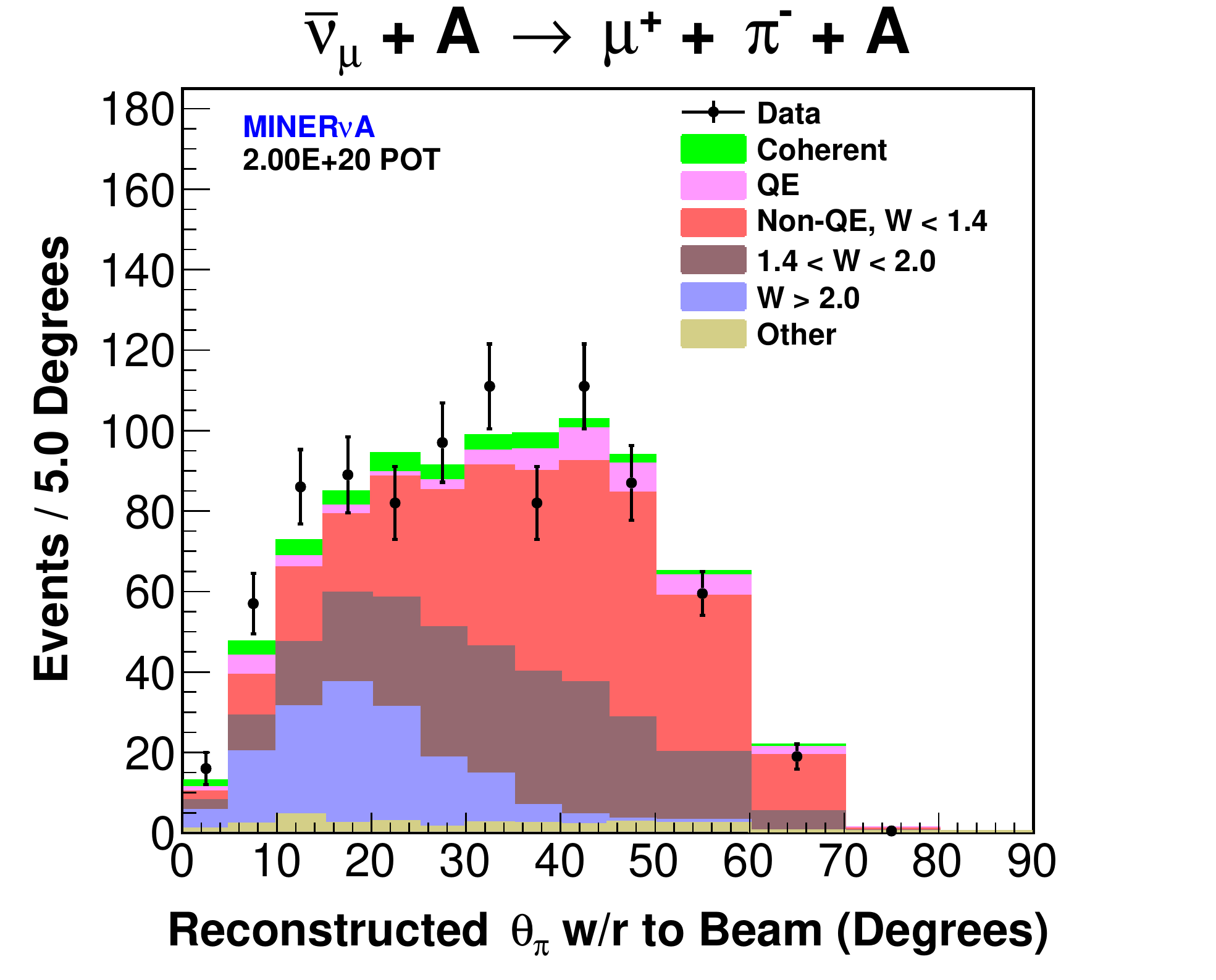}}
\caption{The top plots show the \numu (left) and \numubar (right) sideband \thetapi distributions after background tuning.  The middle plots show the \thetapi weighting applied to the MC background to correct the disagreement between the data and the MC.  The bottom plots show the sideband \thetapi distributions after background tuning and \thetapi weighting.}
\label{fig:thetapi_weighting}
\end{figure*}

Disagreement between data and MC remains in the \numu and \numubar sideband \thetapi distributions after background
tuning (Fig.~\ref{fig:thetapi_weighting}).  There is no model for the origin of this disagreement, and 
correct for it using data-driven weights of the total tuned background as a function of reconstructed
\thetapi (Fig.~\ref{fig:thetapi_weighting}).  For each group of two or more consecutive bins in the sideband \thetapi
distribution with tuned MC above (below) the data at greater than 1$\sigma$ statistical significance in each bin, the weighting
decreases (increases) the total background in this group of bins by an amount required to bring the group's agreement between data and MC to within 1$\sigma$.
The \numu and \numubar sideband \thetapi distributions after background tuning and \thetapi weighting are shown in
Fig.~\ref{fig:thetapi_weighting}.  The \thetapi weighting is applied to the tuned background in the coherent-like sample.
The change in the background prediction from the \thetapi weighting is applied as a systematic uncertainty.

\section{Calculation of the Cross Sections}
\label{sec:xs}

%This section details the calculation of the measured \numu and \numubar coherent cross sections.  The calculation consists of the following operations listed in the order performed:
%Measurements of the \numu and \numubar charged current coherent pion production cross sections were extracted from the data by subtracting the tuned background prediction from the selected data coherent candidates, unfolding to correct distortion of the kinematic distributions from reconstruction, correcting for the efficiency of selecting coherent interactions, normalizing to the neutrino flux, normalizing to the number of carbon nuclei in the fiducial volume, and correcting for non-carbon nuclei in the fiducial volume.  

The measured cross section in true-\enu bin $i$ is calculated as
\begin{equation}
\label{eq:tot_xs}
\sigma_{i} = \frac{\beta\sum_{j}U_{ij}(N^{data}_{j}-N^{bkgd}_{j})}{T\phi_{i}\epsilon_{i}},
\end{equation}
where $N^{data}_{j}$ is the number of data coherent candidates in reconstructed \enu bin $j$, $N^{bkgd}_{j}$ is the
tuned estimate of the number of background coherent candidates, $U_{ij}$ is the unfolding matrix element that estimates
the signal contribution from reconstructed bin $j$ to true bin $i$, $\epsilon_{i}$ is the coherent selection efficiency,
$\phi_{i}$ is the \numu or \numubar flux, $T$ is the number of carbon nuclei targets in the fiducial volume, and $\beta$ is the
correction to the coherent event rate for interactions on non-carbon nuclei (see Sec.~\ref{sec:target_norm}).
Differential cross sections as functions of $\epi$, $\thetapi$, and $\qsq$ are calculated similarly, but the flux is integrated over
2 $< \enu <$ \unit[20]{GeV}.
%Similarly, the measured differential cross section in true bin $i$ of kinematic parameter $\xi$ was calculated as
%\begin{equation}
%\label{eq:diff_xs}
%\left(\frac{d\sigma}{d\xi}\right)_{i} = \frac{\beta\sum_{j} U_{ij} (N^{data}_{j}-N^{bkgd}_{j}) }{ (\Delta\xi)_{i}T\Phi\epsilon_{i} },
%\end{equation}
%where $\Phi$ is the integrated \numu/\numubar flux and $(\Delta\xi)_{i}$ is the bin width.
%
%Cross sections for the \numubar partial and full detector data sets (Section~\ref{sec:data_mc_samples}) were calculated separately and combined to account for the effect of \argoneut on the reconstructed kinematics and selection efficiency.  The combined \numubar cross section was calculated as
%\begin{equation}
%\label{eqn:comb_numubar_xs}
%\sigma_{\numubar} = \frac{P_{p}T_{p}\sigma_{p}+P_{f}T_{f}\sigma_{f}}{P_{p}T_{p}+P_{f}T_{f}} = 0.317\sigma_{p}+0.683\sigma_{f},
%\end{equation}
%where $P$ and $T$ are the POT (Section~\ref{sec:data_mc_samples}) and number of carbon nuclei in the fiducial volume (Section~\ref{sec:target_norm}), respectively, and the subscripts $p$ and $f$ denote the partial and full detector data samples, respectively.  This calculation accounts for the difference in the flux and fiducial mass between the two samples.  The combined \numubar differential cross sections were calculated in the same way.

The unfolding matrices and efficiency corrections for measuring the cross sections were estimated using coherent
events in the MC, where events with \epi\lt 0.5 GeV were weighted by 50\%.  This weighting is referred to as the
signal model weighting, and is discussed in Sec.~\ref{sec:signal_model_weighting}.
%Coherent events in the MC are used to estimate the unfolding matrices and efficiency corrections for the measuring the cross sections.  This introduces dependence of the measured cross sections on the signal model (\ie GENIE implementation of the Rein-Sehgal coherent model), since the unfolding matrices and efficiency corrections are dependent on the signal model kinematics.  Comparisons of the coherent model to the initial measurements of the cross sections found better agreement by weighting the rate of interactions with \epi\lt 0.5 GeV predicted by the model by 50\% (Section~\ref{sec:signal_model_weighting}).  To minimize systematic bias on the measured cross sections from the signal model, this weighting was applied to coherent events in the MC, the unfolding matrices and efficiency corrections were re-estimated, and the cross sections were remeasured.  The effect of the signal model weighting on the measured cross sections is shown in Section~\ref{sec:signal_model_weighting}.  The unfolding matrices and efficiency corrections shown hereafter were estimated from the weighted signal model.
The following sections detail each step of the cross section calculation.

\subsection{Unfolding}
\label{sec:unfolding}

\begin{figure*}[tpb]
\centering
\mbox{
\includegraphics[width=0.49\linewidth]{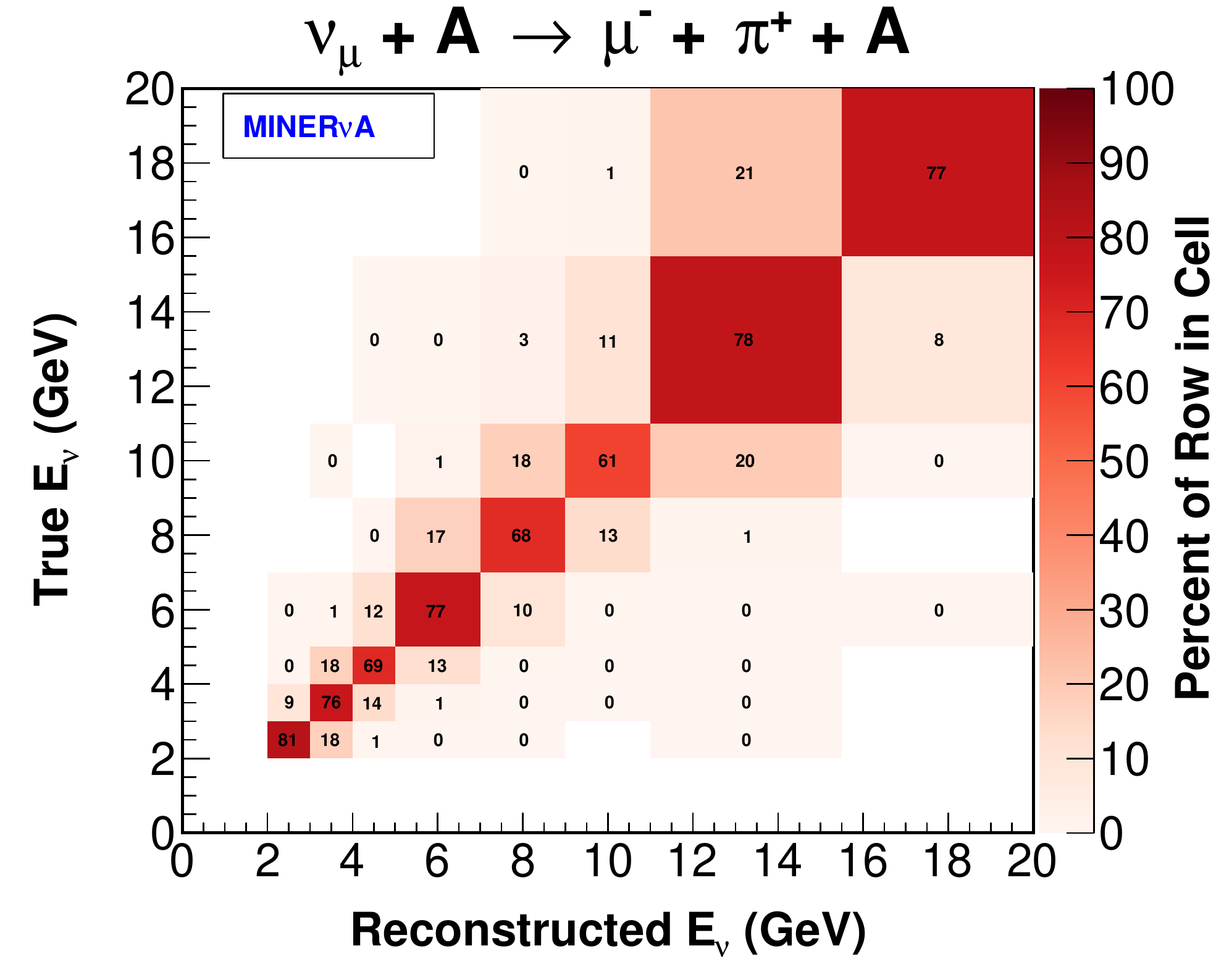}
\includegraphics[width=0.49\linewidth]{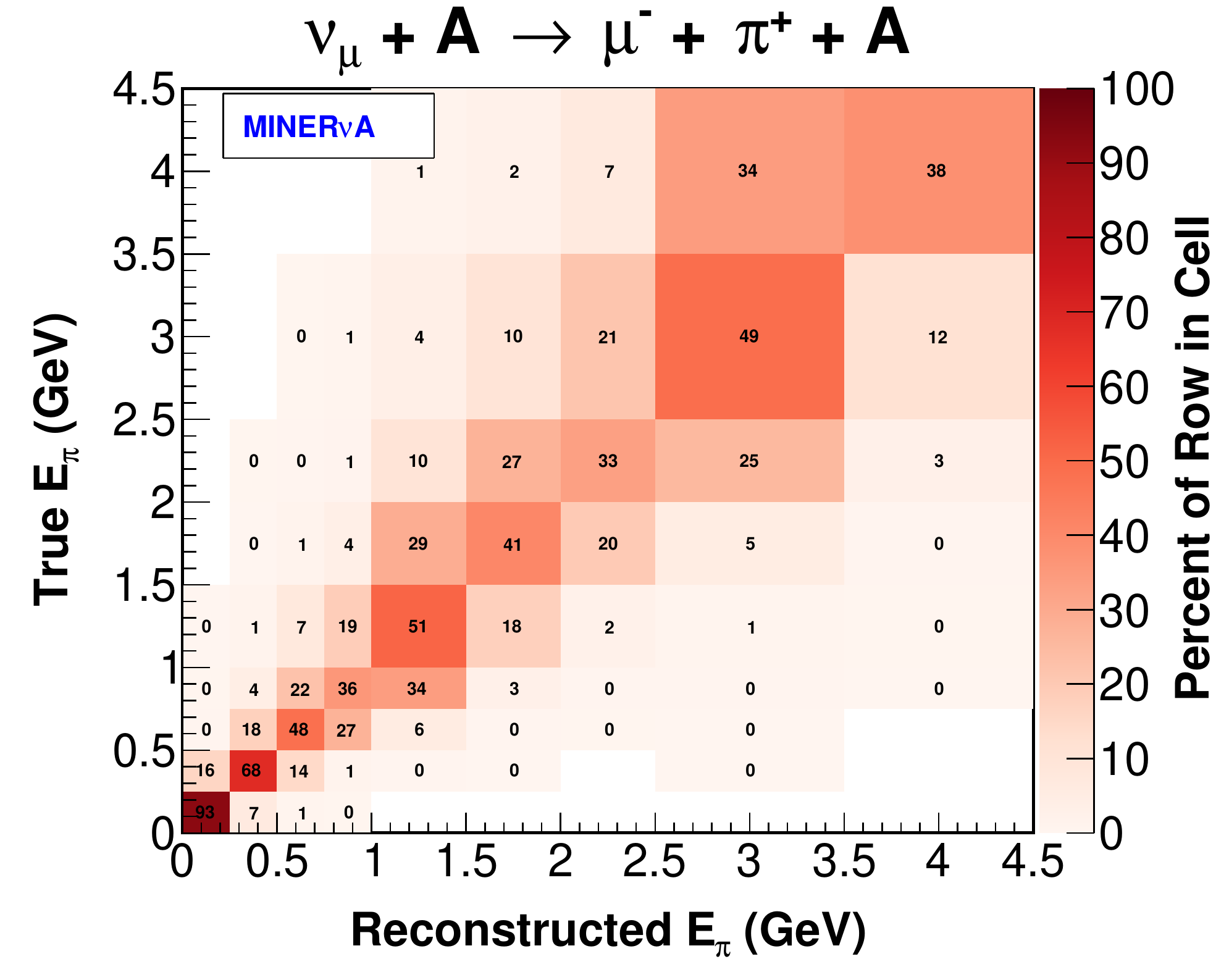}
}
\mbox{
\includegraphics[width=0.49\linewidth]{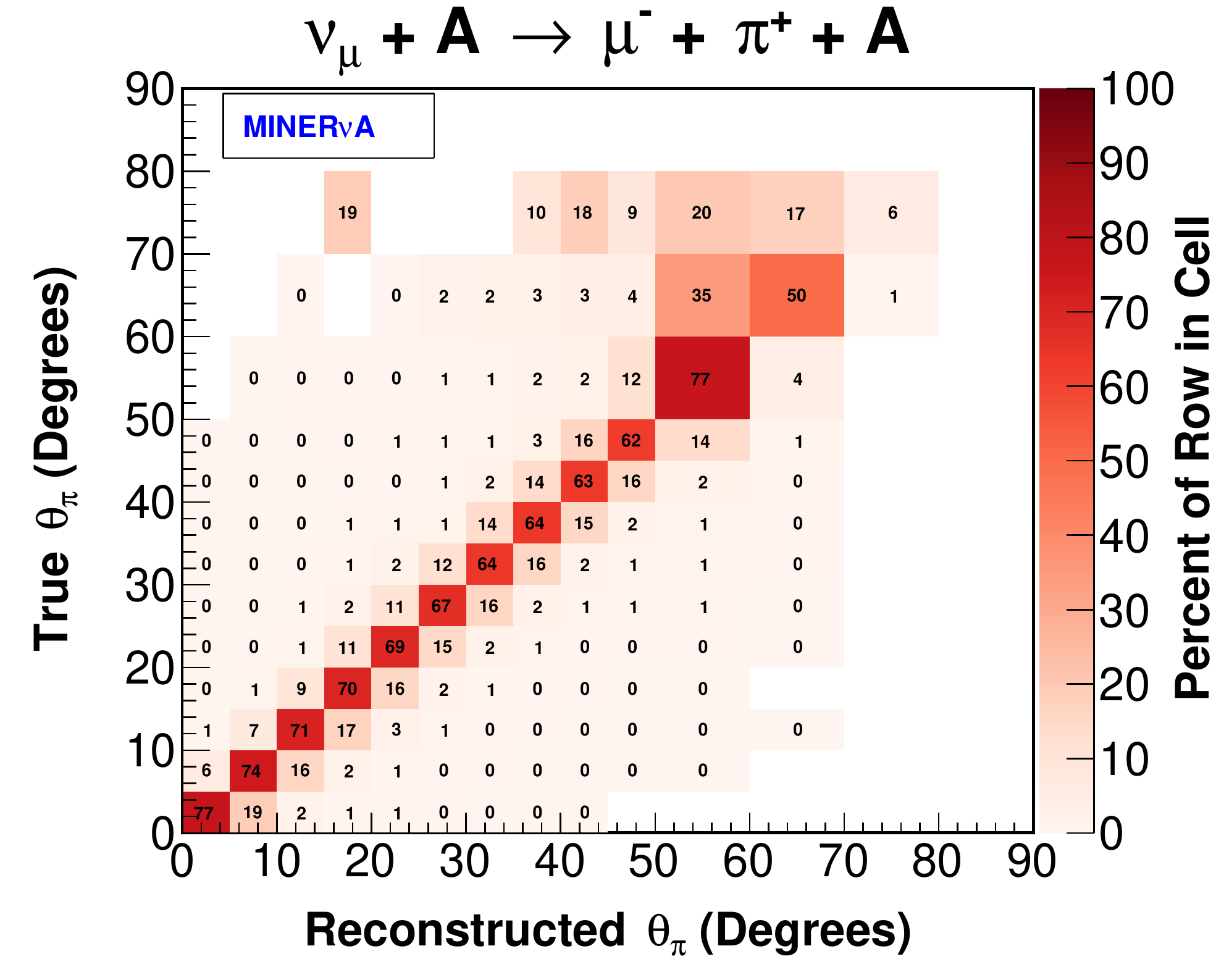}
\includegraphics[width=0.49\linewidth]{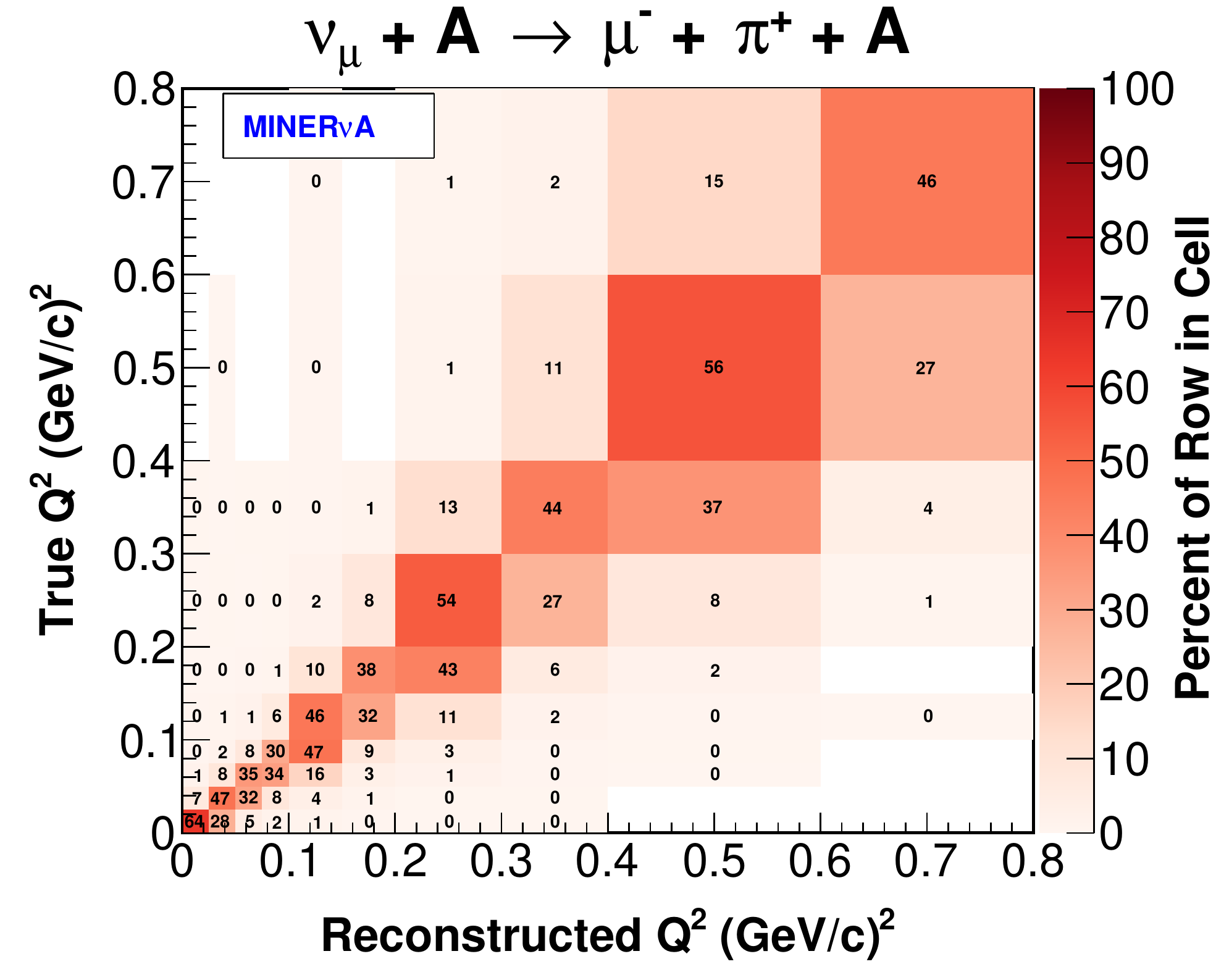}
}
\caption[The unfolding matrices in \enu, \epi, \thetapi and \qsq]{The unfolding matrices in \enu, \epi, \thetapi and \qsq for the \numu sample.}
\label{fig:unfolding}
\end{figure*}

After background subtraction, the \numu (\numubar) sample contains 1411 (481) coherent candidates.  
The distributions of the kinematic variables of these candidates are unfolded to correct for distortions and resolution
effects created in the reconstruction process.  
%Given a distribution of event rates in bins of a reconstructed quantity, the unfolding estimates the event rates in the bins of the true value of that quantity.  
%% IT HAS BEEN POINTED OUT BY SEVERAL AUTHORS THAT THE METHOD IS ACTUALLY FREQUENTIST, NOT BAYESIAN
The distributions were unfolded using the iterative %Bayesian 
unfolding method of D'Agostini~\cite{bib:dagostini}.  
%The method begins by using Bayes' theorem to estimate the unfolded distribution from the reconstructed distribution and an unfolding matrix.  
The unfolding matrix $U_{ij}$ estimates the contribution from true bin $i$ to reconstructed bin $j$.  It was calculated
from the signal-only MC samples as
\begin{equation}
U_{ij} = \frac{N^{sel}_{ij}}{\sum_{j}N^{sel}_{ij}},
\end{equation}
where $N^{sel}_{ij}$ is the number of coherent events passing all selection cuts in bin $ij$.  The signal model
weighting (Sec.~\ref{sec:signal_model_weighting}) was applied in calculating the unfolding matrices.  In each successive
iteration the unfolded distribution is recalculated in the same way from the unfolded distribution resulting from the previous iteration.  
The unfolding matrices for the \numu are shown in Fig.~\ref{fig:unfolding}, and \numubar distributions are similar.

The diagonal entries of the statistical covariance matrix increase with each unfolding iteration.  
The number of unfolding iterations, two, 
was optimized to give adequate correction of reconstruction distortion while minimizing the increase of these diagonals.

Unfolding creates correlations in the statistical uncertainty between bins which are included in the statistical covariance matrix.
%, which is propagated to the measured cross section.

\subsection{Efficiency Correction}
\label{sec:efficiency}

\begin{figure*}[tpb]
\centering
\mbox{
\includegraphics[width=0.49\linewidth]{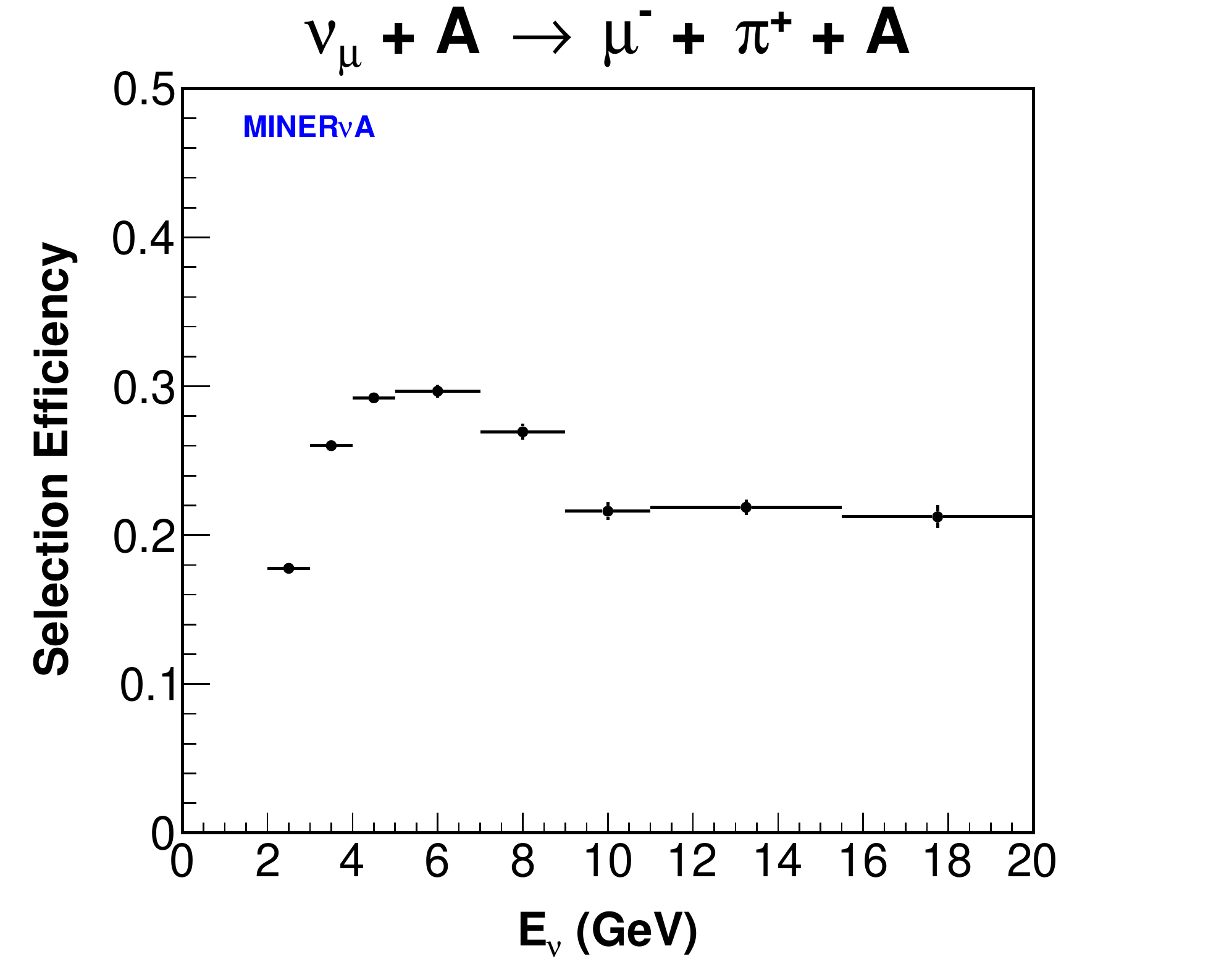}
\includegraphics[width=0.49\linewidth]{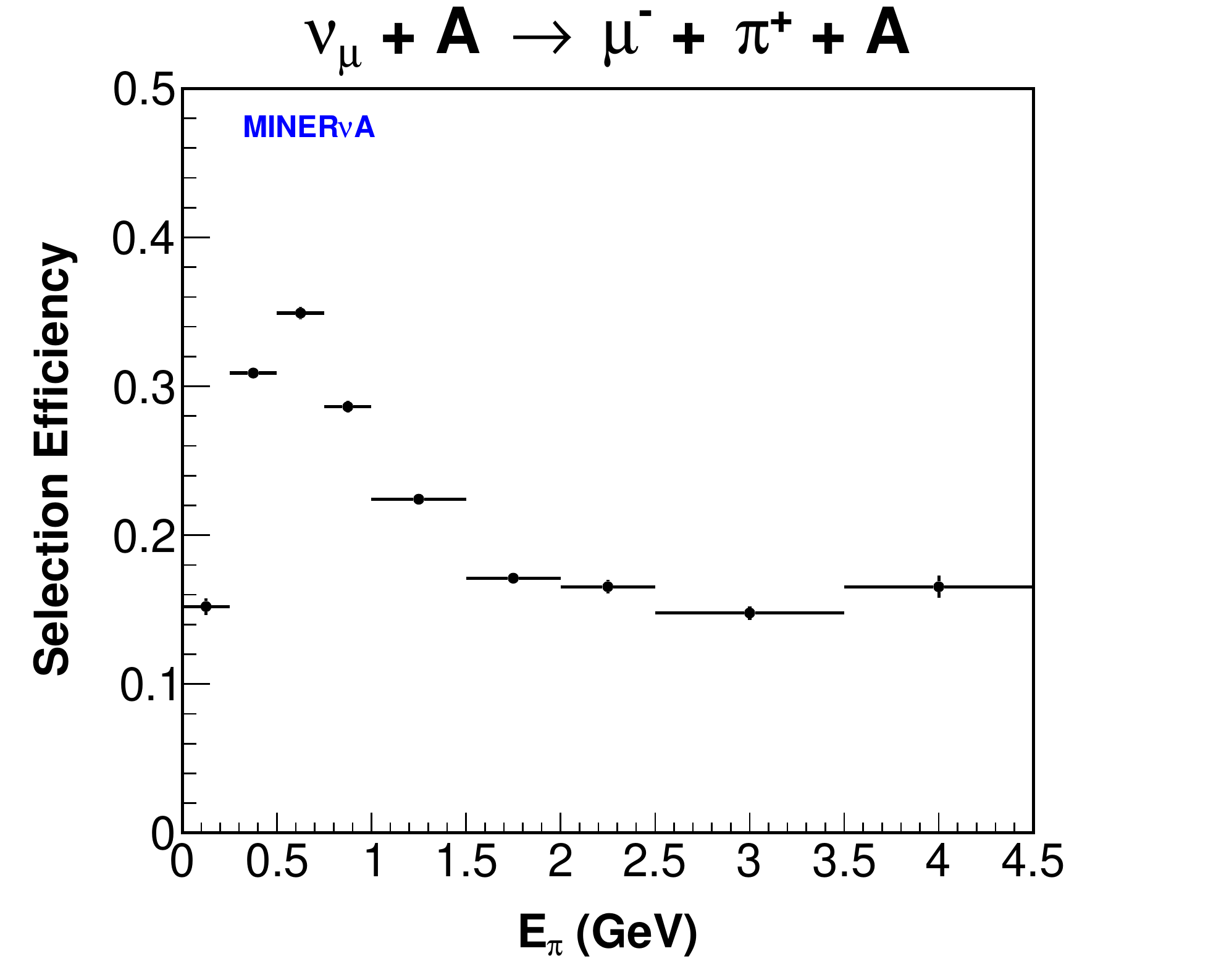}
}
\mbox{
\includegraphics[width=0.49\linewidth]{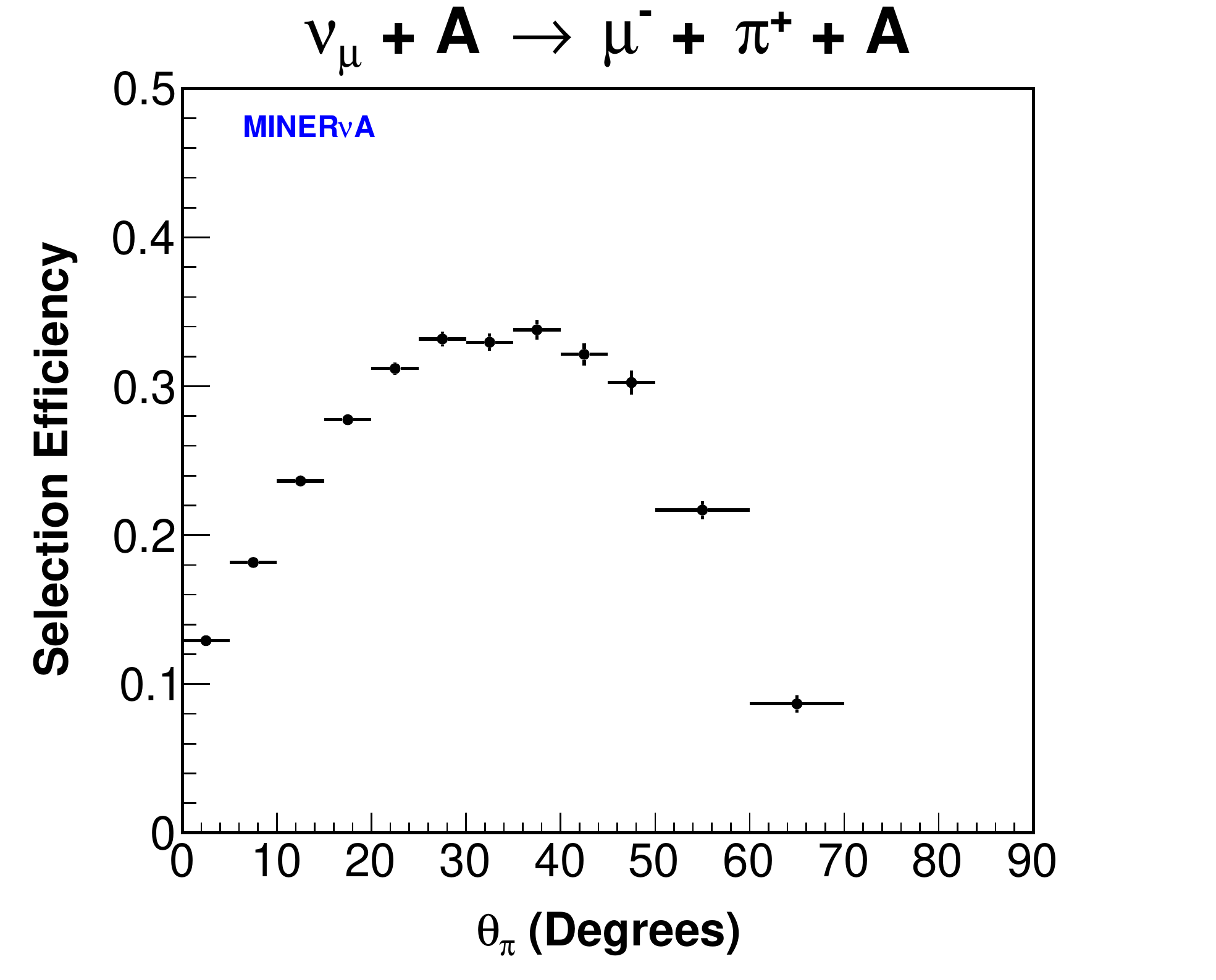}
\includegraphics[width=0.49\linewidth]{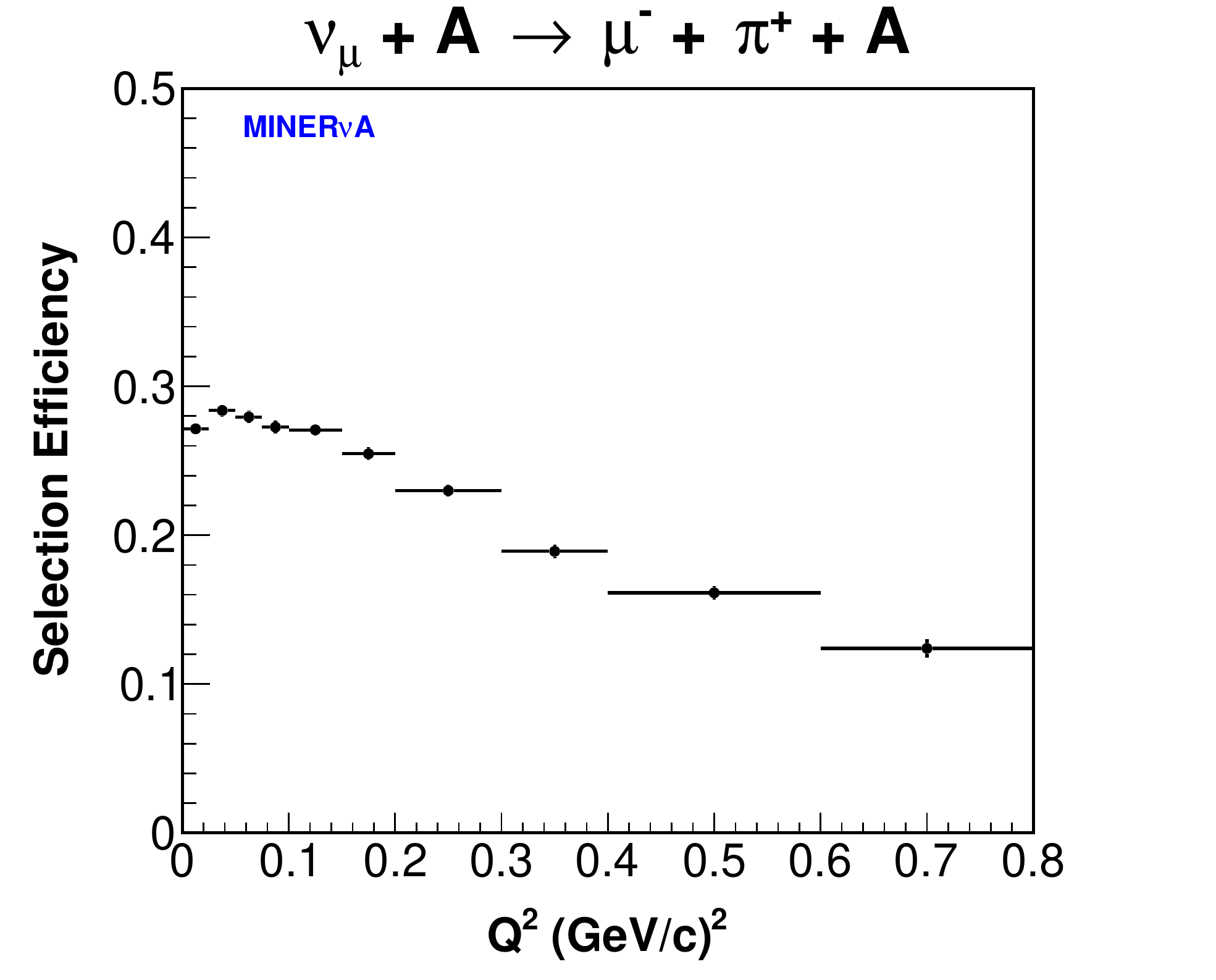}
}
\caption[The selection efficiency as a function of \enu, \epi, \thetapi and \qsq for the \numu sample.]{The selection
efficiency as a function of \enu, \epi, \thetapi and \qsq for the \numu sample.}
\label{fig:efficiency}
\end{figure*}

After unfolding, the kinematic distributions are corrected for signal selection efficiency, which was estimated using coherent events
from the signal MC samples.  The efficiency in each true kinematic bin $\epsilon_{i}$ was calculated as
\begin{equation}
\epsilon_{i} = \frac{N^{sel}_{i}}{N^{gen}_{i}},
\label{eq:efficiency}
\end{equation}
where $N^{gen}_{i}$ is the number of signal events generated inside the fiducial volume with true \enu satisfying \evrange
in bin $i$, and $N^{sel}_{i}$ is the subset of those events that passed all reconstruction and selection cuts.  The signal
event weighting described earlier was applied in calculating the efficiency.  The efficiency as a function of each kinematic
parameter for the \numu samples is shown in Fig.~\ref{fig:efficiency}; the \numubar efficiencies are similar.  The overall
efficiency for the \numu and \numubar samples is 24-25\%.

The selection efficiency includes the acceptance of the reconstruction.
The requirement that the muon be reconstructed in both \minerva and \minos limits \thetamu of accepted events to
\thetamu\lt$20^\circ$, and the minimum number of planes required to form a track limits \thetapi of accepted events to
\thetapi\lt$70^\circ$.  For signal events in the MC occurring inside the fiducial volume, 96\% have \thetamu\lt$20^\circ$
and \thetapi\lt$70^\circ$.  About one third of the signal events are lost because the muon must have high enough energy to be
tracked in the MINOS detector and then matched to a track in the MINERvA detector, and around half of the remaining
signal events are lost by requiring the pion to make a second track in the event.  Pion tracking drops off below $\sim$\unit[200]{MeV}
and, as mentioned earlier, muon tracking fails below \unit[1.5]{MeV}.

\subsection{Flux Normalization}
\label{sec:flux_norm}

%\input{tables/flux_normalization.tex}
%\begingroup
%\squeezetable
%\begin{table}[bp]
%\scriptsize
%\begin{center}
%\tabcolsep=0.11cm
%% \begin{tabular}{c|c|c|c|c}
%% $E_{\nu}$ \ (GeV) & 2.0 - 3.0 & 3.0 - 4.0 & 4.0 - 5.0 & 5.0 - 7.0\\
%% \hline
%% $(\nu_{\mu}/$cm$^{2}/$POT$) \times 10^{-8}$             & 0.740 & 0.765 & 0.312 & 0.189 \\
%% $(\overline{\nu}_{\mu}/$cm$^{2}/$POT$) \times 10^{-8}$ & 0.662 & 0.644 & 0.244 & 0.129 \\
%% \end{tabular}
%% \vspace*{2ex}\\
%% \begin{tabular}{c|c|c|c|c}
%% $E_{\nu}$ \ (GeV) & 7.0 - 9.0 & 9.0 - 11.0 & 11.0 - 15.5 & 15.5 - 20.0 \\
%% \hline
%% $(\nu_{\mu}/$cm$^{2}/$POT$) \times 10^{-8}$             & 0.098 & 0.065 & 0.077 & 0.044 \\
%% $(\overline{\nu}_{\mu}/$cm$^{2}/$POT$) \times 10^{-8}$ & 0.055 & 0.033 & 0.035 & 0.018 \\
%% \end{tabular}
%\begin{tabular}{c|c|c}
%$E_{\nu}$ \ (GeV) & $(\nu_{\mu}/$cm$^{2}/$POT$) \times 10^{-8}$ & $(\overline{\nu}_{\mu}/$cm$^{2}/$POT$) \times 10^{-8}$ \\
%$E_{\nu}$ \ (GeV) & $(\nu_{\mu}/$cm$^{2}/$POT$)$ & $(\overline{\nu}_{\mu}/$cm$^{2}/$POT$)$ \\
%				  & $\times 10^{-8}$			 & $ \times 10^{-8}$ \\
%\hline
%2.0 - 3.0   &  0.740      &   0.662     \\
%3.0 - 4.0   &  0.765      &   0.644     \\
%4.0 - 5.0   &  0.312      &   0.244     \\
%5.0 - 7.0   &  0.189      &   0.129     \\
%7.0 - 9.0   &  0.098      &   0.055     \\
%9.0 - 11.0  &  0.065      &   0.033     \\
%11.0 - 15.5 &  0.077      &   0.035     \\ 
%15.5 - 20.0 &  0.044      &   0.018     
%\end{tabular}
%\caption{ $\nu_{\mu}$ and $\overline{\nu}_{\mu}$ fluxes }
%\label{tab:flux_norm_bins}
%\end{center}
%\end{table}
%s\endgroup

The cross sections were normalized to the flux prediction (see Sec.~\ref{sec:NuMI_N_me}) scaled to the POT for each
sample.  The cross section $\sigma_{i}$ in each \enu bin $i$ was normalized to the flux integrated over the bin range
(Table~\ref{tab:flux_norm}).
 The differential cross section $\left(\frac{d\sigma}{d\xi}\right)_{i}$ in each $\xi$ bin $i$ was normalized to the flux
 integrated over \evrange.

\subsection{Target Number Normalization}
\label{sec:target_norm}

\begin{table}[bp] \small
\begin{center}
\begin{tabular}{ l | c | c }
Nucleus & $A$ & $T$ (units of 10$^{29}$ nuclei) \\
\hline
$^{1}$H & 1.008 & 2.425 \\
$^{12}$C & 12.011 & 2.404\\
$^{16}$O & 15.999 & 0.0655 \\
$^{27}$Al & 26.982 & 0.0032 \\
$^{28}$Si & 28.085 & 0.0032 \\
$^{35}$Cl & 35.453 & 0.0051 \\
$^{48}$Ti & 47.867 & 0.0047 \\
%$^{16}$O & 15.999 & 6.548 $\times$ 10$^{27}$ \\
%$^{27}$Al & 26.982 & 3.175 $\times$ 10$^{26}$ \\
%$^{28}$Si & 28.085 & 3.167 $\times$ 10$^{26}$ \\
%$^{35}$Cl & 35.453 & 5.111 $\times$ 10$^{26}$ \\
%$^{48}$Ti & 47.867 & 4.749 $\times$ 10$^{26}$ \\
\end{tabular}
\end{center}
\caption[The nuclear mass number $A$ and estimated number of nuclei $T$ for each nuclear species in the full detector fiducial volume]{The nuclear mass number $A$ and estimated number of nuclei $T$ for each non-hydrogen nuclear species in the full detector fiducial volume, which spans the central 108 scintillator planes of the tracker and extends to the edges of a hexagon with an 850 mm apothem in the transverse (XY) plane.}
\label{tab:fiducial_nuclei}
\end{table}

The measured cross sections were normalized to the number of carbon nuclei (the ``targets" of the neutrinos and antineutrinos) 
contained in the fiducial volume.
The number of carbon nuclei was estimated using the detector geometry and material models, the latter of which was
informed by direct material assay.  The
%full (partial) detector fiducial volume was estimated to contain 2.404 $\times$ 10$^{29}$ (1.246 $\times$ 10$^{29}$) carbon nuclei.
full (partial) detector fiducial volume contains 2.404 $\times$ 10$^{29}$ (1.246 $\times$ 10$^{29}$) carbon nuclei.

%The number of coherent candidates $N^{coh} = N^{data}-N^{bkgd}$, where $N^{data}$ and $N^{bkgd}$ are the number of selected data and tuned background events, respectively, include coherent interactions on non-carbon nuclei inside the fiducial volume.  
%There are also 
Non-carbon coherent interactions were not included in $N^{bkgd}$ to avoid dependence of the background estimate on the coherent model.  
Instead, the number of coherent candidates ($N^{coh}$) was corrected  to the number of coherent candidates on carbon only ($N^{coh}_{c}$).
The correction $\beta$ was calculated as
\begin{equation}
\beta = \frac{N^{coh}_{c}}{N^{coh}} = \frac{\phi\epsilon_{c}\sigma_{c}T_{c}}{\sum_{i}\phi\epsilon_{i}\sigma_{i}T_{i}},
\end{equation}
where $\phi$ is the flux and $\epsilon_{i}$, $\sigma_{i}$, and $T_{i}$ are the coherent acceptance and selection efficiency,
coherent cross section, and number of nuclei in the fiducial volume for nuclear species $i$.
Assuming that the signal acceptance and selection efficiency is the same for all nuclear species and that the coherent cross section
scales with the nuclear mass number $A$ as $A^{1/3}$ \cite{bib:RS},
\begin{equation}
\beta \approx \frac{A^{1/3}_{c}T_{c}}{\sum_{i}A^{1/3}_{i}T_{i}}.
\end{equation}
The estimated number of nuclei for each nuclear species in the full detector fiducial volume
(Table~\ref{tab:fiducial_nuclei}) gives $\beta$\ =\  0.962.  
%This correction was applied to all measured cross sections (Equations~\ref{eq:tot_xs} and~\ref{eq:diff_xs}).
%Note that the equivalent of coherent scattering on hydrogen, referred to as "diffractive" scattering, occurs at much higher
%\tabs and has different acceptance due to the recoiling proton in the final state.
%Therefore, unlike coherent scattering, diffractive scattering is considered as a background rather than a signal contribution with different cross section than that on carbon.
Diffractive scattering (see Sec.~\ref{sec:diffractiveMeasure}) is considered as a background rather than as a signal contribution.

%The target number in our cross section calculations was the total number of Carbon atoms contained in our fiducial volume.  We calculated the target number using the TargetUtils tool in the Ana/PlotUtils package.  The element mass fractions of the scintillator planes defined in TargetUtils for data and MC were determined from the material assay and the material model, respectively.  The 2.024 g/cm$^{2}$ areal density of the scintillator planes defined in TargetUtils was determined from the material assay.  The number of Carbon atoms in our full \minerva detector fiducial volume was 2.404e29 and 2.401e29 for data and MC, respectively.  The number of Carbon atoms in our partial \minerva detector fiducial volume was 1.246e29 and 1.245e29 for data and MC, respectively.

% The unfolding matrices and efficiency corrections used to measure the cross sections were 

\subsection{Signal Model Weighting}
\label{sec:signal_model_weighting}

\begin{figure*}[tpb]
\centering
\mbox{
\includegraphics[width=0.49\linewidth]{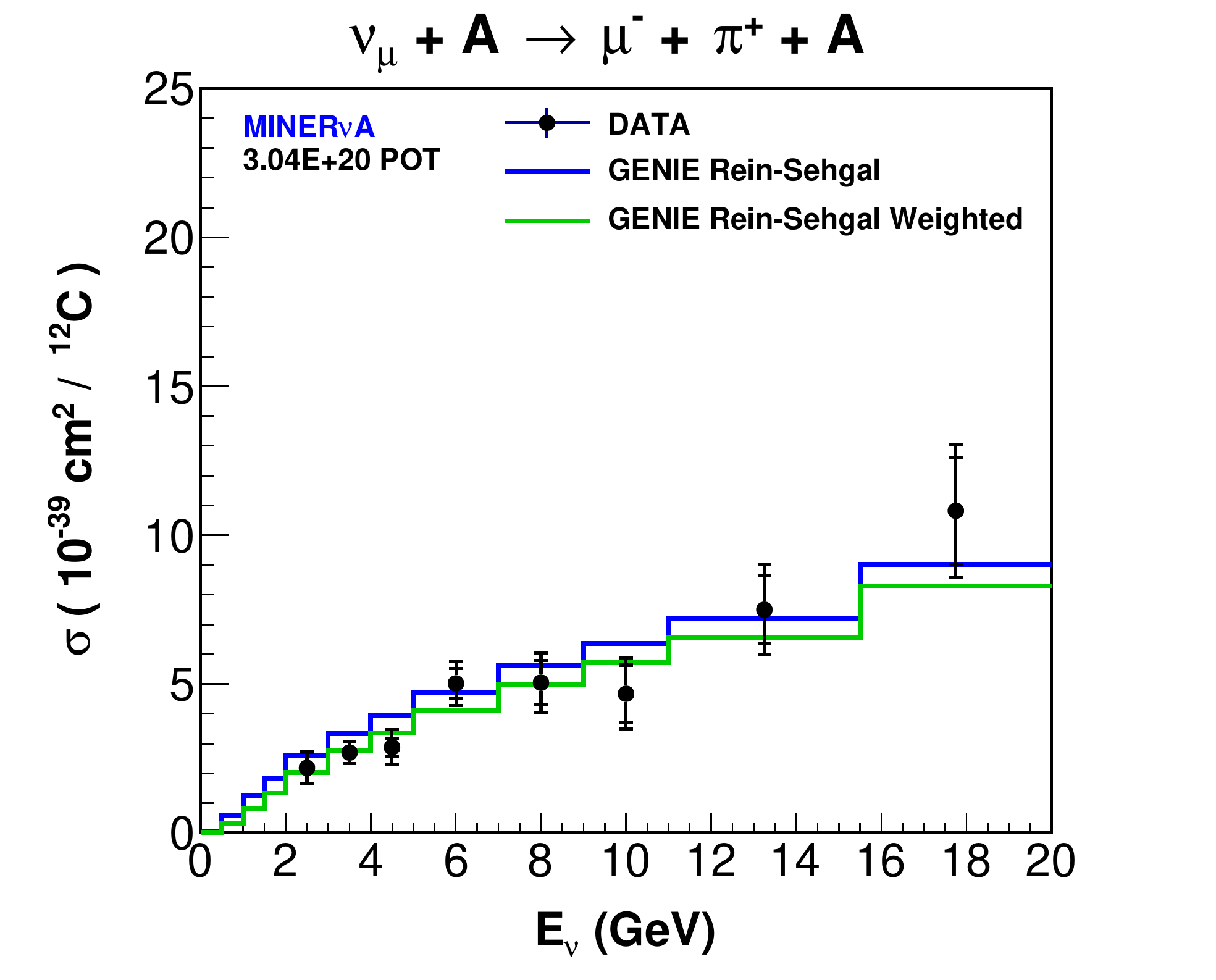}
\includegraphics[width=0.49\linewidth]{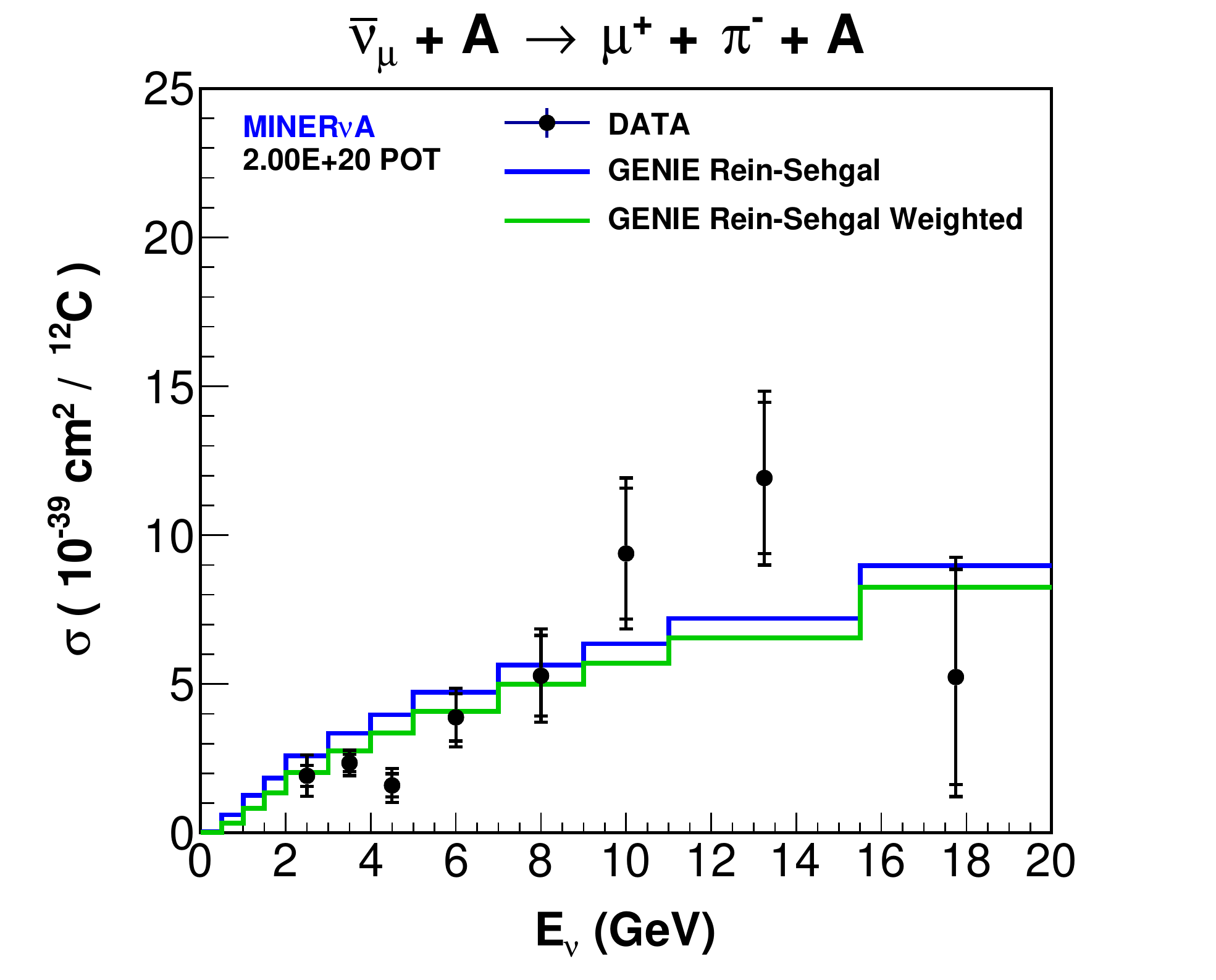}}
\caption[The initial measurements of the \numu and \numubar \sigenu made without the signal model weighting, and the GENIE prediction
with and without the signal model weighting]{The initial measurements of the \numu (left) and \numubar (right) \sigenu made without
the signal model weighting, and the GENIE prediction with and without the signal model weighting.}
\label{fig:initial_meas_sigenu}
\end{figure*}

\begin{figure*}[tpb]
\centering
\mbox{
\includegraphics[width=0.49\linewidth]{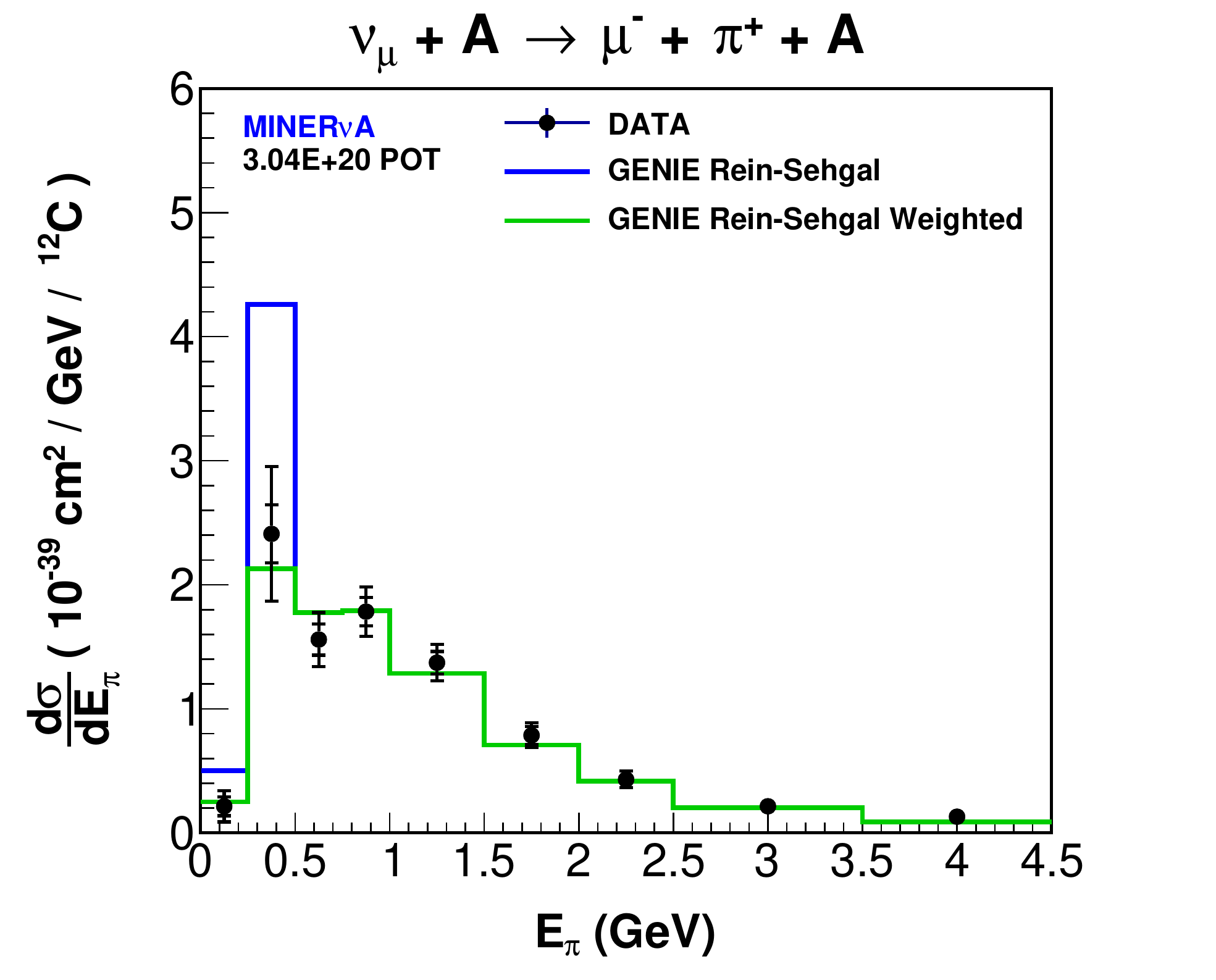}
\includegraphics[width=0.49\linewidth]{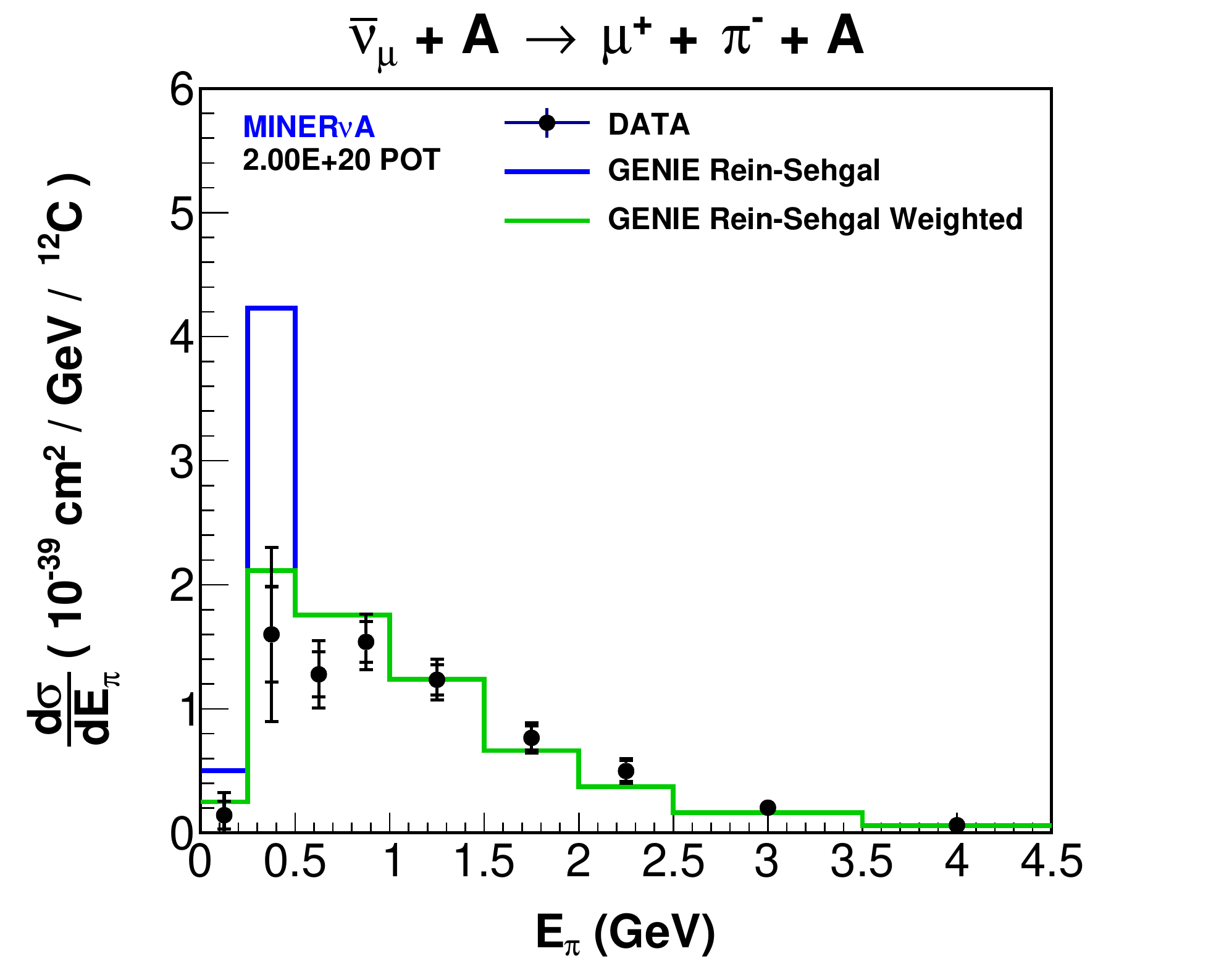}}
\caption[The initial measurements of the \numu and \numubar \dsigdepi made without the signal model weighting, and the GENIE
prediction with and without the signal model weighting]{The initial measurements of the \numu (left) and \numubar (right) \dsigdepi
made without the signal model weighting, and the GENIE prediction with and without the signal model weighting.}
\label{fig:initial_meas_dsigdepi}
\end{figure*}

\begin{figure*}[tpb]
\centering
\mbox{
\includegraphics[width=0.49\linewidth]{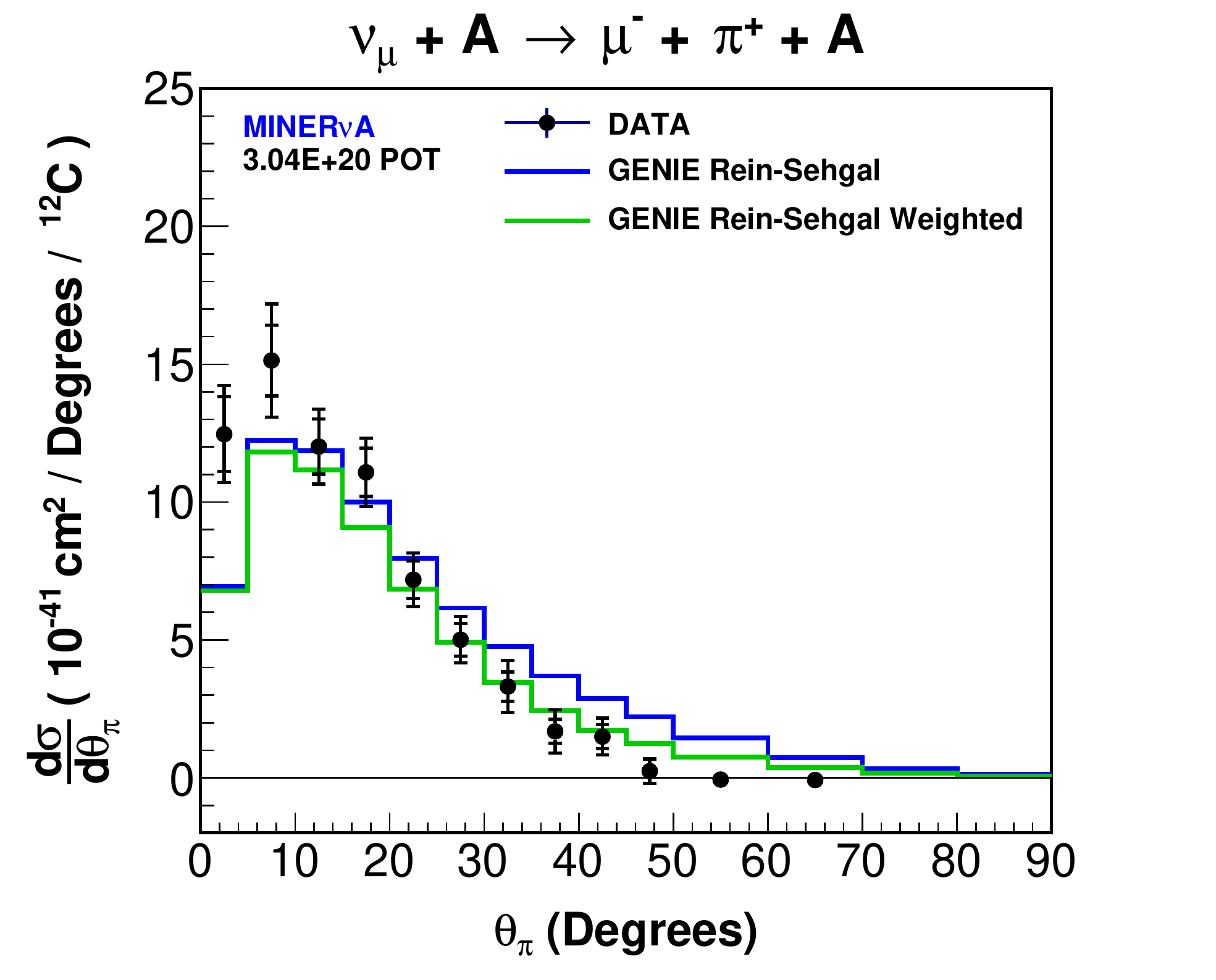}
\includegraphics[width=0.49\linewidth]{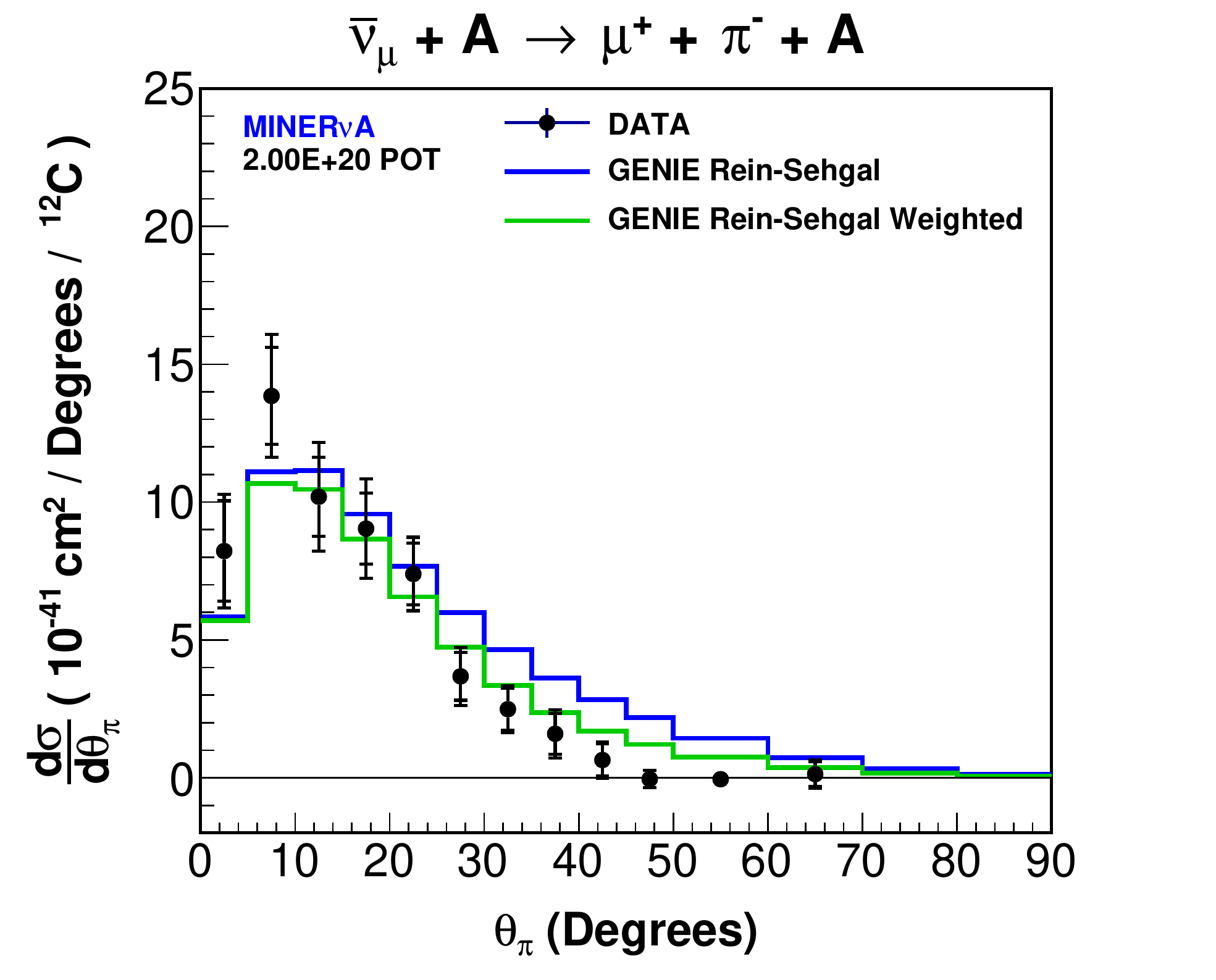}}
\caption[The initial measurements of the \numu and \numubar \dsigdthetapi made without the signal model weighting, and
the GENIE prediction with and without the signal model weighting]{The initial measurements of the \numu (left)
and \numubar (right) \dsigdthetapi made without the signal model weighting, and the GENIE prediction with and without the
signal model weighting.}
\label{fig:initial_meas_dsigdthetapi}
\end{figure*}

\begin{figure*}[tpb]
\centering
\mbox{
\includegraphics[width=0.49\linewidth]{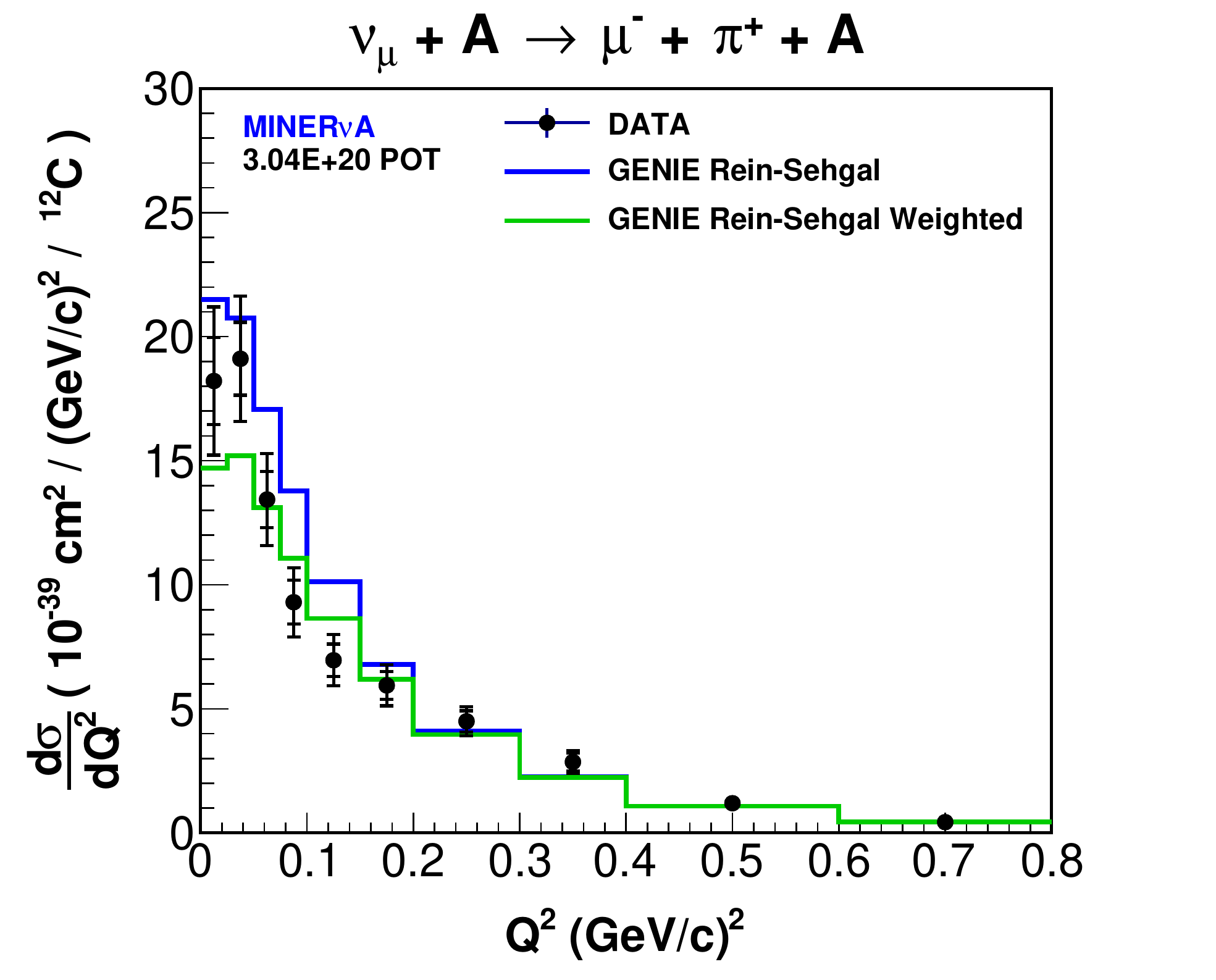}
\includegraphics[width=0.49\linewidth]{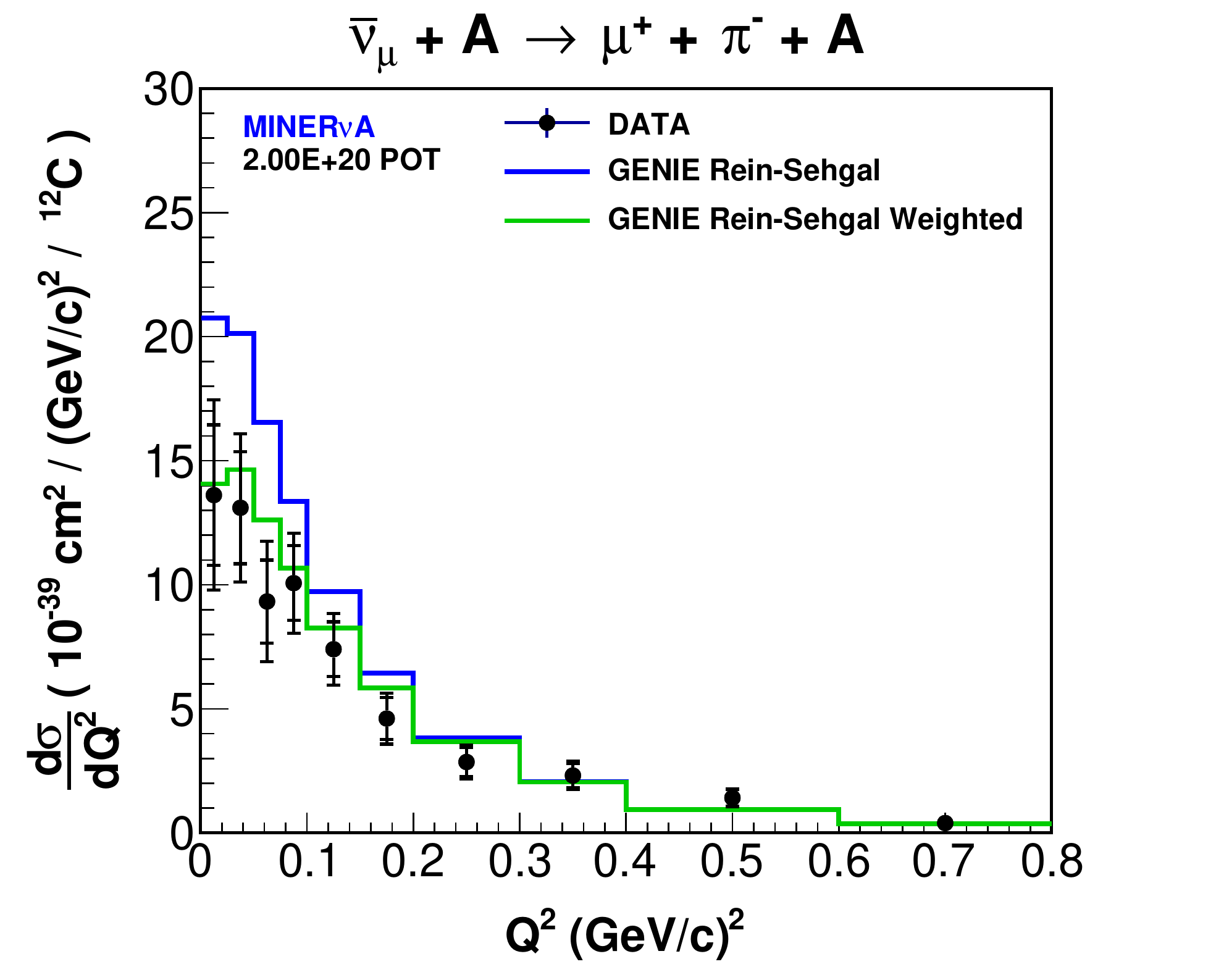}}
\caption[The initial measurements of the \numu and \numubar \dsigdqsq made without the signal model weighting, and the GENIE
prediction with and without the signal model weighting]{The initial measurements of the \numu (left) and \numubar (right) \dsigdqsq
made without the signal model weighting, and the GENIE prediction with and without the signal model weighting.}
\label{fig:initial_meas_dsigdqsq}
\end{figure*}

Initial measurements of the cross sections (Figs.~\ref{fig:initial_meas_sigenu}--\ref{fig:initial_meas_dsigdqsq}) were made
using unfolding matrices and efficiency corrections estimated from the unmodified GENIE Rein-Sehgal coherent model.
These initial measurements revealed that GENIE over-predicts
the production rate at low-\epi and high-\thetapi.  Better agreement between the GENIE prediction and the initial measurements
was achieved by weighting the rate of interactions predicted by GENIE with \epi\lt 0.5 GeV by 50\%
(Fig.s~\ref{fig:initial_meas_sigenu}--\ref{fig:initial_meas_dsigdqsq}).  This is the signal model weighting.  The \chisq
for the comparison of the initial measurement of each cross section to the nominal and weighted GENIE predictions is listed
in Table~\ref{tab:xsec_chisq_rs_wgt}, where the \chisq was calculated per Eq.~(\ref{eq:xsec_chisq}).  

\begingroup
\squeezetable
\begin{table}[bp]
\small
\begin{center}
\begin{tabular}{c|cc|cc|c}
Cross & \multicolumn{2}{c|}{$\nu_{\mu}\ \chi^{2}$} & \multicolumn{2}{c|}{$\overline{\nu}_{\mu}\ \chi^{2}$} & \\
%Cross Section & Rein-Sehgal & Rein-Sehgal Weighted & Rein-Sehgal & Rein-Sehgal Weighted & NDF \\
section & Nominal & Weighted & Nominal & Weighted & NDoF \\
\hline
$\sigma(E_\nu)$                &    11.8 &     7.5 &    27.0 &    17.6 &  8 \\
$d\sigma/dE_\pi$               &    22.4 &    13.5 &    16.6 &     7.1 &  9 \\
$d\sigma/d\theta_\pi$          &  1388.4 &   418.8 &   144.1 &    46.9 & 12 \\
$d\sigma/dQ^{2}$               &    19.2 &    15.4 &    16.8 &    10.0 & 10 \\
\end{tabular}
\end{center}
\caption{ \chisq for the comparisons of the initial measured \numu and \numubar cross sections to the nominal and signal model weighted GENIE Rein-Sehgal predictions }
\label{tab:xsec_chisq_rs_wgt}
\end{table}
\endgroup

\begin{figure*}[tpb]
\centering
\mbox{
\includegraphics[width=0.49\linewidth]{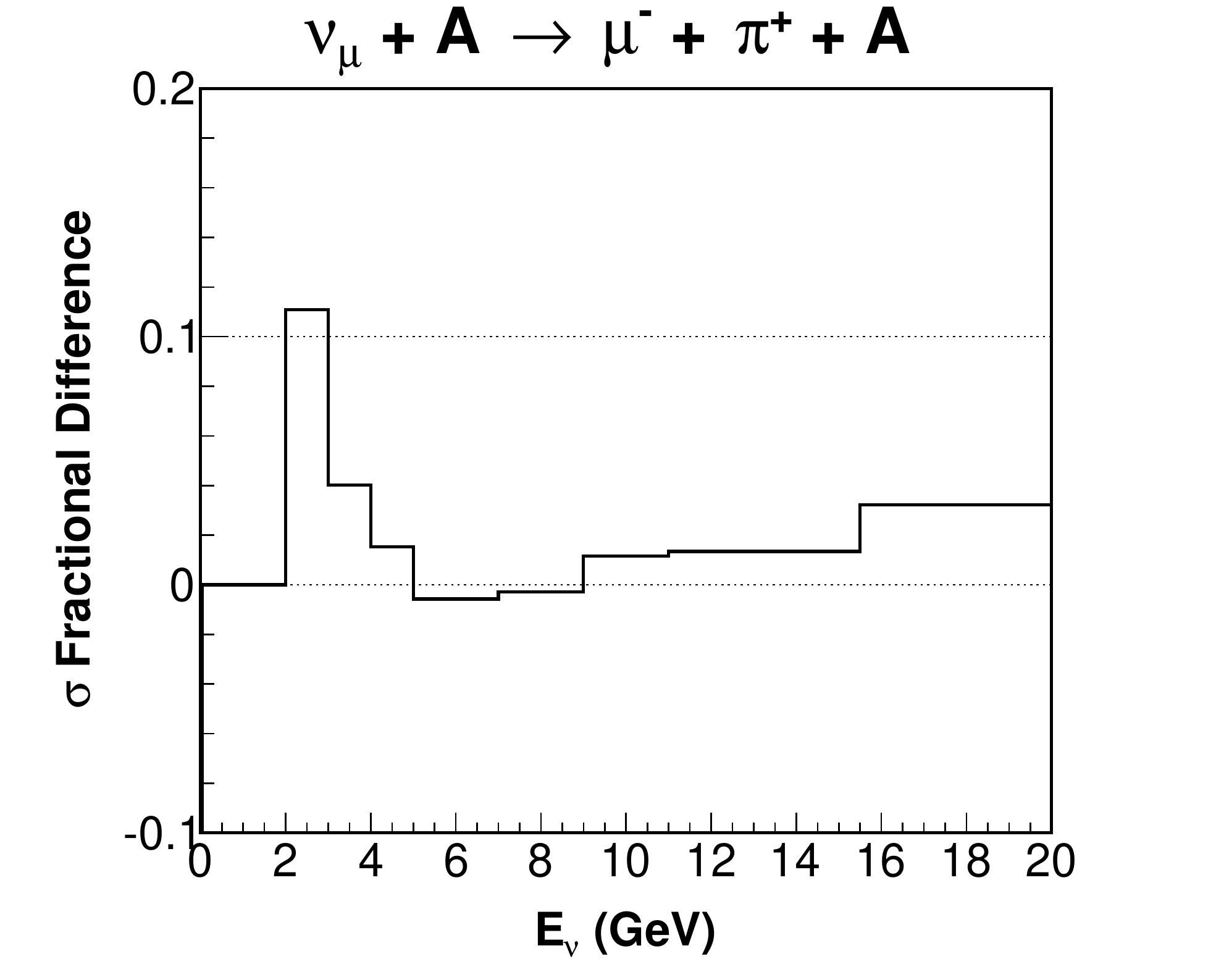}
\includegraphics[width=0.49\linewidth]{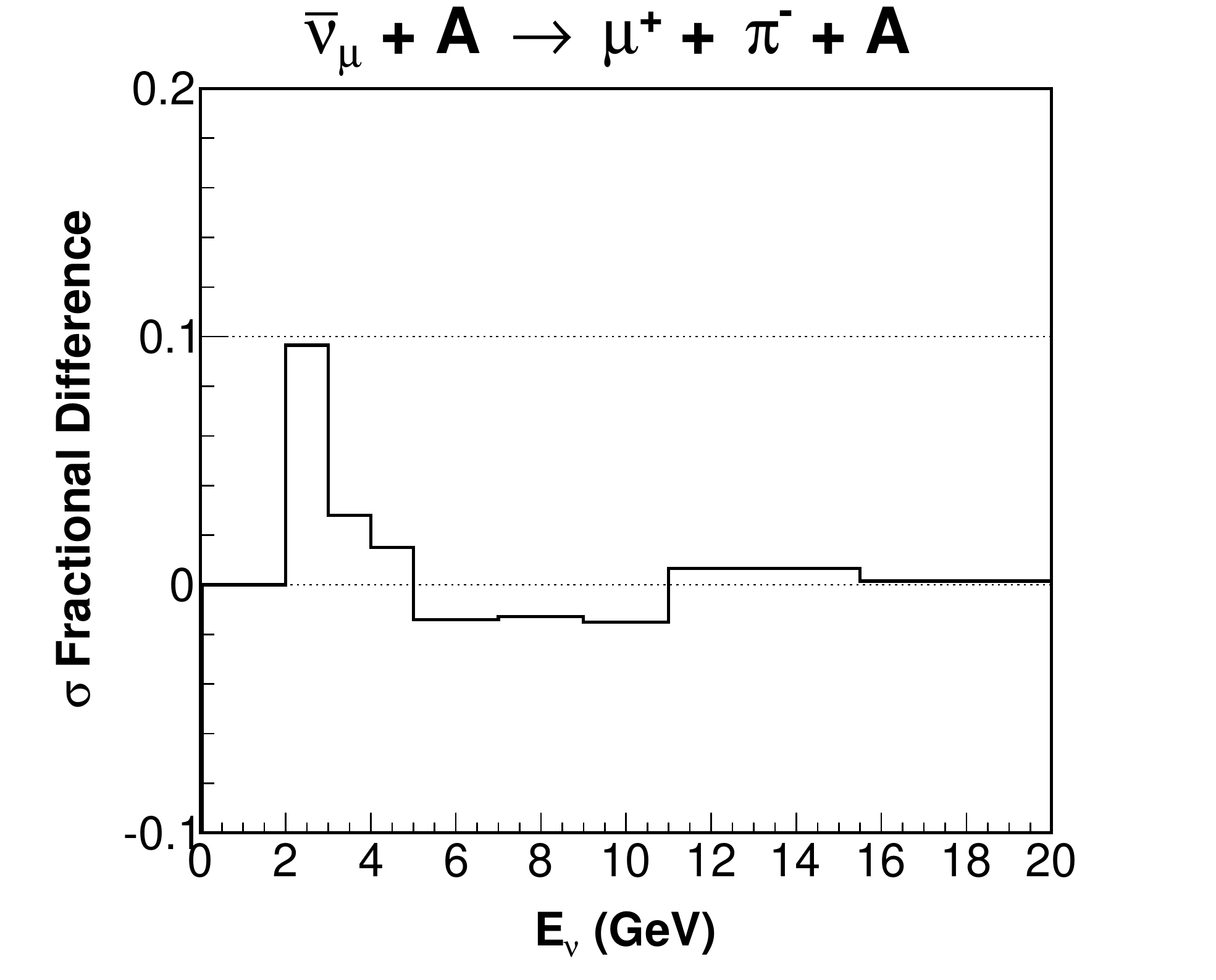}}
\caption[The effect of the signal model weighting on the measured \numu and \numubar \sigenu]
{The effect of the signal model weighting on the measured \numu (left) and \numubar (right) \sigenu.}
\label{fig:signal_model_weighting_sigenu}
\end{figure*}

\begin{figure*}[tpb]
\centering
\mbox{
\includegraphics[width=0.49\linewidth]{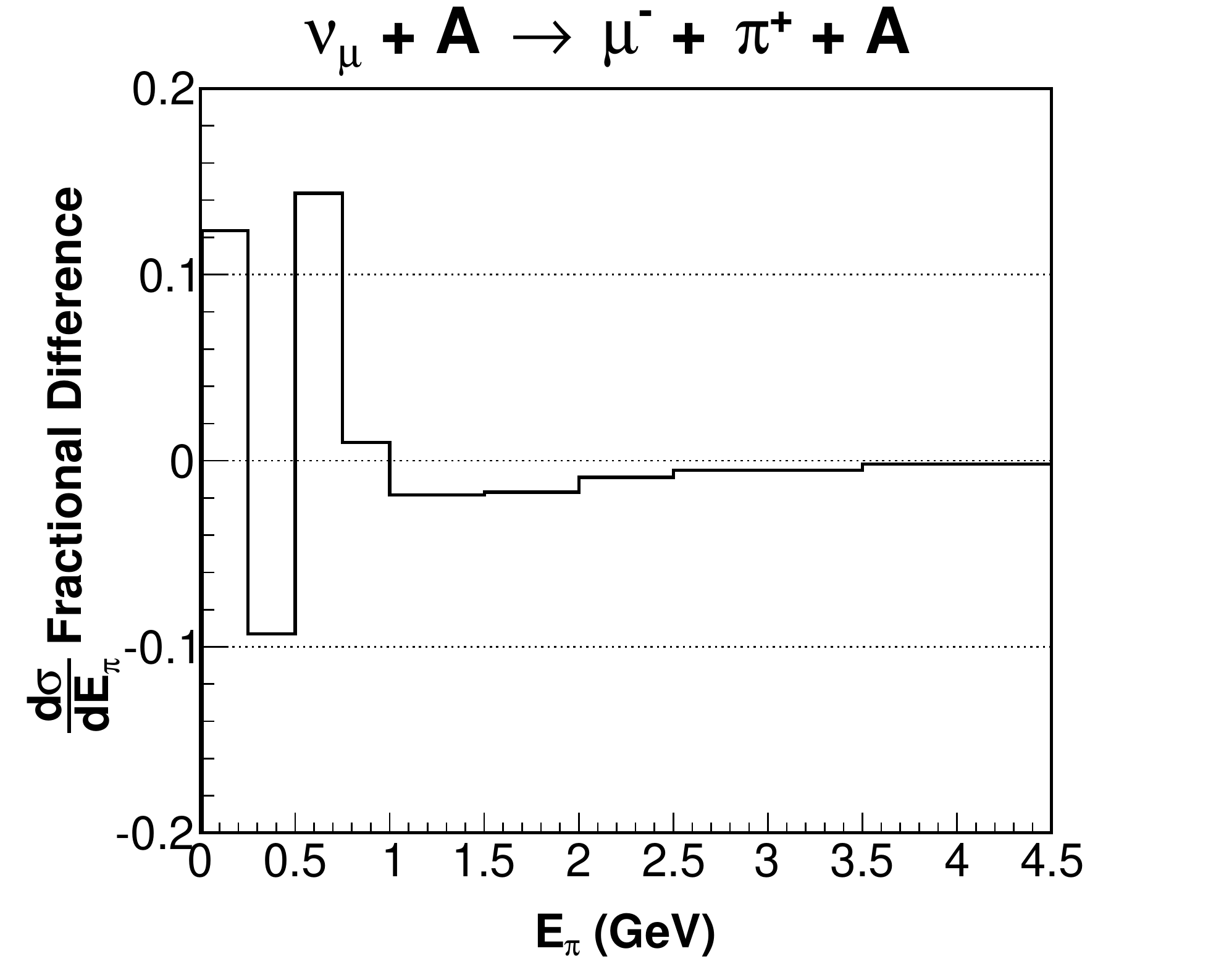}
\includegraphics[width=0.49\linewidth]{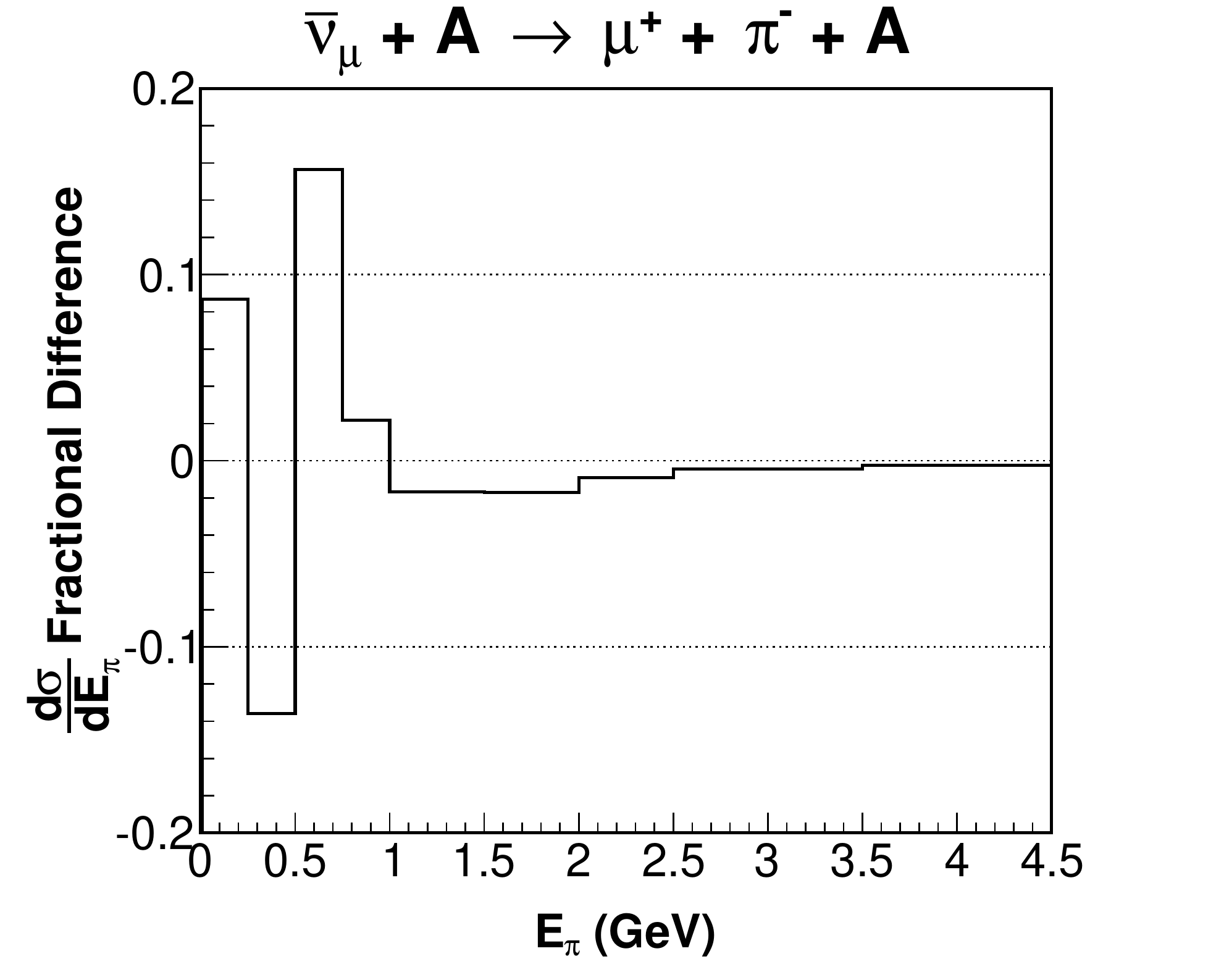}}
\caption[The effect of the signal model weighting on the measured \numu and \numubar \dsigdepi]
{The effect of the signal model weighting on the measured \numu (left) and \numubar (right) \dsigdepi.}
\label{fig:signal_model_weighting_dsigdepi}
\end{figure*}

\begin{figure*}[tpb]
\centering
\mbox{
\includegraphics[width=0.49\linewidth]{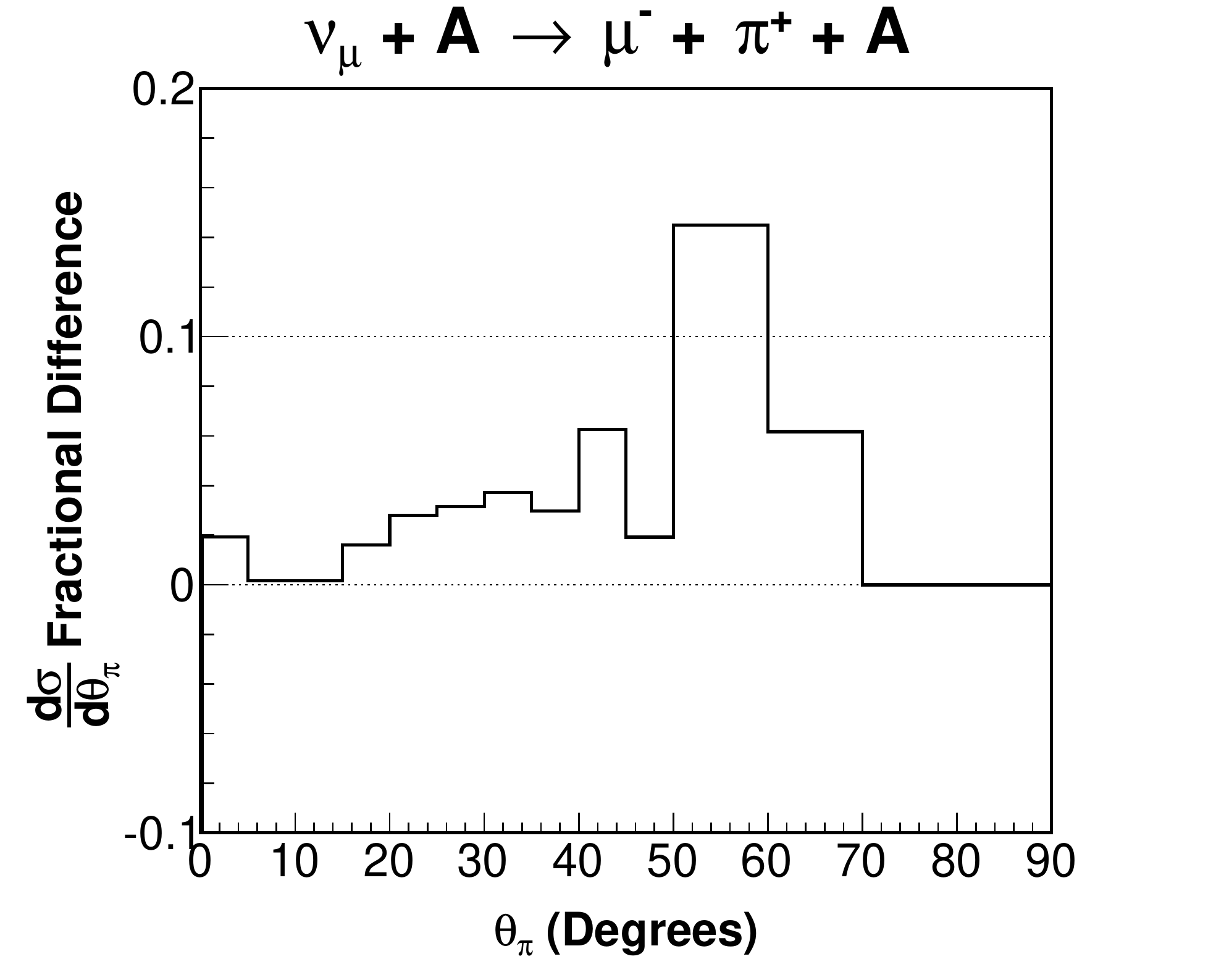}
\includegraphics[width=0.49\linewidth]{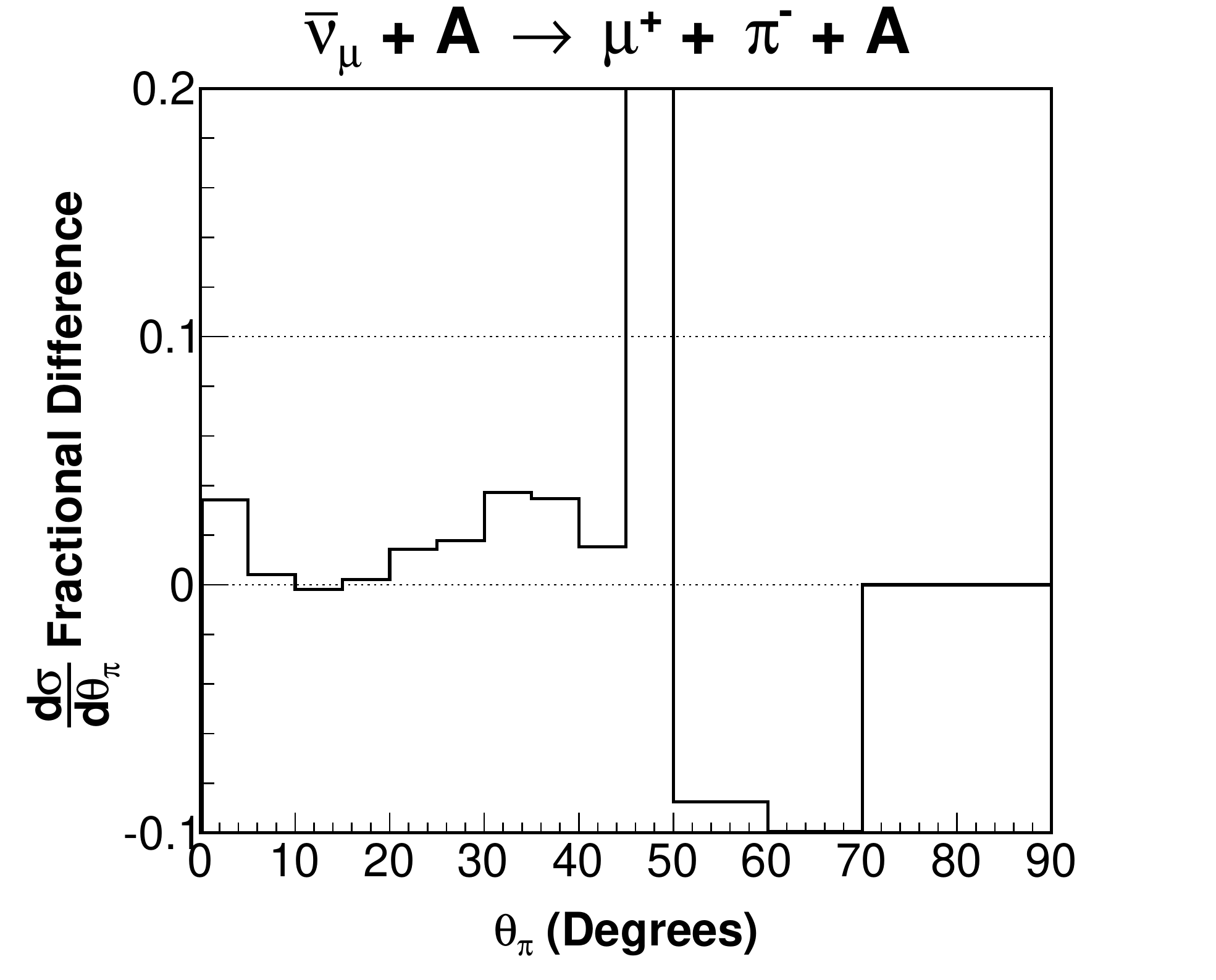}}
\caption[The effect of the signal model weighting on the measured \numu and \numubar \dsigdthetapi]
{The effect of the signal model weighting on the measured \numu (left) and \numubar (right) \dsigdthetapi.}
\label{fig:signal_model_weighting_dsigdthetapi}
\end{figure*}

\begin{figure*}[tpb]
\centering
\mbox{
\includegraphics[width=0.49\linewidth]{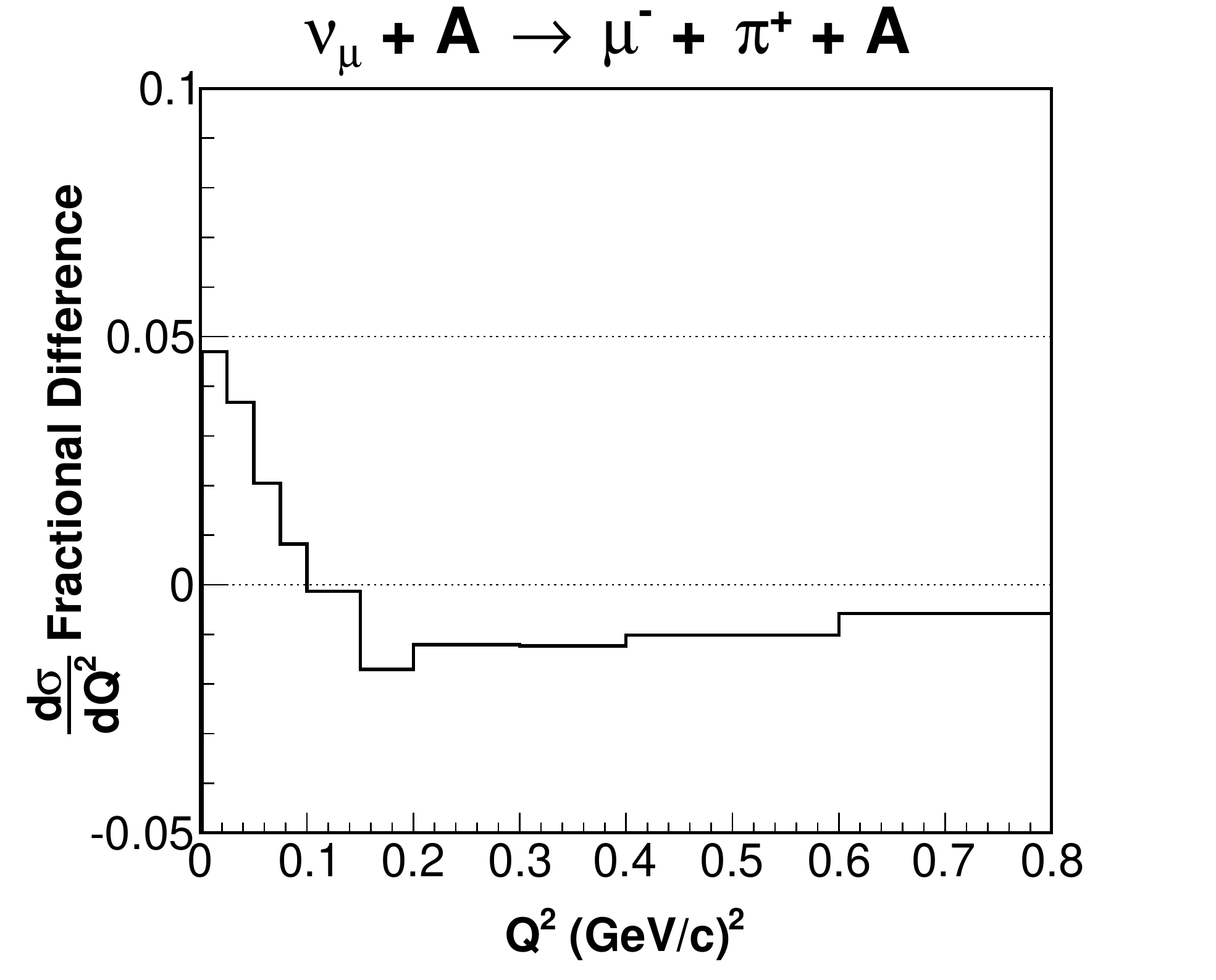}
\includegraphics[width=0.49\linewidth]{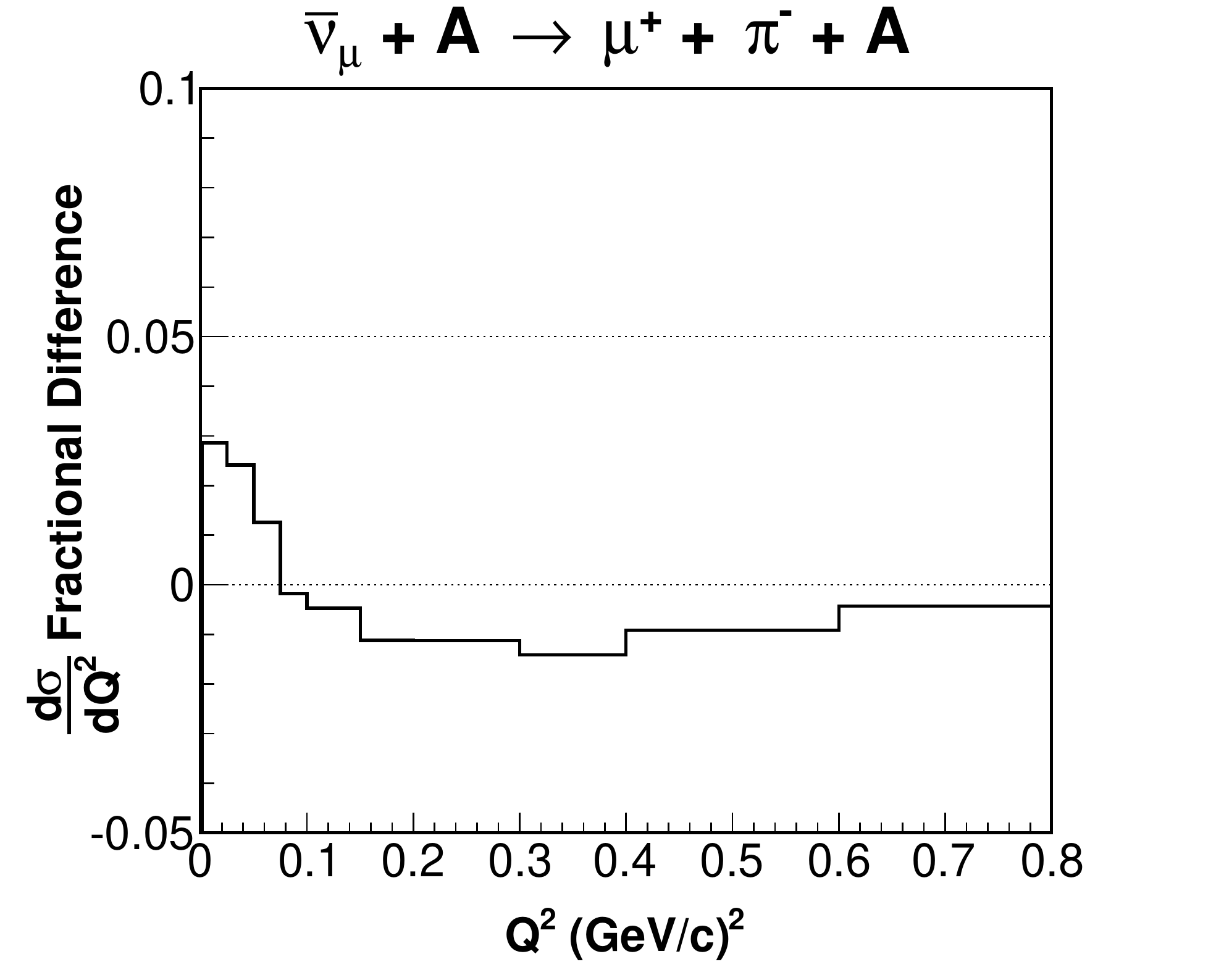}}
\caption[The effect of the signal model weighting on the measured \numu and \numubar \dsigdqsq]
{The effect of the signal model weighting on the measured \numu (left) and \numubar (right) \dsigdqsq.}
\label{fig:signal_model_weighting_dsigdqsq}
\end{figure*}

%The unfolding matrices and efficiency corrections are dependent on the kinematics of the signal model.  
To minimize bias on the measured cross sections from the signal model, the signal model weighting was applied to
coherent events in the MC, the unfolding matrices and efficiency corrections were re-estimated, and the cross sections
were remeasured.  The effect of the signal model weighting on the measured cross sections is shown in
Figs.~\ref{fig:signal_model_weighting_sigenu}--\ref{fig:signal_model_weighting_dsigdqsq}.  
%The signal model weighting increased the measured \numu and \numubar cross sections at low-\enu, high-\thetapi, and low-\qsq.  This kinematic region corresponds to the low-\epi region suppressed by the signal model weighting.  The MC coherent event rate and selection efficiency are maximal near near \epi = 0.5 GeV (Figure~\ref{fig:efficiency_epi}).  Therefore, the signal model weighting reduces the selection efficiency, and thereby the measured cross section, at low-\enu, high-\thetapi, and low-\qsq.
%The signal model weighting reduced the measured \numu and \numubar \dsigdepi by $\sim$10\% in the 0.25-0.5 GeV bin while increasing the cross section in the adjacent bins by a similar amount (Figure~\ref{fig:signal_model_weighting_dsigdepi}).  
The signal model weighting did not actually change the selection efficiency as a function of \epi since events in the numerator and
denominator of the efficiency calculation (Eq.~\ref{eq:efficiency}) in each \epi bin were weighted equally.  Instead,
there was a change to the measured \dsigdepi resulting from the change to the \epi unfolding matrix.  
%The nominal signal model predicts a large peak in the coherent event rate in the 0.25-0.5 GeV \epi bin, and the resulting unfolding matrix predicts a net migration of events from this bin into the adjacent bins.  
The signal model weighting dramatically suppresses the peak in the 0.25-0.5 GeV bin in \epi and thereby the predicted amount of
migration from that bin, resulting in a decrease of the measured \numu and \numubar \dsigdepi in the 0.25-0.5 GeV bin and an increase in the adjacent bins.

\section{Contribution From Diffractive Scattering}
\label{sec:diffractiveMeasure}
Diffractive pion production from free protons may appear in the detector with the same signature as coherent pion production, if the
recoiling proton is undetected.  Although the recoil system is never detected for coherent scattering on heavier nuclei, diffractive
scattering is not the same process and is considered a background.  It was not included in the MC simulations of backgrounds, and
can be constrained from the data.  
An exclusive calculation of the diffractive cross section valid in the kinematic region of the measured
coherent cross sections does not yet exist.  The PCAC-based calculation of diffractive scattering by Rein~\cite{bib:rein_diffractive}
is valid only for $W>2$ GeV, since the interference with $p\pi$ final states from neutrino resonance production must be
calculated for $W<2$ GeV.  For diffractive scattering at small-\tabs, $W>2$ GeV corresponds to \epi\gt 1.5 GeV, which covers
only the high-\epi phase space of the measured coherent cross sections.  
A search for diffractive scattering within the selected coherent candidate sample by looking for ionization
from the recoil proton near the event vertex is presented here.

\subsection{Diffractive Acceptance}

\begin{figure*}[tpb]
\begin{center}
%\mbox{\includegraphics[width=0.5\linewidth]{figures/Diffractive/h_relative_acceptance_neutrino.pdf}
%\includegraphics[width=0.5\linewidth]{figures/Diffractive/h_relative_acceptance_antineutrino.pdf}}
\mbox{\includegraphics[width=0.5\linewidth]{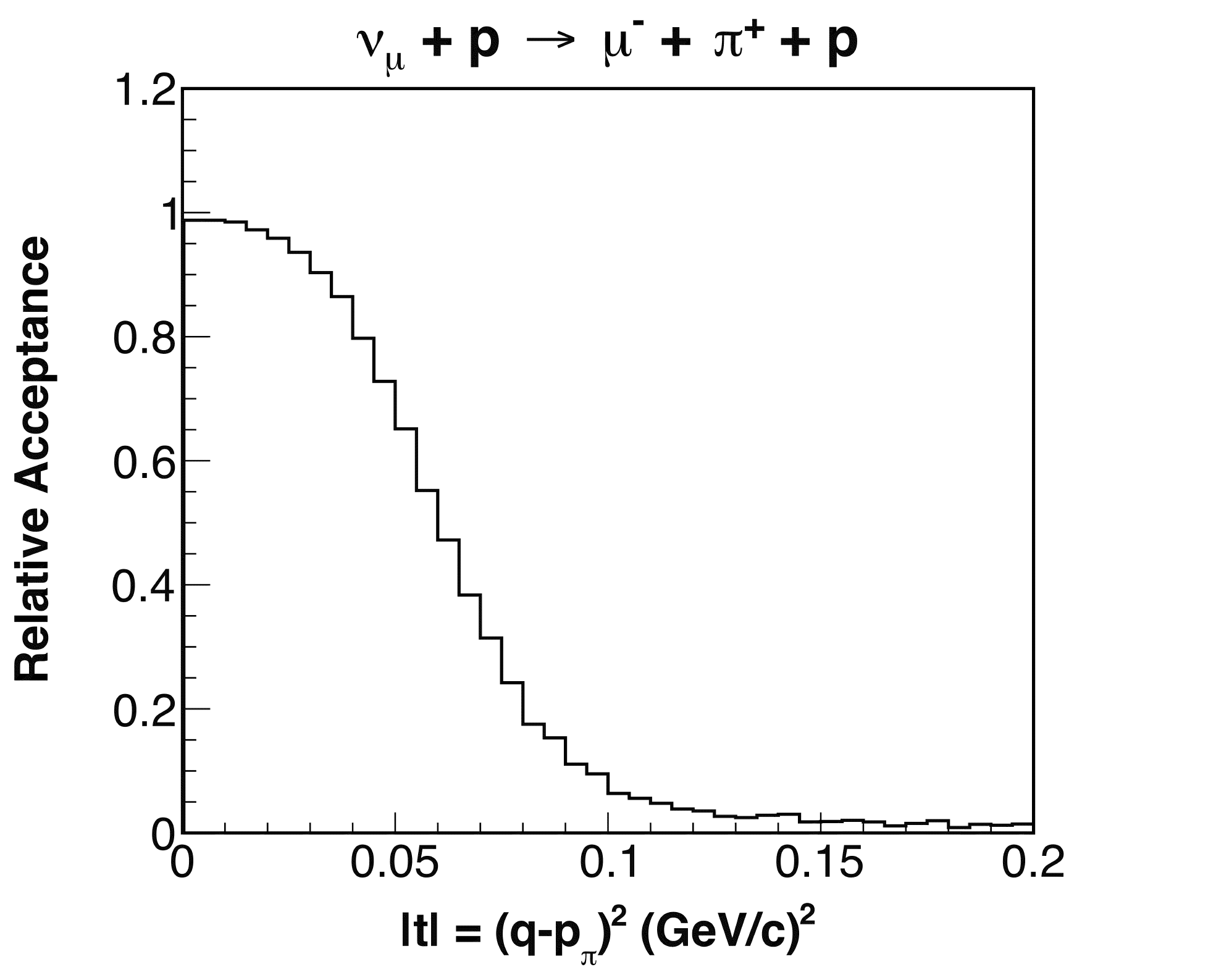}
\includegraphics[width=0.5\linewidth]{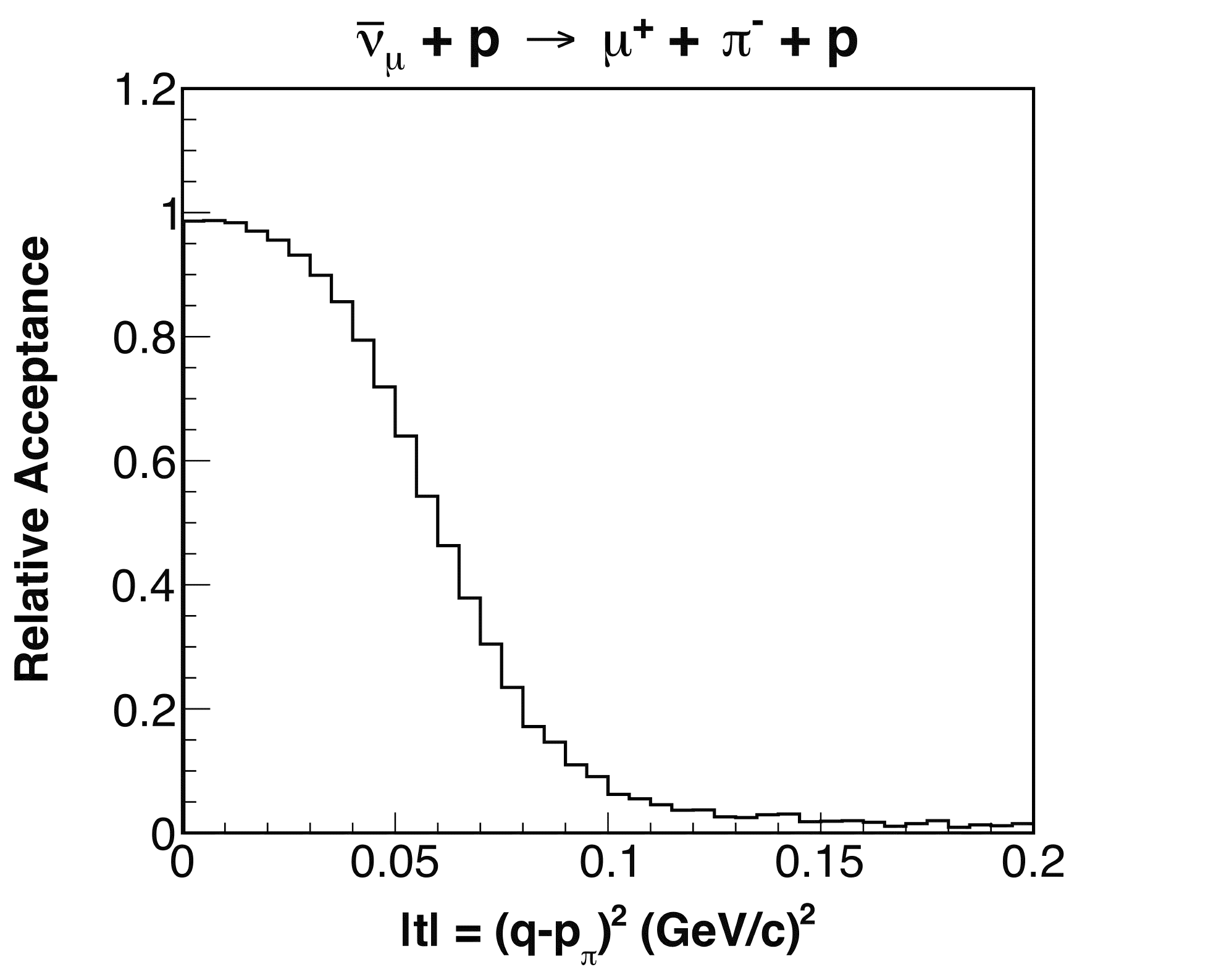}}

%\mbox{\includegraphics[width=0.5\linewidth]{figures/Diffractive/h_coherent_acceptance_neutrino.pdf}
%\includegraphics[width=0.5\linewidth]{figures/Diffractive/h_coherent_acceptance_antineutrino.pdf}}
\mbox{\includegraphics[width=0.5\linewidth]{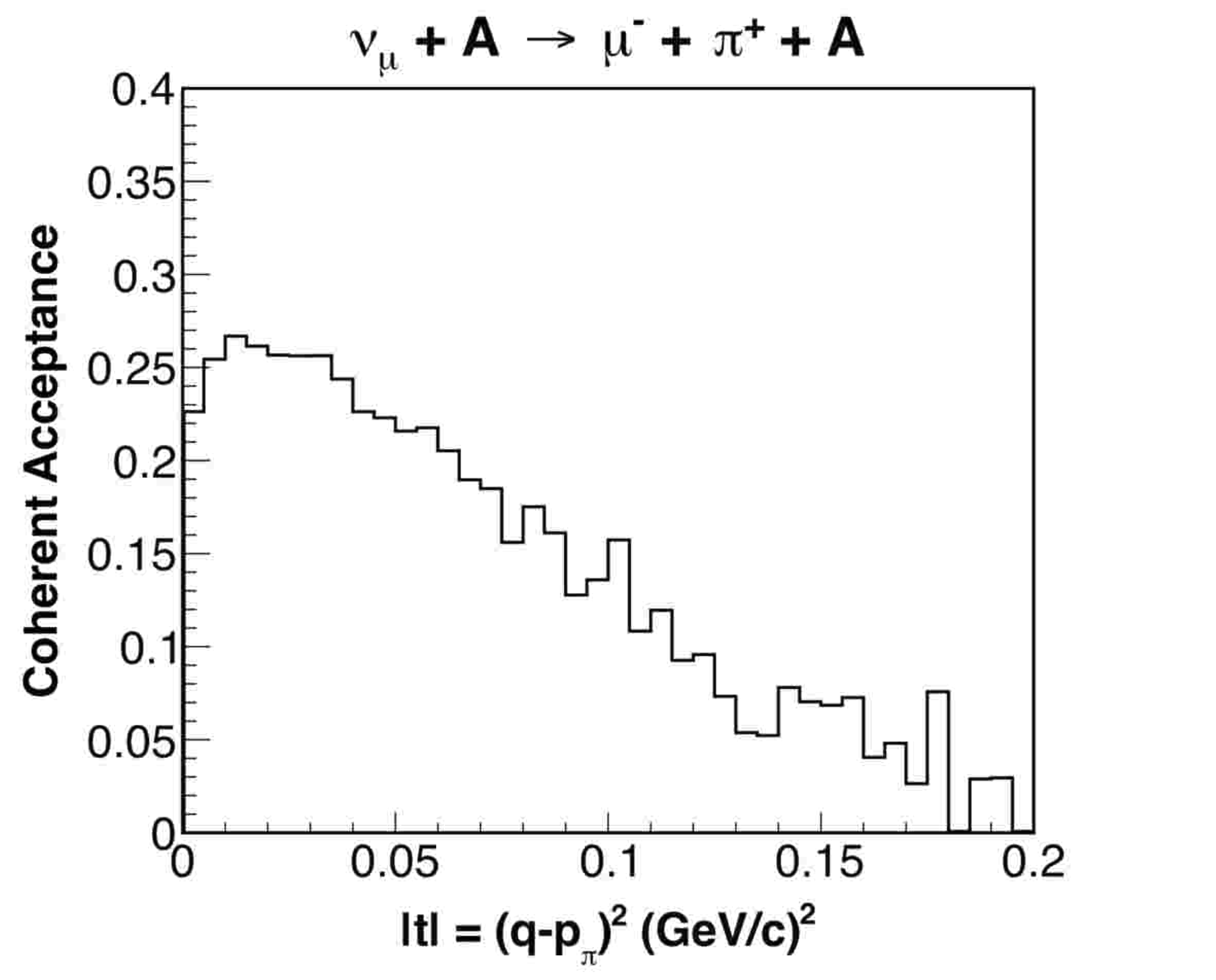}
\includegraphics[width=0.5\linewidth]{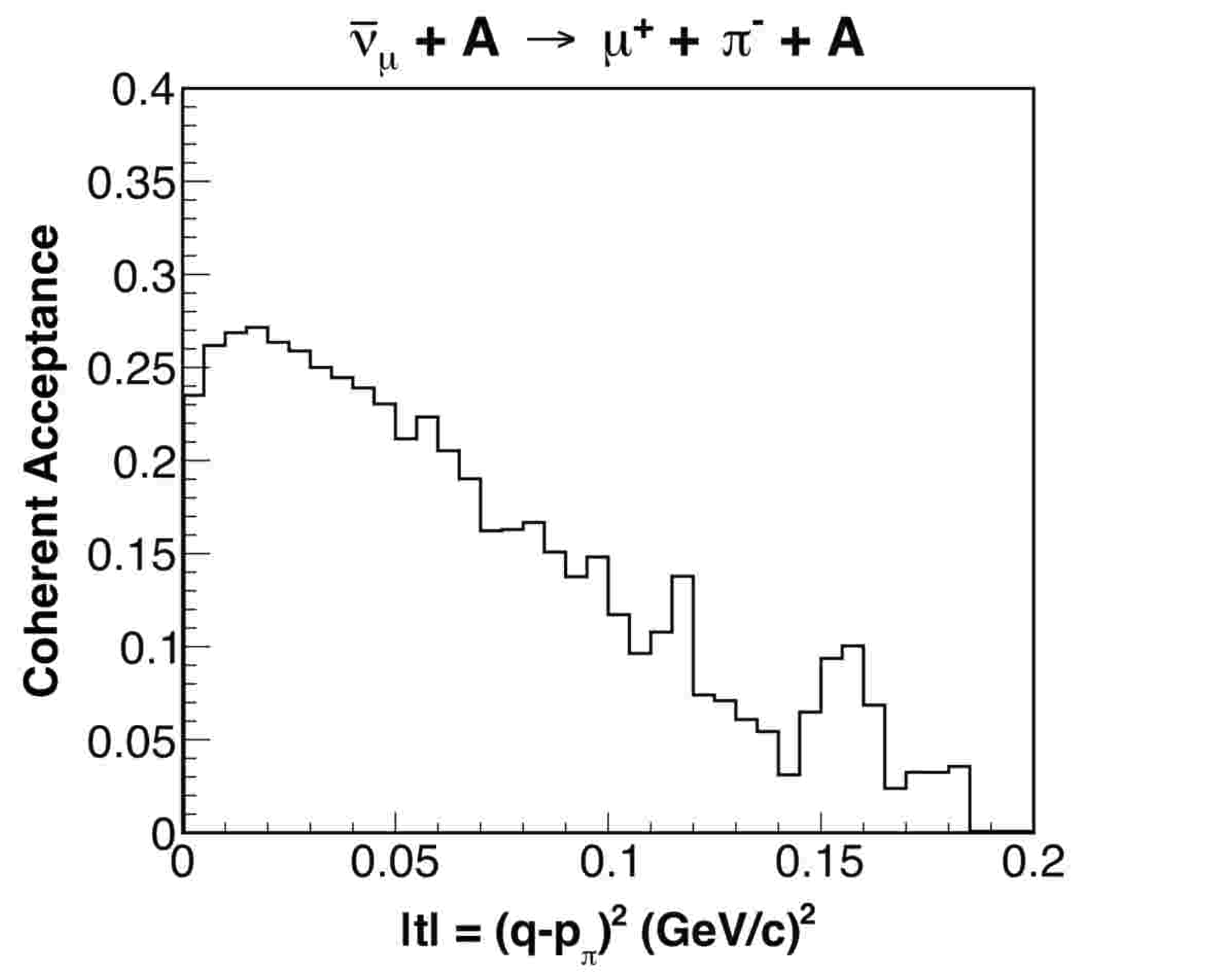}}

%\mbox{\includegraphics[width=0.5\linewidth]{figures/Diffractive/h_diffractive_acceptance_neutrino.pdf}
%\includegraphics[width=0.5\linewidth]{figures/Diffractive/h_diffractive_acceptance_antineutrino.pdf}}
\mbox{\includegraphics[width=0.5\linewidth]{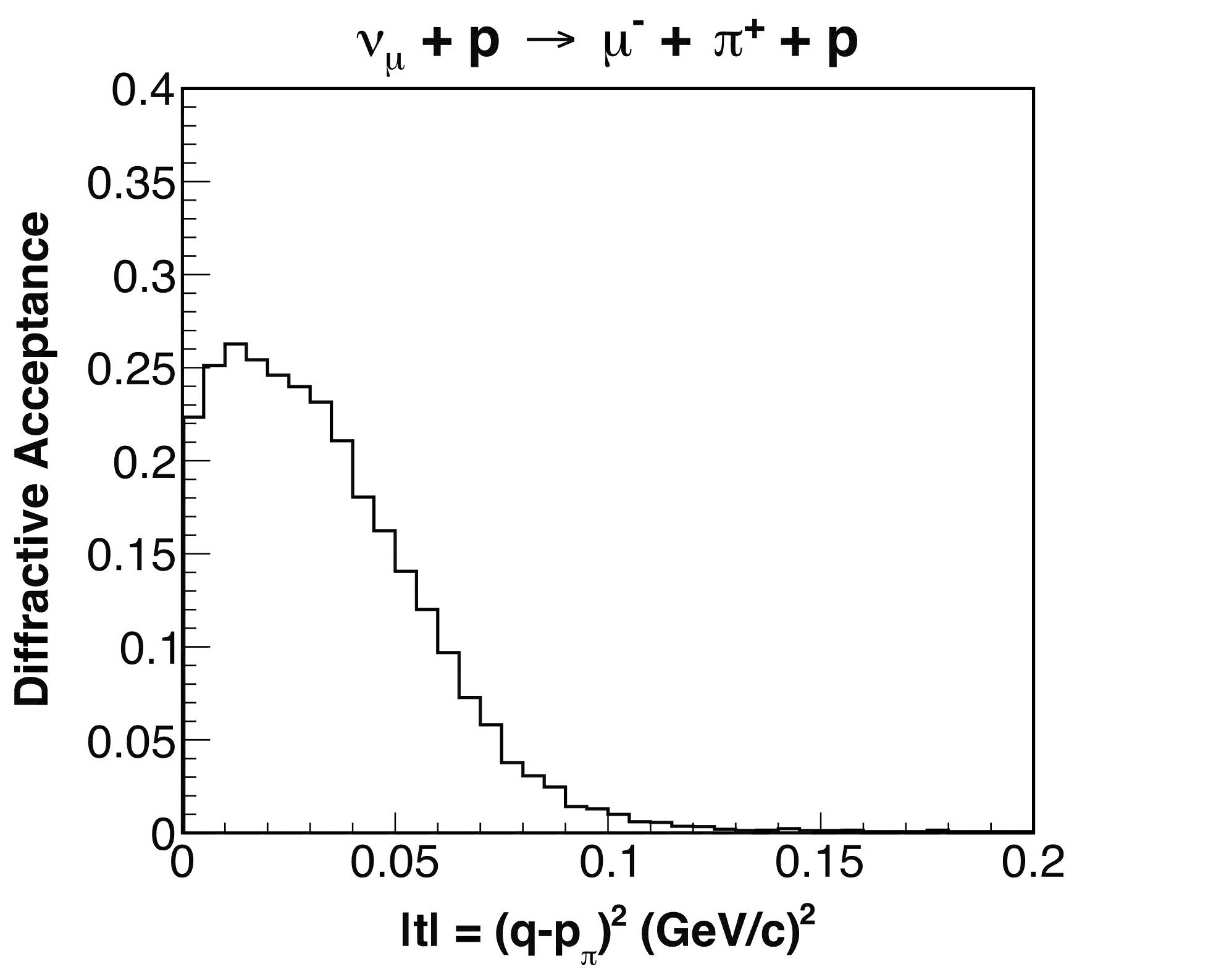}
\includegraphics[width=0.5\linewidth]{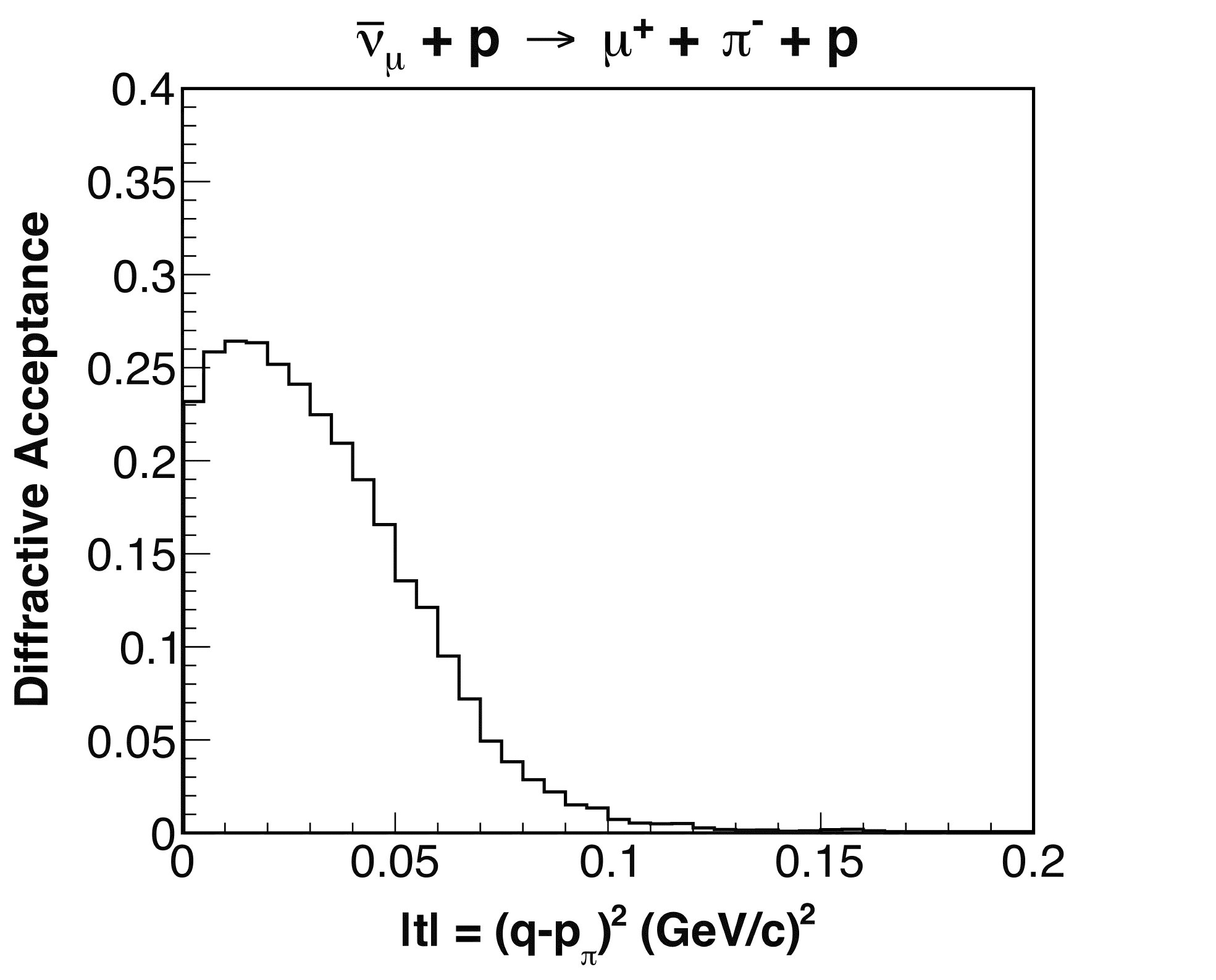}}
\caption[The estimated diffractive acceptance as a function of \tabs]{The \numu (left) and \numubar (right)
estimated relative diffractive-to-coherent acceptance (top), coherent acceptance (middle), and absolute
diffractive acceptance (bottom) as a function of \tabs.  The absolute diffractive acceptance was estimated by
weighting the coherent acceptance by the relative diffractive-to-coherent acceptance.}
\label{fig:diffractive_acceptance}
\end{center}
\end{figure*}

Coherent and diffractive scattering in the detector only differ in acceptance because of the presence of the recoiling proton.
The relative diffractive-to-coherent acceptance of the vertex energy cut is a function of \tabs since the kinetic energy of
the proton, \Tp, is proportional to \tabs.  This relative acceptance was estimated using a distribution of vertex energy
deposited by recoil protons as a function of \Tp, which was estimated from a simulation of single protons
originating in the fiducial volume and isotropic in direction along with a coherent MC sample passing all selection cuts up
to the vertex energy cut.  For each event, the vertex energy from a recoil proton with kinetic energy $T_{p}$ was added to
the vertex energy of the event.  The relative diffractive-to-coherent acceptance (Fig.~\ref{fig:diffractive_acceptance})
was calculated as the ratio of the vertex energy cut acceptance with this added vertex energy to the acceptance without
this added vertex energy, as a function of \tabs, which was calculated event-by-event.
% from \Tp (Equation~\ref{eq:t_diff}).
The vertex energy cut acceptance is insensitive to differences between diffractive and coherent scattering in the muon
and pion kinematics since the energy deposited by the muon and pion is corrected to normal incidence in the vertex energy calculation.

The vertex energy cut rejects nearly all diffractive events with \tabs \gt 0.125 \gevpercsq.  Since the signal sample has
reconstructed \tabs \lt 0.125 \gevpercsq, both types of events have low \tabs, and the acceptance of the \tabs cut is
approximately the same for both of them after cutting on vertex energy.  The  acceptance as a function of true \tabs can
therefore be estimated by weighting the relative diffractive-to-coherent acceptance of the vertex energy cut by the total
selection efficiency for coherent scattering as a function of true \tabs (Fig.~\ref{fig:diffractive_acceptance}).
Note though that the coherent acceptance is non-zero above \tabs = 0.125 \gevpercsq due to the reconstructed \tabs resolution.

The diffractive scattering contribution to the sideband (0.2 \lt \tabs \lt 0.6 \gevpercsq) is negligible
%since the vertex energy cut rejects nearly all diffractive events with \tabs \gt 0.125 \gevpercsq.  Therefore,
and diffractive scattering is neglected in tuning the GENIE prediction of the incoherent backgrounds.

\subsection{Diffractive Cross Section Estimate}

%With the estimated diffractive acceptance as a function of \tabs, the diffractive scattering contribution to the measured
%coherent cross sections can be estimated from the predicted diffractive scattering cross section as a function of \tabs,
%\dsigdt.  There is no microphysical calculation of
%diffractive scattering at \minerva energies.

An estimate of the diffractive cross section can be made from a calculation of inclusive \diffractivenumu and \diffractivenumubar
on free protons by Kopeliovich \etal~\cite{bib:kopeliovich} which uses Adler's PCAC relation and pion-nucleus scattering data.
Relative to the GENIE prediction, the Kopeliovich calculation exhibits a low-\tabs enhancement
%that falls exponentially in \tabs.  The difference includes all low-\tabs enhancements,
from all processes, including that from diffractive scattering, not present in GENIE.
The low-\tabs enhancement as extracted from the Kopeliovich calculation is therefore an estimate of the largest
possible diffractive cross section in this model.

\begin{figure*}[tpb]
\centering
\mbox{\includegraphics[width=0.5\linewidth]{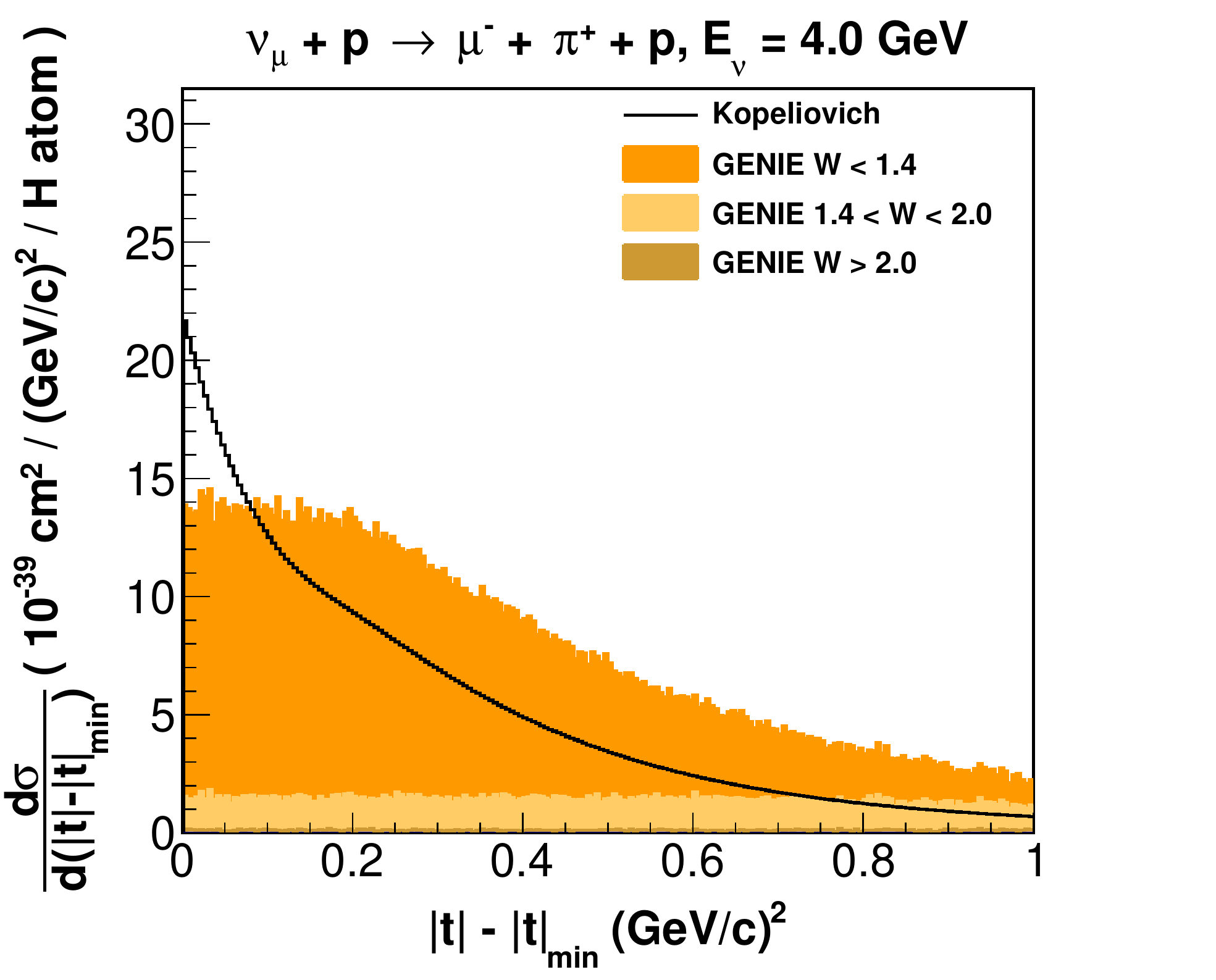}
\includegraphics[width=0.5\linewidth]{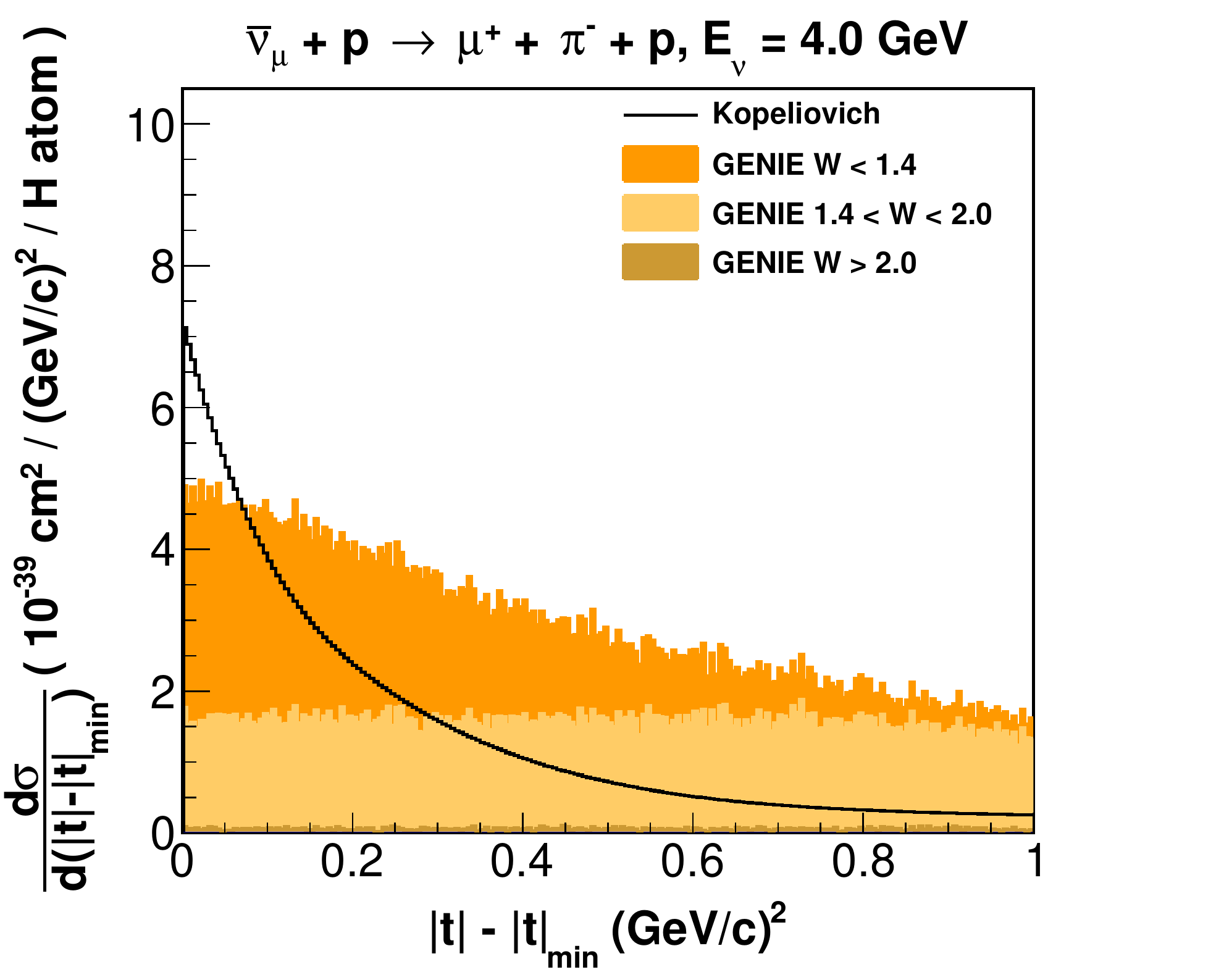}}
\mbox{\includegraphics[width=0.5\linewidth]{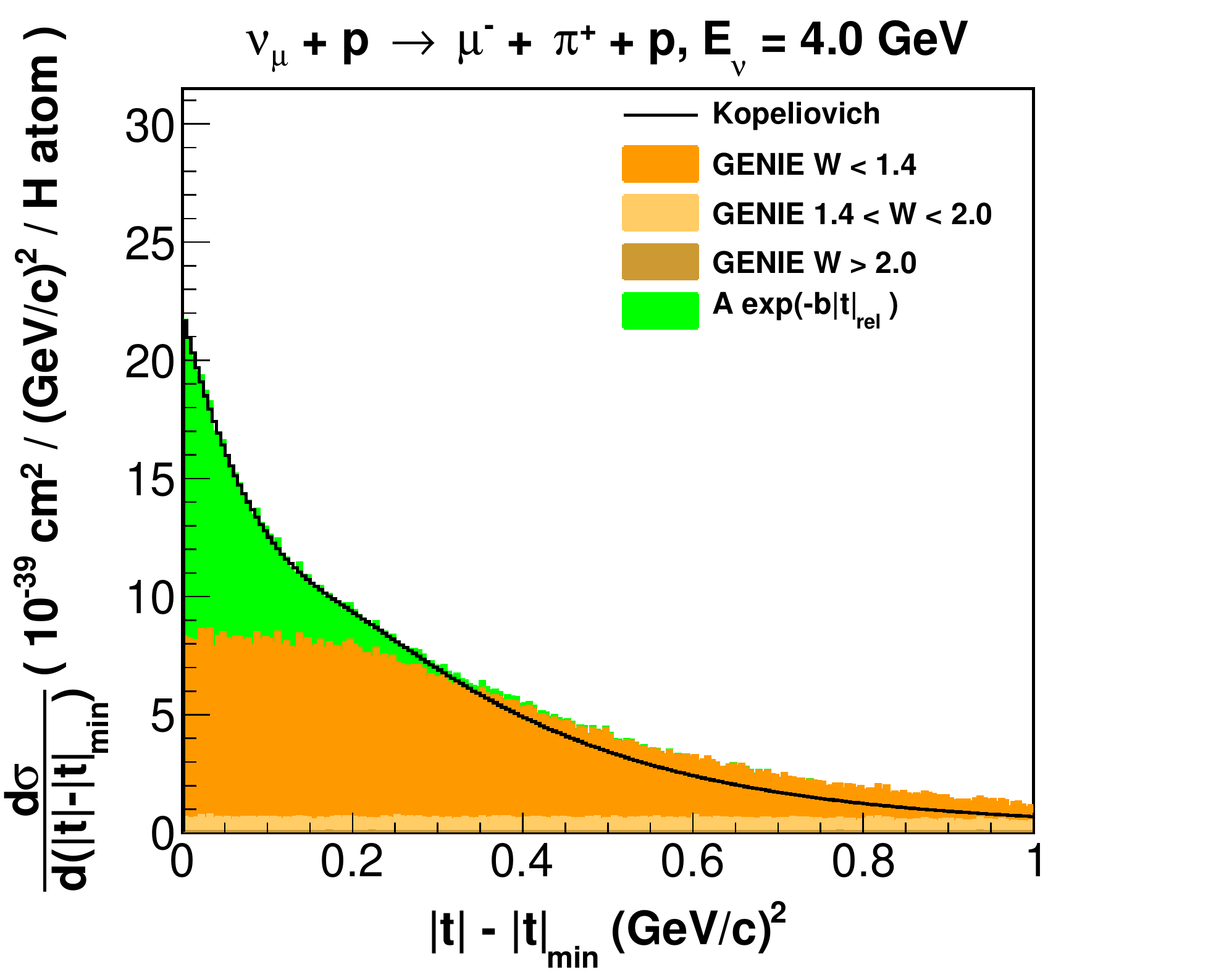}
\includegraphics[width=0.5\linewidth]{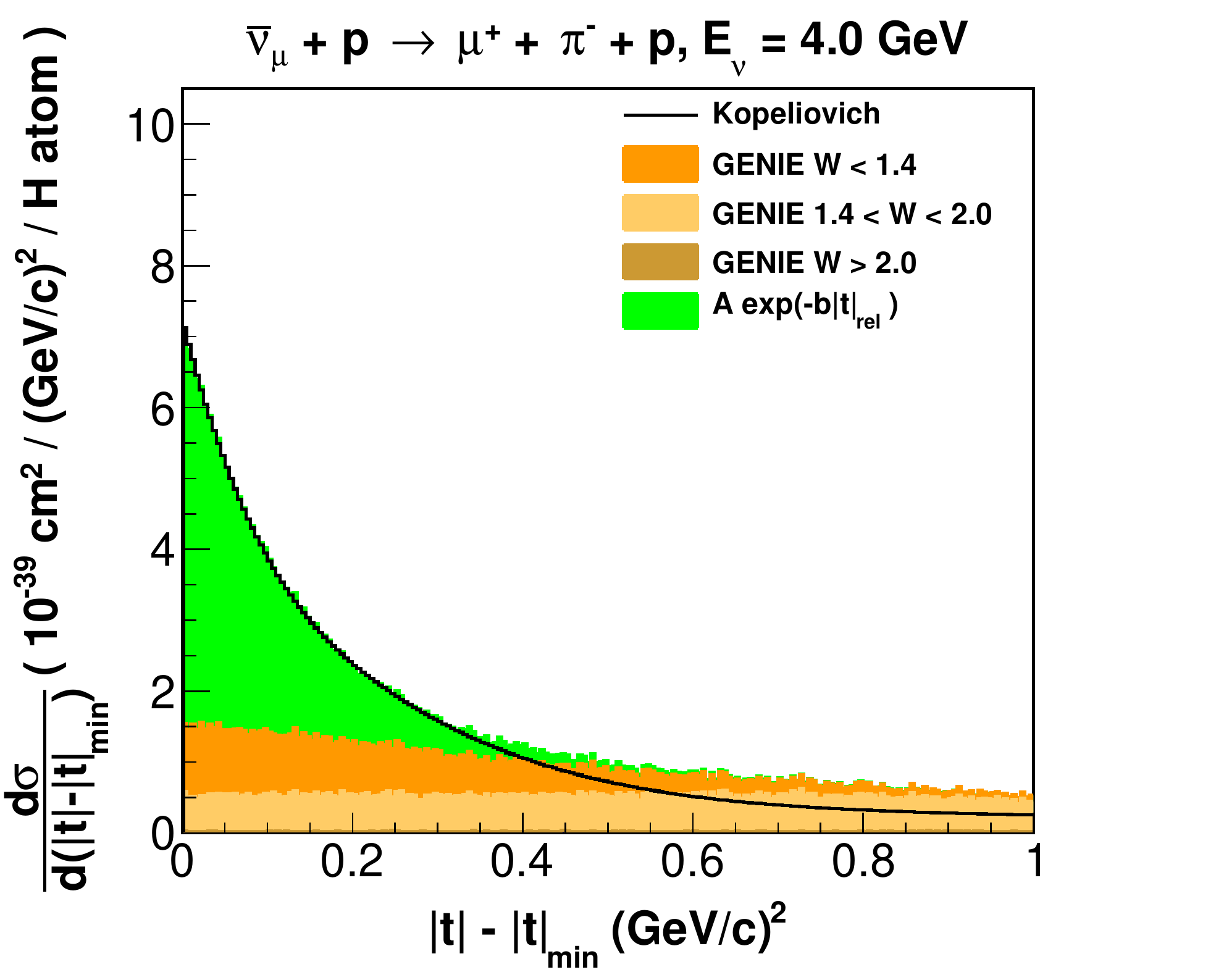}}
\caption[The Kopeliovich and GENIE predictions of \dsigdtrel for inclusive \diffractivenumu and \diffractivenumubar on free protons]{The \numu (left) and \numubar (right) Kopeliovich and GENIE predicted \dsigdtrel for \enu = 4.0 GeV before (top) and after (bottom) fitting the Kopeliovich prediction.  The fit includes an exponential term representing the diffractive \dsigdtrel.}
\label{fig:diff_dsigdtrel_calc}
\end{figure*}

The diffractive and non-diffractive components of the Kopeliovich calculation were estimated by fitting the
GENIE prediction plus an exponential term to the Kopeliovich
\dsigdtrel prediction (Fig.~\ref{fig:diff_dsigdtrel_calc}).
%\tmin is the kinematic minimum \tabs for diffractive scattering.
% to occur and is defined as
%\begin{equation}
%\label{eq:t_min}
%|t|_{min} = \frac{(Q^{2}+m_{\pi}^{2})^{2}-\left[\sqrt{\lambda(W^{2},-Q^{2},m_{n}^{2})}-\sqrt{\lambda(W^{2},m_{\pi}^{2},m_{n}^{2})}\right]^{2}}{4W^{2}},
%\end{equation}
%where $m_{\pi}$ and $m_{n}$ are the pion and nucleon masses and $\lambda(a, b, c) = a^{2}+b^{2}+c^{2}-2ab-2ac-2bc$.
The Kopeliovich cross section was fit as a function of \trel since, for diffractive scattering, \dsigdt will deviate from an
exponential at low \tabs due to \tmin suppression, whereas \dsigdtrel will not.
%have this effect.  Both the normalization and slope of the exponential term were varied in the fit.
%Differences between the Kopeliovich and GENIE predictions of the non-diffractive processes were accounted for by dividing the
%GENIE prediction into regions \invm \lt1.4 GeV and \invm \gt1.4 GeV, due to their shape differences in \tabs, and varying their normalizations in the fit.
The fit range was 0 \lt\trel\lt0.25 \gevpercsq, which covers the range of non-zero diffractive acceptance.
The Kopeliovich \dsigdtrel was fit for \enu = 4.0 GeV, which is near the average \enu of the neutrino flux.
% used for measuring the coherent cross sections.  The exponential normalization, exponential slope, and GENIE normalization scale factors extracted from the fit are listed in Table~\ref{tab:diff_dsigdtrel_fit}.
Further details of the fit may be found in~\cite{bib:Mislivec_thesis}.

%\begin{table}[bp] \small
%\begin{center}
%\begin{tabular}{ l | c | c }
%Fit Parameter & \numu & \numubar \\
%\hline
%Exponential Normalization ( 10$^{-39}$ cm$^{2}$ / \gevpercsq\ / H atom ) &  13.8 & 5.7 \\
%Exponential Slope (GeV/c)$^{-2}$ & 11.1 & 8.3 \\ 
%GENIE \invm \lt 1.4 GeV Normalization Scale Factor & 0.62 & 0.42 \\
%GENIE \invm \gt 1.4 GeV Normalization Scale Factor & 0.30 & 0.33 \\
%\end{tabular}
%\end{center}
%\caption{Kopeliovich \dsigdtrel fit parameters for \enu = 4.0 GeV.}
%\label{tab:diff_dsigdtrel_fit}
%\end{table}

%The diffractive \dsigdt is needed in order to account for the diffractive acceptance, estimated as a function of \tabs,
%in estimating the diffractive contribution to the measured coherent cross sections.  The diffractive acceptance could
%not instead be estimated as a function of \trel since calculating \tmin requires knowing the diffractive \qsq and \invm event-by-event.
The diffractive \dsigdt 
%(Figure~\ref{fig:diff_dsigdt_est})
was estimated by subtracting the GENIE \dsigdt from the Kopeliovich \dsigdt
%for \enu = 4.0 GeV (Figure~\ref{fig:diff_dsigdt_calc})
, where the GENIE \dsigdt was scaled by the normalization scale factors extracted from the fit to the Kopeliovich \dsigdtrel.

The \numu and \numubar diffractive cross sections at \enu = 4 GeV (Table~\ref{tab:diff_sigma}) were obtained by integrating
the exponential extracted from the fit to the Kopeliovich \dsigdtrel.  The \numu (\numubar) diffractive cross section is
34\% (19\%) of the GENIE coherent cross section on carbon at \enu = 4 GeV.
%(Table~\ref{tab:diff_sigma}).
The acceptance-reduced diffractive cross sections
%(Table~\ref{tab:diff_sigma})
were calculated by weighting the diffractive \dsigdt by the relative diffractive-to-coherent acceptance of the vertex energy cut
as a function of \tabs
%(Figure~\ref{fig:diff_dsigdt_est})
and integrating over \tabs.
%When measuring the coherent cross sections, the rate of the accepted diffractive events is corrected by the coherent selection efficiency, where the primary
%difference between the coherent and diffractive selection efficiencies comes from the acceptance of the vertex energy cut.
The diffractive scattering contribution to the measured \numu (\numubar) coherent cross sections is 8\% (4\%).
%, which is consistent with the estimate of~\cite{bib:minerva_coh}.
Again, this is an estimate of the largest possible diffractive contribution.

%The acceptance reduced diffractive cross sections (Table~\ref{tab:diff_sigma}) were calculated by integrating the estimated relative acceptance weighted diffractive \dsigdt.  The diffractive scattering contribution to the measured \numu (\numubar) coherent cross sections is thereby estimated to be 8\% (4\%).  The measured coherent cross sections are not corrected for this possible contribution.

\begin{table}[bp] \small
\begin{center}
\begin{tabular}{ l | c | c }
                       & \multicolumn{2}{c}{$\sigma$ ( 10$^{-39}$ cm$^{2}$ / atom )} \\
					   & \numu & \numubar \\
\hline\hline
Diffractive on H & 1.24 (0.34) & 0.69 (0.19) \\ \hline
Acceptance Reduced & \multirow{2}{*}{0.28 (0.08)} & \multirow{2}{*}{0.15 (0.04)} \\
Diffractive on H & & \\ \hline
GENIE Coherent on $^{12}$C & 3.64 & 3.64 \\
\end{tabular}
\end{center}
\caption{The maximum estimated diffractive cross section, the same cross section with acceptance reduction, and the GENIE coherent cross section for comparison at \enu = 4.0 GeV.  The numbers in the parentheses are the fraction of the coherent cross section.}
\label{tab:diff_sigma}
\end{table}

\subsection{Search for Diffractive Scattering}

The search for diffractive interactions within the selected coherent candidate samples looks for ionization from the
recoil proton near the event vertex.  Accepted diffractive interactions are estimated to have \tabs $\lesssim$ 0.1 \gevpercsq,
corresponding to a recoil proton with \Tp $\lesssim$ 50 MeV and range $\lesssim$ 2 cm in the scintillator.  The search
region for the recoil proton ionization extends $\pm$2 planes (34 mm of scintillator) in the longitudinal direction,
and $\pm$2 strip widths (66 mm of scintillator) in the transverse direction, from the event vertex.  For selected
diffractive interactions, the recoil proton is identified by a large energy deposition in a single strip inside the search region.

\begin{figure*}[tpb]
\centering
\mbox{\includegraphics[width=0.5\linewidth]{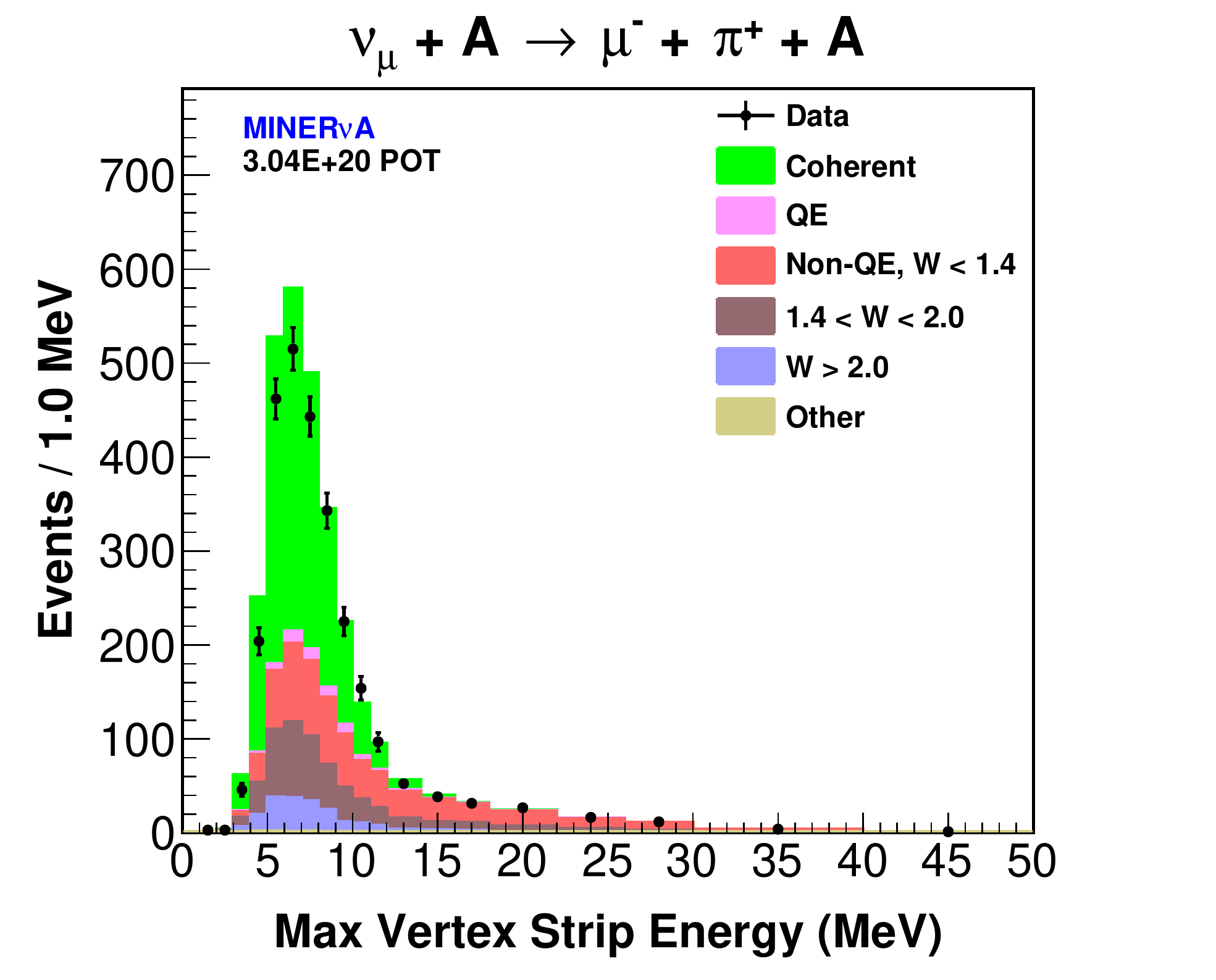}
\includegraphics[width=0.5\linewidth]{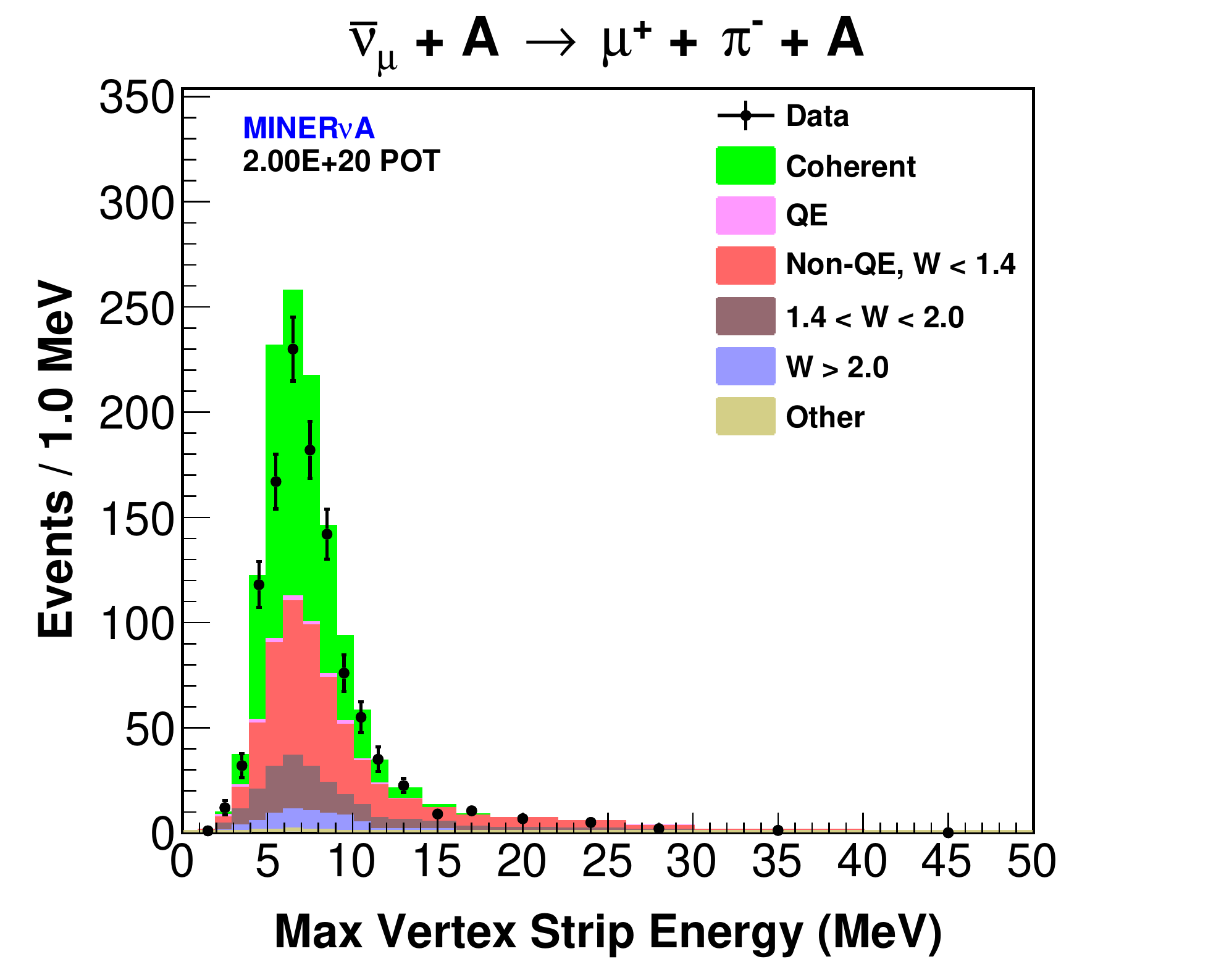}}
\mbox{\includegraphics[width=0.5\linewidth]{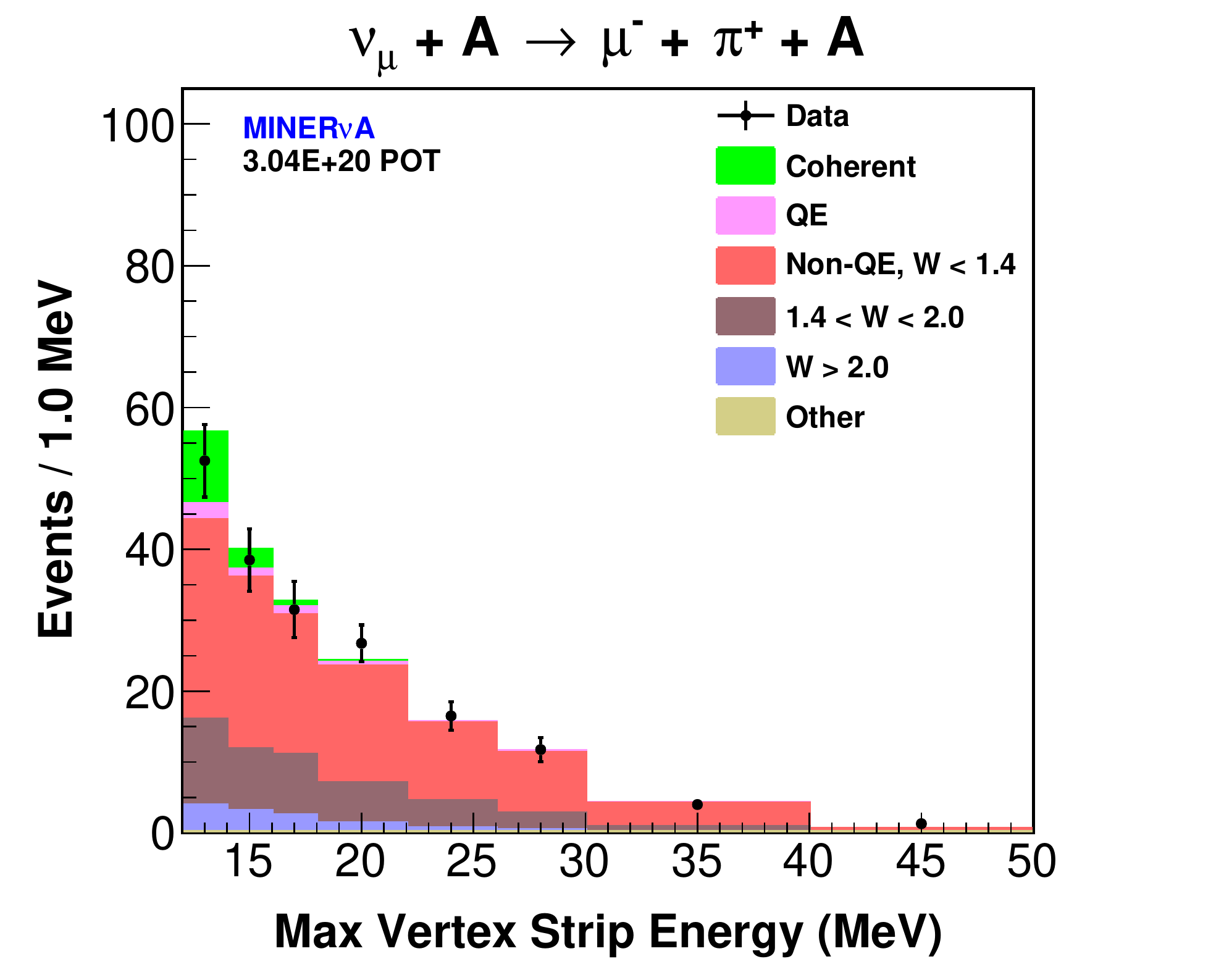}
\includegraphics[width=0.5\linewidth]{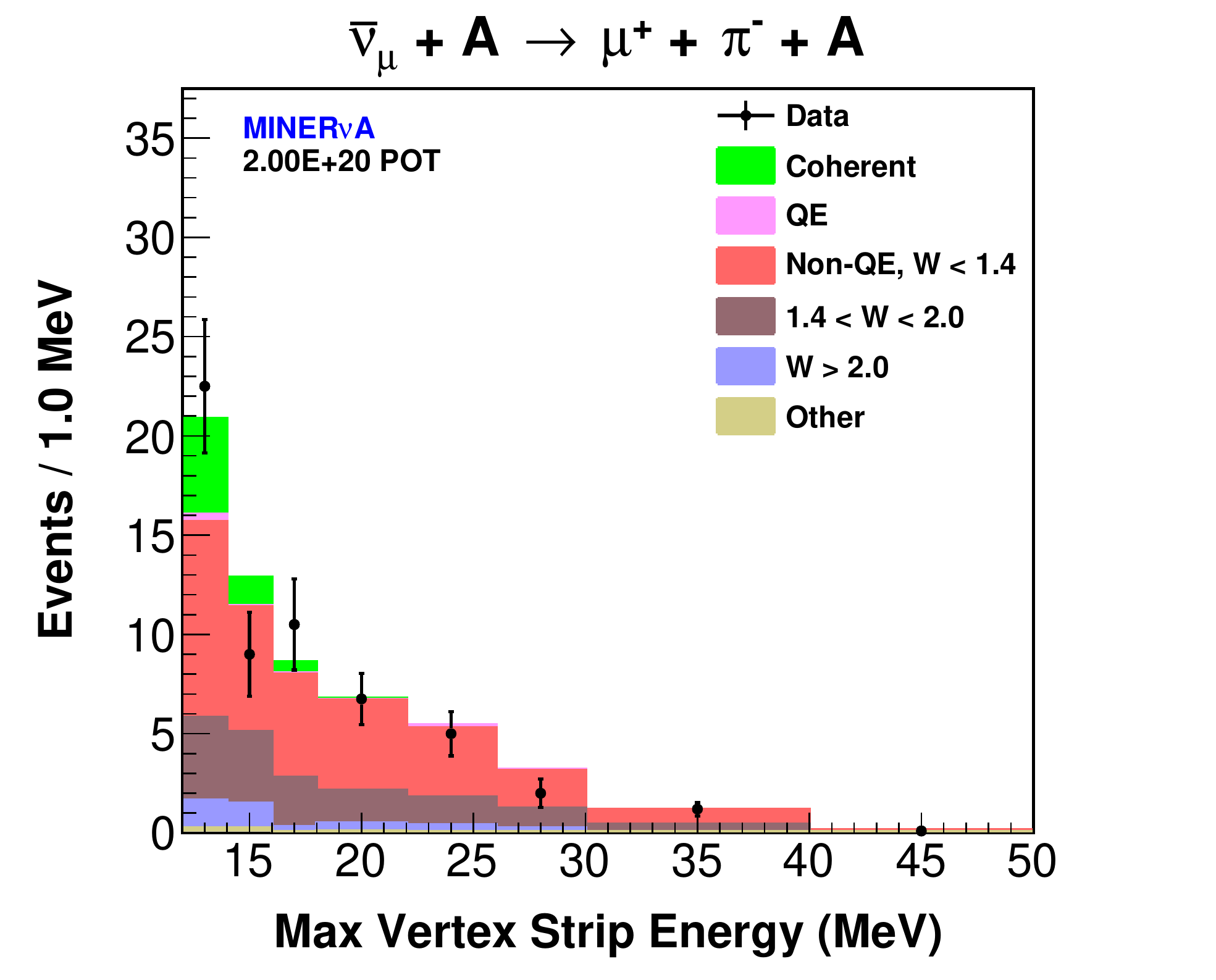}}
\caption[The MVSE distribution for the \numu and \numubar selected coherent candidate samples]{The MVSE for the \numu (left) and \numubar (right) selected coherent candidate samples.  The bottom plots show the high-MVSE region examined for the presence of diffractive scattering.  The non-diffractive background normalizations are tuned.}
\label{fig:mvse}
\end{figure*}

Figure~\ref{fig:mvse} shows the distribution of maximum vertex strip energy (MVSE), defined for each event as the largest amount
of visible energy in a single strip inside the search region, for the \numu and \numubar selected coherent-like samples.
The region of large MVSE where the MC coherent contribution is small is indicative of ionization in addition to that from
a muon and pion only and is analyzed for the presence of diffractive interactions.
%This is done by adding a simulated diffractive sample to the MC and fitting the diffractive sample normalization.

\begin{figure*}[tpb]
\centering
\mbox{\includegraphics[width=0.5\linewidth]{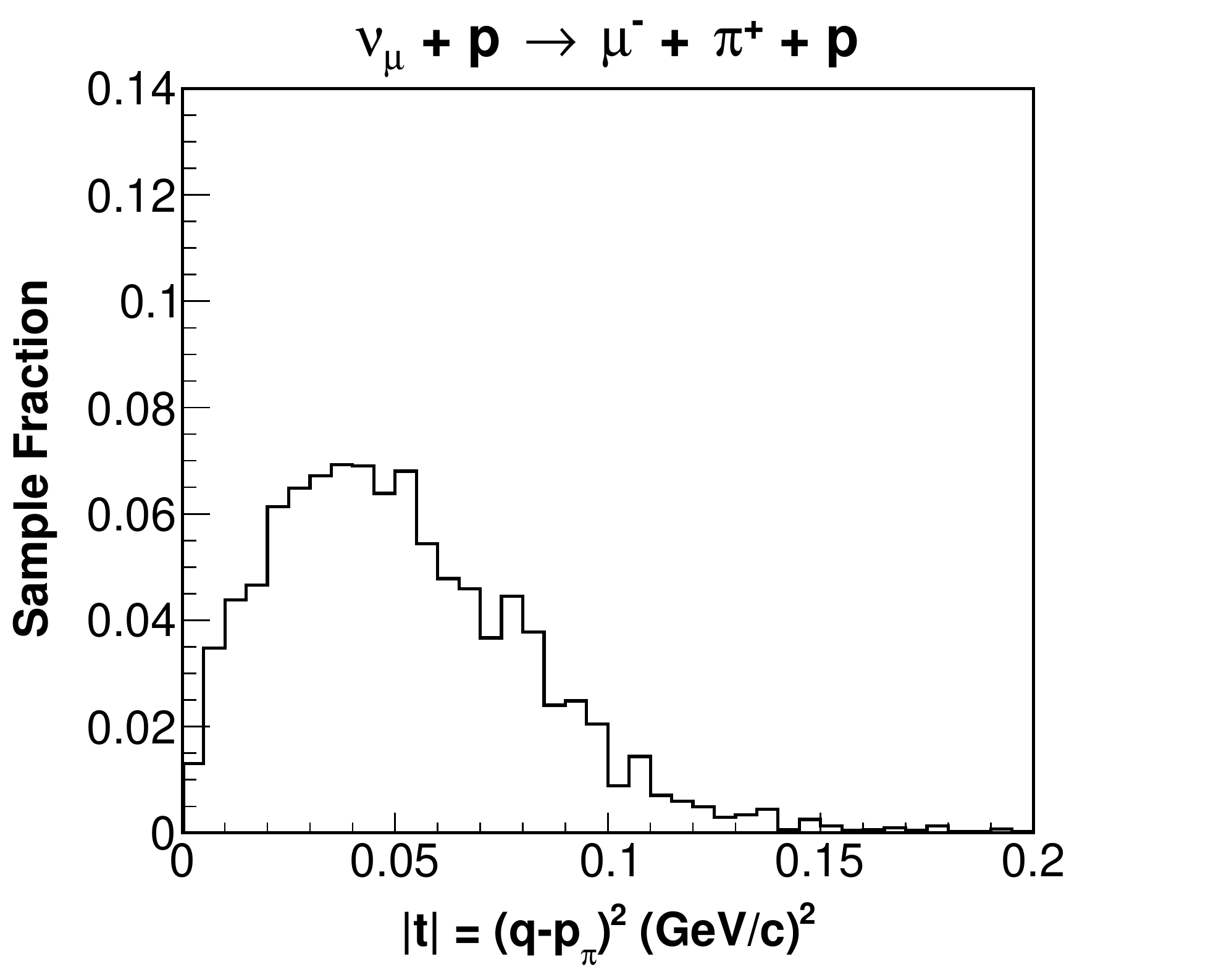}
\includegraphics[width=0.5\linewidth]{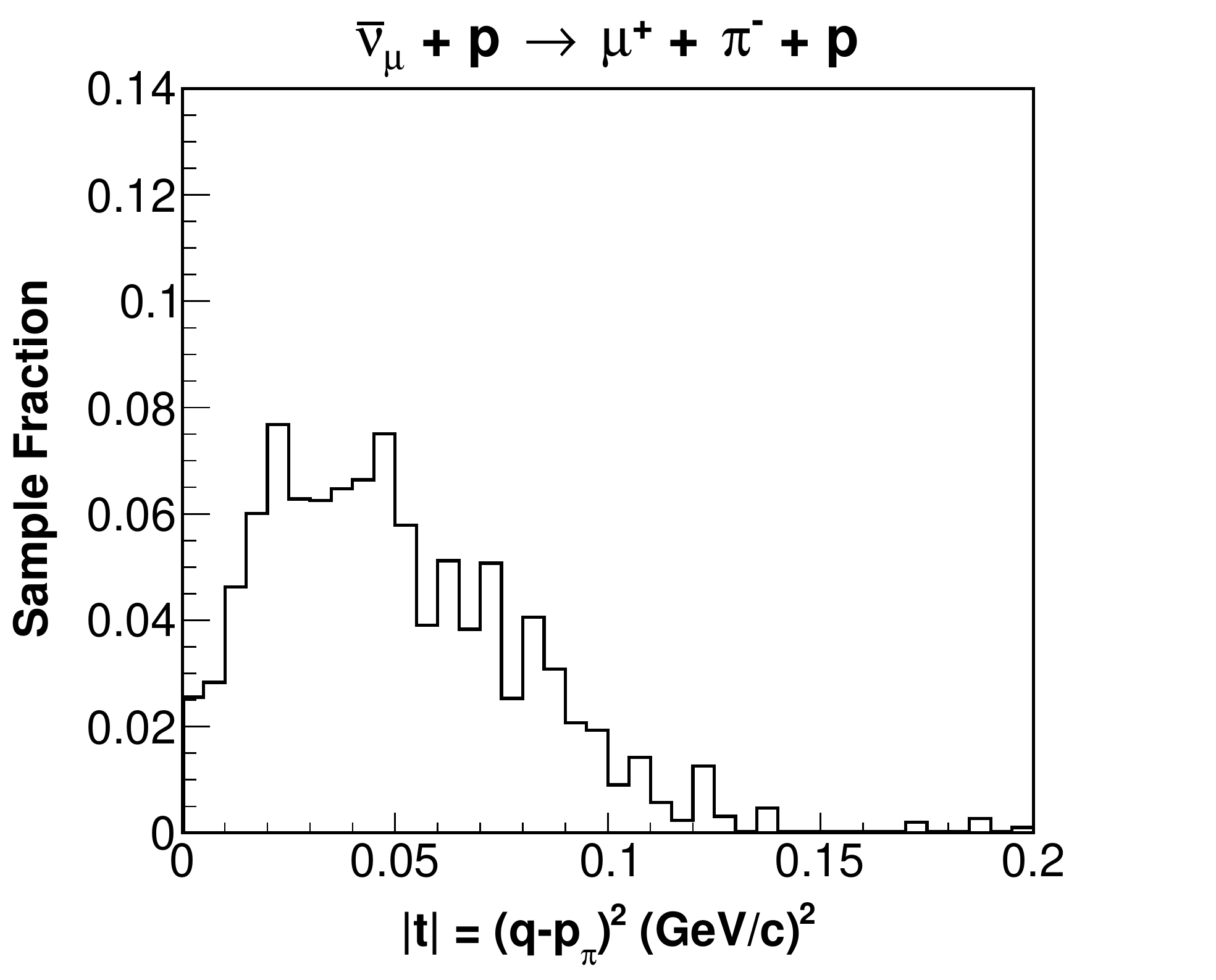}}
\mbox{\includegraphics[width=0.5\linewidth]{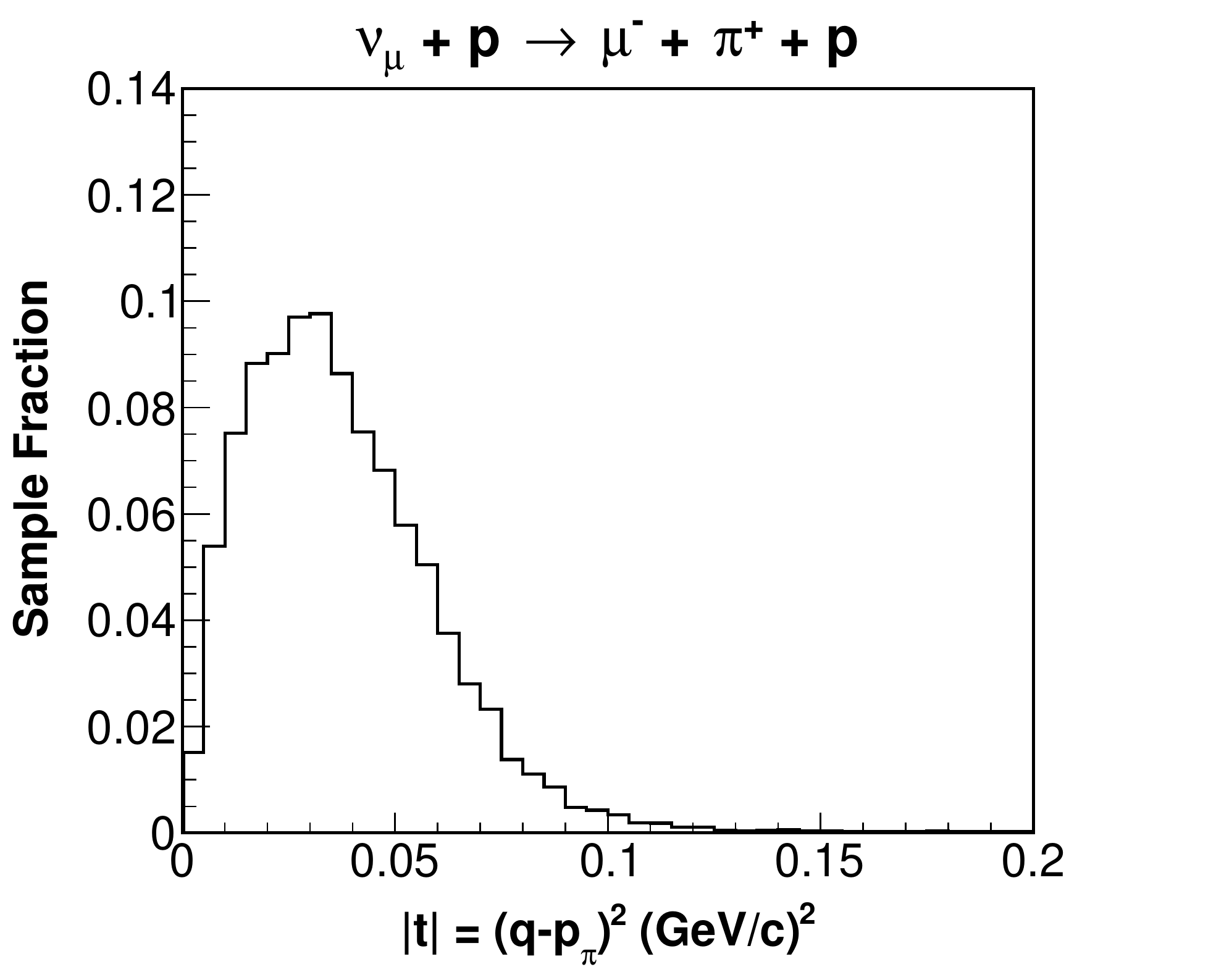}
\includegraphics[width=0.5\linewidth]{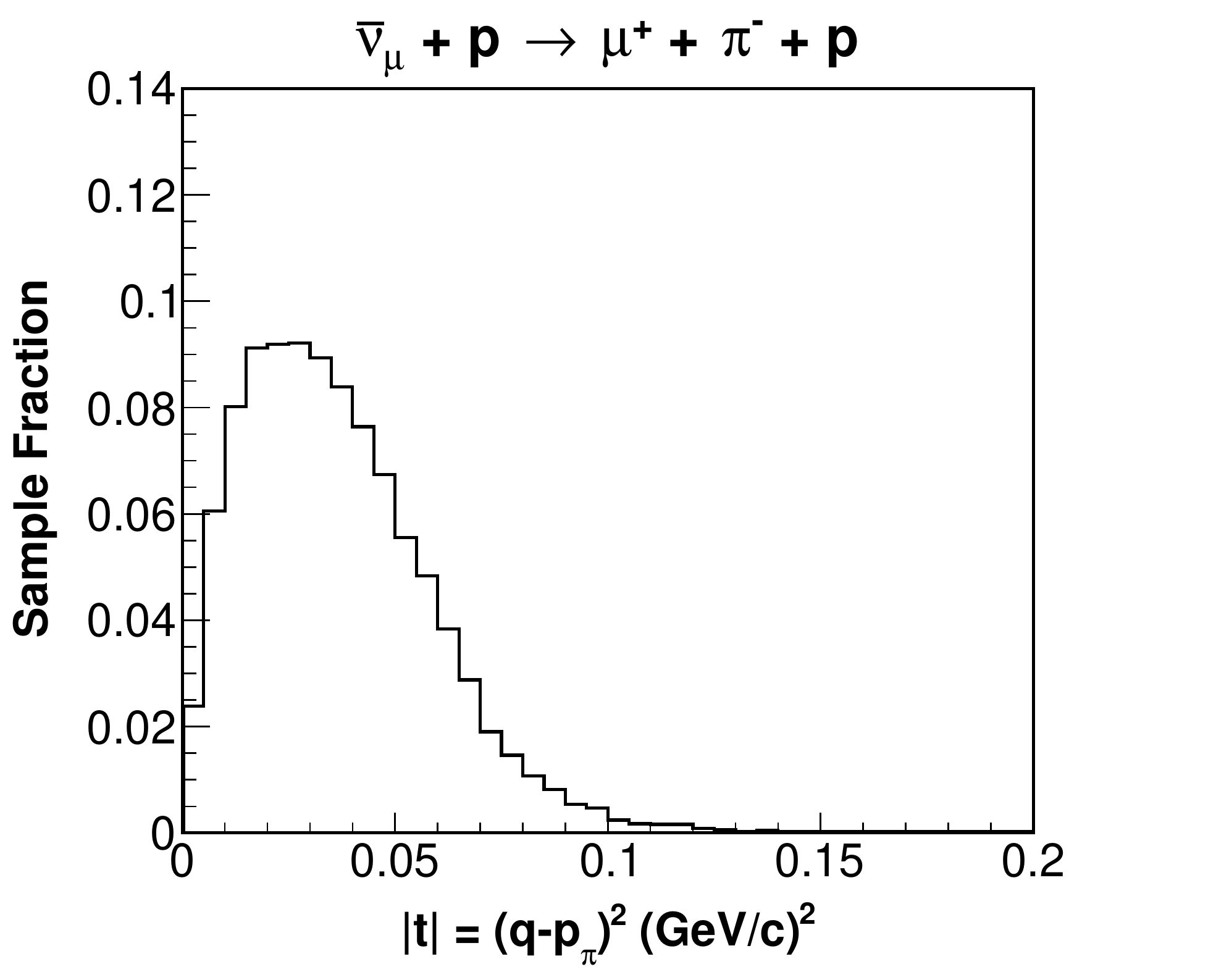}}
\caption[The \numu and \numubar diffractive MC \tabs distributions]{The \numu (left) and \numubar (right) diffractive MC \tabs distributions before (top) and after (bottom) weighting to the shape of the diffractive \dsigdt weighted by the absolute diffractive acceptance.}
\label{fig:diff_sample_t}
\end{figure*}

As mentioned previously, diffractive scattering was not simulated in the MC.  Instead, a stand-in diffractive MC sample
was constructed from MC interactions from other processes that pass all selection cuts and have a final state consisting
of a muon, a charged pion, and a proton.  This sample was then
%To represent the diffractive \tabs-dependence and acceptance, the diffractive MC sample was
weighted (Fig.~\ref{fig:diff_sample_t}) as a function of \tabs (calculated from the proton kinetic energy) to the shape of
the diffractive \dsigdt weighted by the absolute diffractive acceptance (Fig.~\ref{fig:diffractive_acceptance}).
%An important feature of the diffractive MC is the simulation of final state particles depositing energy in the same strip(s) which directly affects the MVSE.

\begin{figure*}[tpb]
\centering
\mbox{\includegraphics[width=0.5\linewidth]{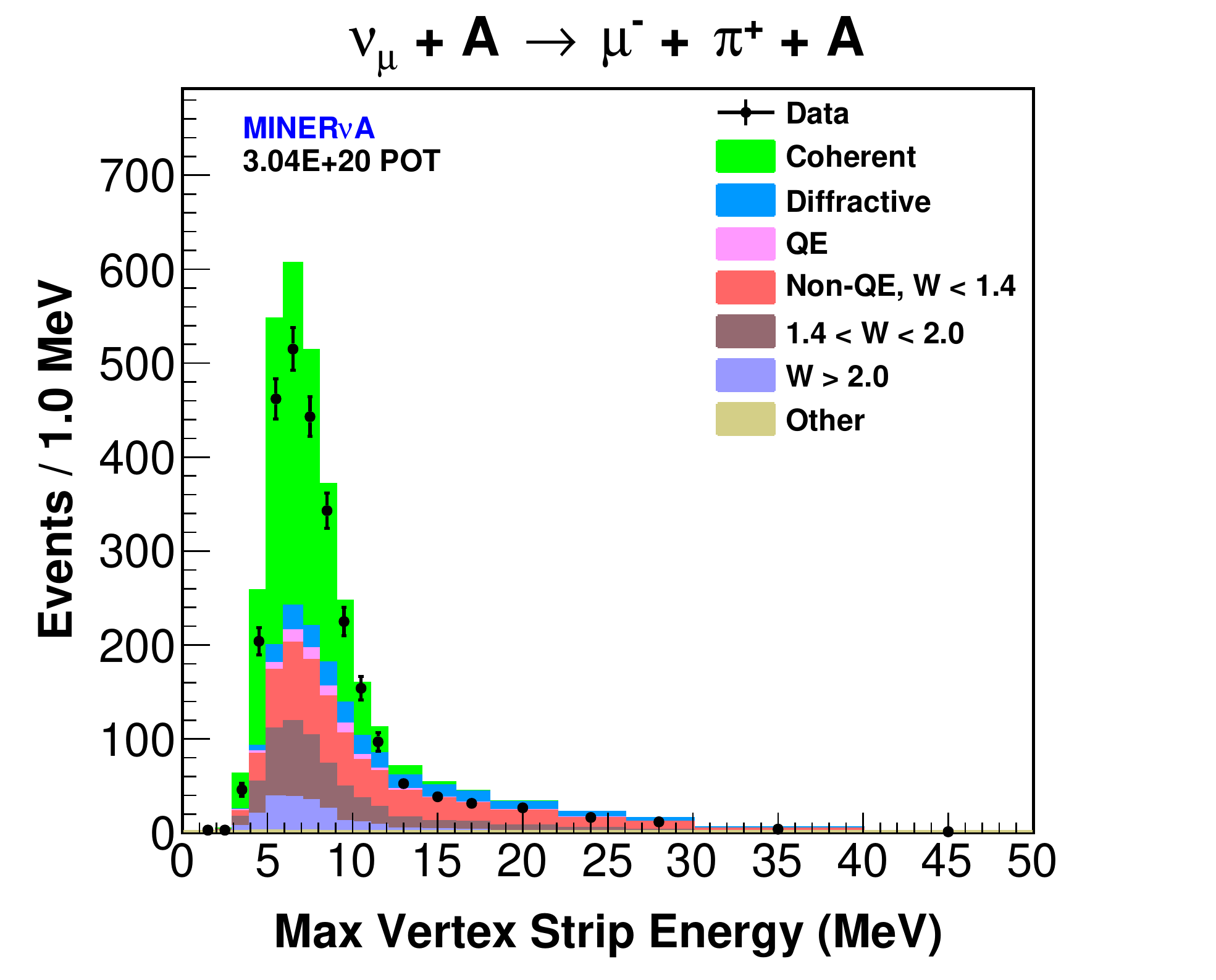}
\includegraphics[width=0.5\linewidth]{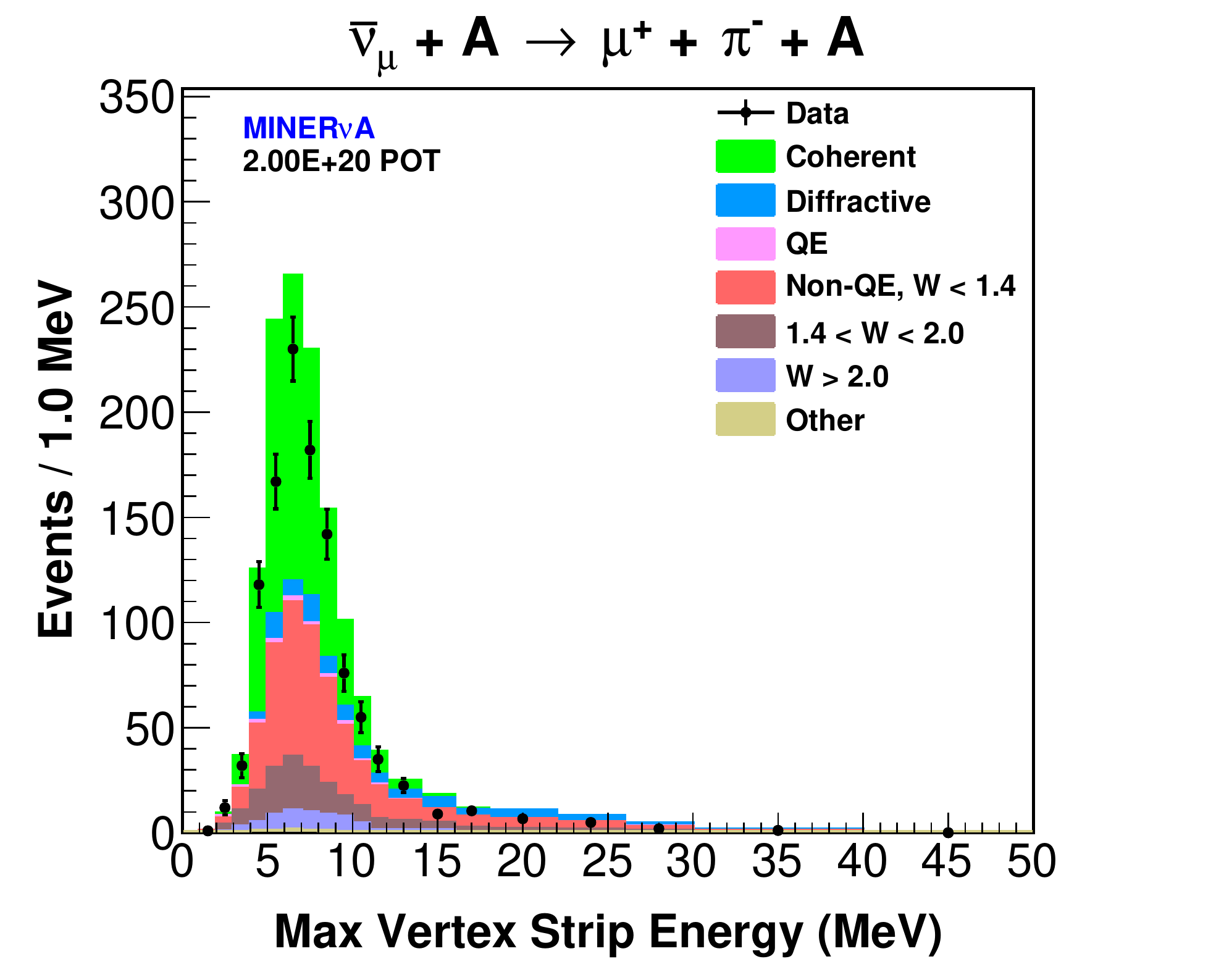}}
\mbox{\includegraphics[width=0.5\linewidth]{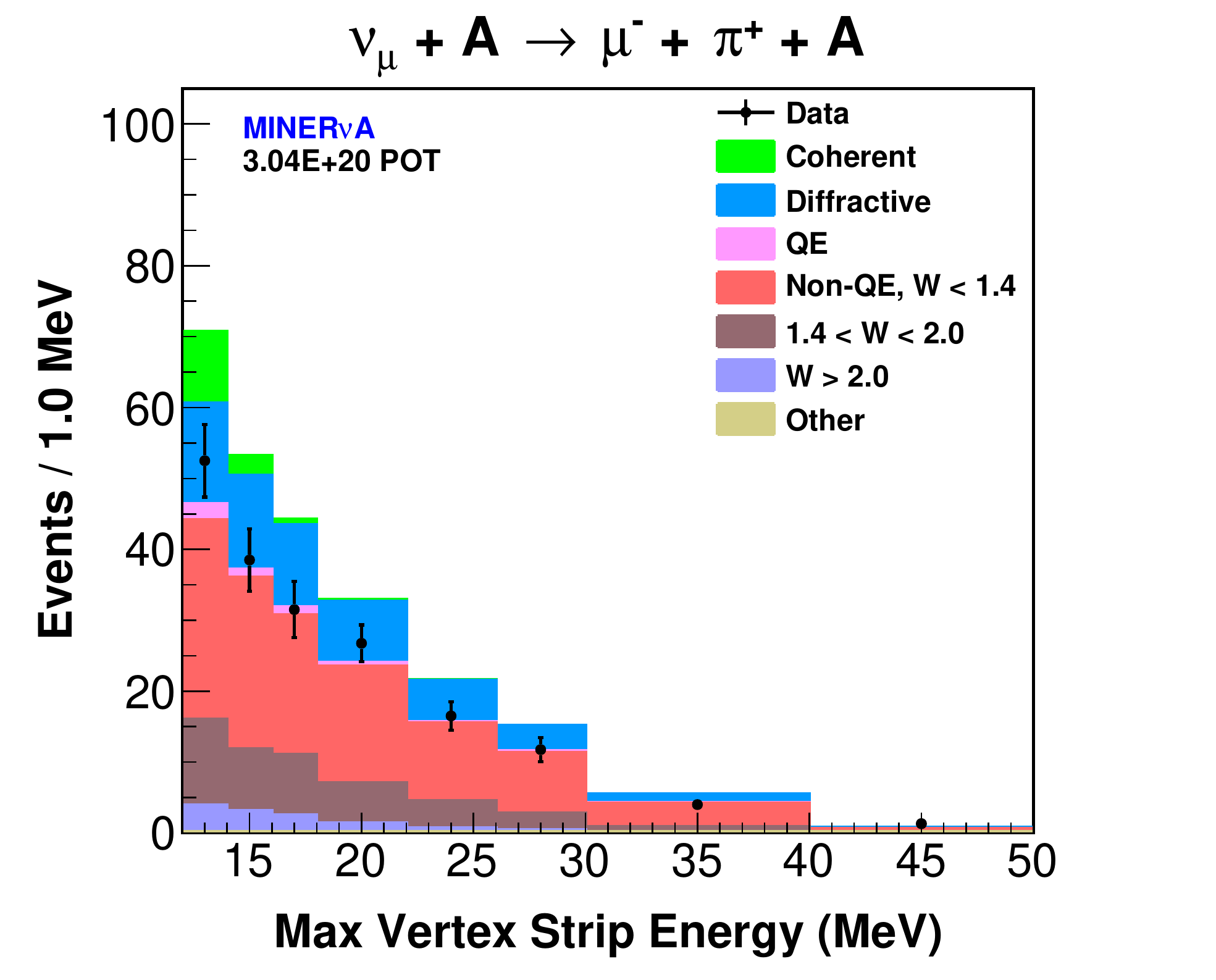}
\includegraphics[width=0.5\linewidth]{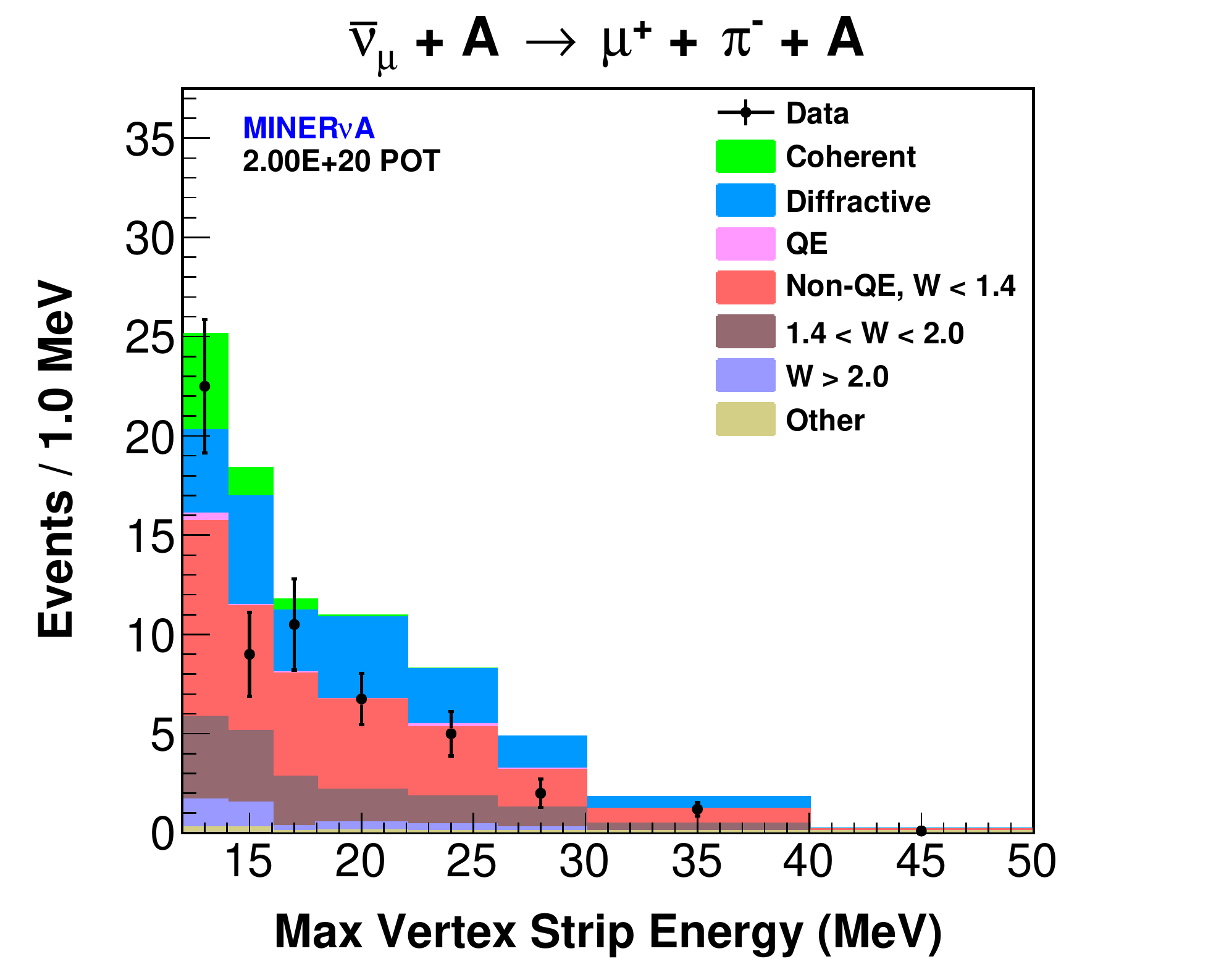}}
\caption[The MVSE distribution for the \numu and \numubar selected coherent candidate samples with the diffractive prediction]{The MVSE distribution for the \numu (left) and \numubar (right) selected coherent-like samples.  The bottom plots show the high-MVSE region examined for the presence of diffractive scattering.  The non-diffractive background normalizations are tuned. The added diffractive MC sample shown is for 0.2 relative diffractive-to-coherent normalization, which is clearly inconsistent with the data.  The fitted \numu (\numubar) relative diffractive-to-coherent normalization is +0.01$\pm$0.08 (-0.03$\pm$0.09).}
\label{fig:mvse-test}
\end{figure*}

The MVSE distribution was tested for the presence of diffractive scattering by adding the diffractive MC to the
existing background tuned MC (Fig.~\ref{fig:mvse-test}) and fitting the diffractive normalization.  The fit was
performed in the region 16 \lt MVSE \lt 40 MeV where the coherent contribution is small.  The $\chi^{2}$ in the fit was calculated as
\begin{equation}
\label{eq:mvse_chi2}
\chi^{2} = AC^{-1}A^{T},
\end{equation}
where $C$ is the total covariance (statistical + systematic) matrix for the MVSE distribution in the fit region, and
\begin{equation}
\label{eq:diff_search_a}
A_{i} = N_{i}^{data} - N_{i}^{MC} - N_{i}^{diff},
\end{equation}
where $N_{i}^{data}$, $N_{i}^{MC}$, and $N_{i}^{diff}$ are the data, non-diffractive MC, and diffractive MC event rates,
respectively, in MVSE bin $i$ within the fit region.  The non-diffractive MC event rate was held constant in the fit.

The \numu (\numubar) 
ratio of diffractive to coherent MC integrated event rates extracted from the fit is +0.01$\pm$0.08 (-0.03$\pm$0.09),
which is consistent with the estimated 8\% (4\%) estimate of the largest possible diffractive contribution to the
measured coherent cross sections.  These results are also consistent with no diffractive scattering contribution to
our measured coherent cross sections, and the measured coherent cross sections are not corrected for the possible contribution from diffractive scattering.

%\subsection{Diffractive Scattering Conclusions}

%The diffractive scattering contribution to the measured \numu (\numubar) coherent cross sections is estimated to be 8\% (4\%), which is supported by the search for diffractive events in the selected coherent-like sample.  The small contribution is the result of the \tabs-dependence of the diffractive scattering cross section in conjunction with the vertex energy cut acceptance for diffractive scattering.  The diffractive scattering cross section falls more slowly with \tabs than the coherent scattering cross section, and the vertex energy cut accepts only low-\tabs diffractive events due to the recoil proton ionization, thereby excluding $\sim$80\% of the diffractive cross section.
%The measured coherent cross sections are not corrected for the possible contribution from diffractive scattering.

\section{Systematic Uncertainties}
\label{sec:systematics}

The cross section measurements relied on the MC to estimate the background, resolution of the kinematic parameters,
signal selection efficiency, and to some extent, the flux.  Uncertainties on the predictions of the MC therefore result
in uncertainties on the measured cross sections.  These uncertainties and their correlations are evaluated by simulating
an ensemble of pseudoexperiments with different assumptions about each systematic uncertainty with altered background and event migration  in each pseudoexperiment.   The results of those
pseudoexperiments then determine a covariance matrix of systematic uncertainties.
% These uncertainties were evaluated by varying the MC predictions and measuring the resulting change to the cross sections.  Each systematic uncertainty was calculated from a covariance matrix $C$ whose elements were were calculated as
%\begin{equation}
%\label{eq:cov}
%C_{ij} = \frac{1}{N}\sum_{k}\Delta\sigma_{ik}\Delta\sigma_{jk},
%\end{equation}
%where $\Delta\sigma_{ik}$ is the change to the cross section in bin $i$ for variation $k$, and $N$ is the number of variations from which the uncertainty was evaluated.  The MC inputs to the measured cross sections were recalculated for each variation to the MC.  
%Varied inputs were the background prediction, the unfolding matrices, the signal selection efficiency, and the flux prediction.

\begin{figure*}[tpb]
\centering
\mbox{
\includegraphics[width=0.49\linewidth]{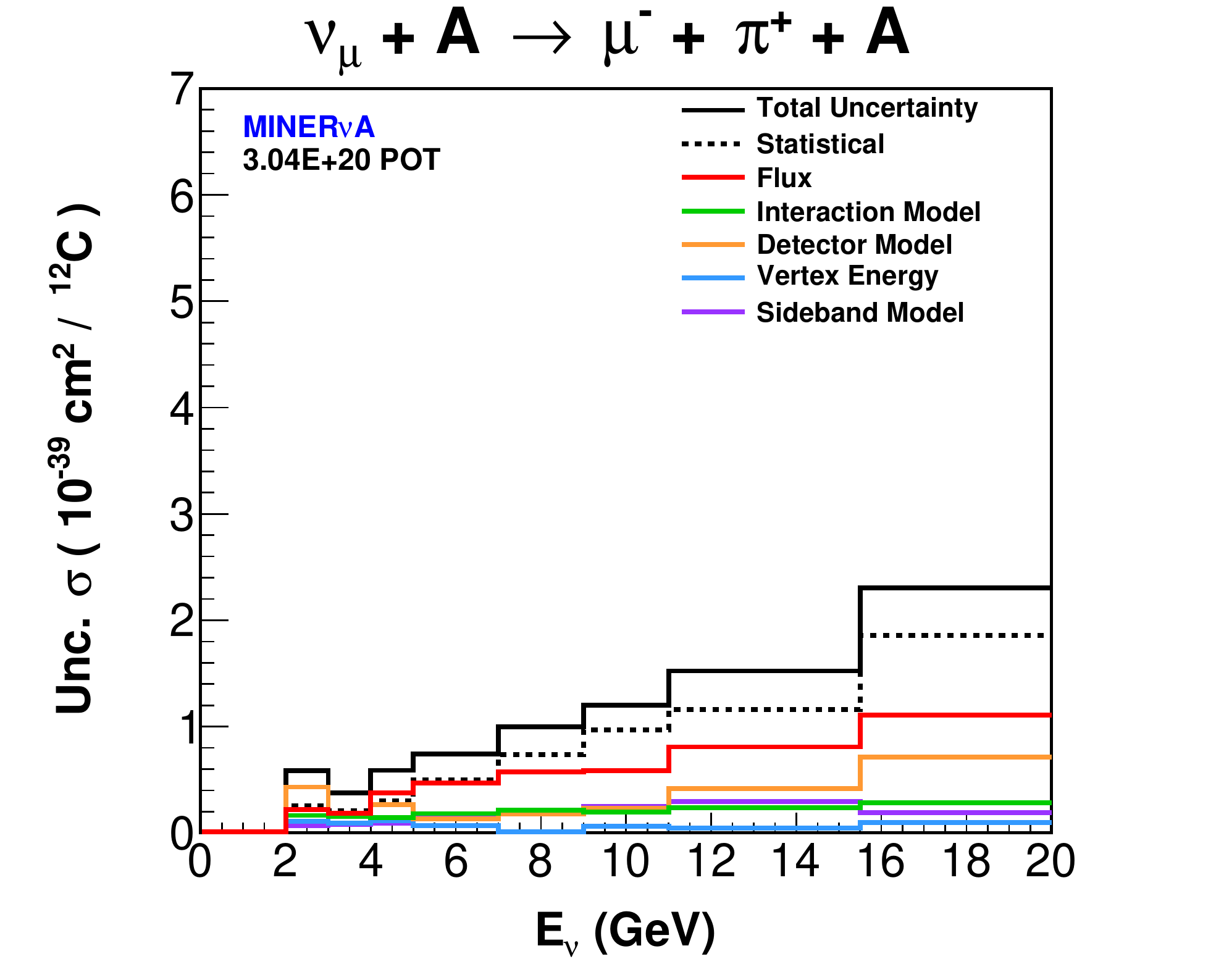}
\includegraphics[width=0.49\linewidth]{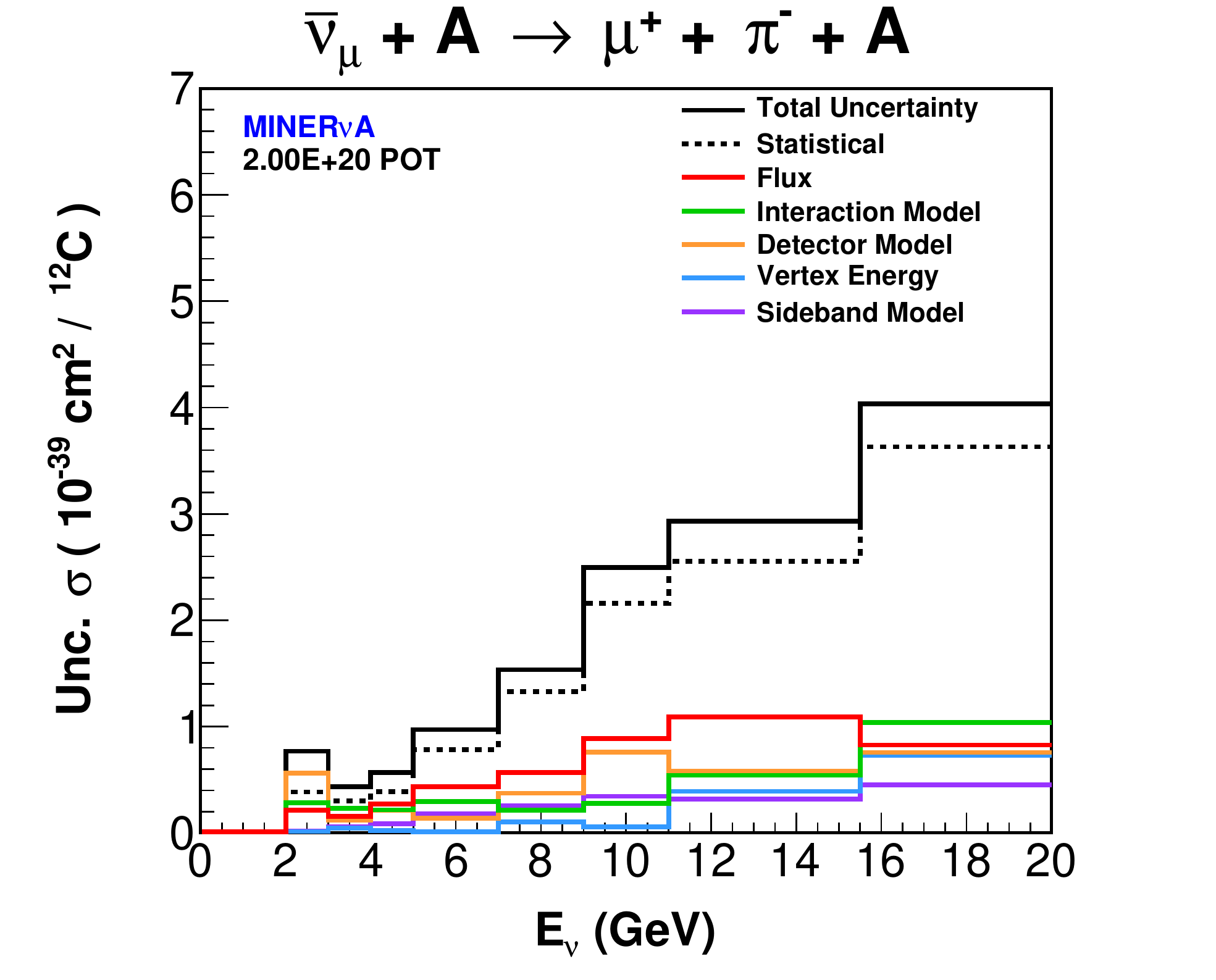}}
\mbox{
\includegraphics[width=0.49\linewidth]{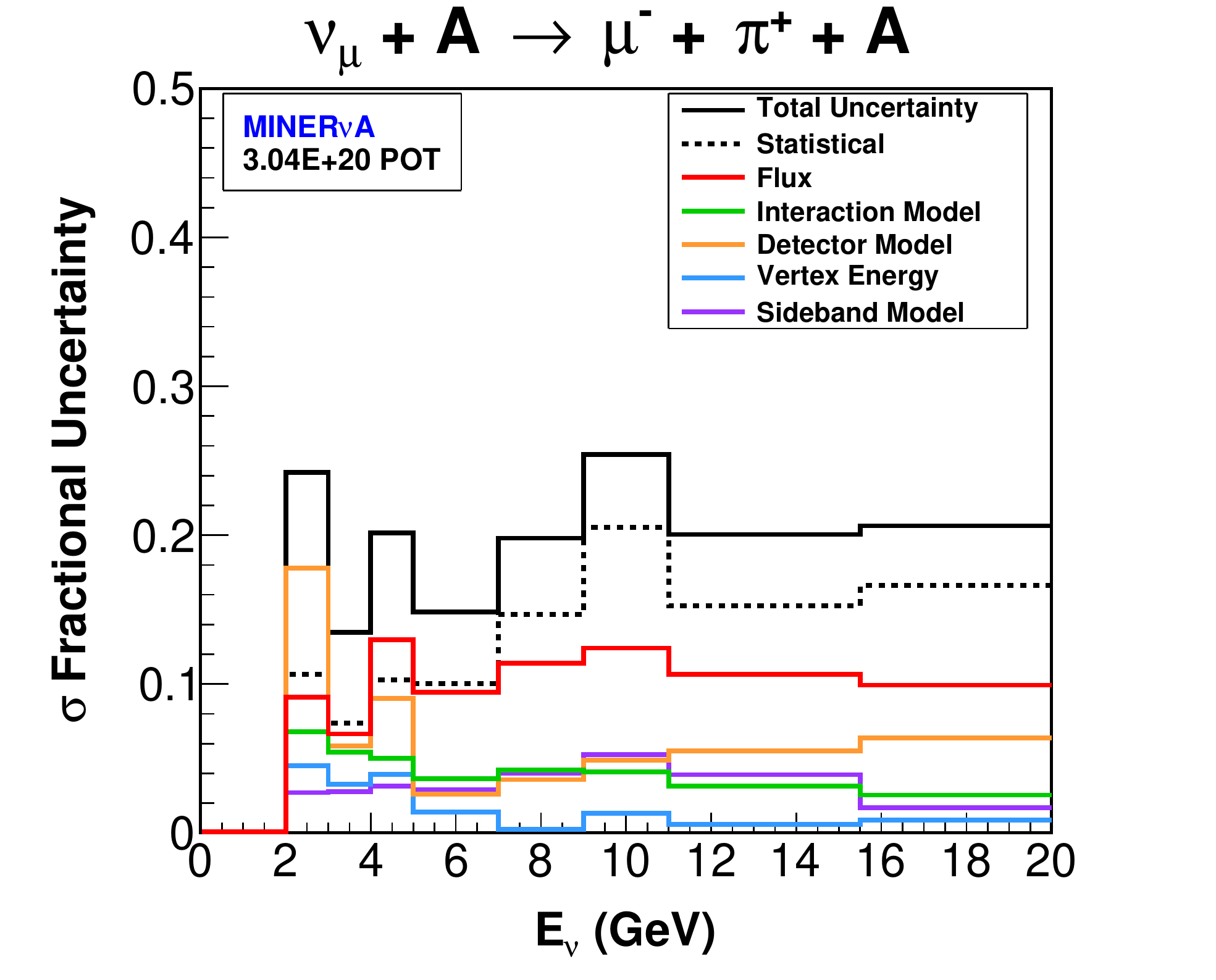}
\includegraphics[width=0.49\linewidth]{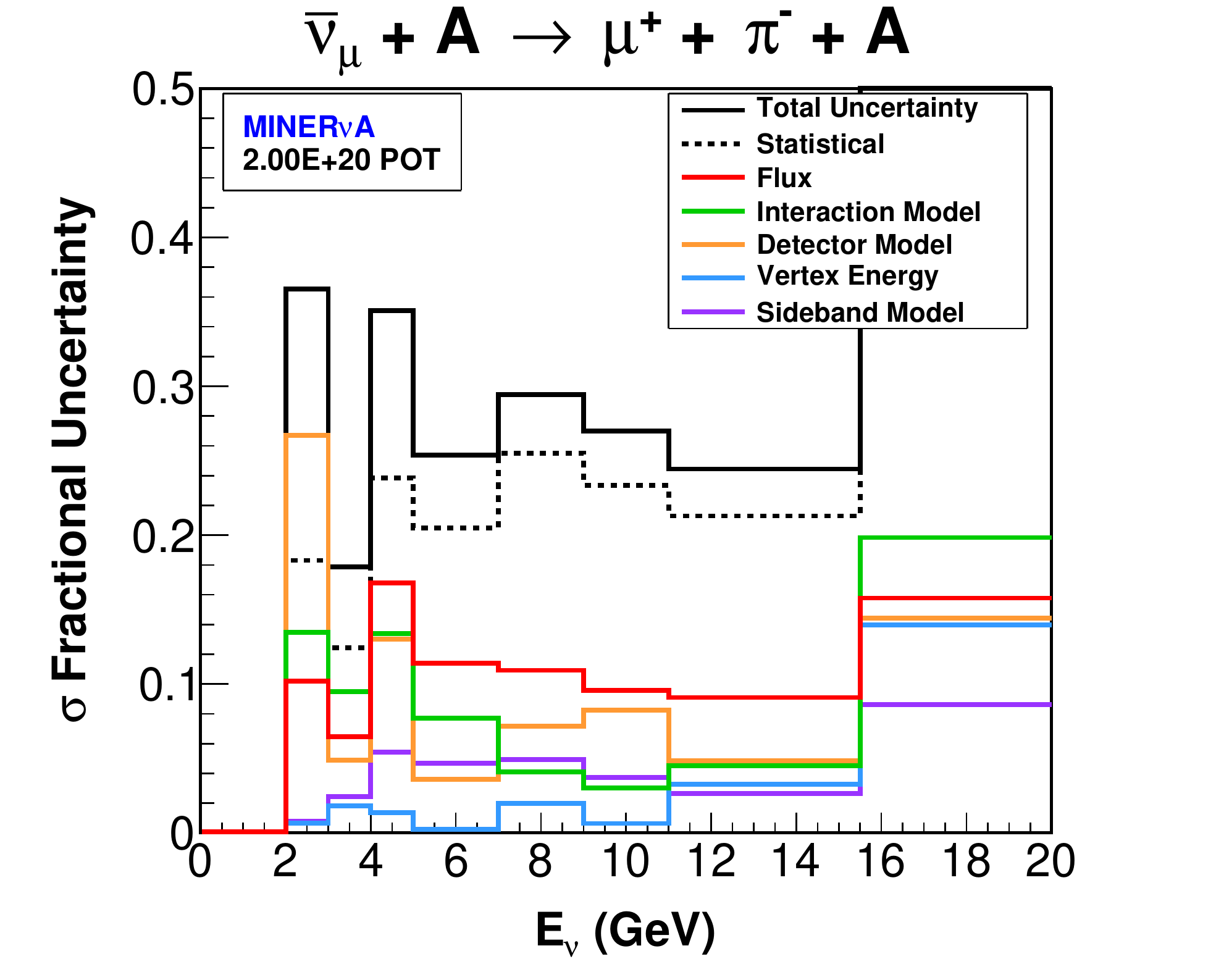}}
\caption[The statistical and systematic uncertainties on the measured \numu and \numubar \sigenu]{The statistical and systematic uncertainties on the measured \sigenu for \numu (left) and \numubar (right).  The top (bottom) plots show the absolute (fractional) uncertainty.}
\label{fig:sys_sigenu}
\end{figure*}

\begin{figure*}[tpb]
\centering
\mbox{
\includegraphics[width=0.49\linewidth]{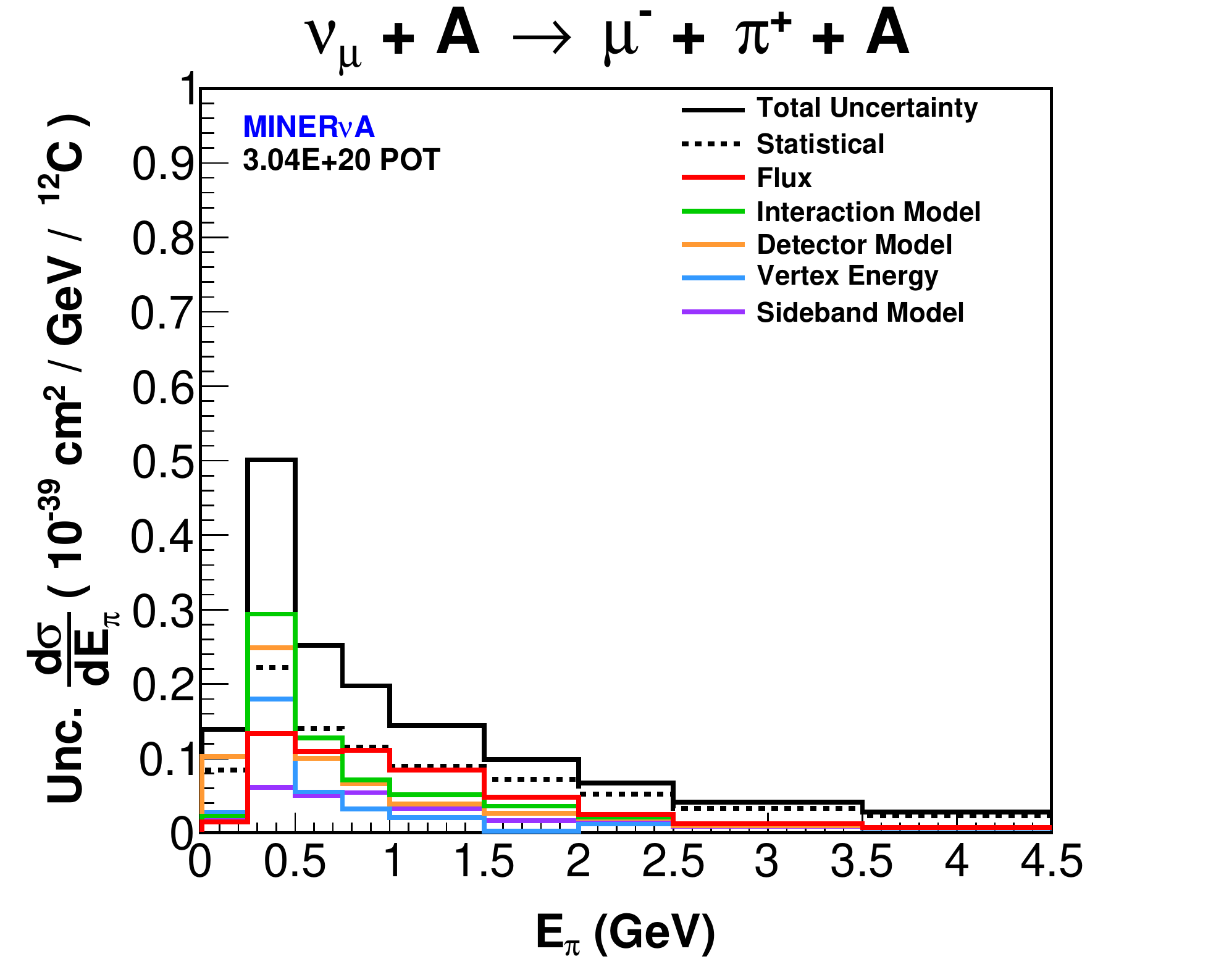}
\includegraphics[width=0.49\linewidth]{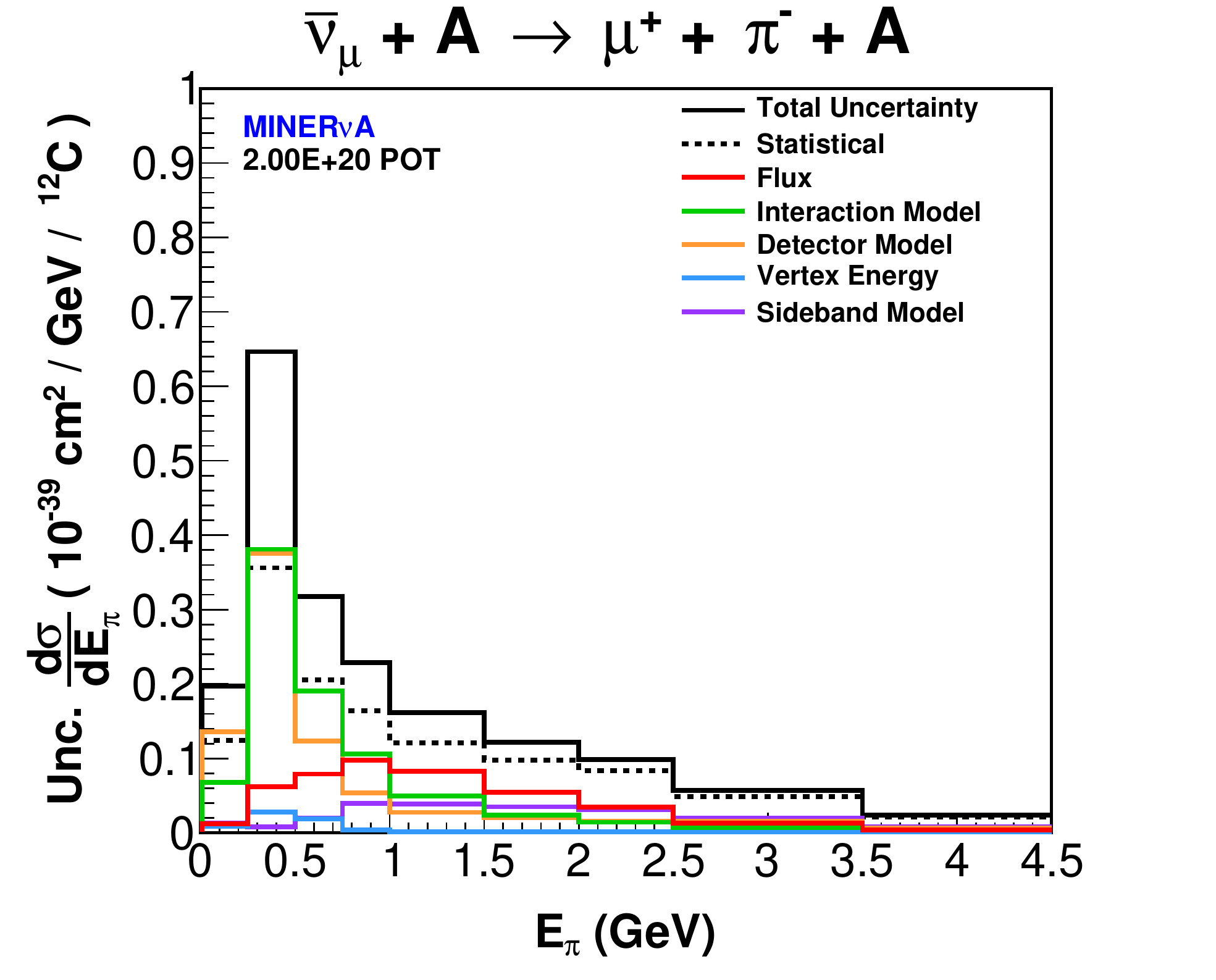}}
\mbox{
\includegraphics[width=0.49\linewidth]{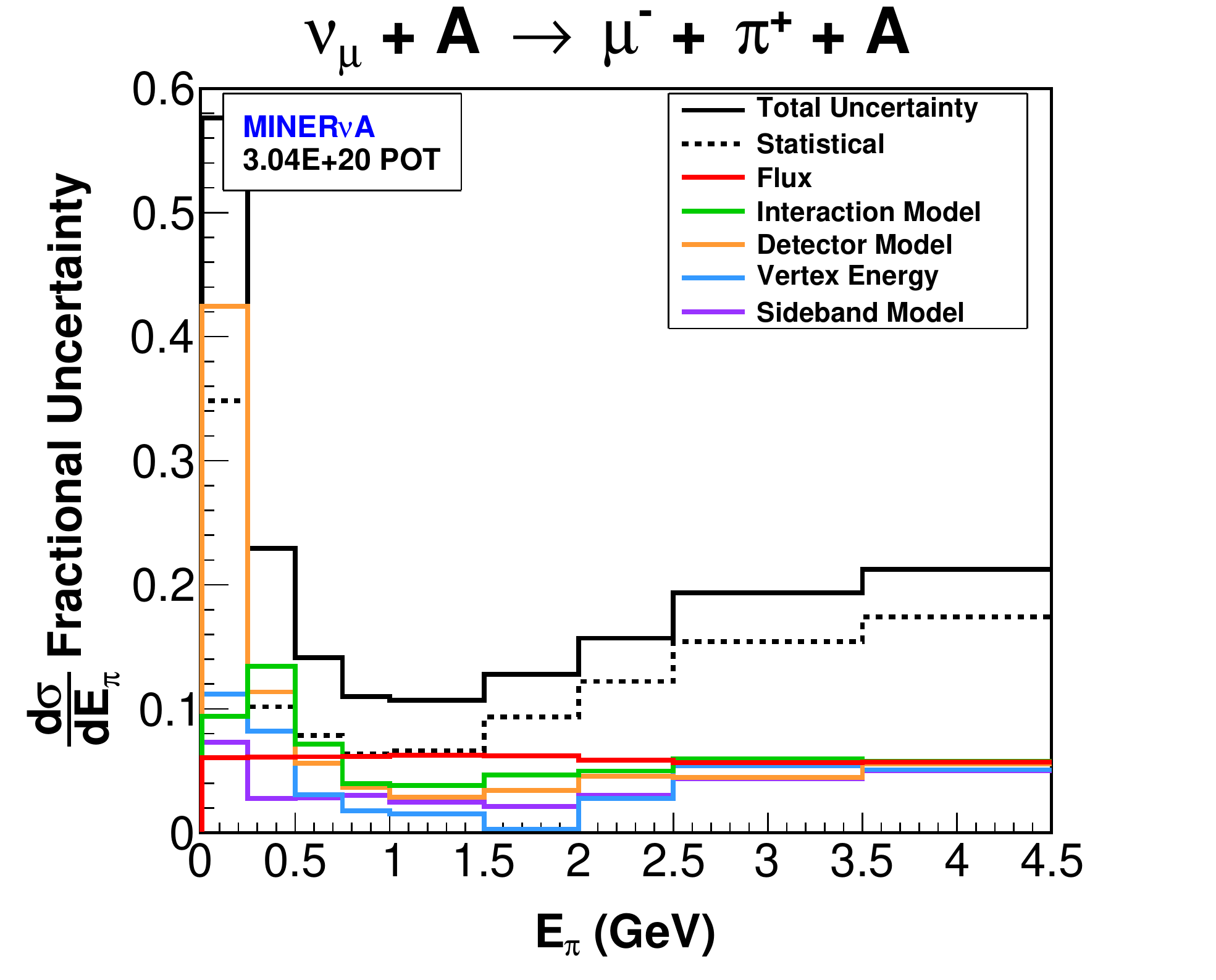}
\includegraphics[width=0.49\linewidth]{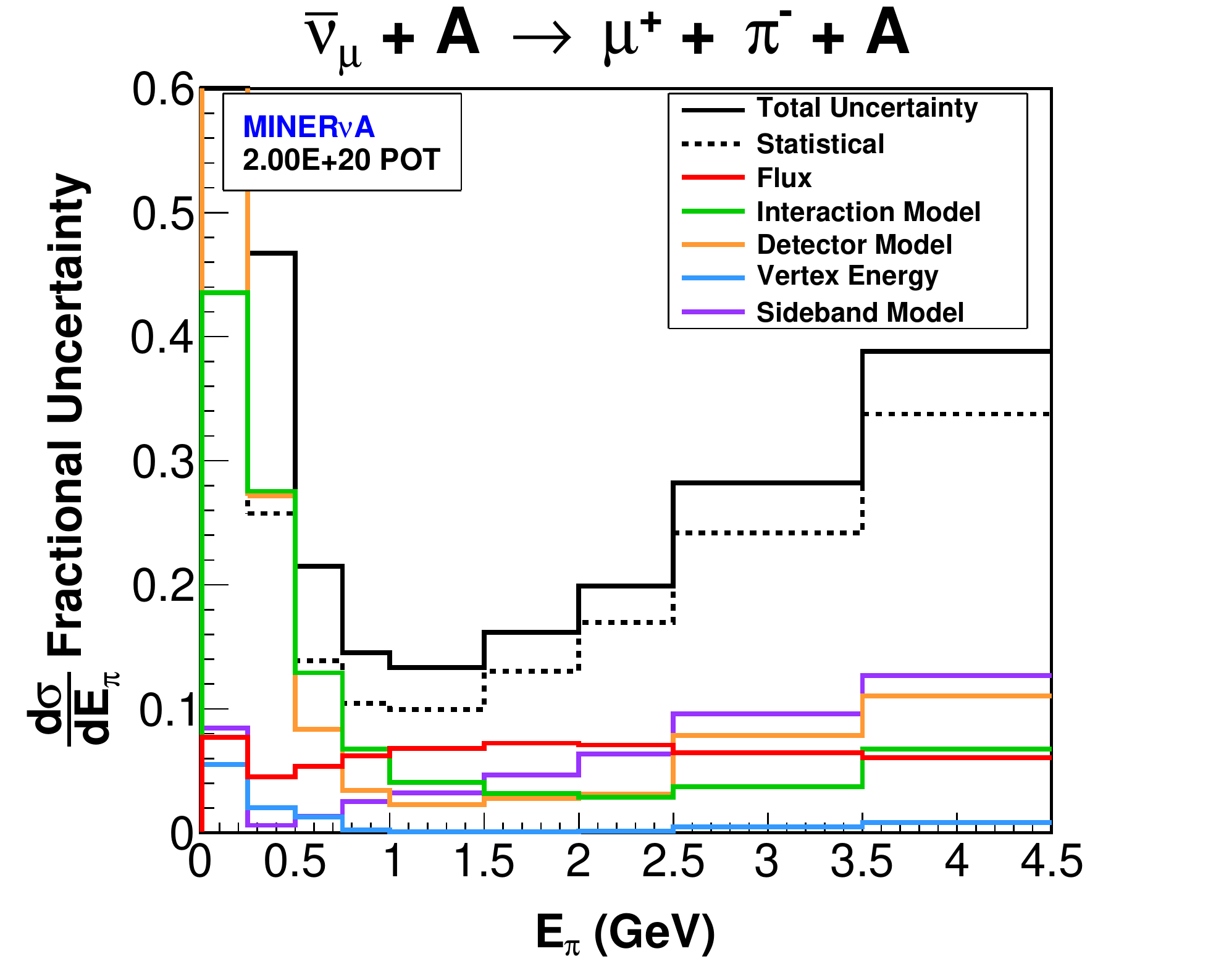}}
\caption[The statistical and systematic uncertainties on the measured \numu and \numubar \dsigdepi]{The statistical and systematic uncertainties on the measured \dsigdepi for \numu (left) and \numubar (right).  The top (bottom) plots show the absolute (fractional) uncertainty.  In the 0-0.25 GeV bin, the fractional uncertainty is large due to the small measured cross section.}
\label{fig:sys_dsigdepi}
\end{figure*}

\begin{figure*}[tpb]
\centering
\mbox{
\includegraphics[width=0.49\linewidth]{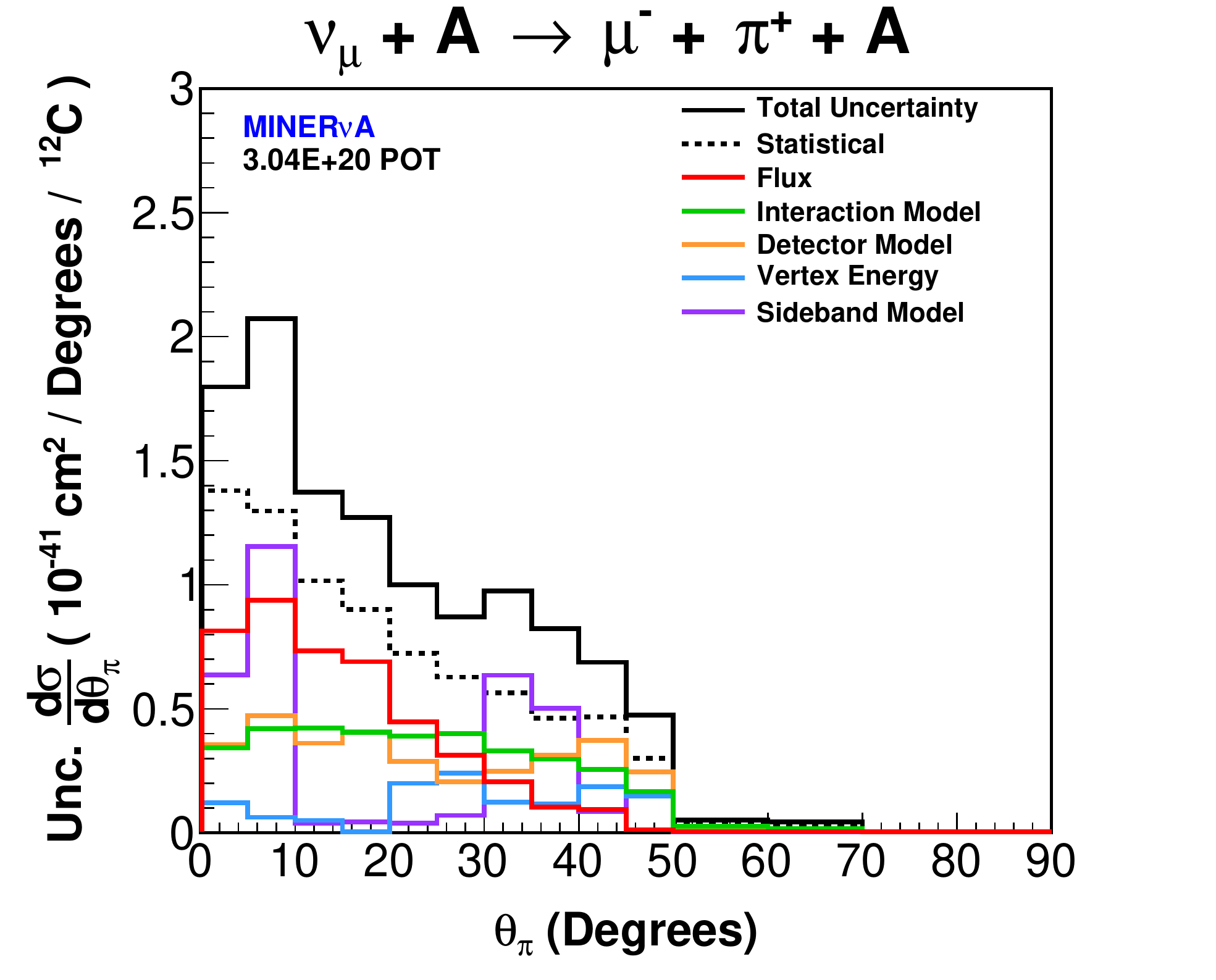}
\includegraphics[width=0.49\linewidth]{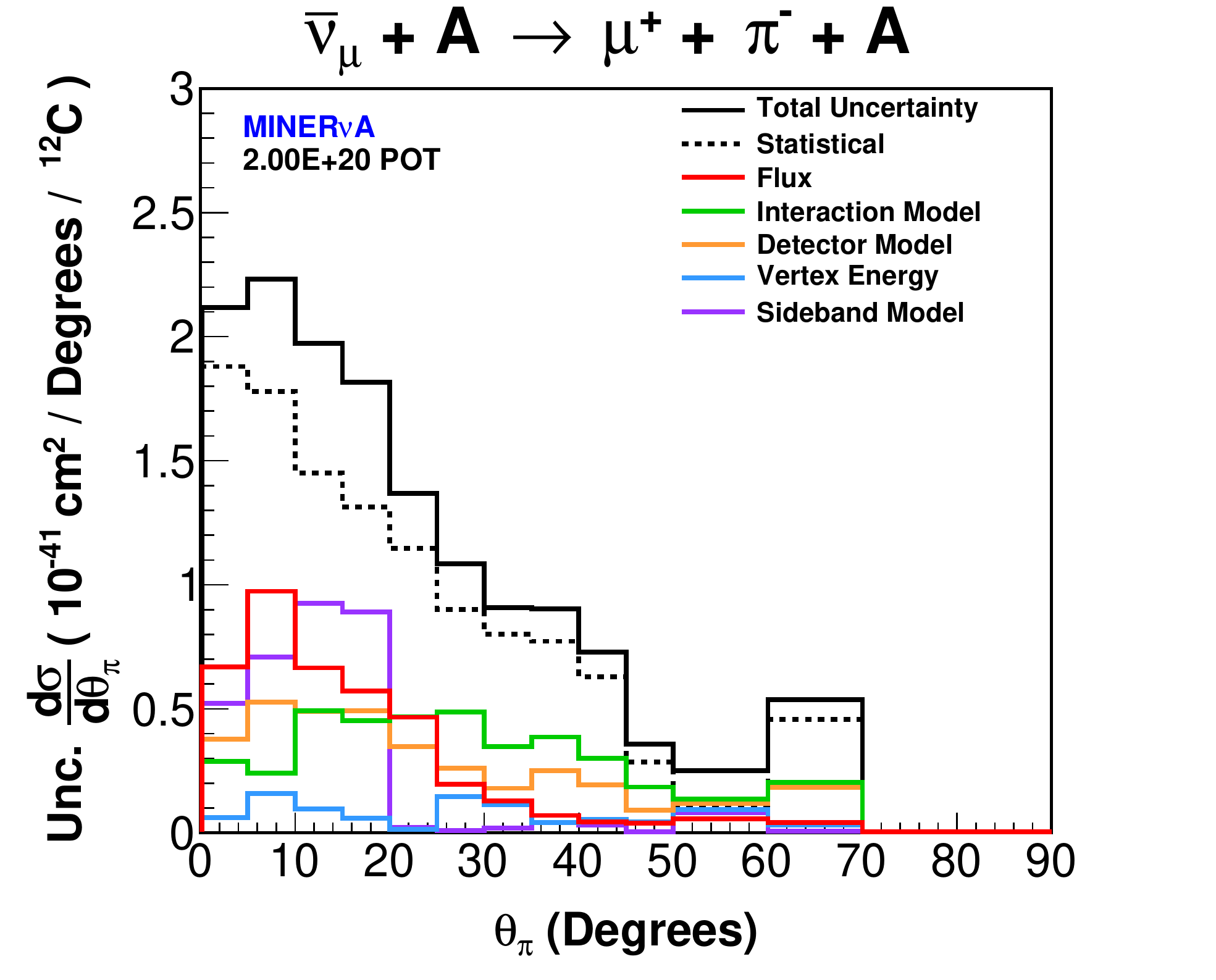}}
\mbox{
\includegraphics[width=0.49\linewidth]{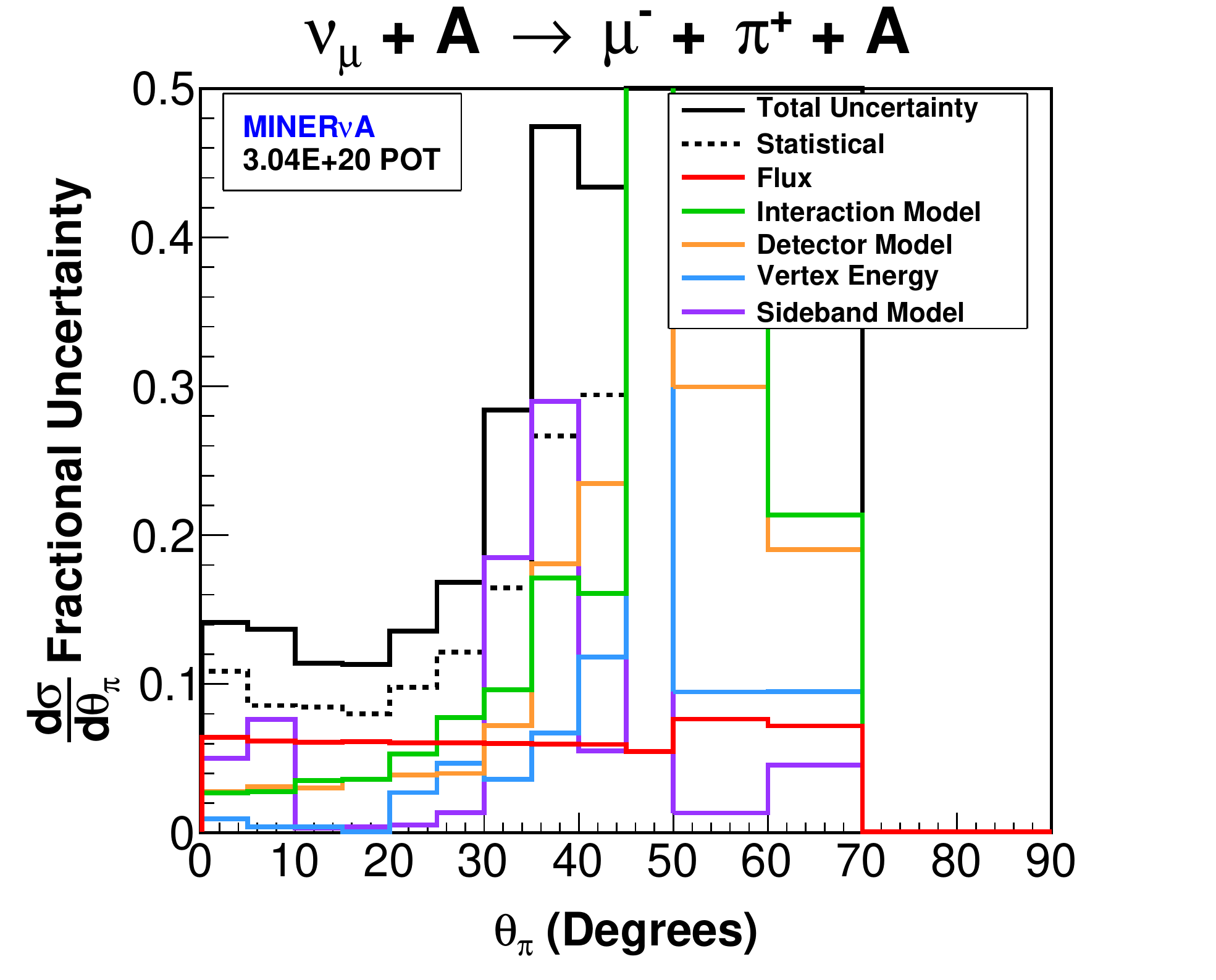}
\includegraphics[width=0.49\linewidth]{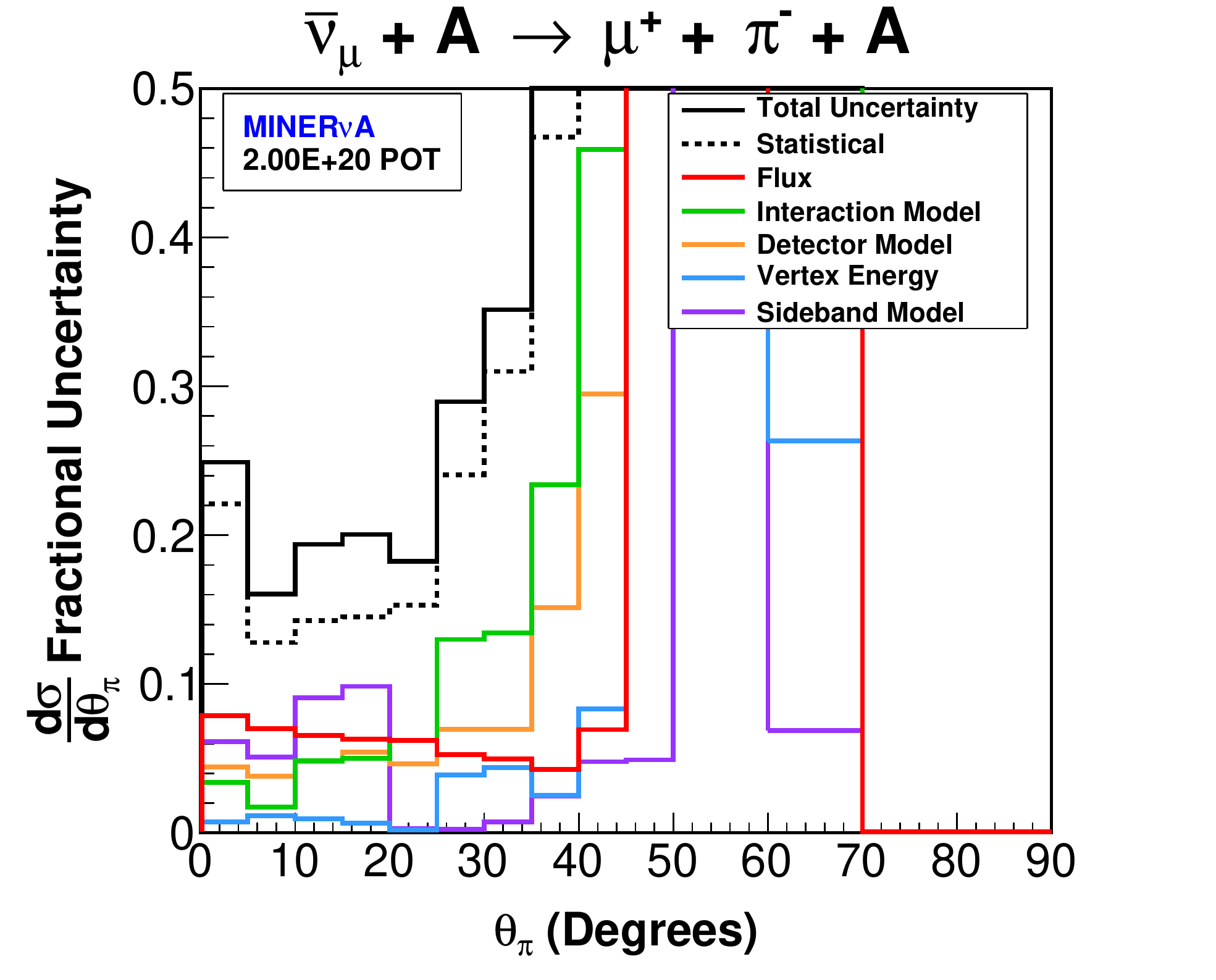}}
\caption[The statistical and systematic uncertainties on the measured \numu and \numubar \dsigdthetapi]{The statistical and systematic uncertainties on the measured \dsigdthetapi for \numu (left) and \numubar (right).  The top (bottom) plots show the absolute (fractional) uncertainty.  At high-\thetapi, the fractional uncertainty is large due to the small measured cross section.}
\label{fig:sys_dsigdthetapi}
\end{figure*}

\begin{figure*}[tpb]
\centering
\mbox{
\includegraphics[width=0.49\linewidth]{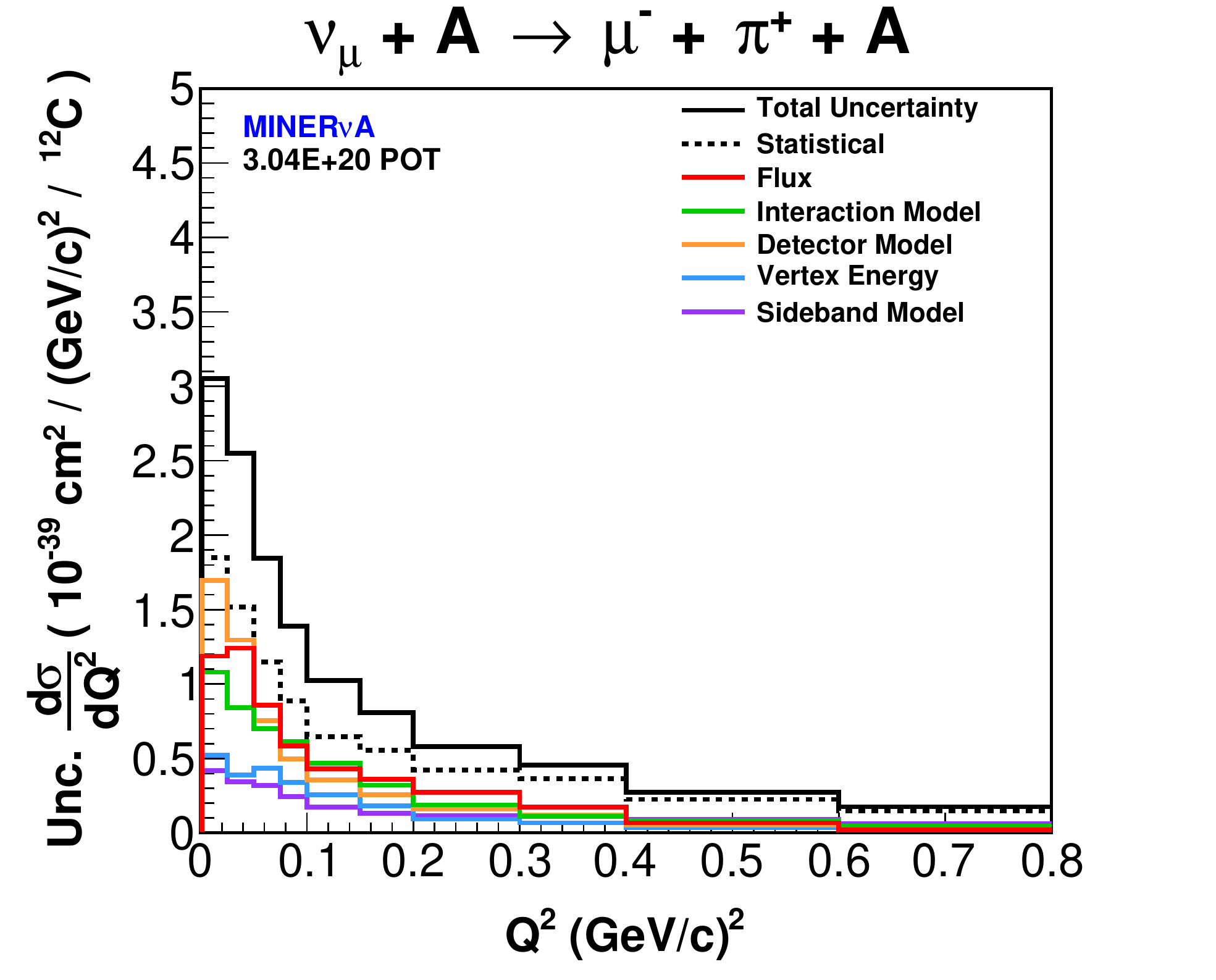}
\includegraphics[width=0.49\linewidth]{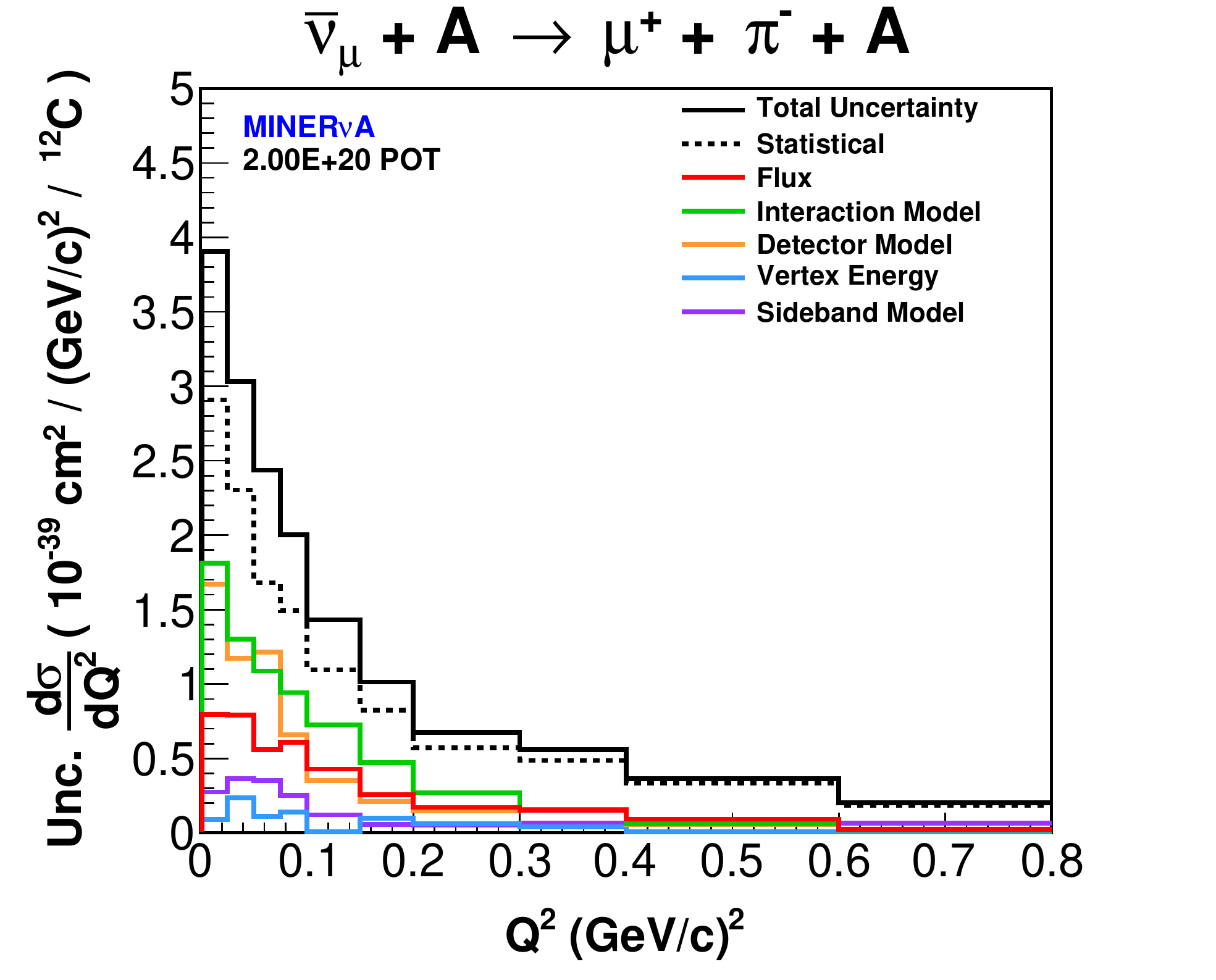}}
\mbox{
\includegraphics[width=0.49\linewidth]{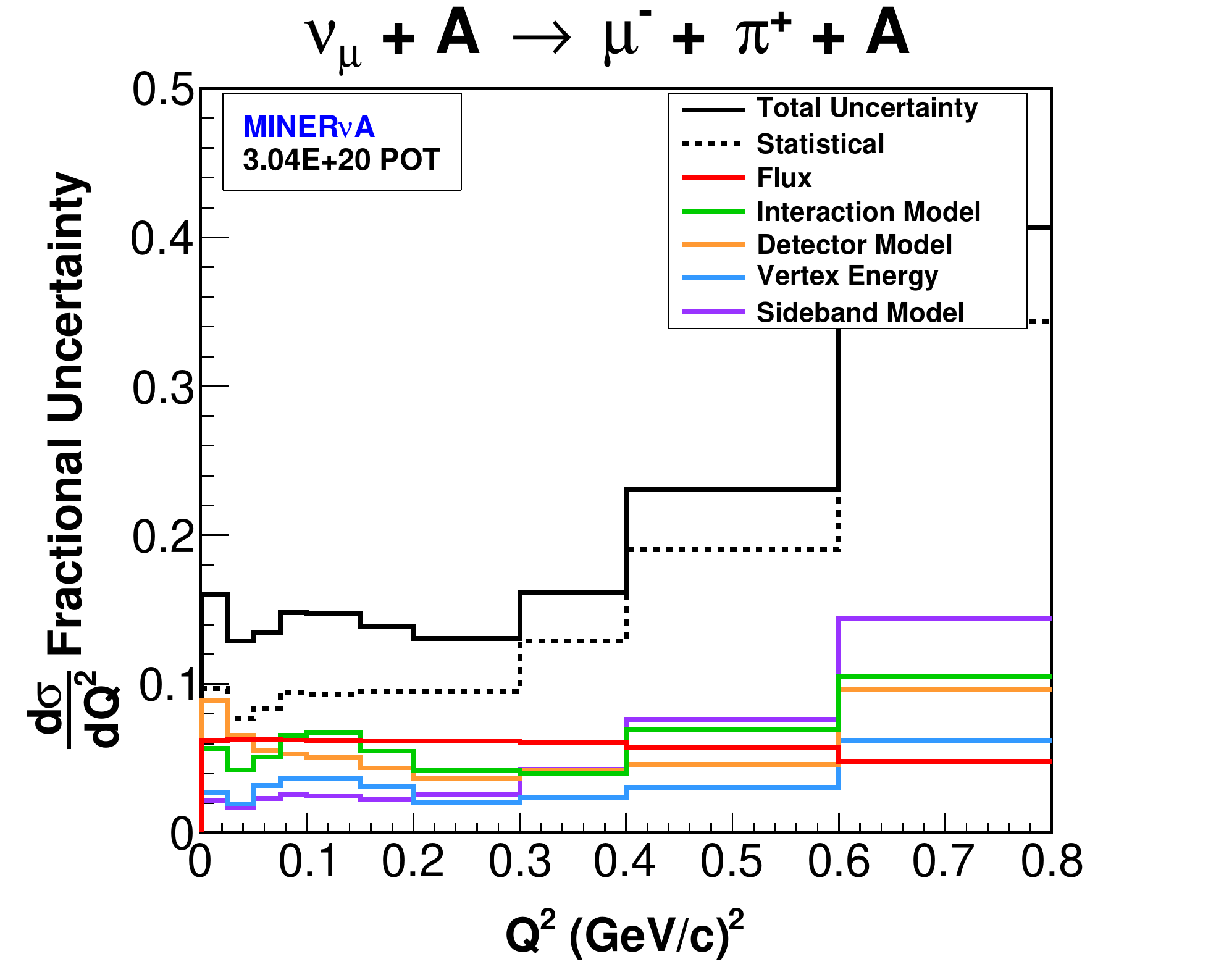}
\includegraphics[width=0.49\linewidth]{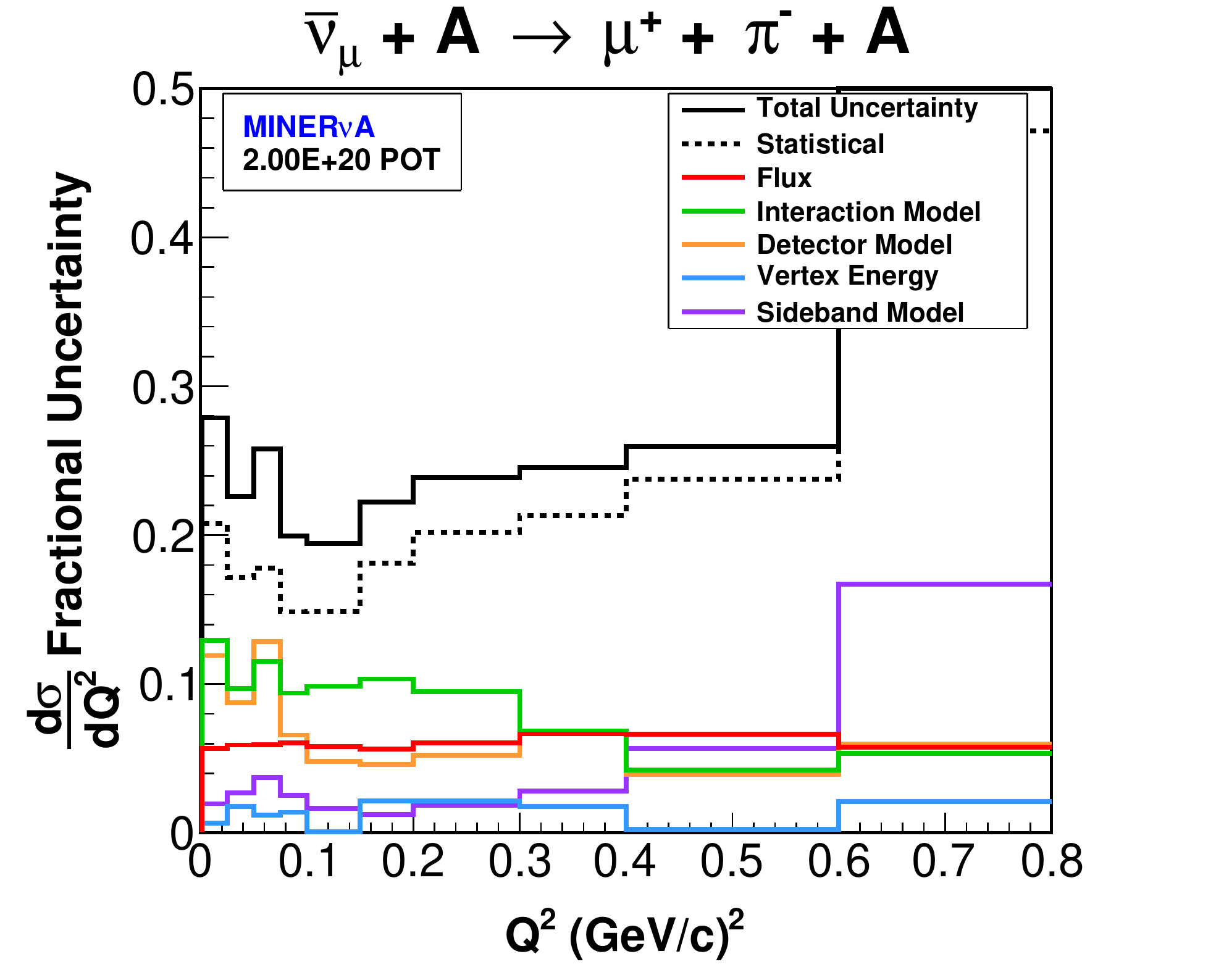}}
\caption[The statistical and systematic uncertainties on the measured \numu and \numubar \dsigdqsq]{The statistical and systematic uncertainties on the measured \dsigdqsq for \numu (left) and \numubar (right).  The top (bottom) plots show the absolute (fractional) uncertainty.}
\label{fig:sys_dsigdqsq}
\end{figure*}

The systematic uncertainties in the measured cross sections are shown in Figs.~\ref{fig:sys_sigenu}--\ref{fig:sys_dsigdqsq}.
The interaction model and detector model uncertainties in these figures are decomposed into their constituent uncertainties.
The fractional systematic uncertainties tend to be larger for the \numubar cross sections than the \numu cross sections.  This
is due to the larger background fraction in the \numubar candidate sample, coupled with the systematic uncertainties on the background prediction.

The evaluation of the systematic uncertainties on the measured cross sections is detailed in the following sections.  The
effects of the individual systematic uncertainties on each measured quantity can be found in Ref.~\cite{bib:Mislivec_thesis}. 

\subsection{Flux}
\label{sec:sys_flux}

%Our flux uncertainty was composed of the total uncertainties on the three primary components of the neutrino beam simulation: hadron production, tertiary interactions, and beam focusing.  The total uncertainty on each of these components was determined from 100 universes, where each universe was a simultaneous variation of the component's parameters within their 1-standard deviation uncertainty.  The background estimate, efficiency, and flux were remeasured for each of these universes.

The uncertainty on the flux prediction, detailed in~\cite{bib:flux}, consists of uncertainties
%on the parameters in the flux MC
that govern hadronic interactions in the NuMI target and downstream beamline materials, uncertainties in the magnetic
focusing of hadrons emerging from the target, and the geometry model of the beamline components.  The flux uncertainty
was evaluated from a set of 100 alternate flux predictions.  Each alternate flux prediction, referred to herein as a
flux variation, was the result of simultaneously varying the parameters in the flux model, where each parameter was varied
by a random amount determined from a gaussian distribution centered on the nominal parameter value with a 1$\sigma$ width
equal to the parameter uncertainty.  To evaluate the flux uncertainty on the cross sections, the cross sections were
remeasured for each flux variation and a covariance matrix for the set of flux variations was calculated for each cross section.

\subsection{Neutrino Interaction Model}
\label{sec:sys_genie}

The cross section measurements rely on the MC to predict the rate of the incoherent backgrounds.  The measured cross
sections are therefore subject to uncertainties on the underlying neutrino-nucleus interaction models of GENIE.  For
the purpose of evaluating the effects of these uncertainties, GENIE provides event-by-event weights.
These weights modify the predictions of the cross section models and FSI models.
%the model of interactions of final state particles within the target nucleus (FSI).
These weights modify both the normalization and  kinematic dependence of the model predictions.
The weights used to evaluate the neutrino-nucleus interaction model uncertainties on the measured cross sections
correspond to $\pm 1\sigma$ uncertainties on the GENIE model parameters.  Table~\ref{tab:genie_knobs} lists the
default $\pm 1\sigma$ uncertainties on these parameters for the \minerva implementation of GENIE used in this analysis.
%~\cite{bib:genie_knobs}.  THIS CITATION IS TO A MINERVA DOCDB ENTRY.

\begin{table}[bp]
\footnotesize
\squeezetable
\begin{center}
\begin{tabular}{ l c }
Parameter & Variation \\
\hline
\multicolumn{2}{c}{\textit{CC Quasielastic}} \\
Normalization & +20\%, -15\% \\
Axial vector mass (shape only) & $\pm$10\% \\
Vector form factor model (shape only) & BBBA to \\
& dipole \\
& parameterization \\
Pauli suppression & $\pm$30\% \\
\hline
\multicolumn{2}{c}{\textit{NC Elastic}} \\
Axial vector mass & $\pm$25\% \\
Strange axial form factor $\eta$ & $\pm$30\% \\
\hline
\multicolumn{2}{c}{\textit{CC Resonance Production}} \\
Normalization & $\pm$20\% \\
Axial vector mass & $\pm$20\% \\
Vector mass & $\pm$10\% \\
\hline
\multicolumn{2}{c}{\textit{CC \& NC Non-Resonant Pion Production}} \\
Normalization of 1$\pi$ final states from $\nu p$ / $\overline{\nu} p$ & $\pm$50\% \\
Normalization of 1$\pi$ final states from $\nu n$ / $\overline{\nu} n$ & $\pm$50\% \\
Normalization of 2$\pi$ final states from $\nu p$ / $\overline{\nu} p$ & $\pm$50\% \\
Normalization of 2$\pi$ final states from $\nu n$ / $\overline{\nu} n$ & $\pm$50\% \\
\hline
\multicolumn{2}{c}{\textit{Deep Inelastic Scattering}} \\
Bodek-Yang model parameter $A_{HT}$ & $\pm$25\% \\
Bodek-Yang model parameter $B_{HT}$ & $\pm$25\% \\
Bodek-Yang model parameter $C_{V1u}$ & $\pm$30\% \\
Bodek-Yang model parameter $C_{V2u}$ & $\pm$40\% \\
\hline

%--> For 1 column double spaced version only
%\end{tabular}
%\end{center}
%\end{table}
%\newpage
%\begin{table}[tp]
%\footnotesize
%\squeezetable
%\begin{center}
%\begin{tabular}{ l c }
%--> to here.

\multicolumn{2}{c}{\textit{Final State Interactions}} \\
Nucleon mean free path & $\pm$20\% \\
Nucleon charge exchange probability & $\pm$50\% \\
Nucleon elastic interaction probability & $\pm$30\% \\
Nucleon inelastic interaction probability & $\pm$40\% \\
Nucleon absorption probability & $\pm$20\% \\
Nucleon $\pi$-production probability & $\pm$20\% \\
$\pi$ mean free path & $\pm$20\% \\
$\pi$ charge exchange probability & $\pm$50\% \\
$\pi$ elastic interaction probability & $\pm$10\% \\
$\pi$ inelastic interaction probability & $\pm$40\% \\
$\pi$ absorption probability & $\pm$30\% \\
$\pi$ $\pi$-production probability & $\pm$20\% \\
\hline
\multicolumn{2}{c}{\textit{Hadronization and Resonance Decay}} \\
$x_{F}$ dependence for $N\pi$ final states&\\ ~~in AGKY hadronization model & $\pm$20\% \\
Resonance $\to X+1\gamma$ branching ratio & $\pm$50\% \\
Pion angular distribution in $\Delta^{++}\to N\pi$ & Isotropic $\to$ \\
& Rein-Sehgal \\
& parameterization \\
\end{tabular}
\end{center}
\caption{The default $\pm1\sigma$ variations to the GENIE model parameters.}
\label{tab:genie_knobs}
\end{table}

The GENIE prediction of single-pion final states was fit to \numu deuterium scattering data~\cite{bib:D2Data,bib:D2Fit}.
This fit resulted in improved values and reduced uncertainties for the axial vector mass for resonant pion production
$M^{RES}_{A}$, and the resonant pion production and non-resonant single pion production normalizations in GENIE.
Furthermore, the uncertainty on the vector mass for resonant pion production $M^{RES}_{V}$ was reduced from the GENIE
default $\pm$10\% to $\pm$3\%.  This reduction was supported by comparisons of predicted and measured helicity amplitudes
for resonance production in electron-nucleus scattering~\cite{bib:mvres_constraint}.  The reduced CC resonance production
and CC non-resonant single pion production uncertainties are listed in Table~\ref{tab:reduced_genie_uncertainties}.

\begin{table}[bp] \footnotesize
\begin{center}
\begin{tabular}{ l c }
Parameter & Variation \\
\hline
\multicolumn{2}{c}{\textit{CC Resonance Production}} \\
Normalization & $\pm$7\% \\
Axial vector mass & $\pm$5\% \\
Vector mass & $\pm$3\% \\
\hline
\multicolumn{2}{c}{\textit{CC \& NC Non-Resonant Pion Production}} \\
Normalization of 1$\pi$ final states from $\nu p$ / $\overline{\nu} p$ & $\pm$4\% \\
Normalization of 1$\pi$ final states from $\nu n$ / $\overline{\nu} n$ & $\pm$4\% \\
\end{tabular}
\end{center}
\caption{The reduced $\pm1\sigma$ variations to the CC resonance production and CC non-resonant single pion production model parameters in GENIE.}
\label{tab:reduced_genie_uncertainties}
\end{table}

As discussed in Sec.~\ref{sec:mc_reweight}, the isotropic $\Delta^{++}\to N\pi$ decay in the MC was weighted to
half the non-isotropy predicted by the Rein-Sehgal resonance production model.  The uncertainty on the decay isotropy
applied to the measured cross sections was half the difference between the isotropic and non-isotropic predictions.
%where the isotropic and full non-isotropic predictions serve as the $\pm 1\sigma$ variations to the half non-isotropic prediction.
This is a reduction of the default GENIE uncertainty, which is the full difference between the isotropic and non-isotropic predictions.

The uncertainty on the measured cross sections does not include uncertainty intrinsic to the Rein-Sehgal model of
coherent production in the MC.  The \minerva implementation of GENIE 2.6.2 did not include event weights for these uncertainties.
%on the Rein-Sehgal coherent model.
In principle, the uncertainty in the measured cross sections should include uncertainty in the signal model due to the signal
model dependence introduced by the unfolding and efficiency correction.
Bias from the signal model was evaluated by remeasuring the cross sections with the signal model weighted to the initial cross
section measurements; the differences in the extracted cross sections were small compared to the total uncertainty.
%However, bias from the signal model was minimized by
%remeasuring the cross sections with the signal model weighted to the initial cross section measurements and
%as a result, remaining bias should be small compared to the total uncertainty on the cross sections.
We found that the remaining bias is small compared to the total uncertainty in the cross sections.

The largest GENIE uncertainties in the measured \numu cross sections are the uncertainties in the pion inelastic
interaction and pion absorption rates of the FSI model, and the uncertainty in the $\Delta^{++}$ decay isotropy.  The
largest GENIE uncertainties on the measured \numubar cross sections are the uncertainties on the neutron elastic interaction
and neutron absorption rates in the FSI model.

\subsection{Vertex Energy}
\label{sec:sys_vertex_energy}

%The amount of visible energy near the event vertex affects the acceptance of the vertex energy cut.  

For non-coherent interactions in the MC, the amount of visible energy near the event vertex is dependent upon the
modeling of initial and final state nuclear effects in GENIE.  Since vertex energy is used to reject non-coherent interactions,
the predicted rate of background is thereby sensitive to the modeling of these effects.  This
sensitivity was minimized by tuning the background prediction to data after cutting on vertex energy.  Uncertainties on
the measured cross sections from modeling final state interactions in the nucleus were evaluated using the GENIE FSI weights.

The modeling of initial state nuclear effects in current neutrino-nucleus event generators is known to be incomplete.  In
particular, the version of GENIE used in generating the MC does not model scattering off correlated nucleon pairs, an effect
observed in electron scattering data.  The \minerva \numu CCQE results~\cite{bib:Arturo} provide evidence for scattering off correlated nucleon
pairs in neutrino-nucleus interactions.  This evidence was the result of an analysis of the visible energy near the event vertex of
a \numu CCQE-enhanced sample.  A fit of the vertex energy distribution predicted by the GENIE-based MC to that of the data
preferred the addition of a final state proton with kinetic energy less than 225 MeV to 25\% of the events in the MC.  This
suggests the presence of events with CC scattering off an initial state neutron-proton correlated pair resulting in the
ejection of two final state protons from the nucleus.

An uncertainty was applied to the measured cross sections to account for the absence of modeling the scattering off
correlated nucleon pairs in GENIE.  Informed by the \minerva \numu CCQE results, this uncertainty was evaluated by adding
the vertex energy from a final state proton with kinetic energy less than 225 MeV to the vertex energy of 25\% of events in
the MC where the neutrino scattered off a neutron.  The additional vertex energy was sampled from a vertex energy distribution
for a sample of simulated single protons originating in the tracker with a flat 25-225 MeV kinetic energy spectrum.  The
uncertainty on the measured cross sections from the additional vertex energy is shown in Figs.~\ref{fig:sys_sigenu}--\ref{fig:sys_dsigdqsq}.

%Our analysis used extra energy near the event vertex to reject background.  However, we knew from MINERvA's initial CCQE results \cite{bib:Arturo} that GENIE mis-modeled vertex activity.  Specifically, the CCQE results reported that the data preferred the addition of a final state proton with momentum less than 225 MeV/c in 25\% of events where the target was a neutron.  To estimate our error due to mis-modeled vertex activity, we measured the change in our result due to these additional protons by adding their visible energy deposited in our vertex region (see Section~\ref{sec:event_selection}) to our vertex energy.  We did not consider the additional protons' affect on reconstructed \epi since we excluded non-muon energy within 200 mm from the event vertex when reconstructing \epi (see Section~\ref{sec:kinematics_reco}).

%To measure the change in our result due to the additional protons we first generated a particle gun sample of protons originating in the tracker region with a flat 20-225 MeV/c momentum spectrum and isotropic in direction.  We then measured the spectrum of visible energy deposited by these protons within our vertex region centered at the proton's origin.  In 25\% of events where the target was a neutron an energy was randomly sampled from this distribution and added to the vertex energy.  The event selection and background tuning were repeated for this variation.  %The fractional errors from the resulting covariance matrices were then applied to the background and propagated to the final measured cross sections.

\subsection{Sideband Constraints}
\label{sec:sys_sideband}

The MC was weighted as a function of reconstructed \thetapi to correct the disagreement between the data and MC in the
sideband reconstructed \thetapi distribution after tuning the background normalizations (Sec.~\ref{sec:bg_tuning}).
An uncertainty was applied to the background prediction to account for the extrapolation of the weighting from the
sideband to the signal region.  The size of the uncertainty was the full difference between the weighted and unweighted
tuned background predictions.  This uncertainty was propagated to the measured coherent cross sections (Figs.~\ref{fig:sys_sigenu}--\ref{fig:sys_dsigdqsq}).

%After tuning our backgrounds in \epi, disagreement remained in the sideband \thetapi distribution that exceeded the data and MC statistical errors.  To account for this disagreement, background in all MC (central value and all systematic variations) were weighted by a discontinuous function of \thetapi, referred to herein as \thetapi weighting, that brought the tuned MC and data sideband \thetapi distributions into agreement within statistical error.  The function was constructed by-hand from the data/MC ratio in the sideband \thetapi distribution after background tuning.

%In addition to the \thetapi weighting, an error was also applied to account for the sideband \thetapi disagreement, referred to herein as the sideband \thetapi error, since it was not known whether or not the disagreement resulted from signal contained in the data sideband.  The error was determined by applying the \thetapi weighting to the central value MC and remeasuring the cross section.  The background estimates were remeasured for this variation.  The efficiency was not remeasured since the \thetapi weighting was not applied to signal.  %The fractional errors from the resulting covariance matrices were then applied to the background and propagated to the final measured cross sections.

%It should be noted that the sideband \thetapi error was determined from a change to the measured signal but was applied to the background.  The sideband \thetapi error enters in the error on the final measured cross sections through background subtraction.

\subsection{Detector Model}
\label{sec:sys_detector}

\begin{figure*}[tpb]
\centering
\mbox{
\includegraphics[width=0.49\linewidth]{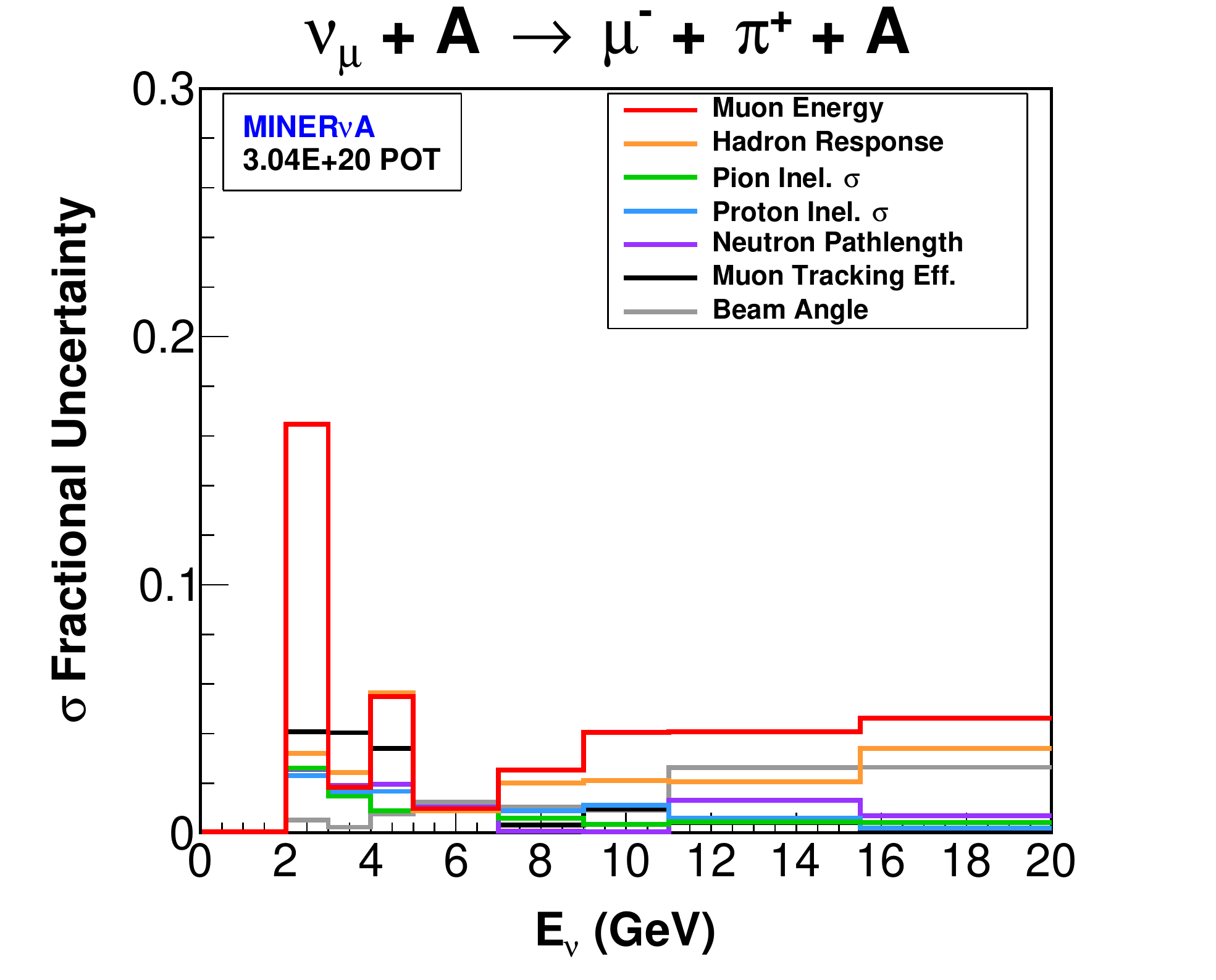}
\includegraphics[width=0.49\linewidth]{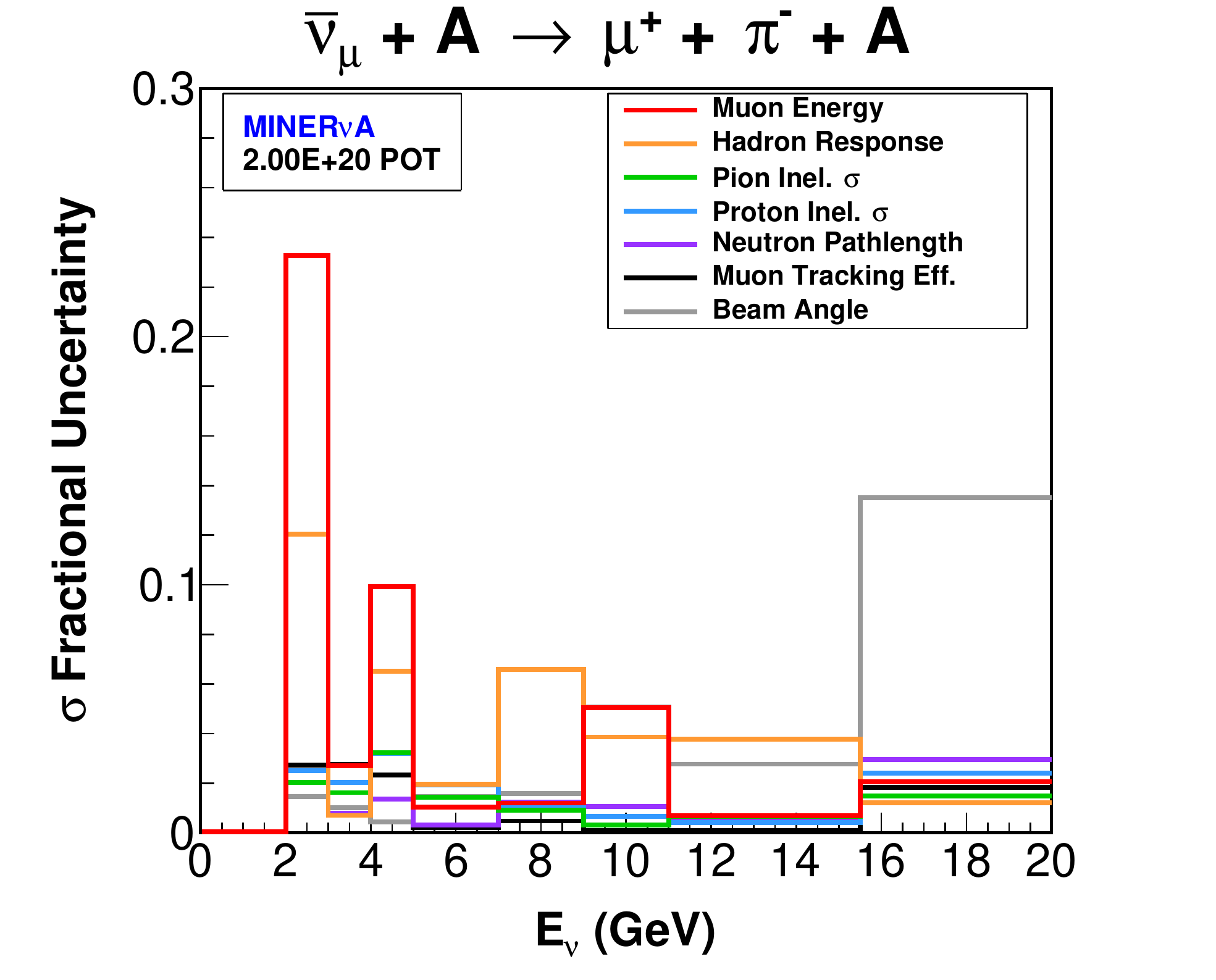}}
\caption[Detector model uncertainties on the measured \numu and \numubar \sigenu]{The fractional detector model uncertainties on the measured \numu (left) and \numubar (right) \sigenu.}
\label{fig:sys_detector_sigenu}
\end{figure*}

\begin{figure*}[tpb]
\centering
\mbox{
\includegraphics[width=0.49\linewidth]{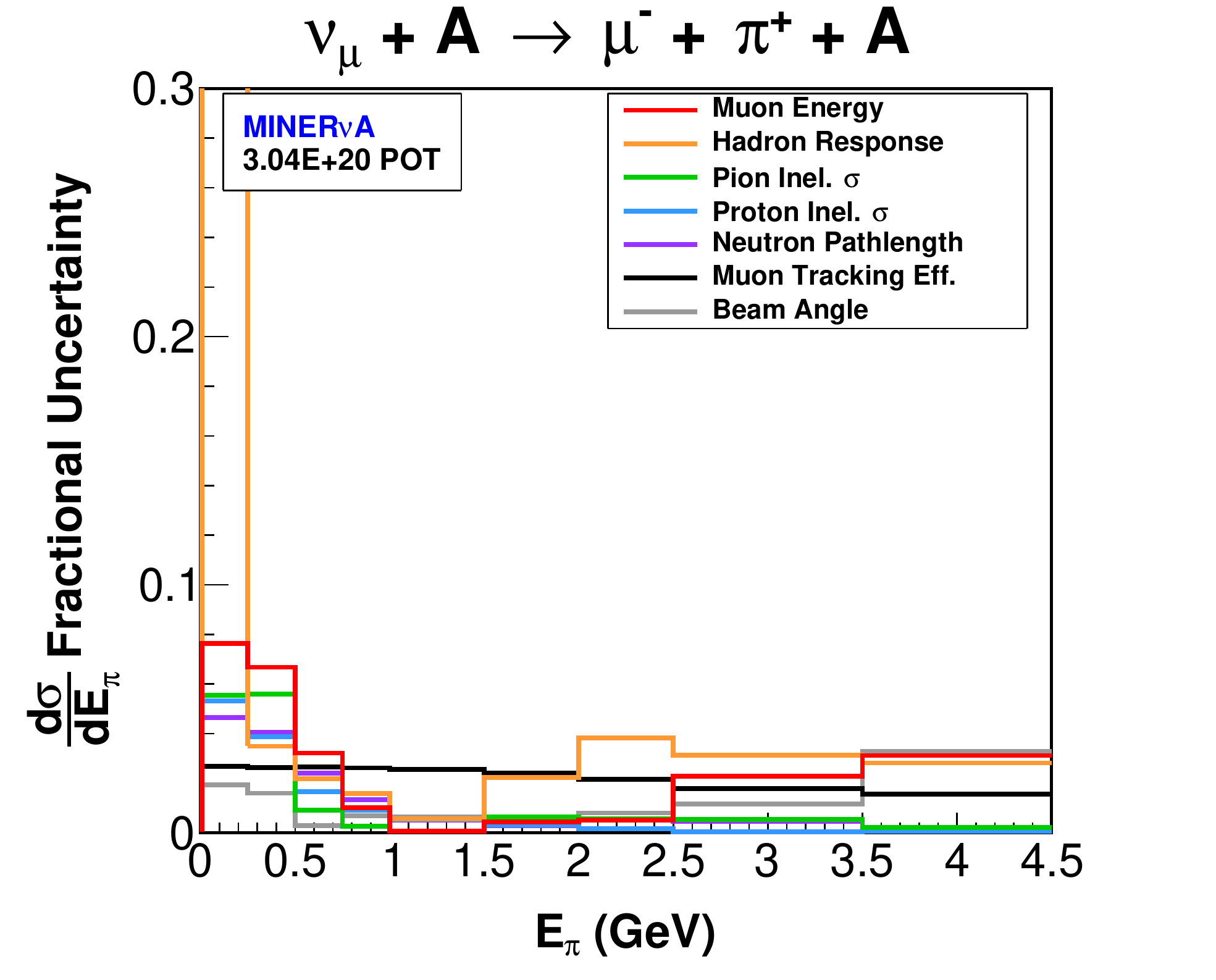}
\includegraphics[width=0.49\linewidth]{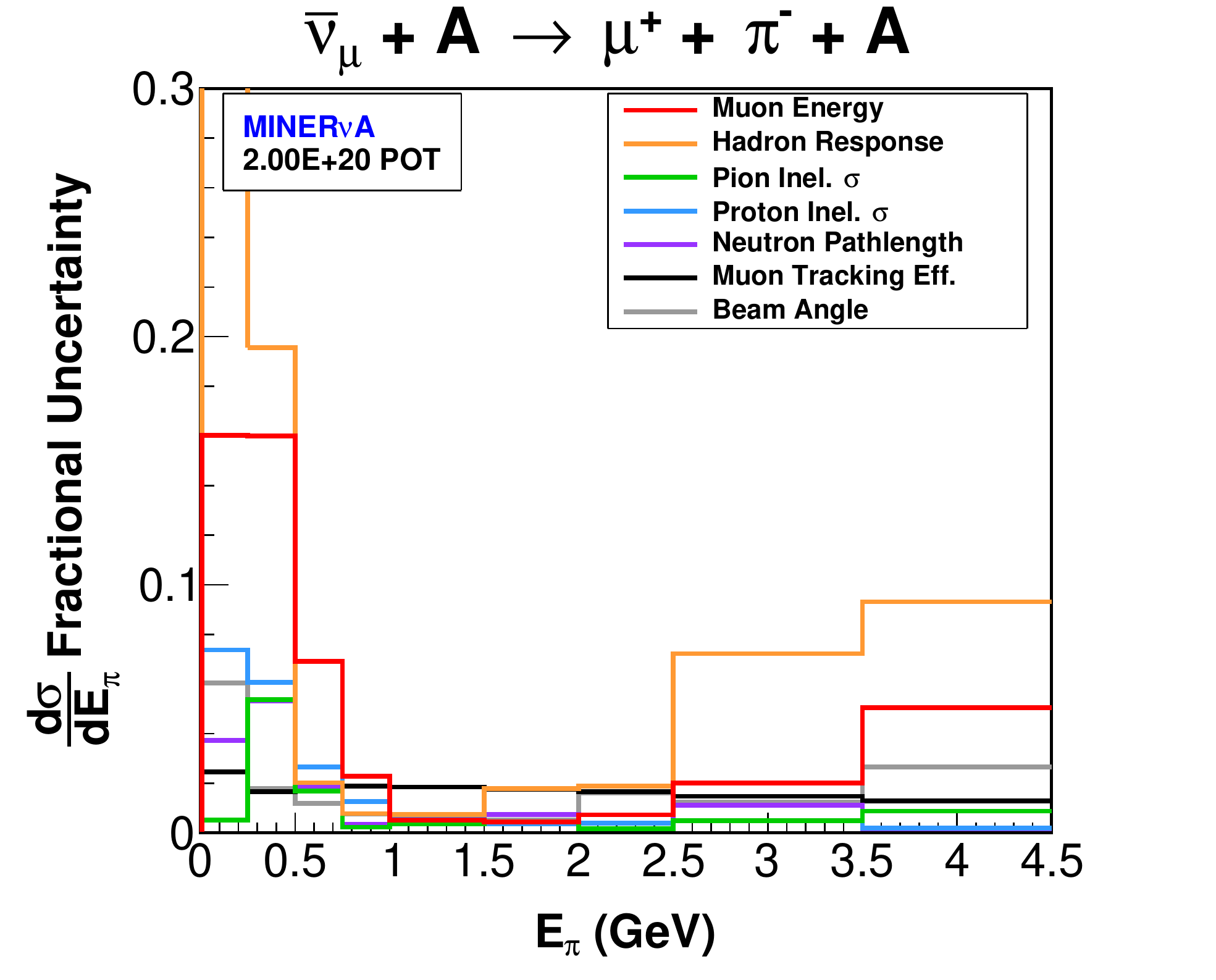}}
\caption[Detector model uncertainties on the measured \numu and \numubar \dsigdepi]{The fractional detector model uncertainties on the measured \numu (left) and \numubar (right) \dsigdepi.}
\label{fig:sys_detector_dsigdepi}
\end{figure*}

\begin{figure*}[tpb]
\centering
\mbox{
\includegraphics[width=0.49\linewidth]{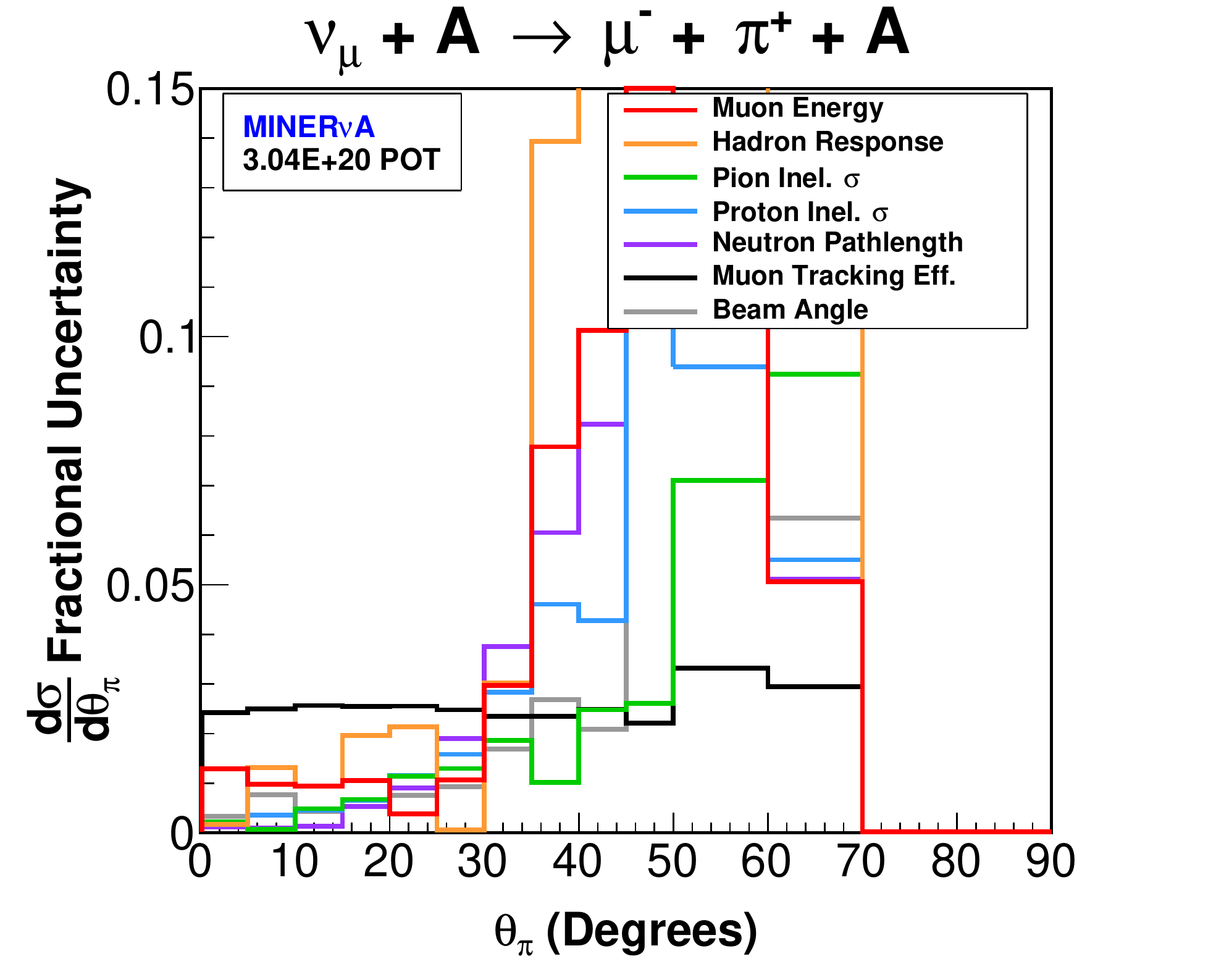}
\includegraphics[width=0.49\linewidth]{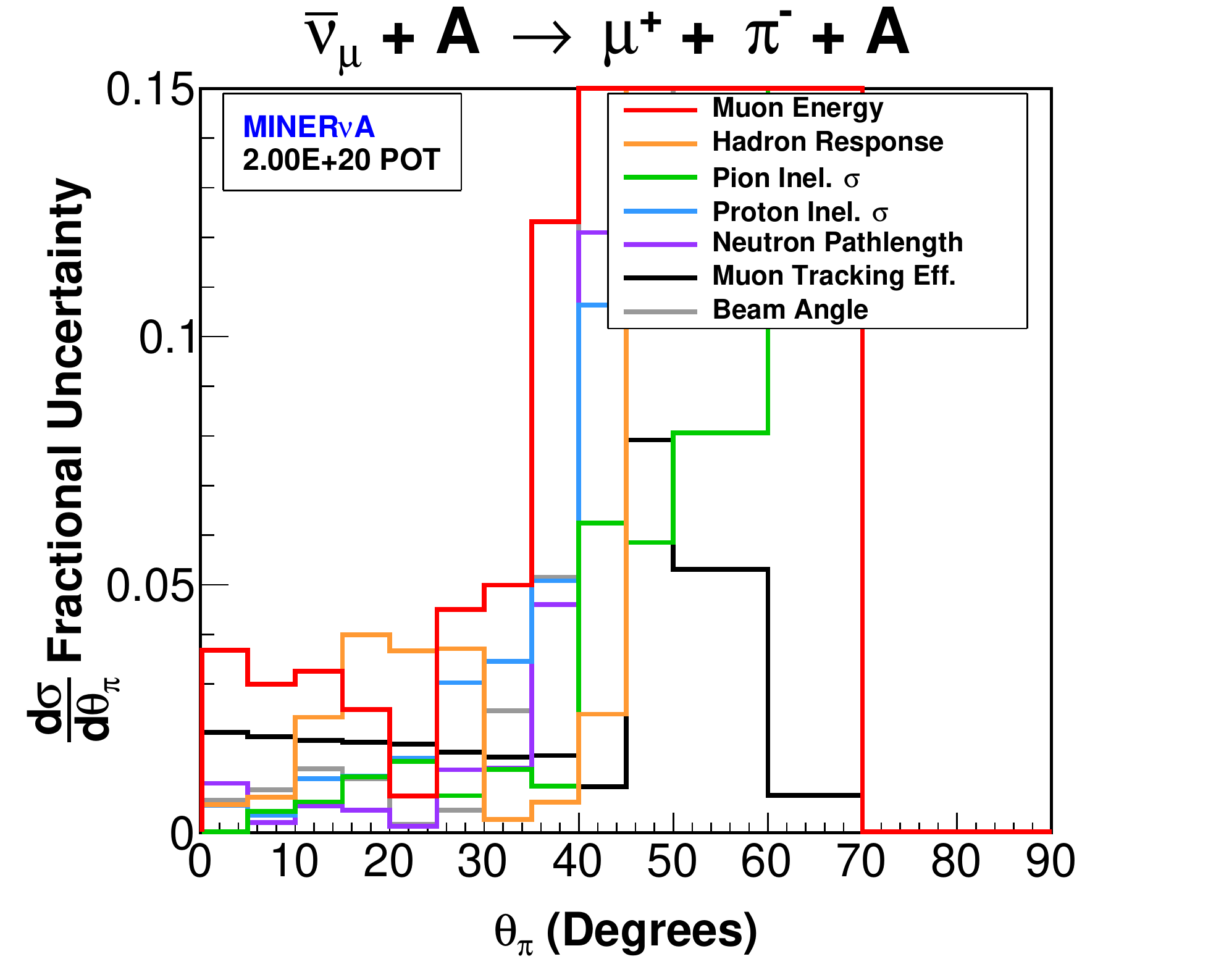}}
\caption[Detector model uncertainties on the measured \numu and \numubar \dsigdthetapi]{The fractional detector model uncertainties on the measured \numu (left) and \numubar (right) \dsigdthetapi.}
\label{fig:sys_detector_dsigdthetapi}
\end{figure*}

\begin{figure*}[tpb]
\centering
\mbox{
\includegraphics[width=0.49\linewidth]{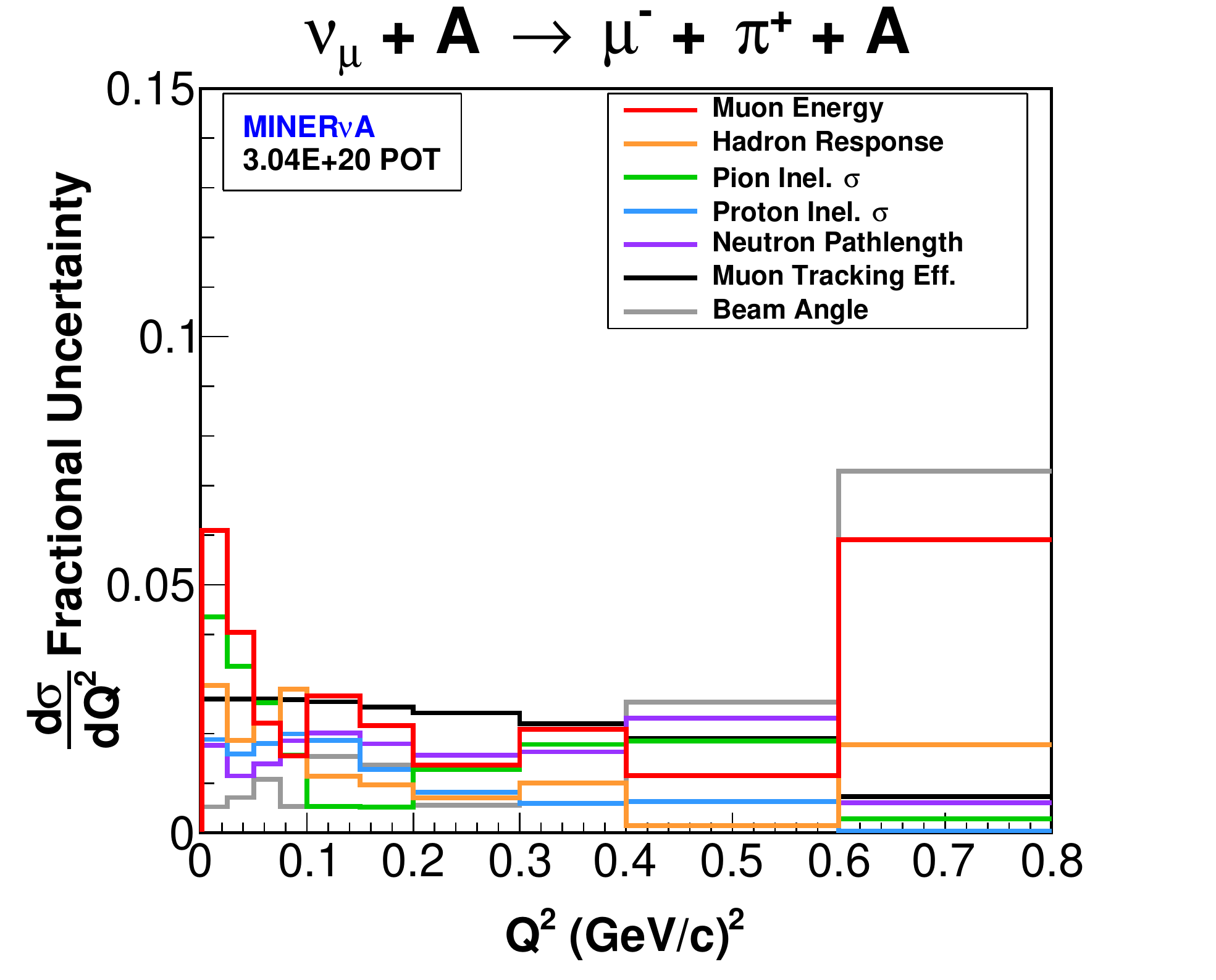}
\includegraphics[width=0.49\linewidth]{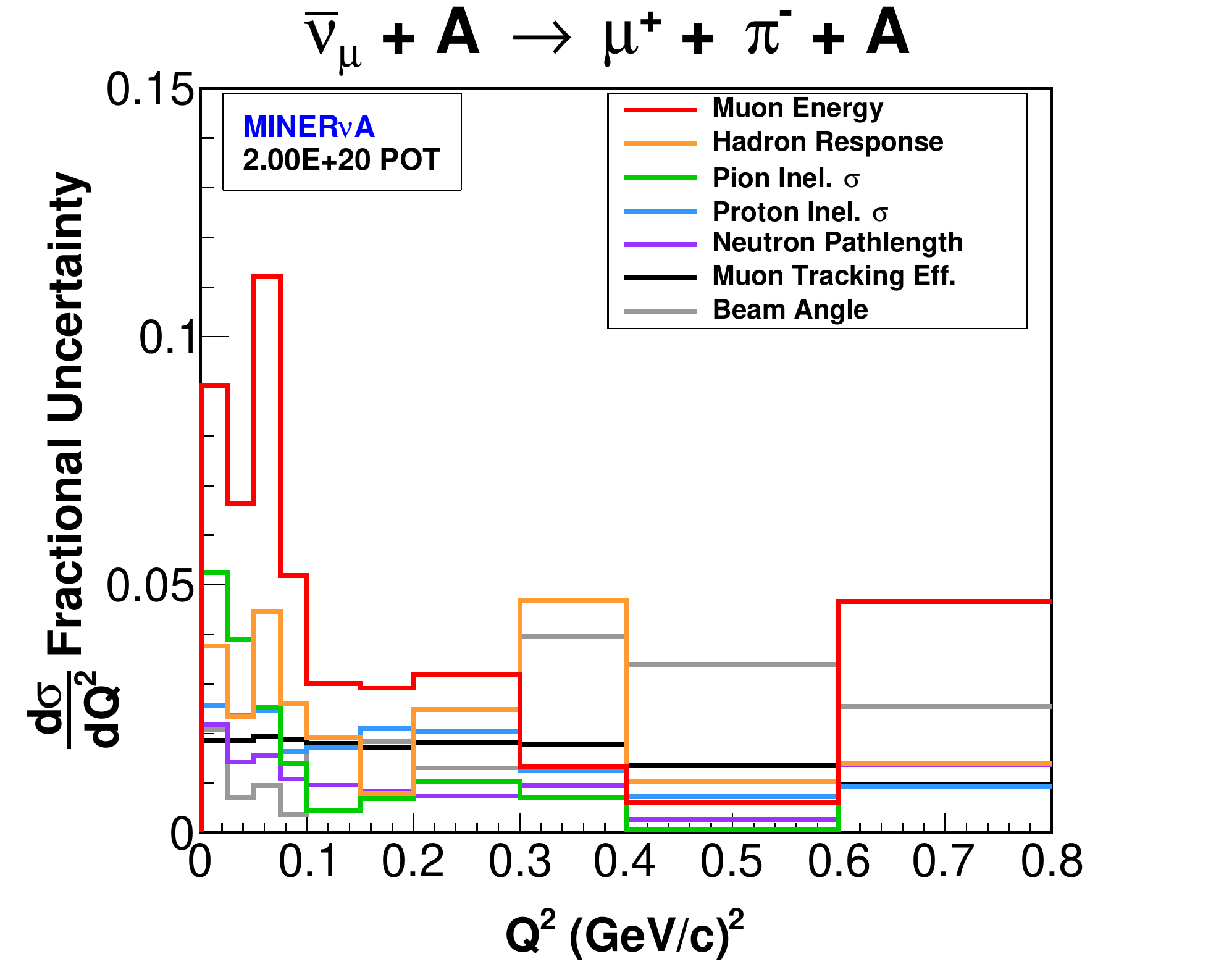}}
\caption[Detector model uncertainties on the measured \numu and \numubar \dsigdqsq]{The fractional detector model uncertainties on the measured \numu (left) and \numubar (right) \dsigdqsq.}
\label{fig:sys_detector_dsigdqsq}
\end{figure*}

The detector model uncertainties on the measured cross sections (Figs.~\ref{fig:sys_detector_sigenu}--\ref{fig:sys_detector_dsigdqsq})
consist of uncertainties on the simulation of particle propagation in the detectors, the particle response of the detectors,
and the particle and kinematics reconstruction.  The following sections detail the evaluation of the detector model uncertainties.

\subsubsection{Muon Tracking Efficiency}
\label{sec:sys_track_eff}

The MC was weighted to correct the difference between MC and the data in the efficiency of tracking the muon in \minos due to pile-up not
being simulated in \minos.  An uncertainty on the correction to the MC tracking efficiency was included in the cross section results, which
was evaluated by varying the size of the corrections by $\pm$50\% of their values.

\subsubsection{Muon Energy Scale}
\label{sec:sys_muon_energy}

Systematic uncertainty on the reconstructed muon energy affects the reconstructed values of other kinematic parameters,
as well as the signal and background coherent candidate event rates via the cuts on \enu and \tabs.  The muon energy uncertainty
consists of uncertainties on the \minos track momentum and the muon energy loss within \minerva.

The momentum of a tracked muon in \minos is measured by either the range of the track or its curvature in the magnetic field,
with the range measurement being more precise.  The range measurement uncertainty was estimated to be 2\%, and consists of
uncertainties on the \minos geometry and material models, muon energy loss calculation, and track vertex
reconstruction~\cite{bib:minos_nim}.  The curvature measurement uncertainty was estimated using a high-statistics sample of
muons produced upstream of the detector that enter the front face of \minos and stop in its fully instrumented region.
%The $\frac{1}{P_{curve}}-\frac{1}{P_{range}}$ distribution (Figure~\ref{fig:minos_muon_p_range_curve_diff}) was measured in both data and MC, where $P_{curve}$ and $P_{range}$ are the curvature and range measurements for a single muon, respectively.  
The difference between the data and MC distribution means established a 0.6\% (2.5\%) uncertainty on the curvature
measurement beyond the 2\% uncertainty on the range measurement for muon momentum above (below) 1 GeV/c.  The uncertainties
were added in quadrature to give an estimated curvature measurement uncertainty of 2.1\% (3.1\%) for muon momentum above (below) 1 GeV/c.

%\begin{figure}[tpb]
%\centering
%\includegraphics[width=1.0\linewidth]{figures/Systematics/minos_muon_p_range_curve_diff.pdf}
%\caption[MINOS muon momentum reconstruction uncertainty]{The $P_{curve} - P_{range}$ distribution (left) and $\frac{1}{P_{curve}}-\frac{1}{P_{range}}$ distribution (right) for \numu CC events in \minerva where the muon stopped in the fully instrumented region of MINOS.  $P_{curve}$ and $P_{range}$ are the momentum of the muon in MINOS reconstructed by the curvature and range of the MINOS muon track, respectively.}
%\label{fig:minos_muon_p_range_curve_diff}
%\end{figure}

The uncertainty on the muon energy loss in \minerva consisted of uncertainties on the \minerva material assay and the
Bethe-Bloch energy loss prediction.  The effects of the material assay uncertainties were estimated by varying the detector mass
in the muon energy loss calculation.  The uncertainty on the Bethe-Bloch energy loss prediction was estimated by comparing the
Bethe-Bloch muon range prediction to the Groom muon range tables~\cite{bib:muon_range_tables} for the materials in the \minerva detector.
For muons originating in the tracker and exiting the back of \minerva, the effects of the material assay and Bethe-Bloch
prediction uncertainties on the muon energy loss were estimated to be 11 MeV and 30 MeV on average, respectively.

For each event, the reconstructed muon energy uncertainty $\Delta$\emu was the quadrature sum of the \minos track momentum
and \minerva muon energy loss uncertainties.  The muon energy uncertainty on the cross sections was evaluated by varying the
muon energy in the MC by $\pm\Delta$\emu.
% Leo asks DOES THIS NEED MORE DETAIL? Kevin doesn't see why.

\subsubsection{Pion and Proton Response}
\label{sec:sys_response}

The calorimetric correction for reconstructing \epi from the visible energy was tuned to the simulated response to pions.
%Uncertainty on the simulated response therefore affects the reconstructed event kinematics and the coherent candidate event rates.
This uncertainty was constrained by measurements of the single pion and proton response in the \minerva test beam data and
MC~\cite{bib:test_beam}.  The response for each pion and proton event was measured as the ratio of the visible energy to
the incoming particle energy, and the detector MC agreed with the data in the mean pion (proton) response to within 5\% (3\%)
over the sampled incoming energy range. This difference was used to evaluate the systematic uncertainty.

%\begin{figure*}[tpb]
%\centering
%\mbox{
%\includegraphics[width=0.49\linewidth]{figures/Systematics/tResponsePiPlus.pdf}
%\includegraphics[width=0.49\linewidth]{figures/Systematics/tResponsePiMinus.pdf}}
%\caption[The test beam detector response to $\pi^{+}$ and $\pi^{-}$ in data and MC]{The mean ratio of visible energy to incoming energy for single $\pi^{+}$ (left) and $\pi^{-}$ (right) events in the \minerva test beam EH configuration data and MC.  The error on the data is the statistical uncertainty, and the error band on the MC is the systematic uncertainty consisting of beam line, detector, and event selection uncertainties.}
%\label{fig:pion_resonse}
%\end{figure*}
%
%\begin{figure*}[tpb]
%\centering
%\includegraphics[width=0.5\linewidth]{figures/Systematics/EH_ratio_mean.pdf}
%\caption[The test beam detector response to protons in data and MC]{The mean ratio of visible energy to incoming energy for single proton events in the \minerva test beam EH configuration data and MC.  The error on the data is the statistical uncertainty, and the error band on the MC is the systematic uncertainty consisting of beam line, detector, and event selection uncertainties.}
%\label{fig:proton_resonse}
%\end{figure*}

\subsubsection{Pion \& Proton Interaction Cross Section}
\label{sec:sys_pionproton_interaction}

The interaction of pions and protons within \minerva affects the tracking efficiency, angular resolution, vertex energy
and calorimetric response for both pions and protons, as well as the proton score.
The uncertainty on the measured coherent cross sections due to the uncertainty in pion and proton interaction rates in the
MC was evaluated by varying the rate of pion and proton inelastic scattering on carbon in the MC by $\pm$10\%.  Here,
inelastic scattering includes charge exchange and absorption processes.  The size
of the variation, referred to as $\delta$ below was determined from comparisons of Geant4 to hadron scattering
data~\cite{Ashery:1981tq,Allardyce:1973ce,Saunders:1996ic,Lee:2002eq,Abfalterer:2001gw,Schimmerling:1973bb,Voss:1956,Slypen:1995fm,Franz:1989cf,Tippawan:2008xk,Bevilacqua:2013rfq,Zanelli:1981zz}.
%~\cite{bib:geant_xsec}. THIS CITATION IS A DOCDB ENTRY.
The inelastic scattering rate was varied using an event-by-event weighting technique
%~\cite{bib:pion_proton_weighting}, THIS CITATION IS ALSO A DOCDB ENTRY
where a weight was calculated for each final state pion and proton.  The weight for a pion or proton that interacted
inelastically was calculated as 
\begin{equation}
\label{eq:weight_inel_xs_int}
W_{inel} = \frac{1-e^{-\rho x \sigma_{inel}(1+\delta)}}{1-e^{-\rho x \sigma_{inel}}},
\end{equation}
and the weight for a pion/proton that did not interact inelastically was calculated as
\begin{equation}
\label{eq:weight_inel_xs_non_int}
W_{non-inel} = \frac{e^{-\rho x \sigma_{inel}(1+\delta)}}{e^{-\rho x \sigma_{inel}}} = e^{-\rho x \sigma_{inel} \delta},
\end{equation}
where $\rho$ is the average density of the scintillator planes, $x$ is the total path length of the particle, $\sigma_{inel}$
is the energy averaged pion or proton inelastic scattering cross section on carbon, and $\delta$ is the variation on the
inelastic scattering rate.  The energy averaged cross section was used in calculating the weights to account for the
dependence of the cross section on pion or proton energy, and was calculated as
\begin{equation}
\label{eq:energy_ave_int_xs}
\sigma_{inel} = \frac{1}{E_{f}-E_{i}}\int_{E_{i}}^{E_{f}}\sigma_{inel}(E)dE,
\end{equation}
where $E_{i}$ and $E_{f}$ are the initial and final kinetic energy of the pion or proton, respectively, and $\sigma_{inel}(E)$
is a parameterization of the pion or proton inelastic scattering cross section on carbon from the hadron scattering data.
The weight for each MC event was the product of the weights for all final state pions and protons.  The event weights were normalized
to preserve the total neutrino interaction rate for each MC event category.

The pion and proton calorimetric response uncertainties from the test beam measurements include uncertainties on the pion and
proton interaction rates.  The separate evaluation of the interaction rate uncertainties here double counts their contribution
to the calorimetric response uncertainties.  Since it is difficult in practice to isolate their contribution to the calorimetric
response uncertainties, the best approach is to conservatively include the pion and proton interaction rate uncertainties on
the calorimetric response along with the calorimetric response uncertainties from the test beam measurements.

%Varying the pion and proton interaction rates affects the simulated calorimetric response.  This evaluation of the pion and proton interaction rate uncertainties therefore double cothe The calorimetric response uncertainty on the measured coherent cross sections was evaluated separately per Sec.~\ref{sec:sys_response}.  Evaluating the 

%Therefore, the component of the calorimetric response uncertainty due to the uncertainty on the pion/proton interaction rate is double counted.  In practice it is difficult to isolate the pion and proton interaction rate component of the calorimetric response uncertainty.  This double counting is a small contribution to the total uncertainty on the measured coherent cross sections and was not corrected.

\subsubsection{Neutron Path Length}
\label{sec:sys_neutron_path}

In \minerva, neutrons tend to either exit the detector without depositing any energy or deposit only a small fraction of
their energy {\it via} a hadronic interaction.  The neutron interaction rate therefore has a small effect on the calorimetric
response.  However, final state neutrons that promptly interact can affect the vertex energy and occasionally produce
a tracked proton or pion.  The neutron interaction rate therefore affects the predicted rate of background coherent candidates.

The uncertainty in the measured coherent cross sections due to the uncertainty on the neutron interaction rate in the MC was
evaluated by varying the neutron mean free path, similar to the procedure in Sec.~\ref{sec:sys_pionproton_interaction}.
% ~\cite{bib:neutron_weighting}.   this is a DocDB; can't cite in PRD
The mean free path was varied $\pm$25\% for
neutron kinetic energy below 40 MeV, $\pm$10\% between 50 and 150 MeV, and $\pm$20\% above 300 MeV with interpolations in the uncertainties
between these regions.  
%The variation in the undefined energy regions was interpolated.  
The amount of these variations was determined from comparisons of GEANT4 and neutron
scattering data.%~\cite{bib:geant_xsec}. this is a DocDB; can't cite in PRD 

%A weight was calculated for each final state neutron, which depended on the path length of the neutron in the tracker region and whether the neutron interacted inside the tracker region.  The weight applied to each MC event was the product of its neutron weights.  The event weights were normalized to preserve the total neutrino interaction rate for each MC event category.

\subsubsection{Beam Direction}
\label{sec:sys_beam_direction}

Reconstruction of event kinematics assumed the incoming neutrino direction was parallel to the neutrino beam axis.
The measured cross sections are therefore sensitive to uncertainty on the beam direction.  The measurement of the beam
direction using the \numu and \numubar CC low-$\nu$ samples agreed with the beam direction in the model of the detector
geometry to within $\sim$3 mrad in both the vertical (YZ) and horizontal (XZ) planes.  The uncertainty in the cross
sections from beam direction uncertainty was evaluated by varying the beam direction $\pm$3 mrad in both the vertical and horizontal planes.

\section{Measured Cross Sections}
\label{sec:measured_cross_sections}

\begin{figure*}[tpb]
\centering
\mbox{
\includegraphics[width=0.49\linewidth]{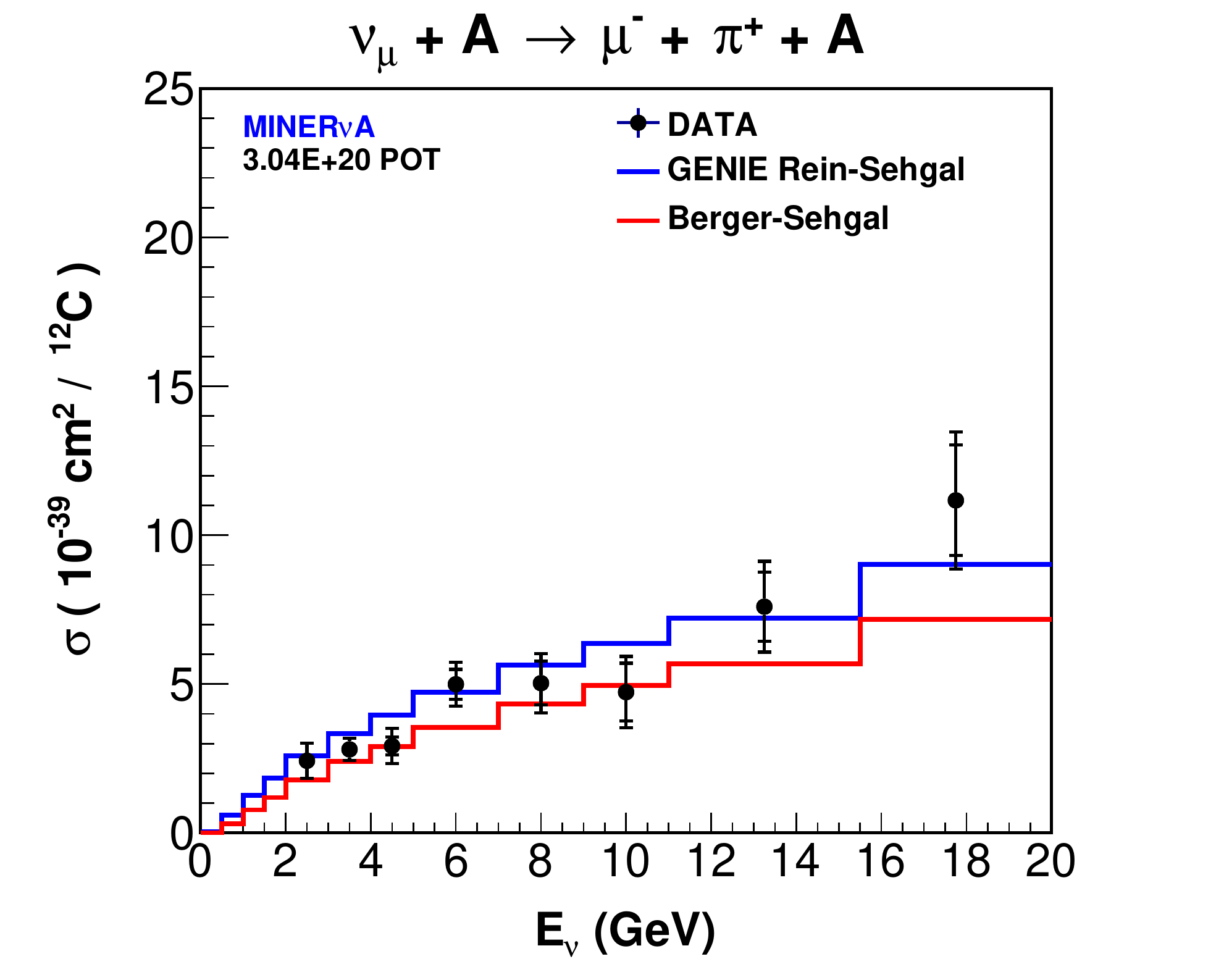}
\includegraphics[width=0.49\linewidth]{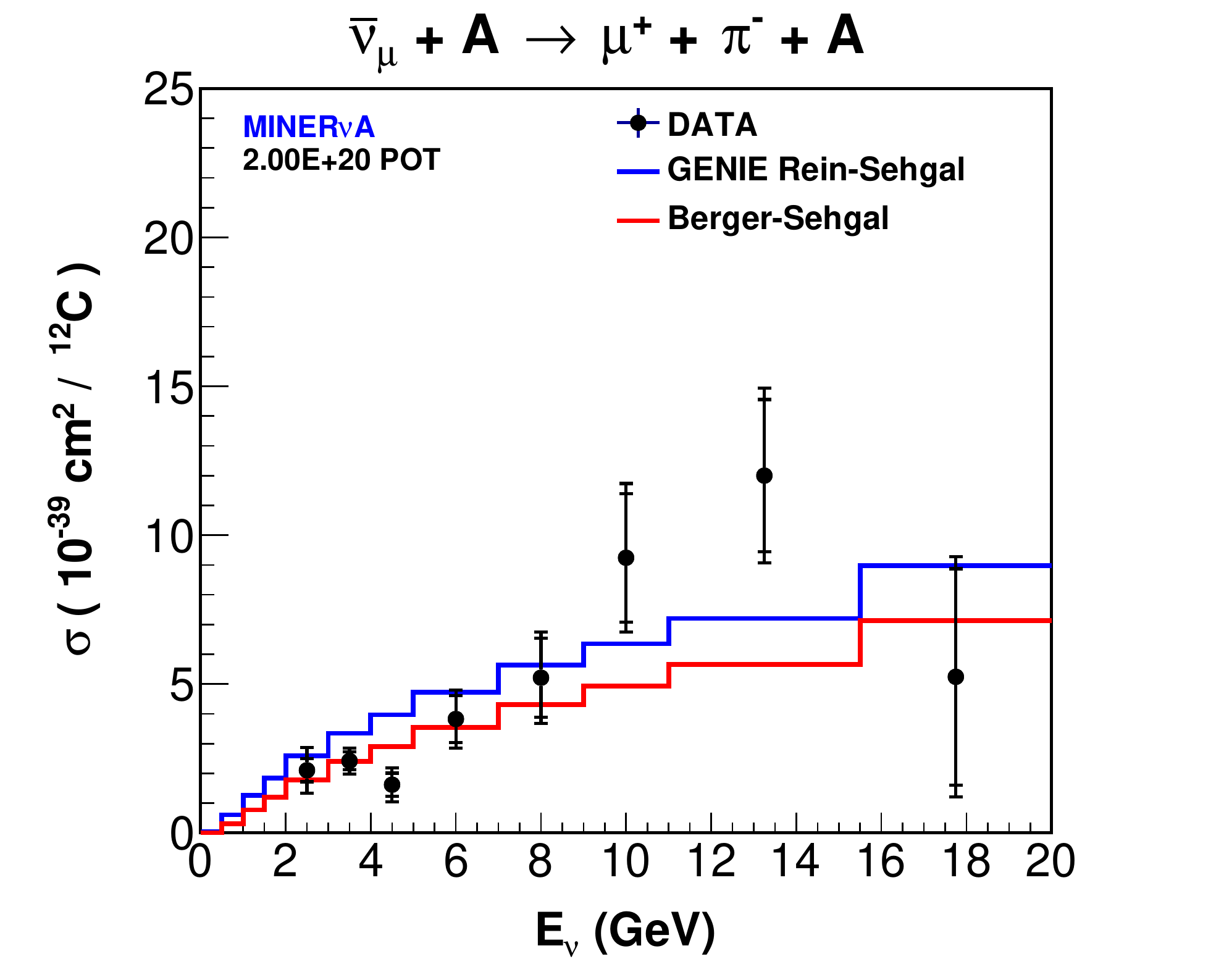}}
%\mbox{
%\includegraphics[width=0.49\linewidth]{figures/CrossSection/h_Ev_Final_XSec_ThetaPiCor_FluxConstrained_WeightedSignalModel_minerva5.pdf}
%\includegraphics[width=0.49\linewidth]{figures/CrossSection/h_Ev_Final_XSec_ThetaPiCor_FluxConstrained_WeightedSignalModel_downstream.pdf}}
\caption[The measured \numu and \numubar \sigenu]{The measured \sigenu for the \numu (left) and \numubar (right), samples.  The inner and outer error bars are the statistical and total (statistical + systematic) uncertainties, respectively.}
\label{fig:meas_sigenu}
\end{figure*}

\begin{figure*}[tpb]
\centering
\mbox{
\includegraphics[width=0.49\linewidth]{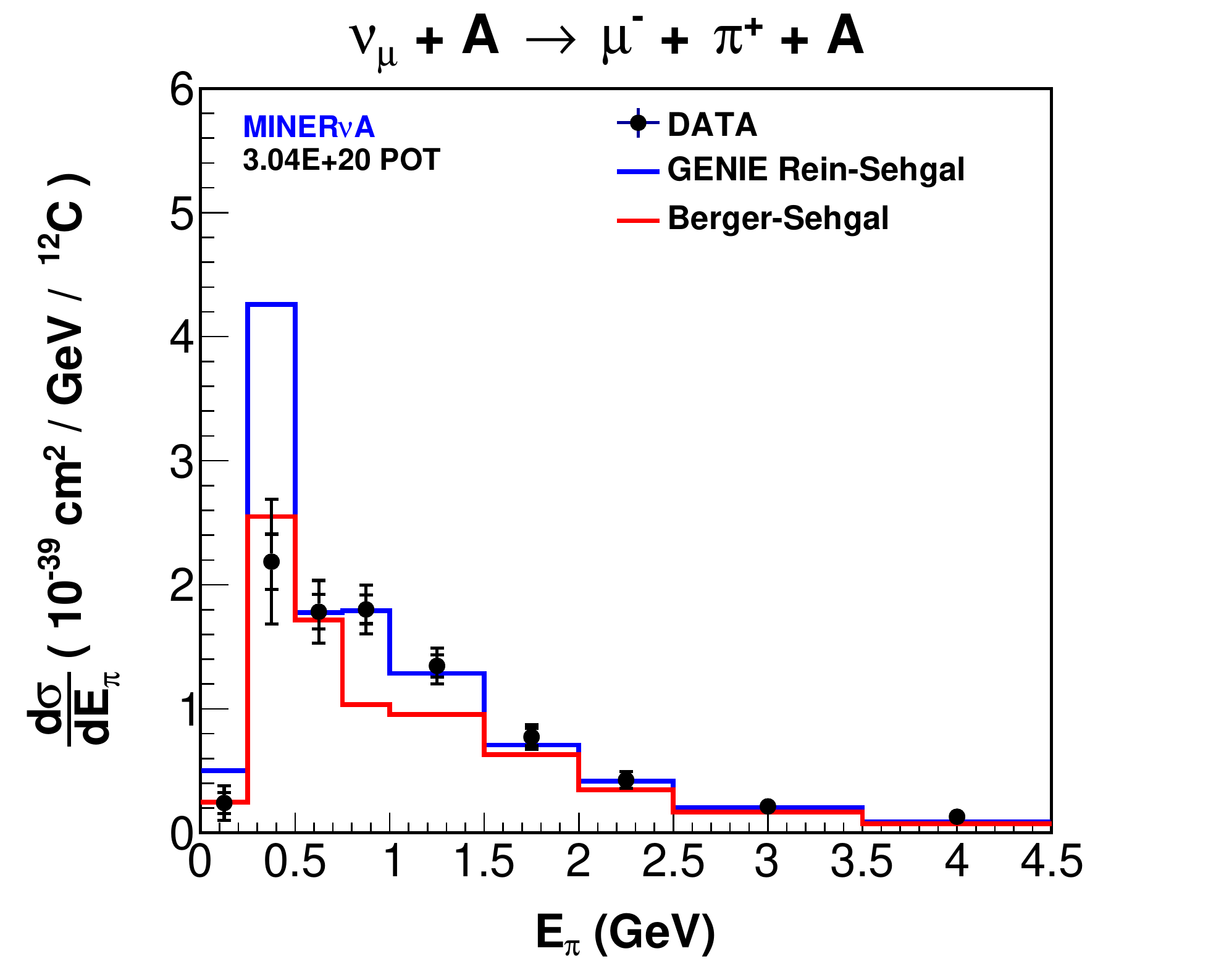}
\includegraphics[width=0.49\linewidth]{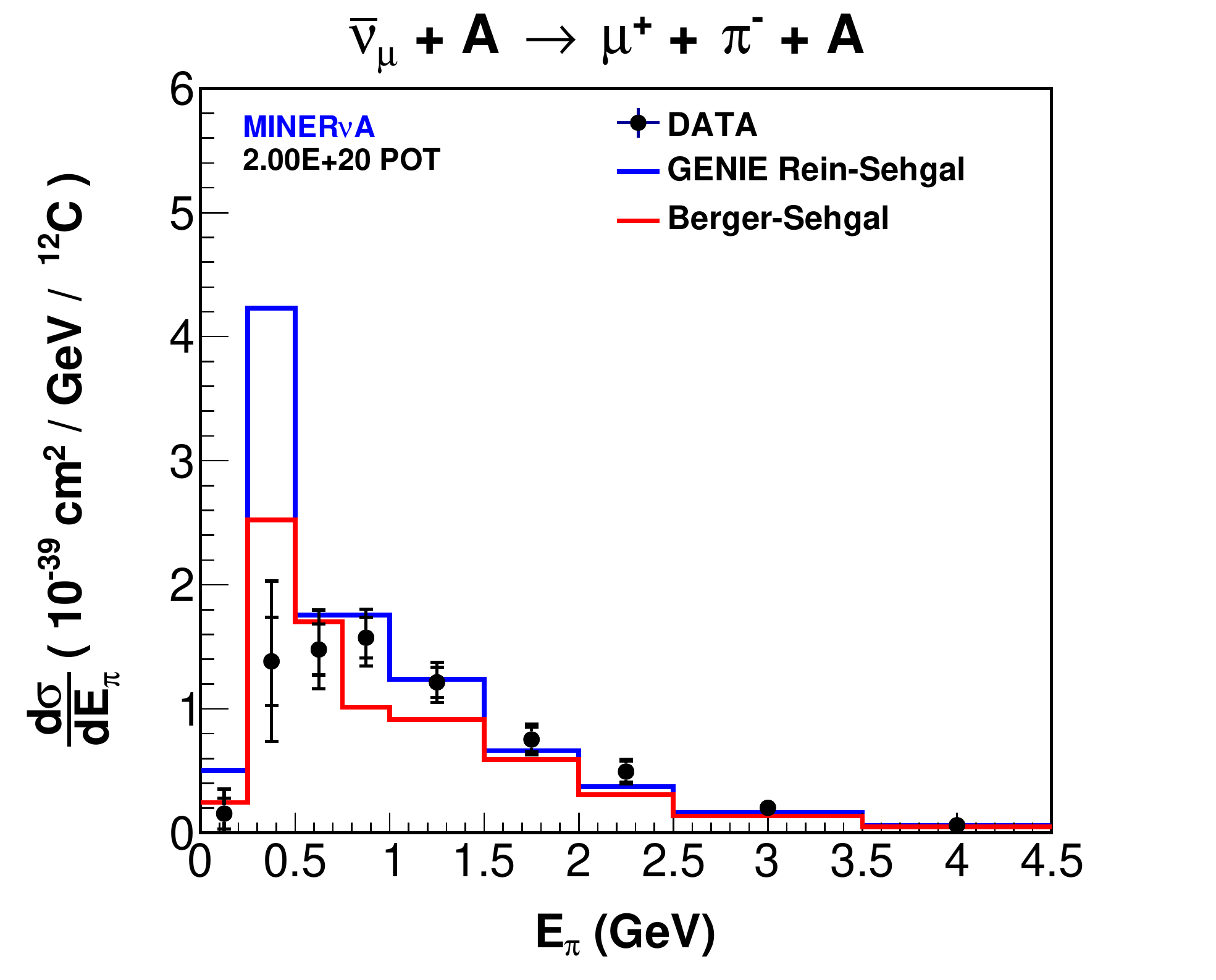}}
%\mbox{
%\includegraphics[width=0.49\linewidth]{figures/CrossSection/h_Epi_Final_XSec_ThetaPiCor_FluxConstrained_WeightedSignalModel_minerva5.pdf}
%\includegraphics[width=0.49\linewidth]{figures/CrossSection/h_Epi_Final_XSec_ThetaPiCor_FluxConstrained_WeightedSignalModel_downstream.pdf}}
\caption[The measured \numu and \numubar \dsigdepi]{The measured \dsigdepi for the \numu (left) and \numubar (right) samples.  The inner and outer error bars are the statistical and total (statistical + systematic) uncertainties, respectively.}
\label{fig:meas_disgdepi}
\end{figure*}

\begin{figure*}[tpb]
\centering
\mbox{
\includegraphics[width=0.49\linewidth]{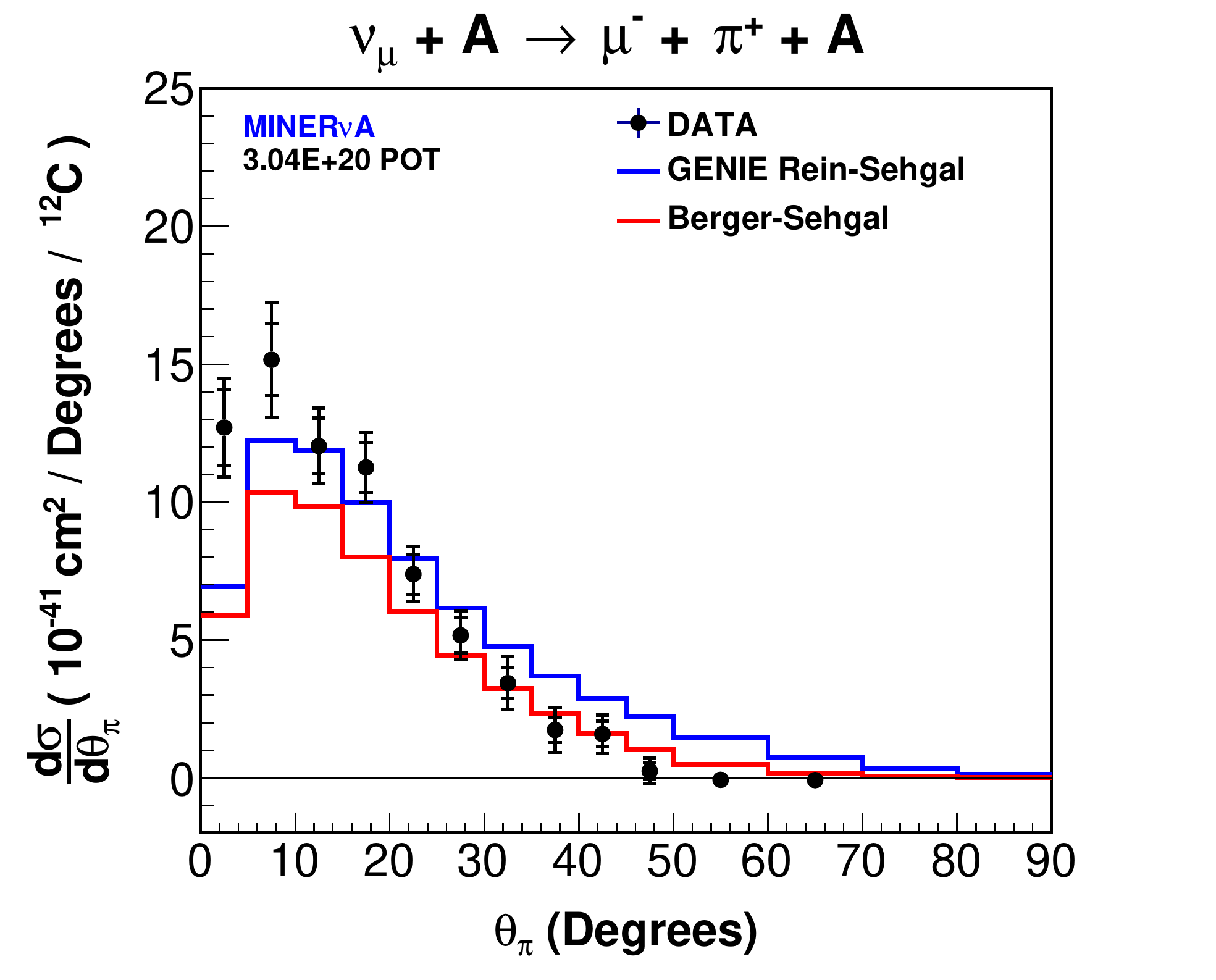}
\includegraphics[width=0.49\linewidth]{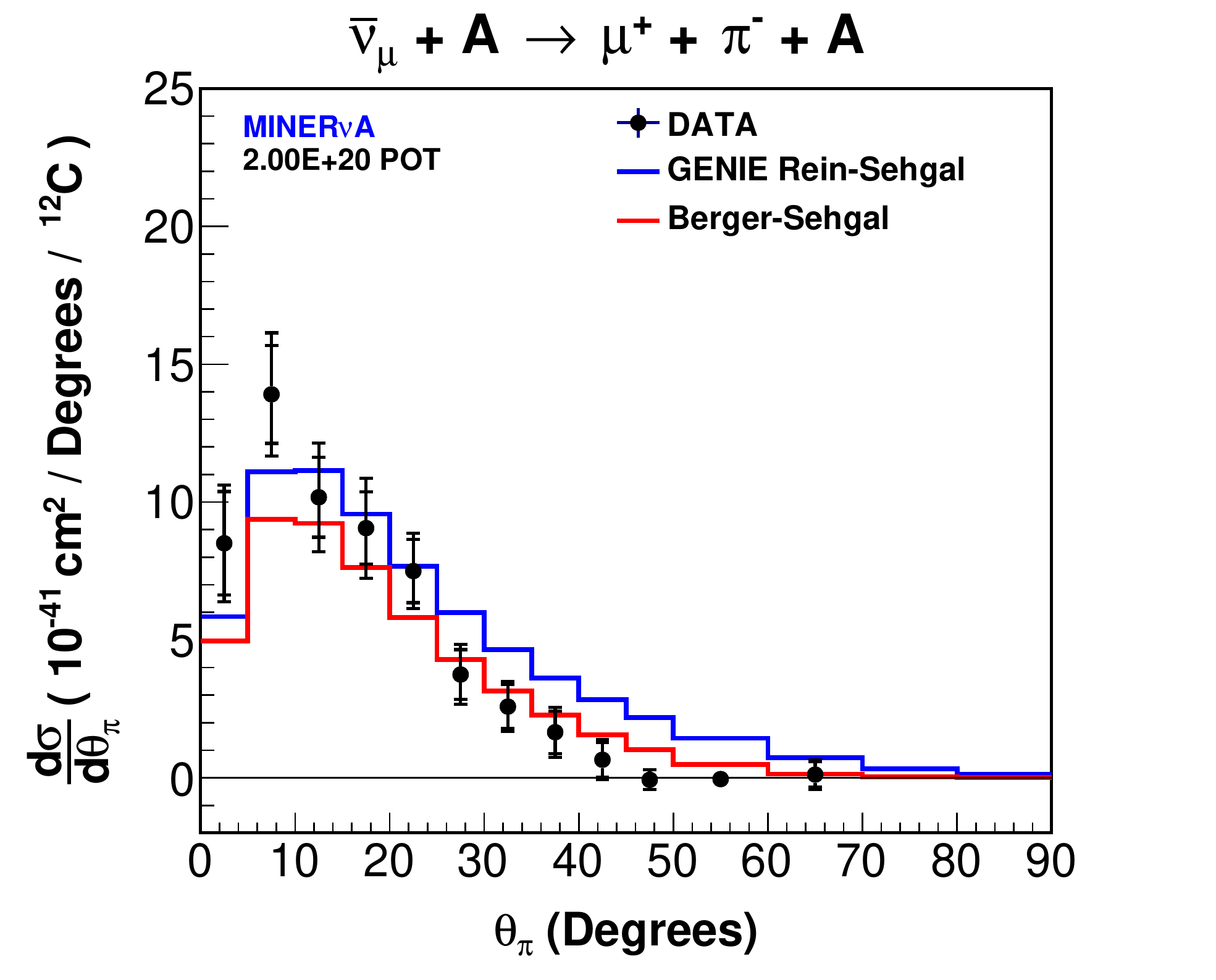}}
%\mbox{
%\includegraphics[width=0.49\linewidth]{figures/CrossSection/h_ThetaPi_Final_XSec_ThetaPiCor_FluxConstrained_WeightedSignalModel_minerva5.pdf}
%\includegraphics[width=0.49\linewidth]{figures/CrossSection/h_ThetaPi_Final_XSec_ThetaPiCor_FluxConstrained_WeightedSignalModel_downstream.pdf}}
\caption[The measured \numu and \numubar \dsigdthetapi]{The measured \dsigdthetapi for the \numu (left) and \numubar (right) samples.  The inner and outer error bars are the statistical and total (statistical + systematic) uncertainties, respectively.}
\label{fig:meas_disgdthetapi}
\end{figure*}

\begin{figure*}[tpb]
\centering
\mbox{
\includegraphics[width=0.49\linewidth]{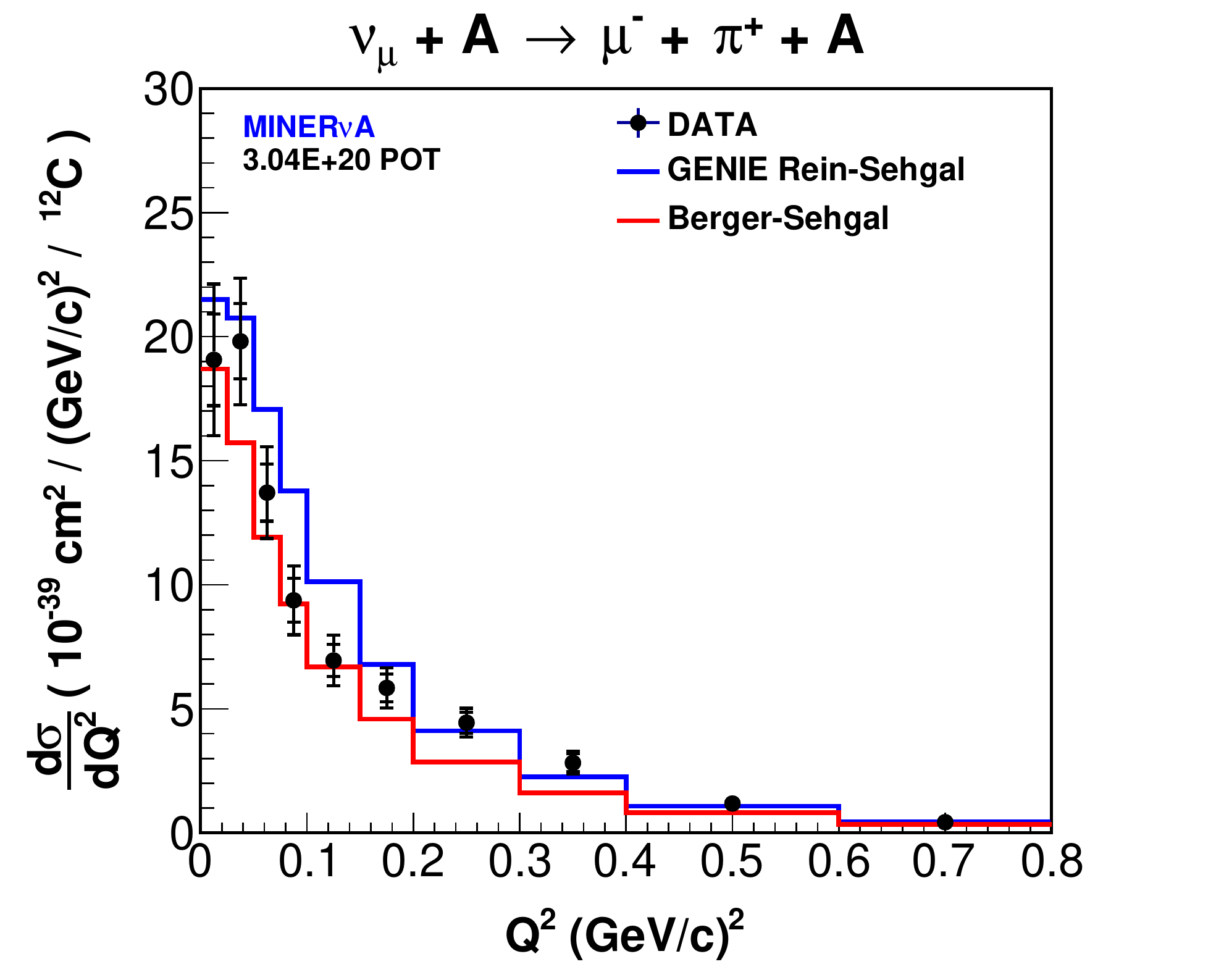}
\includegraphics[width=0.49\linewidth]{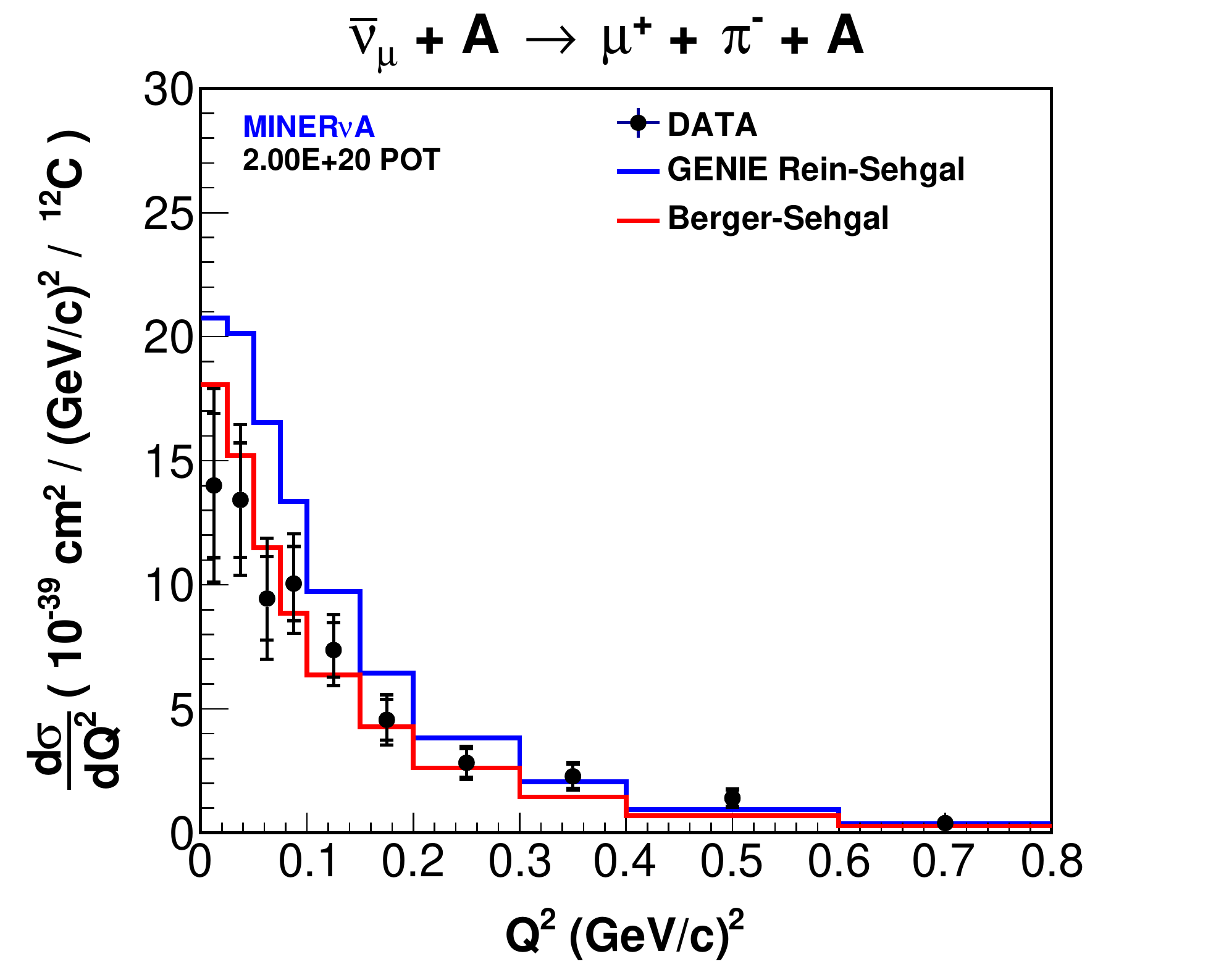}}
%\mbox{
%\includegraphics[width=0.49\linewidth]{figures/CrossSection/h_Q2_Final_XSec_ThetaPiCor_FluxConstrained_WeightedSignalModel_minerva5.pdf}
%\includegraphics[width=0.49\linewidth]{figures/CrossSection/h_Q2_Final_XSec_ThetaPiCor_FluxConstrained_WeightedSignalModel_downstream.pdf}}
\caption[The measured \numu and \numubar \dsigdqsq]{The measured \dsigdqsq for the \numu (left) and \numubar (right) samples.  The inner and outer error bars are the statistical and total (statistical + systematic) uncertainties, respectively.}
\label{fig:meas_disgdqsq}
\end{figure*}

The measured \numu and \numubar cross sections are shown in Figs.~\ref{fig:meas_sigenu}--\ref{fig:meas_disgdqsq} along
with cross section predictions from GENIE 2.8.4\footnote{Recall that GENIE 2.6.2 was used as the base model in this measurement.  GENIE 2.8.4 was used for these comparisons because of the availability of the Berger-Sehgal prediction in this version.} in the default Rein-Sehgal model and the alternate Berger-Sehgal model. 
%The signal model weighting (Section~\ref{sec:signal_model_weighting}) was applied in calculating the unfolding matrices and efficiency corrections used to measure the cross sections.  
The \chisq for the comparison of each measured cross section to the GENIE Rein-Sehgal and Berger-Sehgal predictions is
listed in Table~\ref{tab:xsec_chisq}.  The \chisq was calculated as
\begin{equation}
\chisq = AC^{-1}A^{T},
\label{eq:xsec_chisq}
\end{equation}
where $C$ is the bin-to-bin covariance matrix for the total uncertainty (statistical + total systematic) on the measured
cross section, and the elements of the vector $A$ are
\begin{equation}
A_{i} = \sigma^{meas}_{i} - \sigma^{pred}_{i},
\end{equation}
where $\sigma^{meas}_{i}$ and $\sigma^{pred}_{i}$ are the measured and predicted cross sections in bin $i$, respectively.

Both the GENIE Rein-Sehgal and Berger-Sehgal predictions are consistent with the measured cross sections in overall rate.
The Berger-Sehgal overall rate is lower than the Rein-Sehgal overall because of the lower Berger-Sehgal pion-carbon
elastic scattering cross section in the delta resonance region.  Both predictions differ in shape from the measured
\dsigdepi and \dsigdthetapi.  Assuming the tested validity of PCAC, this suggests
that both models mis-model the pion-carbon elastic scattering cross section.  Both predictions agree in shape with
the measured \dsigdqsq, which supports the axial vector dipole parameterization of the \qsq dependence of the coherent cross section. 

The measured \dsigdthetapi for large \thetapi are below zero because of background subtraction; however the size of
the uncertainties and bin-to-bin correlations is such that these results are fully consistent with zero.

Tables with the cross section results and the relevant covariance matrices describing their uncertainties may be found \SuppLocation.

\begingroup
%\squeezetable
\begin{table}[h]
\small
\begin{center}
\begin{tabular}{c|cc|cc|c}
& \multicolumn{2}{c|}{$\nu_{\mu}\ \chi^{2}$} & \multicolumn{2}{c|}{$\overline{\nu}_{\mu}\ \chi^{2}$} & \\
%Cross Section & Rein-Sehgal & Berger-Sehgal & Rein-Sehgal & Berger-Sehgal & NDF \\
Cross section & RS & BS & RS & BS & NDF \\
\hline
$\sigma(E_\nu)$                &    10.6 &     9.6 &    25.6 &    14.7 &  8 \\
$d\sigma/dE_\pi$               &    46.2 &    58.7 &    35.8 &    40.4 &  9 \\
$d\sigma/d\theta_\pi$          &  1164.5 &   171.9 &   122.2 &    29.1 & 12 \\
$d\sigma/dQ^{2}$               &    19.3 &    13.5 &    16.2 &    11.4 & 10 \\
\end{tabular}
\end{center}
\caption{ \chisq for the comparisons of the measured \numu and \numubar cross sections to the GENIE 2.8.4 Rein-Sehgal (RS) and Berger-Sehgal (BS) predictions }
\label{tab:xsec_chisq}
\end{table}
\endgroup

\section{\numu-\numubar Cross Section Comparisons}
\label{sec:xsec_comparisons}

\begin{figure*}[tpb]
\centering
\mbox{
\includegraphics[width=0.49\linewidth]{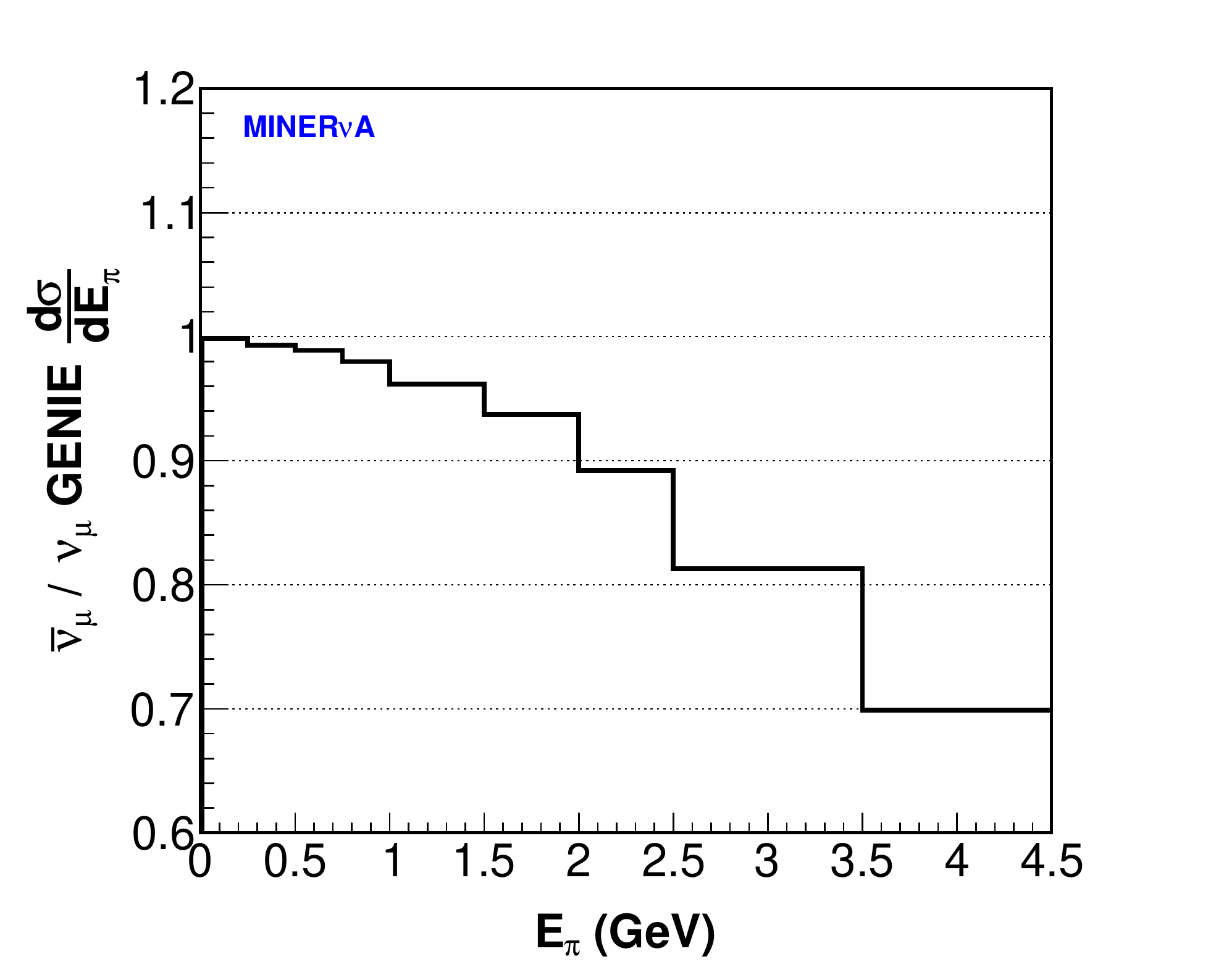}
\includegraphics[width=0.49\linewidth]{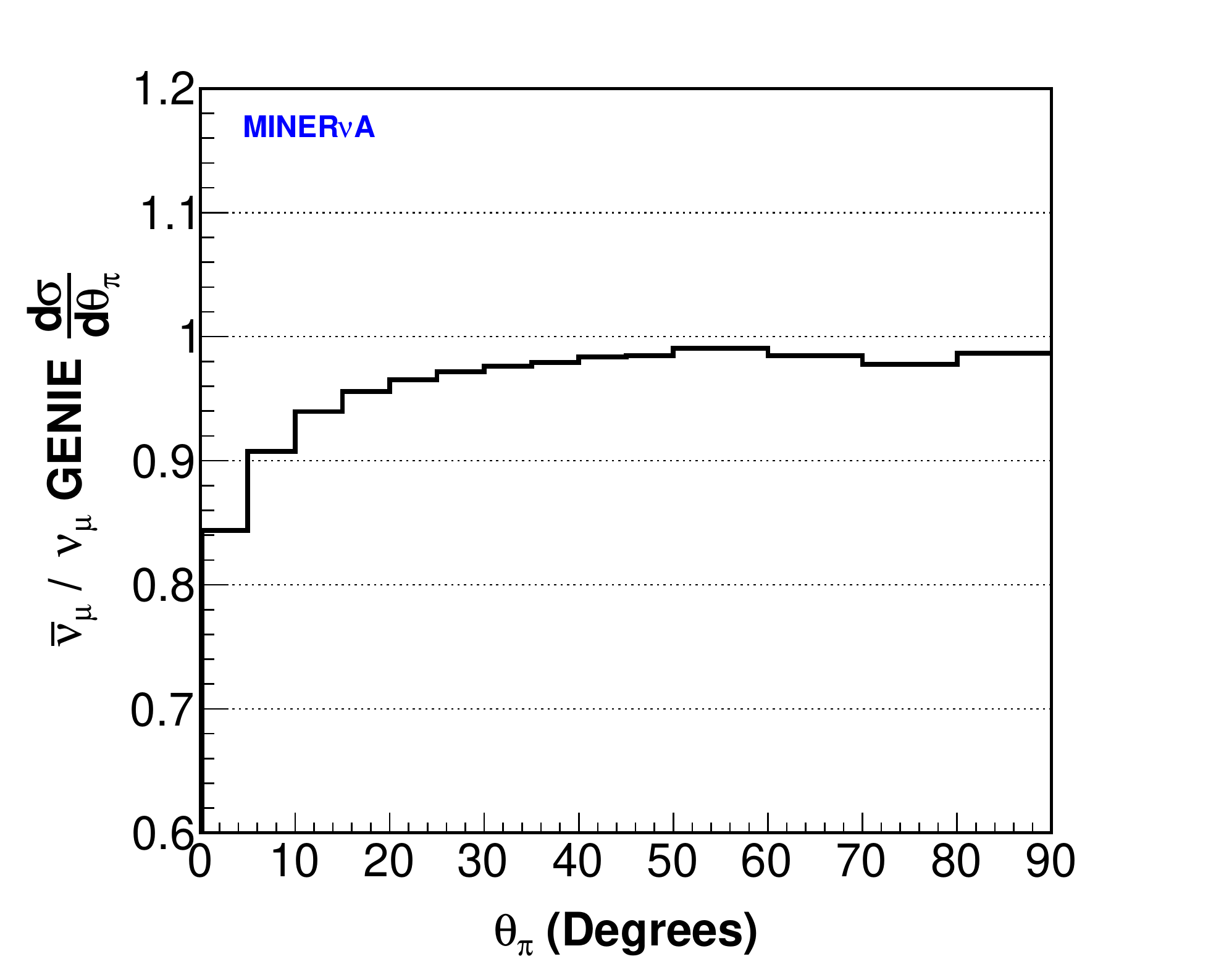}}
\mbox{
\includegraphics[width=0.49\linewidth]{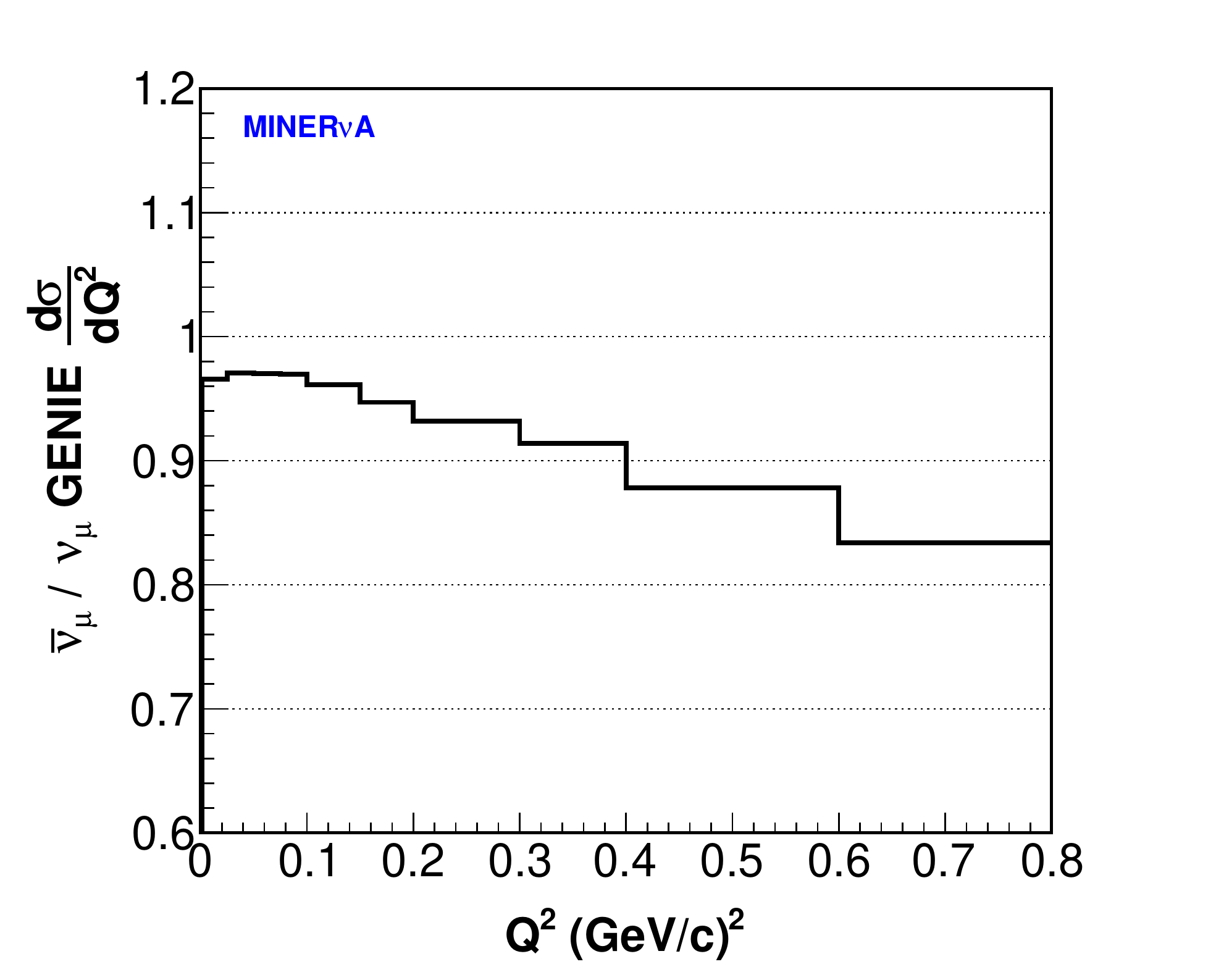}}
\caption[The ratio of the flux integrated coherent pion differential cross sections predicted by GENIE for the \numubar flux to that for the \numu flux]{The ratio of the coherent differential cross sections predicted by GENIE for the \numubar flux to that for the \numu flux}
\label{fig:flux_diff_xsec}
\end{figure*}

The measured \numu and \numubar coherent cross sections may be compared to test the prediction of the PCAC coherent
model (Sec.~\ref{sec:Rein-Sehgal}) that the neutrino and antineutrino cross sections are equal for a particular \enu.
%This prediction results from the assumption of no V-A interference in the scattering amplitude.
The measured \numu and \numubar \sigenu can be compared directly in each \enu bin.  The measured \numu and \numubar flux
integrated differential cross sections cannot be compared directly since they are integrated over different \enu spectra.
%(Figure~\ref{fig:flux_ratio}).
The effect of the different \enu spectra on the \numu and \numubar differential cross sections predicted by the GENIE
implementation of the Rein-Sehgal coherent model is illustrated in Fig.~\ref{fig:flux_diff_xsec}.  The \numubar flux
is $\sim\frac{1}{2}$ the \numu flux for 5 \lt\enu\lt 20 GeV, which results in a lower \numubar cross section at high-\epi,
low-\thetapi, and high-\qsq.

%\begin{figure*}[tpb]
%\centering
%\mbox{\includegraphics[width=0.5\linewidth]{figures/NeutrinoAntineutrinoComparisons/Flux_Ratio.pdf}}
%\caption[\numubar-to-\numu flux prediction ratio]{\numubar-to-\numu flux prediction ratio}
%\label{fig:flux_ratio}
%\end{figure*}

To compare the measured \numu and \numubar differential cross sections, the \numu cross sections are weighted to the \numubar
flux spectrum.  The \numu cross sections are weighted instead of the \numubar cross sections to minimize amplifying statistical
fluctuations.
%Ideally, the weighted \numu cross sections would be obtained by measuring the two-dimensional cross sections $\frac{d\sigma(\enu)}{d\epi}$, $\frac{d\sigma(\enu)}{d\thetapi}$, and $\frac{d\sigma(\enu)}{d\qsq}$, weighting the two-dimensional
%cross sections as a function of \enu by the flux ratio $\phi^{\numubar}(\enu) / \phi^{\numu}(\enu)$, and integrating over \enu.  However, measuring the two-dimensional cross sections is precluded by the statistics of the coherent candidate event samples.
%The alternative
The approach employed was to weight the \numu background subtracted coherent candidate event rates, unfolding matrices,
efficiency corrections, and flux normalization to the \numubar flux and remeasure the \numu cross sections.  The \numu background
subtracted coherent candidate event rate in reconstructed kinematic parameter bin $j$ was weighted as a function of reconstructed \enu as
\begin{equation}
N_{j}^{data} - N_{j}^{bkgd} = \displaystyle\sum_{k}W_{k}^{reco}(N_{jk}^{data} - N_{jk}^{bkgd}),
\end{equation}
where $W_{k}^{reco}$ is the weight for reconstructed \enu bin $k$.  $W_{k}^{reco}$ was calculated as
\begin{equation}
W_{k}^{reco} = \displaystyle\sum_{m}U_{km}^{\numu}\frac{\phi_{m}^{\numubar}}{\phi_{m}^{\numu}},
\end{equation}
where $\phi_{m}^{\numubar}$ and $\phi_{m}^{\numu}$ are the \numubar and \numu fluxes per POT in true \enu bin $m$, respectively,
and $U_{km}^{\numu}$ is the \numu \enu unfolding matrix element normalized to give the estimated fraction of selected coherent
events in reconstructed bin $k$ originating from true bin $m$.  The \numu unfolding matrices and efficiency corrections were
weighted to the \numubar flux by weighting each event in the \numu signal-only MC sample by the flux ratio
$\phi^{\numubar} / \phi^{\numu}$ for the event \enu and recalculating the unfolding matrices and efficiency corrections.
The \numu flux was weighted as a function of \enu by the flux ratio $\phi^{\numubar}(\enu) / \phi^{\numu}(\enu)$ and integrated
to give the flux normalization for the weighted \numu differential cross sections.  The fractional systematic uncertainty on the
weighted \numu differential cross sections was forced to be equal to that on the unweighted cross sections by preserving the
fractional difference of each systematic variation in the weighting.  The ratio of the weighted to unweighted \numu differential
cross sections is shown in Fig.~\ref{fig:flux_diff_xsec}.  For the remainder of this section, \numu differential cross section
refers to the \numu differential cross section weighted to the \numubar flux.

\begin{figure*}[tpb]
\centering
\mbox{
\includegraphics[width=0.49\linewidth]{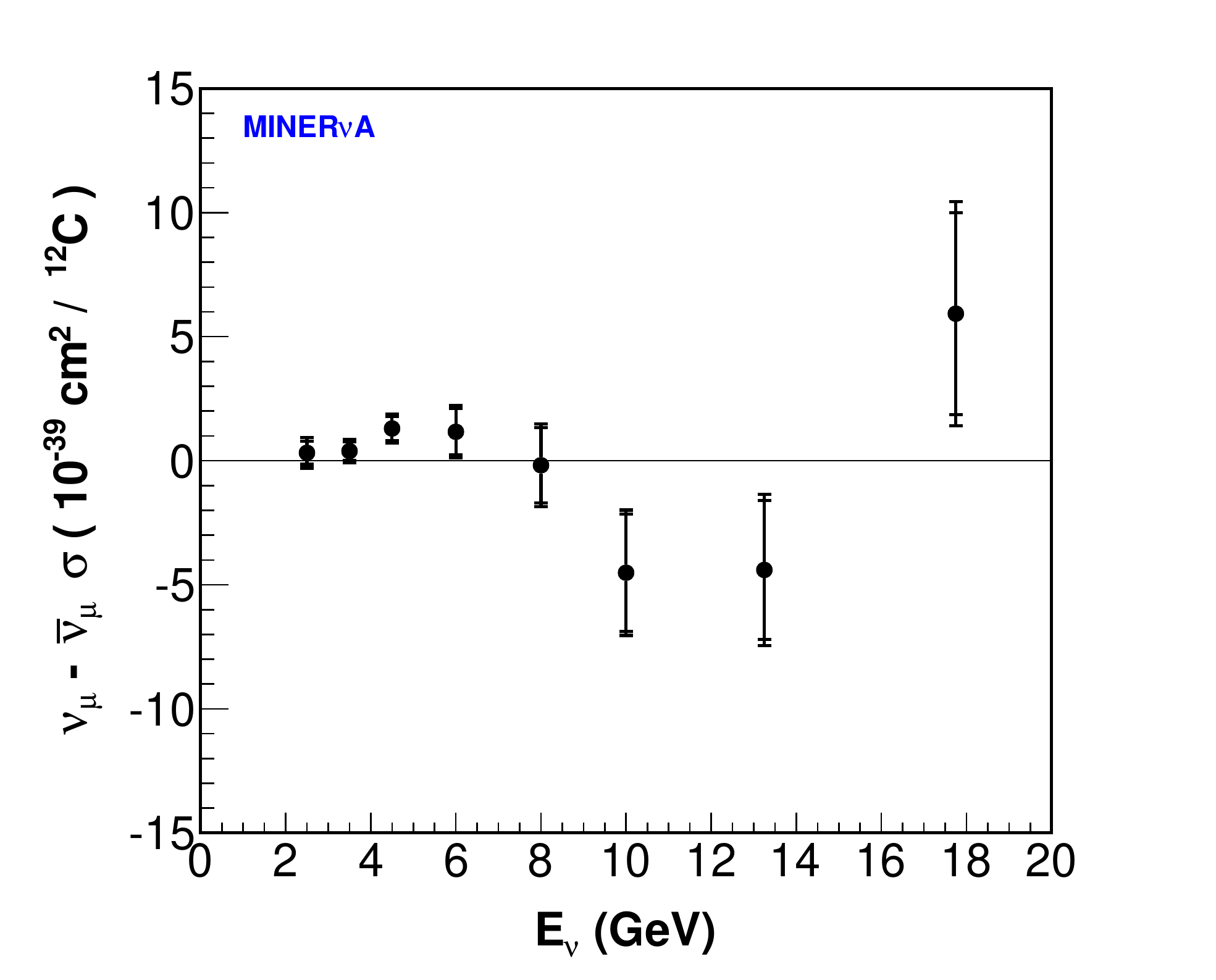}
\includegraphics[width=0.49\linewidth]{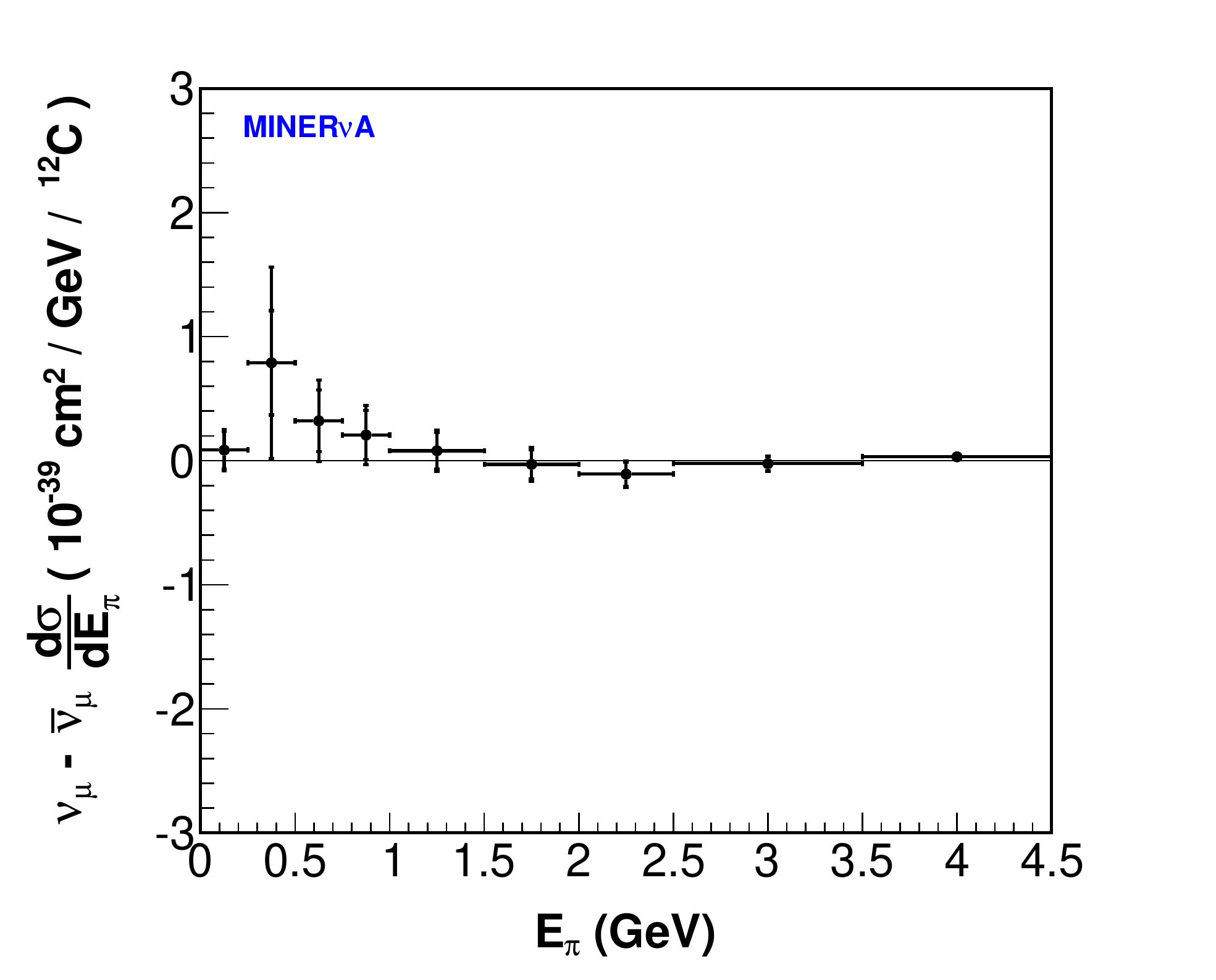}}
\mbox{
\includegraphics[width=0.49\linewidth]{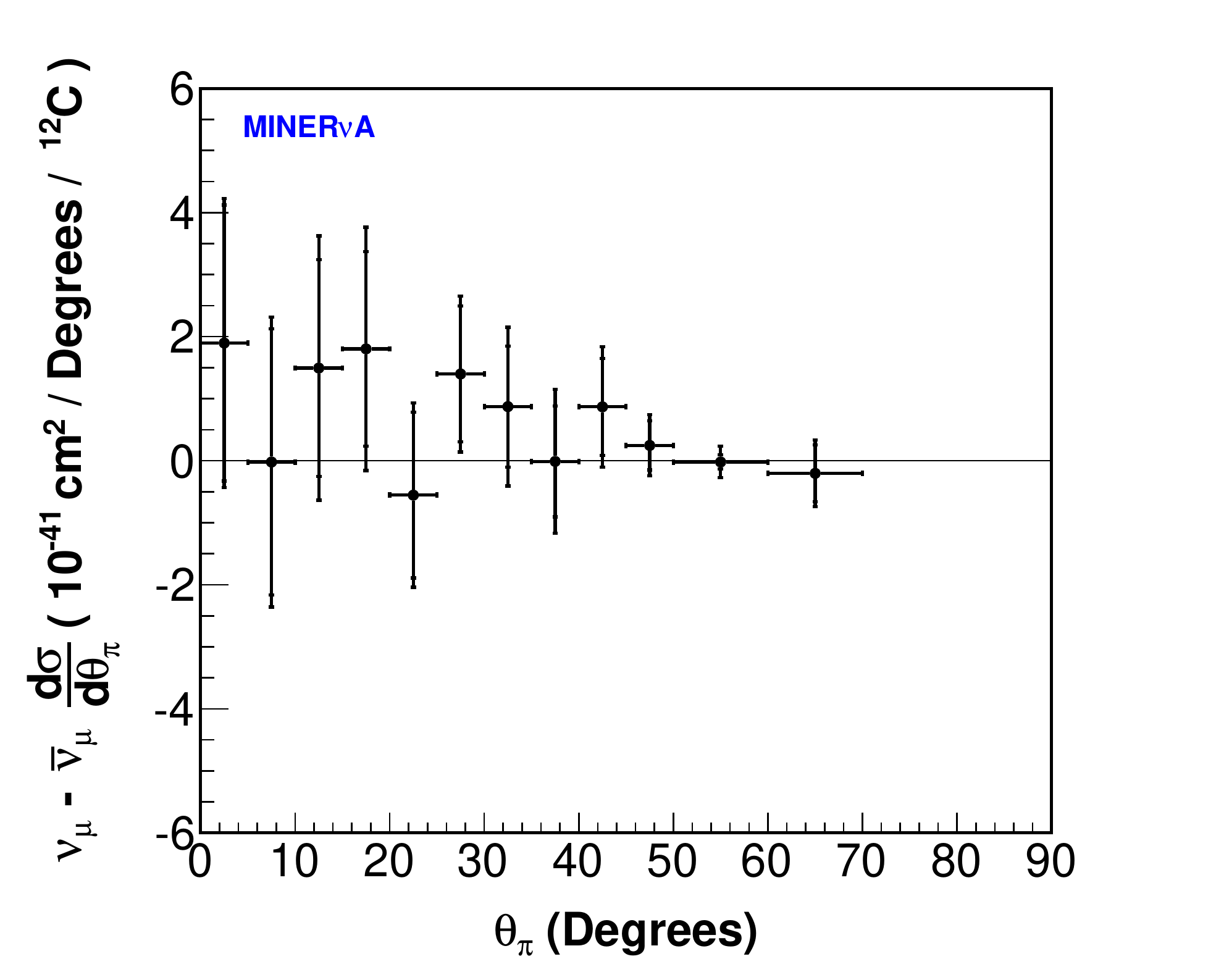}
\includegraphics[width=0.49\linewidth]{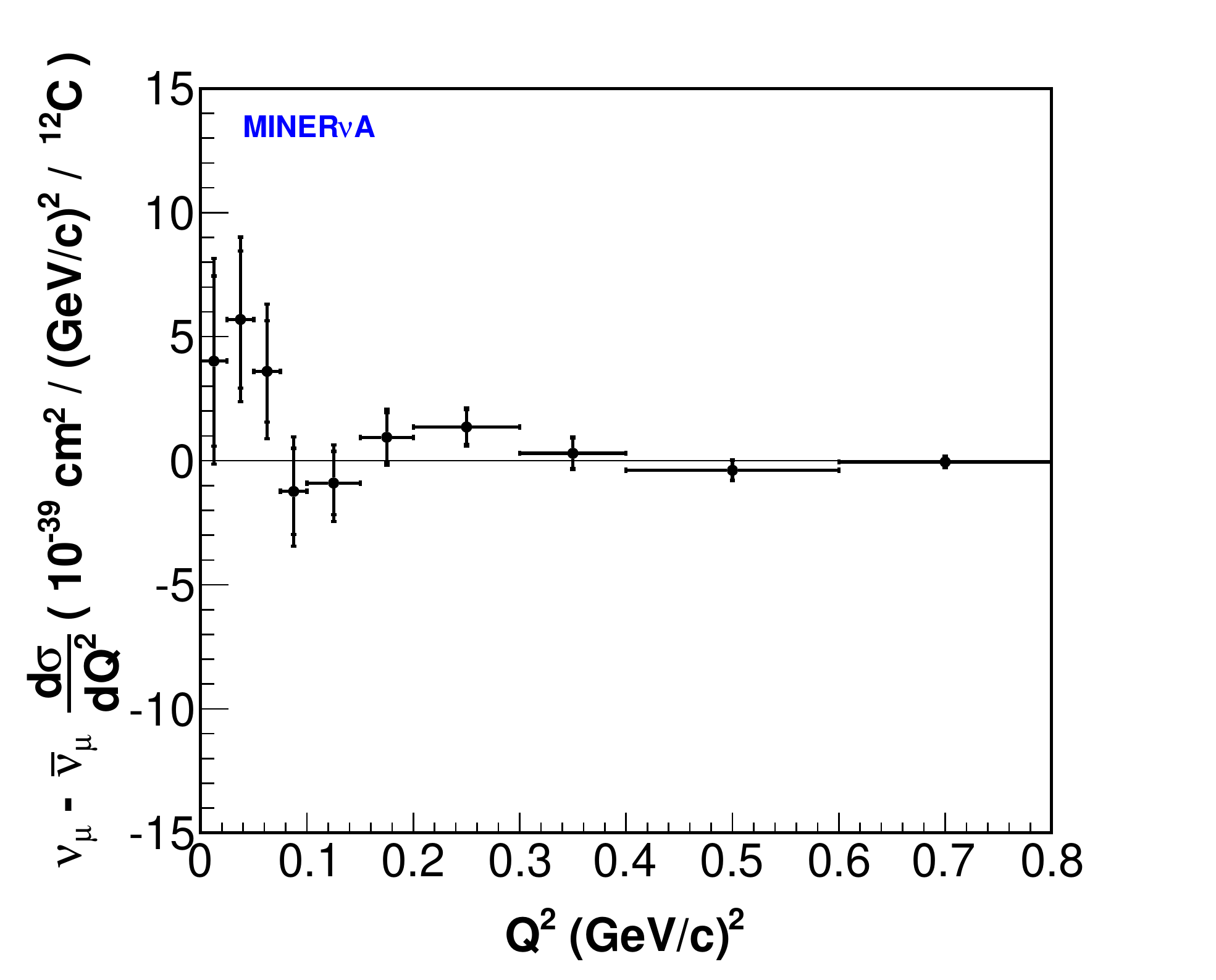}}
\caption[The measured $\numubar - \numu$ cross-section differences]{The measured $\numubar - \numu$ cross-section differences}
\label{fig:numu_numubar_xsec_diff}
\end{figure*}
The measured $\numu - \numubar$ cross-section differences are shown in Fig.~\ref{fig:numu_numubar_xsec_diff}.  The uncertainties
in the cross-section differences were calculated from the combined statistical and systematic covariance matrices for the \numu and
\numubar cross sections.  The combined statistical covariance matrix is the sum of the \numu and \numubar statistical covariance
matrices.  The elements of the combined covariance matrix for each systematic uncertainty $C^{sys}$ were calculated as
\begin{equation}
C^{sys}_{ij} = C^{\numu}_{ij} + C^{\numubar}_{ij} - C^{\numu\numubar}_{ij} - C^{\numu\numubar}_{ji},
\end{equation}
where $C^{\numu}$ and $C^{\numubar}$ are the systematic covariance matrices for the \numu and \numubar cross sections,
respectively, and $C^{\numu\numubar}$ is the joint systematic covariance matrix for the \numu and \numubar cross sections.
The elements of $C^{\numu\numubar}$ were calculated as
\begin{equation}
C^{\numu\numubar}_{ij} = \frac{1}{N}\displaystyle\sum_{k}\Delta\sigma^{\numu}_{ik}\Delta\sigma^{\numubar}_{jk},
\end{equation}
where $\Delta\sigma^{\numu}_{ik}$ ($\Delta\sigma^{\numubar}_{jk}$) is the change to the \numu (\numubar) cross section in bin
$i$ ($j$) for variation $k$, and $N$ is the number of variations from which the systematic uncertainty was evaluated.
The \chisq for the deviation of each $\numu - \numubar$ cross-section difference from zero
%(the prediction of the PCAC coherent models)
are listed in Table~\ref{tab:xsec_diff} and were calculated as
\begin{equation}
\label{eq:diff_search_chi2}
\chi^{2} = AC^{-1}A^{T},
\end{equation}
where $C$ is the total covariance (statistical + total systematic) matrix for the cross-section difference and the elements
of the vector $A$ are defined as
\begin{equation}
A_{i} = \sigma^{\numu}_{i} - \sigma^{\numubar}_{i},
\end{equation}
where $\sigma^{\numu}_{i}$ and $\sigma^{\numubar}_{i}$ are the \numu and \numubar cross sections in bin $i$.  The \chisq
probabilities in Table~\ref{tab:xsec_diff} show that the deviation of the $\numu - \numubar$ cross-section differences
from zero are not significant and are consistent with the prediction of the PCAC coherent models
that the neutrino and antineutrino cross sections are equal.

\begingroup
%\squeezetable
\begin{table}[bp]
\small
\begin{center}
\begin{tabular}{c|ccc}
Cross section & \chisq & NDF & \chisq Probability \\
\hline
$\sigma{\enu}$ & 11.4 & 8 & 0.181 \\
$d\sigma/d\epi$ & 5.5 & 9 & 0.791 \\
$d\sigma/d\thetapi$ & 5.3 & 12 & 0.948 \\
$d\sigma/d\qsq$ & 11.2 & 10 & 0.341 \\
\end{tabular}
\end{center}
\caption{The \chisq, number of degrees of freedom (NDF), and \chisq probability for the deviation of the measured $\numu - \numubar$
cross-section differences from zero.}
\label{tab:xsec_diff}
\end{table}
\endgroup

The primary sources of systematic uncertainties in the \numu - \numubar cross-section differences are the flux, neutron FSI,
and pion response.  Flux predictions can be improved by measuring hadron production from a neutrino production target.  The
electron \vs photon separation capability of liquid argon detectors can also improve the flux constraint from an \insitu
measurement of neutrino-electron elastic scattering.  Reductions in cross section and FSI uncertainties on the incoherent
single pion production backgrounds can be achieved by leveraging existing and future exclusive measurements.

\section{Conclusions}
\label{sec:conclusions}

This paper has reported measurements of \numu and \numubar CC coherent pion production on carbon from \minerva data.
The cross sections were measured for neutrino energies in the range 2.0 $<$ \enu $<$ 20 GeV, where
$\langle\enu\rangle\approx$ 4 GeV.  The measurements were made by isolating coherent interactions using the model
independent experimental signature of CC coherent scattering, which consists of a charged lepton and pion in the
forward direction, no evidence of nuclear breakup, and small \tabs.
A data-driven constraint on the background prediction was designed to minimize dependence on modeling nuclear
effects which are poorly understood.  Unambiguous signals of \numu and \numubar CC coherent pion production above
the predicted background were observed at small \tabs.  The cross sections \sigenu, \dsigdqsq, \dsigdepi
and \dsigdthetapi were measured for both \numu and \numubar CC coherent pion production.  \enu, \qsq, \epi, \thetapi
and \tabs characterize the kinematics of coherent scattering completely.

The measured cross sections were compared to the Rein-Sehgal and Berger-Sehgal coherent models, which are both
based on Adler's PCAC theorem.  The Rein-Sehgal model calculates the pion-nucleus elastic scattering cross section
from pion-nucleon scattering data.  The Berger-Sehgal model calculates the pion-nucleus scattering cross section
from pion-carbon elastic scattering data and scales the cross section to other nuclei.  Both the Rein-Sehgal and
Berger Sehgal predictions agree with the measured total cross-section as a function of neutrino energy in the \numu and \numubar samples.  For both
\numu and \numubar, the measured \dsigdepi (\dsigdthetapi) exhibits a harder (more forward) spectrum than the
Rein-Sehgal and Berger-Sehgal predictions, which suggests that both models miscalculate the pion-nucleus elastic
scattering cross section.  The Rein-Sehgal predictions were brought into better agreement with the measured
\dsigdepi and \dsigdthetapi by weighting the predicted rate of interactions with $\epi<$ 500 MeV by 50\%.  For
both \numu and \numubar, the Rein-Sehgal and Berger-Sehgal predictions are similar in shape to the measured \dsigdqsq,
which supports the axial vector dipole parameterization of the \qsq dependence of the coherent cross section.

PCAC coherent models assume coherent scattering is a purely axial vector process.  PCAC coherent models therefore
predict equal cross sections for neutrinos and antineutrinos.  To test this prediction, the measured \numu and
\numubar cross sections were compared, where the \numu differential cross sections were weighted to the \numubar flux.
No significant differences between the \numu and \numubar cross sections were observed.

Since the \minerva scintillator contains equal numbers of carbon and hydrogen atoms, diffractive pion production on
hydrogen is a possible contribution to the measured cross sections.  Diffractive pion production is indistinguishable
from coherent pion production when the recoil proton is undetected.  A search for diffractive pion production within
the coherent pion production candidate samples was performed by looking for ionization from a recoil proton near the
interaction vertex.  Neither the \numu nor \numubar coherent candidate samples exhibited evidence for such a diffractive
contribution.
%The absence of diffractive pion production in the coherent pion candidate samples is understood from the diffractive acceptance of the vertex energy cut and the \tabs dependence of the diffractive cross section.

\begin{figure}[tpb]
\centering
%\includegraphics[width=0.9\columnwidth]{figures/CrossSection/h_Ev_Final_XSec_WorldData_ThetaPiCor_FluxConstrained_WeightedSignalModel_neutrino.pdf}
%\includegraphics[width=0.9\columnwidth]{figures/CrossSection/h_Ev_Final_XSec_WorldData_ThetaPiCor_FluxConstrained_WeightedSignalModel_antineutrino.pdf}
%\caption[Compilation of measurements of CC coherent pion production including this result]{Compilation of measurements of the \numu (top) and \numubar (bottom) CC
%	coherent pion production cross sections as a function of energy.  All results are scaled to A=12 (carbon) assuming $\sigma\propto A^{1/3}$. }
\includegraphics[width=0.9\columnwidth]{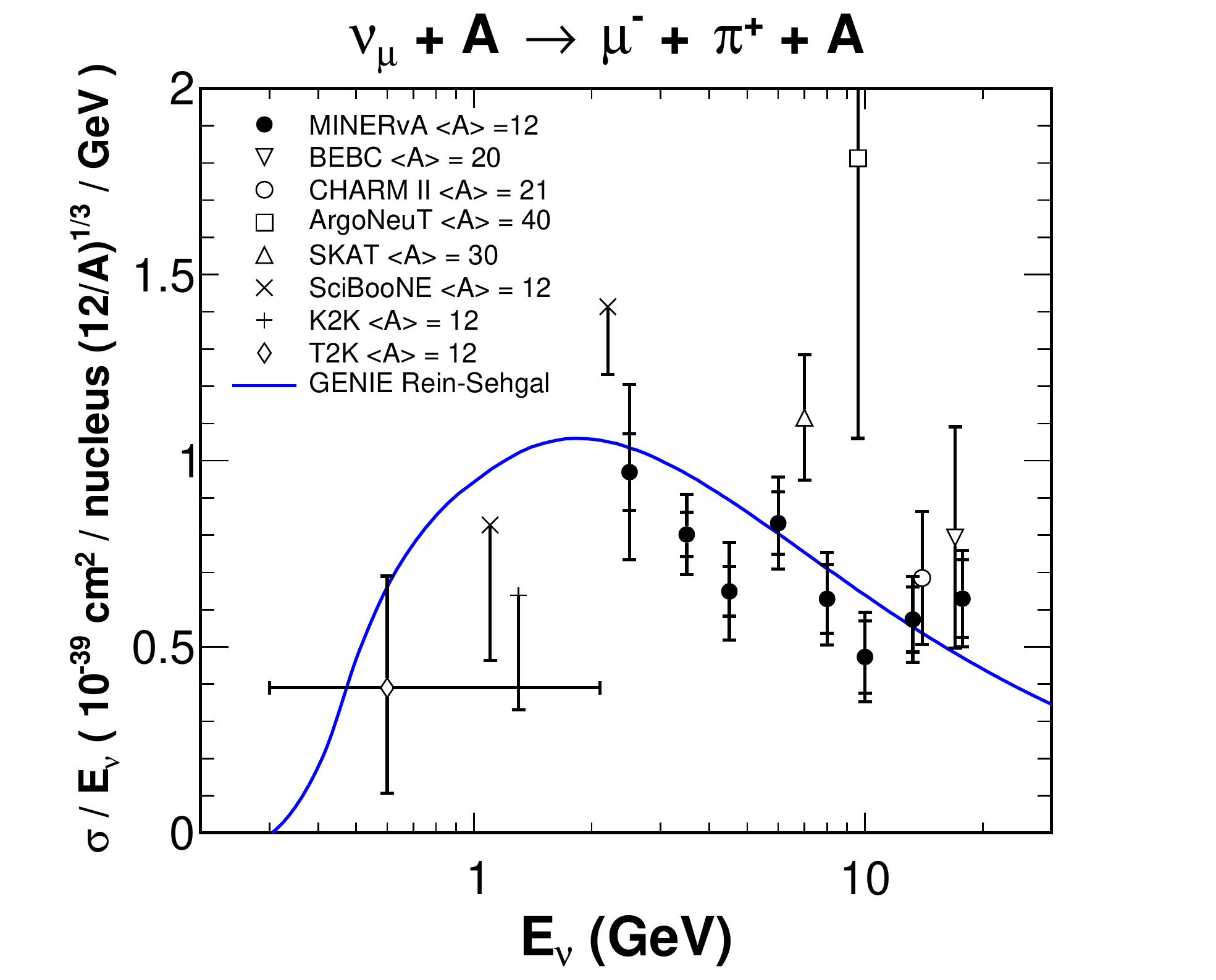}
\includegraphics[width=0.9\columnwidth]{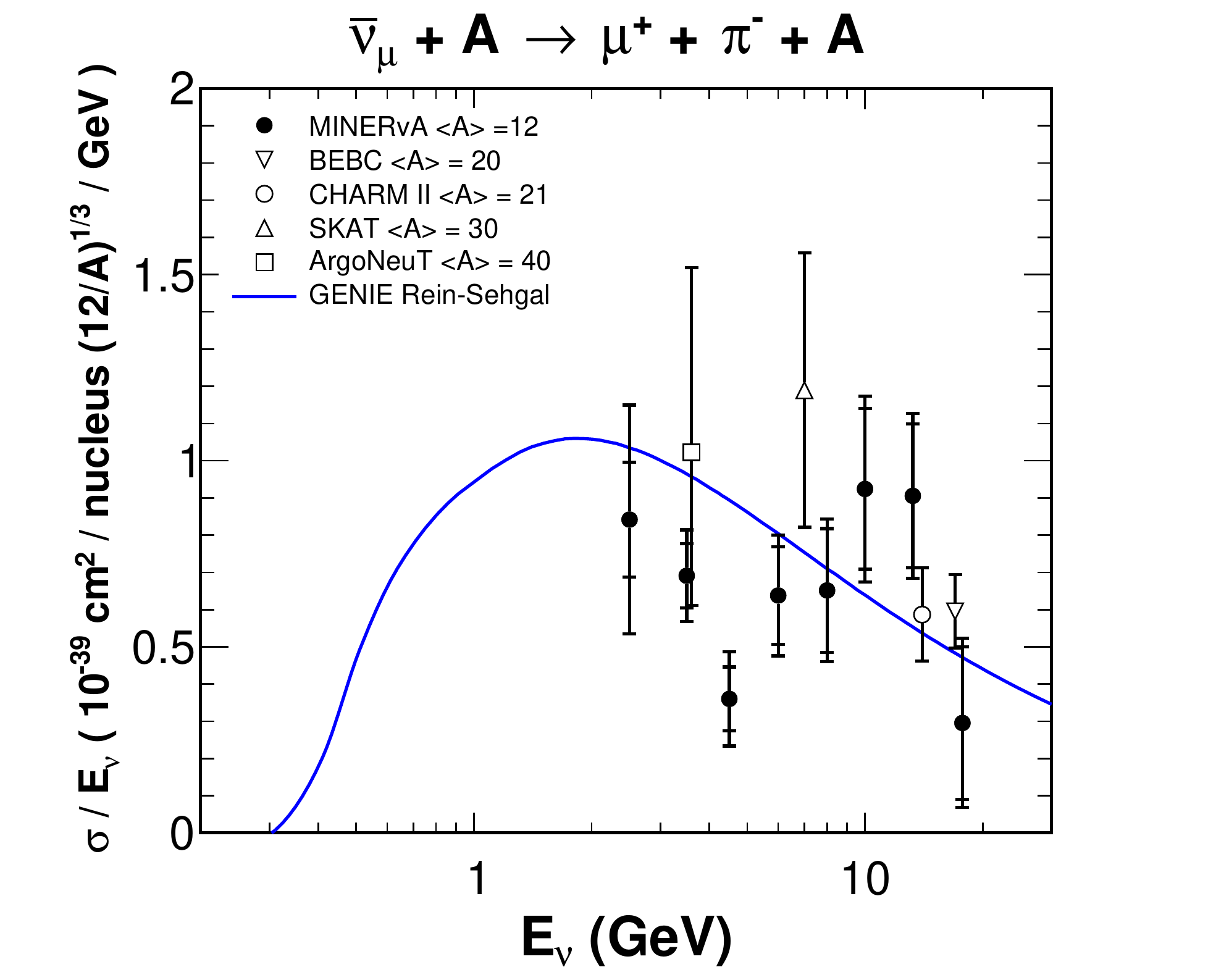}
\caption[Compilation of measurements of CC coherent pion production including this result]{Compilation of measurements of the \numu (top) and \numubar (bottom) CC
	coherent pion production cross sections as a function of energy.  Cross sections divided by neutrino energy is shown for display convenience.  All results are scaled to A=12 (carbon) assuming $\sigma\propto A^{1/3}$. }
\label{fig:coh_cc_compilation}
\end{figure}

The measurements reported in this paper are a significant addition to the world data set of neutrino-nucleus
coherent pion production.  Cross sections as a function of neutrino energy from the world data set are shown in Fig.~\ref{fig:coh_cc_compilation}.
This includes recent measurements from ArgoNEUT~\cite{bib:coh_cc_argo} and T2K~\cite{bib:coh_cc_t2k} which are significantly less precise than the
MINERvA measurements and do not report differential cross sections.  The T2K measurement, however, plays a special role here in that it shows
consistency of the cross section at lower energies than accessible by MINERvA with the $d\sigma/dE_\pi$ at low pion energies below the prediction
of the GENIE Rein-Sehgal model. 

The MINERvA measurements presented in this paper provide the most precise information about charged-current coherent pion production below neutrino
energies of 15 GeV.  The measurements provide constraints on not only the total rate, but also on the
kinematics of coherent pion production.  These measurements can be used to benchmark coherent pion production models
from production threshold up to neutrino energies of 20 GeV, which is the most important energy regime for current and future neutrino oscillation
experiments.

\ifnum\sizecheck=0
\begin{acknowledgments}

This document was prepared by members of the MINERvA collaboration using the resources of the Fermi National Accelerator
Laboratory (Fermilab), a U.S. Department of Energy, Office of Science, HEP User Facility.  Fermilab is managed by Fermi
Research Alliance, LLC (FRA), acting under Contract No. DE-AC02-07CH11359.  These resources included support for the \minerva
construction project, and support for construction also was granted by the United States National Science Foundation under
Award PHY-0619727 and by the University of Rochester.  Support for participating scientists was provided by NSF and DOE (USA) by CAPES
and CNPq (Brazil), by CoNaCyT (Mexico), by Proyecto Basal FB 0821, CONICYT PIA ACT1413, Fondecyt 3170845 and 11130133 (Chile),
by CONCYTEC, DGI-PUCP and UDI/IGI-UNI (Peru), and by the Latin American Center for Physics (CLAF).  We thank the MINOS Collaboration
for use of its near detector data. Finally, we thank the staff of Fermilab for support of the beamline, the detector and computing infrastructure.
%
% FNAL requirement, 28 August 2017
%
%Acknowledgement
%
%This must be on all Scientific/Technical Publications:
%
%Either:
%
%This manuscript has been authored by Fermi Research Alliance, LLC under Contract No. DE-%AC02-07CH11359 with the U.S. Department of Energy, Office of Science, Office of High Energy Physics.
%
%Or:
%
%This document was prepared by [COLLABORATION NAME] using the resources of the Fermi %National Accelerator Laboratory (Fermilab), a U.S. Department of Energy, Office of %Science, HEP User Facility. Fermilab is managed by Fermi Research Alliance, LLC (FRA), %acting under Contract No. DE-AC02-07CH11359.
%
%Copyright Statement
%
%This can be covered when the Fermilab Technical Publications Officer signs the publisher's copyright form:
%
%The U.S. Government retains and the publisher, by accepting the article for publication, acknowledges that the U.S. Government retains a non-exclusive, paid-up, irrevocable, world-wide license to publish or reproduce the published form of this manuscript, or allow others to do so, for U.S. Government purposes.
%
\end{acknowledgments}

\fi

%\bigskip

\ifnum\sizecheck=0
\bibliographystyle{apsrev4-1}
%\bibliography{PRD_include/PRD_Bibliography}
\clearpage
\thispagestyle{myheadings}

\fi

\ifnum\APPsupp=1
  \clearpage
% To expedite debugging LaTex on sections above.
\appendix{Cross Section Results and Covariance Matrices}
\label{appendixResults}

\onecolumngrid\footnotesize
\begingroup
\endgroup

\begingroup
%\squeezetable
\begin{table}[h]
\tabcolsep=0.11cm
\begin{tabular}{c|c|c}
& $\sigma(E_{\nu})\ \  (\text{cm}^{2}/^{12}\text{C})\times 10^{-39}$ & $\sigma(E_{\nu})\ \  (\text{cm}^{2}/^{12}\text{C})\times 10^{-39}$ \\
$E_{\nu}$ \ (GeV) & \numu & \numubar \\
\hline
  2.0 - 3.0 &  2.424 $\pm$ 0.258 $\pm$ 0.221 $\pm$ 0.479 &  2.105 $\pm$ 0.385 $\pm$ 0.214 $\pm$ 0.630 \\
  3.0 - 4.0 &  2.806 $\pm$ 0.207 $\pm$ 0.187 $\pm$ 0.255 &  2.418 $\pm$ 0.301 $\pm$ 0.156 $\pm$ 0.268 \\
  4.0 - 5.0 &  2.920 $\pm$ 0.300 $\pm$ 0.379 $\pm$ 0.336 &  1.621 $\pm$ 0.386 $\pm$ 0.272 $\pm$ 0.316 \\
  5.0 - 7.0 &  4.994 $\pm$ 0.501 $\pm$ 0.472 $\pm$ 0.276 &  3.828 $\pm$ 0.785 $\pm$ 0.436 $\pm$ 0.372 \\
  7.0 - 9.0 &  5.031 $\pm$ 0.738 $\pm$ 0.573 $\pm$ 0.344 &  5.212 $\pm$ 1.330 $\pm$ 0.569 $\pm$ 0.512 \\
 9.0 - 11.0 &  4.728 $\pm$ 0.970 $\pm$ 0.587 $\pm$ 0.397 &  9.241 $\pm$ 2.158 $\pm$ 0.886 $\pm$ 0.883 \\
11.0 - 15.5 &  7.599 $\pm$ 1.159 $\pm$ 0.810 $\pm$ 0.567 & 12.000 $\pm$ 2.555 $\pm$ 1.092 $\pm$ 0.941 \\
15.5 - 20.0 & 11.169 $\pm$ 1.856 $\pm$ 1.109 $\pm$ 0.796 &  5.243 $\pm$ 3.631 $\pm$ 0.827 $\pm$ 1.547 \\
\end{tabular}
\caption{ The measurement of the cross section as a function of $E_\nu$ with statistical, flux systematic, and non-flux systematic uncertainties }
\end{table}
\endgroup

\begingroup
%\squeezetable
\begin{table}[h]
\tabcolsep=0.11cm
\begin{tabular}{c|c|c}
& $d\sigma/dE_{\pi}\ \  (\text{cm}^{2}/\text{GeV}/^{12}\text{C})\times 10^{-39}$ & $d\sigma/dE_{\pi}\ \  (\text{cm}^{2}/\text{GeV}/^{12}\text{C})\times 10^{-39}$ \\
$E_{\pi}$ \ (GeV) & \numu & \numubar \\
\hline
  0 - 0.25 &  0.242 $\pm$ 0.084 $\pm$ 0.015 $\pm$ 0.110 &  0.156 $\pm$ 0.125 $\pm$ 0.012 $\pm$ 0.153 \\
0.25 - 0.5 &  2.186 $\pm$ 0.222 $\pm$ 0.133 $\pm$ 0.429 &  1.383 $\pm$ 0.356 $\pm$ 0.062 $\pm$ 0.536 \\
0.5 - 0.75 &  1.783 $\pm$ 0.140 $\pm$ 0.110 $\pm$ 0.179 &  1.478 $\pm$ 0.205 $\pm$ 0.079 $\pm$ 0.229 \\
0.75 - 1.0 &  1.802 $\pm$ 0.115 $\pm$ 0.111 $\pm$ 0.116 &  1.574 $\pm$ 0.164 $\pm$ 0.098 $\pm$ 0.126 \\
 1.0 - 1.5 &  1.347 $\pm$ 0.089 $\pm$ 0.084 $\pm$ 0.075 &  1.214 $\pm$ 0.121 $\pm$ 0.083 $\pm$ 0.069 \\
 1.5 - 2.0 &  0.773 $\pm$ 0.072 $\pm$ 0.048 $\pm$ 0.048 &  0.753 $\pm$ 0.098 $\pm$ 0.054 $\pm$ 0.047 \\
 2.0 - 2.5 &  0.427 $\pm$ 0.052 $\pm$ 0.025 $\pm$ 0.034 &  0.495 $\pm$ 0.084 $\pm$ 0.035 $\pm$ 0.038 \\
 2.5 - 3.5 &  0.214 $\pm$ 0.033 $\pm$ 0.012 $\pm$ 0.022 &  0.203 $\pm$ 0.049 $\pm$ 0.013 $\pm$ 0.026 \\
 3.5 - 4.5 &  0.131 $\pm$ 0.023 $\pm$ 0.007 $\pm$ 0.014 &  0.062 $\pm$ 0.021 $\pm$ 0.004 $\pm$ 0.011 \\
\end{tabular}
\caption{ The measurement of the differential cross section as a function of pion energy, $d\sigma/dE_{\pi}$, with statistical, flux systematic, and non-flux systematic uncertainties }
\end{table}
\endgroup

\begingroup
%\squeezetable
\begin{table}[h]
\tabcolsep=0.11cm
\begin{tabular}{c|c|c}
& $d\sigma/d\theta_{\pi}\ \  (\text{cm}^{2}/\text{Degree}/^{12}\text{C})\times 10^{-41}$ & $d\sigma/d\theta_{\pi}\ \  (\text{cm}^{2}/\text{Degree}/^{12}\text{C})\times 10^{-41}$ \\
$\theta_{\pi}$\ (Degrees) & \numu & \numubar \\
\hline
  0 - 5 & 12.703 $\pm$ 1.380 $\pm$ 0.814 $\pm$ 0.815 &  8.505 $\pm$ 1.880 $\pm$ 0.668 $\pm$ 0.710 \\
 5 - 10 & 15.161 $\pm$ 1.298 $\pm$ 0.938 $\pm$ 1.317 & 13.908 $\pm$ 1.779 $\pm$ 0.974 $\pm$ 0.930 \\
10 - 15 & 12.032 $\pm$ 1.016 $\pm$ 0.734 $\pm$ 0.559 & 10.175 $\pm$ 1.450 $\pm$ 0.665 $\pm$ 1.160 \\
15 - 20 & 11.256 $\pm$ 0.900 $\pm$ 0.690 $\pm$ 0.576 &  9.056 $\pm$ 1.313 $\pm$ 0.571 $\pm$ 1.116 \\
20 - 25 &  7.385 $\pm$ 0.723 $\pm$ 0.447 $\pm$ 0.527 &  7.498 $\pm$ 1.146 $\pm$ 0.466 $\pm$ 0.584 \\
25 - 30 &  5.166 $\pm$ 0.628 $\pm$ 0.312 $\pm$ 0.515 &  3.746 $\pm$ 0.901 $\pm$ 0.197 $\pm$ 0.571 \\
30 - 35 &  3.434 $\pm$ 0.565 $\pm$ 0.207 $\pm$ 0.768 &  2.582 $\pm$ 0.800 $\pm$ 0.129 $\pm$ 0.408 \\
35 - 40 &  1.735 $\pm$ 0.462 $\pm$ 0.103 $\pm$ 0.673 &  1.653 $\pm$ 0.772 $\pm$ 0.071 $\pm$ 0.464 \\
40 - 45 &  1.586 $\pm$ 0.467 $\pm$ 0.094 $\pm$ 0.496 &  0.655 $\pm$ 0.630 $\pm$ 0.045 $\pm$ 0.363 \\
45 - 50 &  0.245 $\pm$ 0.301 $\pm$ 0.013 $\pm$ 0.368 & -0.072 $\pm$ 0.286 $\pm$ 0.039 $\pm$ 0.211 \\
50 - 60 & -0.074 $\pm$ 0.036 $\pm$ 0.006 $\pm$ 0.036 & -0.046 $\pm$ 0.107 $\pm$ 0.056 $\pm$ 0.220 \\
60 - 70 & -0.085 $\pm$ 0.036 $\pm$ 0.006 $\pm$ 0.026 &  0.117 $\pm$ 0.458 $\pm$ 0.042 $\pm$ 0.276 \\
\end{tabular}
\caption{ The measurement of the differential cross section as a function of pion angle, $d\sigma/d\theta_{\pi}$, with statistical, flux systematic, and non-flux systematic uncertainties }
\end{table}
\endgroup

\begingroup
%\squeezetable
\begin{table}[h]
\tabcolsep=0.11cm
\begin{tabular}{c|c|c}
& $d\sigma/dQ^{2}\ \  (\text{cm}^{2}/(\text{GeV}/\text{c})^{2}/^{12}\text{C})\times 10^{-39}$ & $d\sigma/dQ^{2}\ \  (\text{cm}^{2}/(\text{GeV}/\text{c})^{2}/^{12}\text{C})\times 10^{-39}$ \\
$Q^{2}$\ (GeV/c)$^{2}$ & \numu & \numubar \\
\hline
   0 - 0.025 & 19.068 $\pm$ 1.848 $\pm$ 1.188 $\pm$ 2.119 & 14.004 $\pm$ 2.909 $\pm$ 0.794 $\pm$ 2.481 \\
0.025 - 0.05 & 19.812 $\pm$ 1.519 $\pm$ 1.242 $\pm$ 1.629 & 13.421 $\pm$ 2.303 $\pm$ 0.792 $\pm$ 1.805 \\
0.05 - 0.075 & 13.713 $\pm$ 1.149 $\pm$ 0.857 $\pm$ 1.163 &  9.445 $\pm$ 1.681 $\pm$ 0.558 $\pm$ 1.672 \\
 0.075 - 0.1 &  9.375 $\pm$ 0.886 $\pm$ 0.585 $\pm$ 0.896 & 10.049 $\pm$ 1.493 $\pm$ 0.609 $\pm$ 1.187 \\
  0.1 - 0.15 &  6.951 $\pm$ 0.647 $\pm$ 0.431 $\pm$ 0.665 &  7.370 $\pm$ 1.096 $\pm$ 0.428 $\pm$ 0.816 \\
  0.15 - 0.2 &  5.844 $\pm$ 0.554 $\pm$ 0.361 $\pm$ 0.467 &  4.559 $\pm$ 0.825 $\pm$ 0.258 $\pm$ 0.528 \\
   0.2 - 0.3 &  4.444 $\pm$ 0.422 $\pm$ 0.274 $\pm$ 0.289 &  2.827 $\pm$ 0.571 $\pm$ 0.171 $\pm$ 0.316 \\
   0.3 - 0.4 &  2.825 $\pm$ 0.364 $\pm$ 0.172 $\pm$ 0.214 &  2.282 $\pm$ 0.486 $\pm$ 0.152 $\pm$ 0.232 \\
   0.4 - 0.6 &  1.186 $\pm$ 0.226 $\pm$ 0.068 $\pm$ 0.138 &  1.402 $\pm$ 0.333 $\pm$ 0.093 $\pm$ 0.114 \\
   0.6 - 0.8 &  0.432 $\pm$ 0.149 $\pm$ 0.021 $\pm$ 0.092 &  0.396 $\pm$ 0.187 $\pm$ 0.023 $\pm$ 0.074 \\
\end{tabular}
\caption{ The measurement of the differential cross section as a function of $Q^{2}$, $d\sigma/dQ^{2}$, with statistical, flux systematic, and non-flux systematic uncertainties }
\end{table}
\endgroup

% These numbers differ insignificantly from the ones in Aaron's thesis; but in a phone call on 26 Oct 2017
% Aaron said he thinks these are the right ones.  These are the ones he used in the cross-section calculation
\begingroup
%\squeezetable
\begin{table}[h]
\tabcolsep=0.11cm
\begin{tabular}{c|c|c}
$E_{\nu}$ \ (GeV) & $(\numu/\text{cm}^{2}/\text{POT}) \times 10^{-8}$ & $(\numubar/\text{cm}^{2}/\text{POT}) \times 10^{-8}$ \\
\hline
  2.0 - 3.0 & 0.742 & 0.663 \\
  3.0 - 4.0 & 0.767 & 0.645 \\
  4.0 - 5.0 & 0.313 & 0.244 \\
  5.0 - 7.0 & 0.189 & 0.130 \\
  7.0 - 9.0 & 0.098 & 0.055 \\
 9.0 - 11.0 & 0.065 & 0.033 \\
11.0 - 15.5 & 0.077 & 0.035 \\
15.5 - 20.0 & 0.044 & 0.018 \\
\end{tabular}
\caption{ \numu and \numubar fluxes }
\label{tab:flux_norm}
\end{table}
\endgroup

\begingroup
\squeezetable
\begin{table}[h]
\tabcolsep=0.11cm
\begin{tabular}{c|c|c|c}
& $\sigma(E_{\nu})\ \  (\text{cm}^{2}/^{12}\text{C})\times 10^{-39}$ & $\sigma(E_{\nu})\ \  (\text{cm}^{2}/^{12}\text{C})\times 10^{-39}$& $\sigma(E_{\nu})\ \  (\text{cm}^{2}/^{12}\text{C})\times 10^{-39}$ \\
$E_{\nu}$ \ (GeV) & \numu & \numubar & $\numu - \numubar$ \\
\hline
  2.0 - 3.0 &  2.424 $\pm$ 0.258 $\pm$ 0.221 $\pm$ 0.479 &  2.105 $\pm$ 0.385 $\pm$ 0.214 $\pm$ 0.630 &  0.319 $\pm$ 0.464 $\pm$ 0.146 $\pm$ 0.381 \\
  3.0 - 4.0 &  2.806 $\pm$ 0.207 $\pm$ 0.187 $\pm$ 0.255 &  2.418 $\pm$ 0.301 $\pm$ 0.156 $\pm$ 0.268 &  0.388 $\pm$ 0.365 $\pm$ 0.132 $\pm$ 0.259 \\
  4.0 - 5.0 &  2.920 $\pm$ 0.300 $\pm$ 0.379 $\pm$ 0.336 &  1.621 $\pm$ 0.386 $\pm$ 0.272 $\pm$ 0.316 &  1.299 $\pm$ 0.489 $\pm$ 0.170 $\pm$ 0.260 \\
  5.0 - 7.0 &  4.994 $\pm$ 0.501 $\pm$ 0.472 $\pm$ 0.276 &  3.828 $\pm$ 0.785 $\pm$ 0.436 $\pm$ 0.372 &  1.166 $\pm$ 0.931 $\pm$ 0.347 $\pm$ 0.344 \\
  7.0 - 9.0 &  5.031 $\pm$ 0.738 $\pm$ 0.573 $\pm$ 0.344 &  5.212 $\pm$ 1.330 $\pm$ 0.569 $\pm$ 0.512 & -0.182 $\pm$ 1.521 $\pm$ 0.401 $\pm$ 0.539 \\
 9.0 - 11.0 &  4.728 $\pm$ 0.970 $\pm$ 0.587 $\pm$ 0.397 &  9.241 $\pm$ 2.158 $\pm$ 0.886 $\pm$ 0.883 & -4.513 $\pm$ 2.366 $\pm$ 0.582 $\pm$ 0.668 \\
11.0 - 15.5 &  7.599 $\pm$ 1.159 $\pm$ 0.810 $\pm$ 0.567 & 12.000 $\pm$ 2.555 $\pm$ 1.092 $\pm$ 0.941 & -4.401 $\pm$ 2.805 $\pm$ 0.763 $\pm$ 0.894 \\
15.5 - 20.0 & 11.169 $\pm$ 1.856 $\pm$ 1.109 $\pm$ 0.796 &  5.243 $\pm$ 3.631 $\pm$ 0.827 $\pm$ 1.547 &  5.925 $\pm$ 4.078 $\pm$ 0.956 $\pm$ 1.669 \\
\end{tabular}
\caption{ The measurement of the cross section as a function of $E_\nu$ with statistical, flux systematic, and non-flux systematic uncertainties for \numu, \numubar, and the difference between \numu and \numubar }
\end{table}
\endgroup

\begingroup
\squeezetable
\begin{table}[h]
\tabcolsep=0.11cm
\begin{tabular}{c|c|c|c}
& $d\sigma/dE_{\pi}\ \  (\text{cm}^{2}/\text{GeV}/^{12}\text{C})\times 10^{-39}$ & $d\sigma/dE_{\pi}\ \  (\text{cm}^{2}/\text{GeV}/^{12}\text{C})\times 10^{-39}$& $d\sigma/dE_{\pi}\ \  (\text{cm}^{2}/\text{GeV}/^{12}\text{C})\times 10^{-39}$ \\
$E_{\pi}$ \ (GeV) & \numu & \numubar & $\numu - \numubar$ \\
\hline
  0 - 0.25 &  0.241 $\pm$ 0.085 $\pm$ 0.015 $\pm$ 0.110 &  0.156 $\pm$ 0.125 $\pm$ 0.012 $\pm$ 0.153 &  0.086 $\pm$ 0.151 $\pm$ 0.011 $\pm$ 0.071 \\
0.25 - 0.5 &  2.173 $\pm$ 0.224 $\pm$ 0.132 $\pm$ 0.426 &  1.383 $\pm$ 0.356 $\pm$ 0.062 $\pm$ 0.536 &  0.790 $\pm$ 0.421 $\pm$ 0.112 $\pm$ 0.637 \\
0.5 - 0.75 &  1.800 $\pm$ 0.141 $\pm$ 0.110 $\pm$ 0.180 &  1.478 $\pm$ 0.205 $\pm$ 0.079 $\pm$ 0.229 &  0.321 $\pm$ 0.249 $\pm$ 0.084 $\pm$ 0.199 \\
0.75 - 1.0 &  1.781 $\pm$ 0.114 $\pm$ 0.110 $\pm$ 0.114 &  1.574 $\pm$ 0.164 $\pm$ 0.098 $\pm$ 0.126 &  0.207 $\pm$ 0.200 $\pm$ 0.086 $\pm$ 0.102 \\
 1.0 - 1.5 &  1.295 $\pm$ 0.086 $\pm$ 0.081 $\pm$ 0.073 &  1.214 $\pm$ 0.121 $\pm$ 0.083 $\pm$ 0.069 &  0.081 $\pm$ 0.149 $\pm$ 0.066 $\pm$ 0.040 \\
 1.5 - 2.0 &  0.724 $\pm$ 0.068 $\pm$ 0.045 $\pm$ 0.048 &  0.753 $\pm$ 0.098 $\pm$ 0.054 $\pm$ 0.047 & -0.029 $\pm$ 0.119 $\pm$ 0.040 $\pm$ 0.049 \\
 2.0 - 2.5 &  0.386 $\pm$ 0.047 $\pm$ 0.023 $\pm$ 0.033 &  0.495 $\pm$ 0.084 $\pm$ 0.035 $\pm$ 0.038 & -0.108 $\pm$ 0.096 $\pm$ 0.026 $\pm$ 0.033 \\
 2.5 - 3.5 &  0.179 $\pm$ 0.027 $\pm$ 0.010 $\pm$ 0.019 &  0.203 $\pm$ 0.049 $\pm$ 0.013 $\pm$ 0.026 & -0.024 $\pm$ 0.056 $\pm$ 0.011 $\pm$ 0.020 \\
 3.5 - 4.5 &  0.094 $\pm$ 0.017 $\pm$ 0.005 $\pm$ 0.011 &  0.062 $\pm$ 0.021 $\pm$ 0.004 $\pm$ 0.011 &  0.031 $\pm$ 0.027 $\pm$ 0.004 $\pm$ 0.008 \\
\end{tabular}
\caption{ The measurement of the differential cross section as a function of pion energy, $d\sigma/dE_{\pi}$, with statistical, flux systematic, and non-flux systematic uncertainties for \numu, \numubar, and the difference between \numu and \numubar.  Note that the \numu cross section is weighted to the \numubar flux for calculating the cross section difference. }
\end{table}
\endgroup

\begingroup
\squeezetable
\begin{table}[h]
\tabcolsep=0.11cm
\begin{tabular}{c|c|c|c}
& $d\sigma/d\theta_{\pi}\ \  (\text{cm}^{2}/\text{Degree}/^{12}\text{C})\times 10^{-41}$ & $d\sigma/d\theta_{\pi}\ \  (\text{cm}^{2}/\text{Degree}/^{12}\text{C})\times 10^{-41}$& $d\sigma/d\theta_{\pi}\ \  (\text{cm}^{2}/\text{Degree}/^{12}\text{C})\times 10^{-41}$ \\
$\theta_{\pi}$\ (Degrees) & \numu & \numubar & $\numu - \numubar$ \\
\hline
  0 - 5 & 10.402 $\pm$ 1.190 $\pm$ 0.663 $\pm$ 0.682 &  8.505 $\pm$ 1.880 $\pm$ 0.668 $\pm$ 0.710 &  1.897 $\pm$ 2.225 $\pm$ 0.537 $\pm$ 0.445 \\
 5 - 10 & 13.887 $\pm$ 1.197 $\pm$ 0.859 $\pm$ 1.206 & 13.908 $\pm$ 1.779 $\pm$ 0.974 $\pm$ 0.930 & -0.021 $\pm$ 2.144 $\pm$ 0.742 $\pm$ 0.559 \\
10 - 15 & 11.669 $\pm$ 0.969 $\pm$ 0.712 $\pm$ 0.553 & 10.175 $\pm$ 1.450 $\pm$ 0.665 $\pm$ 1.160 &  1.494 $\pm$ 1.744 $\pm$ 0.562 $\pm$ 1.087 \\
15 - 20 & 10.859 $\pm$ 0.860 $\pm$ 0.666 $\pm$ 0.565 &  9.056 $\pm$ 1.313 $\pm$ 0.571 $\pm$ 1.116 &  1.803 $\pm$ 1.570 $\pm$ 0.517 $\pm$ 1.055 \\
20 - 25 &  6.943 $\pm$ 0.686 $\pm$ 0.420 $\pm$ 0.502 &  7.498 $\pm$ 1.146 $\pm$ 0.466 $\pm$ 0.584 & -0.555 $\pm$ 1.336 $\pm$ 0.365 $\pm$ 0.537 \\
25 - 30 &  5.145 $\pm$ 0.618 $\pm$ 0.311 $\pm$ 0.515 &  3.746 $\pm$ 0.901 $\pm$ 0.197 $\pm$ 0.571 &  1.398 $\pm$ 1.093 $\pm$ 0.238 $\pm$ 0.577 \\
\end{tabular}
\caption{ The measurement of the differential cross section as a function of pion angle, $d\sigma/d\theta_{\pi}$, with statistical, flux systematic, and non-flux systematic uncertainties for \numu, \numubar, and the difference between \numu and \numubar.  Note that the \numu cross section is weighted to the \numubar flux for calculating the cross section difference. }
\end{table}
\endgroup

\begingroup
\squeezetable
\begin{table}[h]
\tabcolsep=0.11cm
\begin{tabular}{c|c|c|c}
& $d\sigma/dQ^{2}\ \  (\text{cm}^{2}/(\text{GeV}/\text{c})^{2}/^{12}\text{C})\times 10^{-39}$ & $d\sigma/dQ^{2}\ \  (\text{cm}^{2}/(\text{GeV}/\text{c})^{2}/^{12}\text{C})\times 10^{-39}$& $d\sigma/dQ^{2}\ \  (\text{cm}^{2}/(\text{GeV}/\text{c})^{2}/^{12}\text{C})\times 10^{-39}$ \\
$Q^{2}$\ (GeV/c)$^{2}$ & \numu & \numubar & $\numu - \numubar$ \\
\hline
   0 - 0.025 & 18.019 $\pm$ 1.812 $\pm$ 1.121 $\pm$ 2.022 & 14.004 $\pm$ 2.909 $\pm$ 0.794 $\pm$ 2.481 &  4.015 $\pm$ 3.427 $\pm$ 0.843 $\pm$ 2.175 \\
0.025 - 0.05 & 19.114 $\pm$ 1.529 $\pm$ 1.196 $\pm$ 1.576 & 13.421 $\pm$ 2.303 $\pm$ 0.792 $\pm$ 1.805 &  5.694 $\pm$ 2.765 $\pm$ 0.897 $\pm$ 1.599 \\
0.05 - 0.075 & 13.047 $\pm$ 1.163 $\pm$ 0.814 $\pm$ 1.109 &  9.445 $\pm$ 1.681 $\pm$ 0.558 $\pm$ 1.672 &  3.602 $\pm$ 2.044 $\pm$ 0.616 $\pm$ 1.679 \\
 0.075 - 0.1 &  8.812 $\pm$ 0.886 $\pm$ 0.548 $\pm$ 0.849 & 10.049 $\pm$ 1.493 $\pm$ 0.609 $\pm$ 1.187 & -1.237 $\pm$ 1.736 $\pm$ 0.476 $\pm$ 1.266 \\
  0.1 - 0.15 &  6.471 $\pm$ 0.647 $\pm$ 0.401 $\pm$ 0.627 &  7.370 $\pm$ 1.096 $\pm$ 0.428 $\pm$ 0.816 & -0.899 $\pm$ 1.273 $\pm$ 0.337 $\pm$ 0.802 \\
  0.15 - 0.2 &  5.504 $\pm$ 0.555 $\pm$ 0.339 $\pm$ 0.447 &  4.559 $\pm$ 0.825 $\pm$ 0.258 $\pm$ 0.528 &  0.945 $\pm$ 0.995 $\pm$ 0.253 $\pm$ 0.460 \\
   0.2 - 0.3 &  4.186 $\pm$ 0.420 $\pm$ 0.258 $\pm$ 0.271 &  2.827 $\pm$ 0.571 $\pm$ 0.171 $\pm$ 0.316 &  1.359 $\pm$ 0.709 $\pm$ 0.194 $\pm$ 0.251 \\
   0.3 - 0.4 &  2.583 $\pm$ 0.360 $\pm$ 0.158 $\pm$ 0.202 &  2.282 $\pm$ 0.486 $\pm$ 0.152 $\pm$ 0.232 &  0.301 $\pm$ 0.605 $\pm$ 0.125 $\pm$ 0.241 \\
   0.4 - 0.6 &  1.016 $\pm$ 0.219 $\pm$ 0.058 $\pm$ 0.121 &  1.402 $\pm$ 0.333 $\pm$ 0.093 $\pm$ 0.114 & -0.386 $\pm$ 0.399 $\pm$ 0.067 $\pm$ 0.107 \\
   0.6 - 0.8 &  0.347 $\pm$ 0.134 $\pm$ 0.016 $\pm$ 0.074 &  0.396 $\pm$ 0.187 $\pm$ 0.023 $\pm$ 0.074 & -0.050 $\pm$ 0.230 $\pm$ 0.019 $\pm$ 0.056 \\
\end{tabular}
\caption{ The measurement of the differential cross section as a function of $Q^{2}$, $d\sigma/dQ^{2}$, with statistical, flux systematic, and non-flux systematic uncertainties for \numu, \numubar, and the difference between \numu and \numubar.  Note that the \numu cross section is weighted to the \numubar flux for calculating the cross section difference. }
\end{table}
\endgroup

%\twocolumngrid\footnotesize
\clearpage

\begingroup
%\squeezetable
\begin{turnpage}
\begin{table}[h]
\tabcolsep=0.11cm
\begin{tabular}{cc|cccccccc|cccccccc}
 &  & \multicolumn{8}{c|}{\numu} & \multicolumn{8}{c}{\numubar} \\
%& $E_{\nu}$ \ (GeV) & 2.0 - 3.0 & 3.0 - 4.0 & 4.0 - 5.0 & 5.0 - 7.0 & 7.0 - 9.0 & 9.0 - 11.0 & 11.0 - 15.5 & 15.5 - 20.0 & 2.0 - 3.0 & 3.0 - 4.0 & 4.0 - 5.0 & 5.0 - 7.0 & 7.0 - 9.0 & 9.0 - 11.0 & 11.0 - 15.5 & 15.5 - 20.0 \\
 & $E_{\nu}$ \ (GeV) & 2.0 - & 3.0 - & 4.0 - & 5.0 - & 7.0 - & 9.0 - & 11.0 - & 15.5 - & 2.0 - & 3.0 - & 4.0 - & 5.0 - & 7.0 - & 9.0 - & 11.0 - & 15.5 - \\
 &                   & 3.0   & 4.0   & 5.0   & 7.0   & 9.0   & 11.0  & 15.5   & 20.0   & 3.0   & 4.0   & 5.0   & 7.0   & 9.0   & 11.0  & 15.5	& 20.0   \\
\hline
\multirow{8}{*}{\numu}
 &   2.0 - 3.0 &  1.000 & -0.004 & -0.088 & -0.014 & -0.001 & -0.000 & -0.000 & -0.000 &  0.000 &  0.000 &  0.000 &  0.000 &  0.000 &  0.000 &  0.000 &  0.000 \\
 &   3.0 - 4.0 & -0.004 &  1.000 &  0.117 & -0.083 & -0.016 & -0.001 & -0.000 & -0.000 &  0.000 &  0.000 &  0.000 &  0.000 &  0.000 &  0.000 &  0.000 &  0.000 \\
 &   4.0 - 5.0 & -0.088 &  0.117 &  1.000 &  0.097 & -0.061 & -0.016 & -0.002 & -0.000 &  0.000 &  0.000 &  0.000 &  0.000 &  0.000 &  0.000 &  0.000 &  0.000 \\
 &   5.0 - 7.0 & -0.014 & -0.083 &  0.097 &  1.000 &  0.078 & -0.073 & -0.023 & -0.001 &  0.000 &  0.000 &  0.000 &  0.000 &  0.000 &  0.000 &  0.000 &  0.000 \\
 &   7.0 - 9.0 & -0.001 & -0.016 & -0.061 &  0.078 &  1.000 &  0.231 & -0.066 & -0.026 &  0.000 &  0.000 &  0.000 &  0.000 &  0.000 &  0.000 &  0.000 &  0.000 \\
\end{tabular}
\caption{ Correlation matrix for the $\sigma(E_{\nu})$ statistical uncertainty }
\end{table}
\end{turnpage}
\endgroup

\begingroup
%\squeezetable
\begin{turnpage}
\begin{table}[h]
\tabcolsep=0.11cm
\begin{tabular}{cc|cccccccc|cccccccc}
 &  & \multicolumn{8}{c|}{\numu} & \multicolumn{8}{c}{\numubar} \\
%& $E_{\nu}$ \ (GeV) & 2.0 - 3.0 & 3.0 - 4.0 & 4.0 - 5.0 & 5.0 - 7.0 & 7.0 - 9.0 & 9.0 - 11.0 & 11.0 - 15.5 & 15.5 - 20.0 & 2.0 - 3.0 & 3.0 - 4.0 & 4.0 - 5.0 & 5.0 - 7.0 & 7.0 - 9.0 & 9.0 - 11.0 & 11.0 - 15.5 & 15.5 - 20.0 \\
 & $E_{\nu}$ \ (GeV) & 2.0 - & 3.0 - & 4.0 - & 5.0 - & 7.0 - & 9.0 - & 11.0 - & 15.5 - & 2.0 - & 3.0 - & 4.0 - & 5.0 - & 7.0 - & 9.0 - & 11.0 - & 15.5 - \\
 &                   & 3.0   & 4.0   & 5.0   & 7.0   & 9.0   & 11.0  & 15.5   & 20.0   & 3.0   & 4.0   & 5.0   & 7.0   & 9.0   & 11.0  & 15.5	& 20.0   \\
\hline
\multirow{8}{*}{\numu}
 &   2.0 - 3.0 &  1.000 &  0.481 & -0.226 &  0.213 &  0.534 &  0.495 &  0.431 &  0.359 &  0.777 &  0.200 & -0.419 & -0.042 &  0.388 &  0.424 &  0.397 &  0.308 \\
 &   3.0 - 4.0 &  0.481 &  1.000 &  0.678 &  0.503 &  0.126 & -0.031 & -0.044 & -0.054 &  0.174 &  0.717 &  0.460 &  0.295 &  0.067 &  0.110 &  0.134 & -0.060 \\
 &   4.0 - 5.0 & -0.226 &  0.678 &  1.000 &  0.624 & -0.095 & -0.277 & -0.257 & -0.222 & -0.443 &  0.576 &  0.916 &  0.606 & -0.025 & -0.042 & -0.004 & -0.221 \\
 &   5.0 - 7.0 &  0.213 &  0.503 &  0.624 &  1.000 &  0.674 &  0.491 &  0.455 &  0.411 & -0.044 &  0.227 &  0.402 &  0.711 &  0.566 &  0.543 &  0.541 &  0.267 \\
 &   7.0 - 9.0 &  0.534 &  0.126 & -0.095 &  0.674 &  1.000 &  0.963 &  0.906 &  0.807 &  0.422 & -0.150 & -0.339 &  0.258 &  0.754 &  0.784 &  0.751 &  0.593 \\
\end{tabular}
\caption{ Correlation matrix for the $\sigma(E_{\nu})$ flux systematic uncertainty }
\end{table}
\end{turnpage}
\endgroup

\begingroup
%\squeezetable
\begin{turnpage}
\begin{table}[h]
\tabcolsep=0.11cm
\begin{tabular}{cc|cccccccc|cccccccc}
 &  & \multicolumn{8}{c|}{\numu} & \multicolumn{8}{c}{\numubar} \\
%& $E_{\nu}$ \ (GeV) & 2.0 - 3.0 & 3.0 - 4.0 & 4.0 - 5.0 & 5.0 - 7.0 & 7.0 - 9.0 & 9.0 - 11.0 & 11.0 - 15.5 & 15.5 - 20.0 & 2.0 - 3.0 & 3.0 - 4.0 & 4.0 - 5.0 & 5.0 - 7.0 & 7.0 - 9.0 & 9.0 - 11.0 & 11.0 - 15.5 & 15.5 - 20.0 \\
 & $E_{\nu}$ \ (GeV) & 2.0 - & 3.0 - & 4.0 - & 5.0 - & 7.0 - & 9.0 - & 11.0 - & 15.5 - & 2.0 - & 3.0 - & 4.0 - & 5.0 - & 7.0 - & 9.0 - & 11.0 - & 15.5 - \\
 &                   & 3.0   & 4.0   & 5.0   & 7.0   & 9.0   & 11.0  & 15.5   & 20.0   & 3.0   & 4.0   & 5.0   & 7.0   & 9.0   & 11.0  & 15.5	& 20.0   \\
\hline
\multirow{8}{*}{\numu}
 &   2.0 - 3.0 &  1.000 &  0.614 & -0.114 &  0.260 & -0.166 &  0.554 &  0.624 & -0.544 &  0.797 &  0.472 & -0.314 &  0.095 &  0.292 &  0.591 & -0.030 & -0.012 \\
 &   3.0 - 4.0 &  0.614 &  1.000 &  0.698 &  0.791 &  0.365 &  0.248 &  0.270 &  0.133 &  0.205 &  0.512 &  0.296 &  0.218 &  0.018 &  0.164 &  0.160 & -0.060 \\
 &   4.0 - 5.0 & -0.114 &  0.698 &  1.000 &  0.785 &  0.645 & -0.155 & -0.162 &  0.658 & -0.418 &  0.274 &  0.685 &  0.252 & -0.223 & -0.318 &  0.249 & -0.053 \\
 &   5.0 - 7.0 &  0.260 &  0.791 &  0.785 &  1.000 &  0.767 &  0.347 &  0.258 &  0.503 & -0.001 &  0.502 &  0.593 &  0.467 &  0.256 &  0.141 &  0.233 &  0.161 \\
 &   7.0 - 9.0 & -0.166 &  0.365 &  0.645 &  0.767 &  1.000 &  0.405 &  0.200 &  0.734 & -0.201 &  0.380 &  0.751 &  0.593 &  0.255 & -0.068 &  0.498 &  0.417 \\
\end{tabular}
\caption{ Correlation matrix for the $\sigma(E_{\nu})$ non-flux systematic uncertainty }
\end{table}
\end{turnpage}
\endgroup

\begingroup
\squeezetable
\begin{turnpage}
\begin{table}[h]
\tabcolsep=0.11cm
\begin{tabular}{cc|ccccccccc|ccccccccc}
 &  & \multicolumn{9}{c|}{\numu} & \multicolumn{9}{c}{\numubar} \\
%& $E_{\pi}$ \ (GeV) & 0 - 0.25 & 0.25 - 0.5 & 0.5 - 0.75 & 0.75 - 1.0 & 1.0 - 1.5 & 1.5 - 2.0 & 2.0 - 2.5 & 2.5 - 3.5 & 3.5 - 4.5 & 0 - 0.25 & 0.25 - 0.5 & 0.5 - 0.75 & 0.75 - 1.0 & 1.0 - 1.5 & 1.5 - 2.0 & 2.0 - 2.5 & 2.5 - 3.5 & 3.5 - 4.5 \\
 & $E_{\pi}$ \ (GeV) & 0 -  & 0.25 - & 0.5 - & 0.75 - & 1.0 - & 1.5 - & 2.0 - & 2.5 - & 3.5 - & 0 -  & 0.25 - & 0.5 - & 0.75 - & 1.0 - & 1.5 - & 2.0 - & 2.5 - & 3.5 -  \\
 &                   & 0.25 & 0.5    & 0.75  & 1.0    & 1.5   & 2.0   & 2.5   & 3.5   & 4.5   & 0.25 & 0.5    & 0.75  & 1.0    & 1.5   & 2.0   & 2.5   & 3.5   & 4.5    \\
\hline
\multirow{9}{*}{\numu}
 &   0 - 0.25 &  1.000 &  0.257 & -0.135 & -0.050 & -0.013 &  0.001 &  0.000 &  0.000 &  0.000 &  0.000 &  0.000 &  0.000 &  0.000 &  0.000 &  0.000 &  0.000 &  0.000 &  0.000 \\
 & 0.25 - 0.5 &  0.257 &  1.000 &  0.157 & -0.103 & -0.105 & -0.038 & -0.011 & -0.004 & -0.001 &  0.000 &  0.000 &  0.000 &  0.000 &  0.000 &  0.000 &  0.000 &  0.000 &  0.000 \\
 & 0.5 - 0.75 & -0.135 &  0.157 &  1.000 &  0.568 &  0.012 & -0.137 & -0.075 & -0.030 & -0.009 &  0.000 &  0.000 &  0.000 &  0.000 &  0.000 &  0.000 &  0.000 &  0.000 &  0.000 \\
 & 0.75 - 1.0 & -0.050 & -0.103 &  0.568 &  1.000 &  0.620 & -0.013 & -0.121 & -0.074 & -0.032 &  0.000 &  0.000 &  0.000 &  0.000 &  0.000 &  0.000 &  0.000 &  0.000 &  0.000 \\
 &  1.0 - 1.5 & -0.013 & -0.105 &  0.012 &  0.620 &  1.000 &  0.479 &  0.064 & -0.084 & -0.072 &  0.000 &  0.000 &  0.000 &  0.000 &  0.000 &  0.000 &  0.000 &  0.000 &  0.000 \\
\end{tabular}
\caption{ Correlation matrix for the $d\sigma/dE_{\pi}$ statistical uncertainty }
\end{table}
\end{turnpage}
\endgroup

\begingroup
\squeezetable
\begin{turnpage}
\begin{table}[h]
\tabcolsep=0.11cm
\begin{tabular}{cc|ccccccccc|ccccccccc}
 &  & \multicolumn{9}{c|}{\numu} & \multicolumn{9}{c}{\numubar} \\
%& $E_{\pi}$ \ (GeV) & 0 - 0.25 & 0.25 - 0.5 & 0.5 - 0.75 & 0.75 - 1.0 & 1.0 - 1.5 & 1.5 - 2.0 & 2.0 - 2.5 & 2.5 - 3.5 & 3.5 - 4.5 & 0 - 0.25 & 0.25 - 0.5 & 0.5 - 0.75 & 0.75 - 1.0 & 1.0 - 1.5 & 1.5 - 2.0 & 2.0 - 2.5 & 2.5 - 3.5 & 3.5 - 4.5 \\
 & $E_{\pi}$ \ (GeV) & 0 -  & 0.25 - & 0.5 - & 0.75 - & 1.0 - & 1.5 - & 2.0 - & 2.5 - & 3.5 - & 0 -  & 0.25 - & 0.5 - & 0.75 - & 1.0 - & 1.5 - & 2.0 - & 2.5 - & 3.5 -  \\
 &                   & 0.25 & 0.5    & 0.75  & 1.0    & 1.5   & 2.0   & 2.5   & 3.5   & 4.5   & 0.25 & 0.5    & 0.75  & 1.0    & 1.5   & 2.0   & 2.5   & 3.5   & 4.5    \\
\hline
\multirow{9}{*}{\numu}
 &   0 - 0.25 &  1.000 &  0.998 &  0.996 &  0.992 &  0.987 &  0.980 &  0.985 &  0.953 &  0.943 &  0.661 &  0.555 &  0.658 &  0.659 &  0.652 &  0.639 &  0.644 &  0.653 &  0.697 \\
 & 0.25 - 0.5 &  0.998 &  1.000 &  0.999 &  0.997 &  0.995 &  0.991 &  0.991 &  0.948 &  0.934 &  0.653 &  0.529 &  0.652 &  0.661 &  0.660 &  0.652 &  0.657 &  0.660 &  0.698 \\
 & 0.5 - 0.75 &  0.996 &  0.999 &  1.000 &  0.999 &  0.996 &  0.991 &  0.995 &  0.960 &  0.947 &  0.652 &  0.537 &  0.648 &  0.654 &  0.651 &  0.641 &  0.647 &  0.653 &  0.698 \\
 & 0.75 - 1.0 &  0.992 &  0.997 &  0.999 &  1.000 &  0.998 &  0.995 &  0.995 &  0.953 &  0.939 &  0.648 &  0.523 &  0.647 &  0.657 &  0.657 &  0.649 &  0.654 &  0.656 &  0.699 \\
 &  1.0 - 1.5 &  0.987 &  0.995 &  0.996 &  0.998 &  1.000 &  0.999 &  0.993 &  0.937 &  0.919 &  0.644 &  0.498 &  0.647 &  0.665 &  0.670 &  0.666 &  0.670 &  0.666 &  0.700 \\
\end{tabular}
\caption{ Correlation matrix for the $d\sigma/dE_{\pi}$ flux systematic uncertainty }
\end{table}
\end{turnpage}
\endgroup

\begingroup
\squeezetable
\begin{turnpage}
\begin{table}[h]
\tabcolsep=0.11cm
\begin{tabular}{cc|ccccccccc|ccccccccc}
 &  & \multicolumn{9}{c|}{\numu} & \multicolumn{9}{c}{\numubar} \\
%& $E_{\pi}$ \ (GeV) & 0 - 0.25 & 0.25 - 0.5 & 0.5 - 0.75 & 0.75 - 1.0 & 1.0 - 1.5 & 1.5 - 2.0 & 2.0 - 2.5 & 2.5 - 3.5 & 3.5 - 4.5 & 0 - 0.25 & 0.25 - 0.5 & 0.5 - 0.75 & 0.75 - 1.0 & 1.0 - 1.5 & 1.5 - 2.0 & 2.0 - 2.5 & 2.5 - 3.5 & 3.5 - 4.5 \\
 & $E_{\pi}$ \ (GeV) & 0 -  & 0.25 - & 0.5 - & 0.75 - & 1.0 - & 1.5 - & 2.0 - & 2.5 - & 3.5 - & 0 -  & 0.25 - & 0.5 - & 0.75 - & 1.0 - & 1.5 - & 2.0 - & 2.5 - & 3.5 -  \\
 &                   & 0.25 & 0.5    & 0.75  & 1.0    & 1.5   & 2.0   & 2.5   & 3.5   & 4.5   & 0.25 & 0.5    & 0.75  & 1.0    & 1.5   & 2.0   & 2.5   & 3.5   & 4.5    \\
\hline
\multirow{9}{*}{\numu}
 &   0 - 0.25 &  1.000 &  0.152 &  0.100 &  0.503 &  0.316 & -0.199 & -0.400 & -0.286 & -0.236 &  0.905 &  0.659 &  0.146 &  0.154 &  0.303 &  0.413 & -0.059 & -0.347 & -0.306 \\
 & 0.25 - 0.5 &  0.152 &  1.000 &  0.887 &  0.438 &  0.106 & -0.123 & -0.205 & -0.190 & -0.086 & -0.188 &  0.138 &  0.430 &  0.247 & -0.001 &  0.073 &  0.243 &  0.307 &  0.289 \\
 & 0.5 - 0.75 &  0.100 &  0.887 &  1.000 &  0.706 &  0.390 &  0.147 &  0.085 &  0.085 &  0.150 & -0.155 &  0.168 &  0.550 &  0.491 &  0.286 &  0.268 &  0.435 &  0.462 &  0.400 \\
 & 0.75 - 1.0 &  0.503 &  0.438 &  0.706 &  1.000 &  0.857 &  0.456 &  0.216 &  0.162 &  0.175 &  0.404 &  0.419 &  0.471 &  0.629 &  0.734 &  0.698 &  0.535 &  0.312 &  0.210 \\
 &  1.0 - 1.5 &  0.316 &  0.106 &  0.390 &  0.857 &  1.000 &  0.792 &  0.439 &  0.217 &  0.168 &  0.357 &  0.308 &  0.364 &  0.624 &  0.856 &  0.754 &  0.579 &  0.320 &  0.186 \\
\end{tabular}
\caption{ Correlation matrix for the $d\sigma/dE_{\pi}$ non-flux systematic uncertainty }
\end{table}
\end{turnpage}
\endgroup

\begingroup
\squeezetable
\begin{turnpage}
\begin{table}[h]
\tabcolsep=0.08cm
\begin{tabular}{cc|cccccccccccc|cccccccccccc}
 &  & \multicolumn{12}{c|}{\numu} & \multicolumn{12}{c}{\numubar} \\
%& $\theta_{\pi}$\ (Degrees) & 0 - 5 & 5 - 10 & 10 - 15 & 15 - 20 & 20 - 25 & 25 - 30 & 30 - 35 & 35 - 40 & 40 - 45 & 45 - 50 & 50 - 60 & 60 - 70 & 0 - 5 & 5 - 10 & 10 - 15 & 15 - 20 & 20 - 25 & 25 - 30 & 30 - 35 & 35 - 40 & 40 - 45 & 45 - 50 & 50 - 60 & 60 - 70 \\
 & $\theta_{\pi}$ & 0 - & 5 - & 10 - & 15 - & 20 - & 25 - & 30 - & 35 - & 40 - & 45 - & 50 - & 60 - & 0 - & 5 - & 10 - & 15 - & 20 - & 25 - & 30 - & 35 - & 40 - & 45 - & 50 - & 60 - \\
 & (Degrees)      & 5   & 10  & 15   & 20   & 25   & 30   & 35   & 40   & 45   & 50   & 60   & 70   & 5   & 10  & 15   & 20   & 25	& 30   & 35   & 40   & 45	& 50   & 60   & 70   \\
\hline
\multirow{12}{*}{\numu}
 &   0 - 5 &  1.000 &  0.002 & -0.055 & -0.020 & -0.006 & -0.005 & -0.004 & -0.002 & -0.001 & -0.002 &  0.003 &  0.002 &  0.000 &  0.000 &  0.000 &  0.000 &  0.000 &  0.000 &  0.000 &  0.000 &  0.000 &  0.000 &  0.000 &  0.000 \\
 &  5 - 10 &  0.002 &  1.000 &  0.051 & -0.064 & -0.025 & -0.010 & -0.005 & -0.004 & -0.003 &  0.002 & -0.001 &  0.001 &  0.000 &  0.000 &  0.000 &  0.000 &  0.000 &  0.000 &  0.000 &  0.000 &  0.000 &  0.000 &  0.000 &  0.000 \\
 & 10 - 15 & -0.055 &  0.051 &  1.000 &  0.088 & -0.063 & -0.026 & -0.014 & -0.007 & -0.005 & -0.002 &  0.003 &  0.001 &  0.000 &  0.000 &  0.000 &  0.000 &  0.000 &  0.000 &  0.000 &  0.000 &  0.000 &  0.000 &  0.000 &  0.000 \\
 & 15 - 20 & -0.020 & -0.064 &  0.088 &  1.000 &  0.105 & -0.065 & -0.031 & -0.014 & -0.009 & -0.001 &  0.003 &  0.008 &  0.000 &  0.000 &  0.000 &  0.000 &  0.000 &  0.000 &  0.000 &  0.000 &  0.000 &  0.000 &  0.000 &  0.000 \\
 & 20 - 25 & -0.006 & -0.025 & -0.063 &  0.105 &  1.000 &  0.124 & -0.063 & -0.024 & -0.016 &  0.008 & -0.002 &  0.008 &  0.000 &  0.000 &  0.000 &  0.000 &  0.000 &  0.000 &  0.000 &  0.000 &  0.000 &  0.000 &  0.000 &  0.000 \\
\end{tabular}
\caption{ Correlation matrix for the $d\sigma/d\theta_{\pi}$ statistical uncertainty }
\end{table}
\end{turnpage}
\endgroup

\begingroup
\squeezetable
\begin{turnpage}
\begin{table}[h]
\tabcolsep=0.08cm
\begin{tabular}{cc|cccccccccccc|cccccccccccc}
 &  & \multicolumn{12}{c|}{\numu} & \multicolumn{12}{c}{\numubar} \\
%& $\theta_{\pi}$\ (Degrees) & 0 - 5 & 5 - 10 & 10 - 15 & 15 - 20 & 20 - 25 & 25 - 30 & 30 - 35 & 35 - 40 & 40 - 45 & 45 - 50 & 50 - 60 & 60 - 70 & 0 - 5 & 5 - 10 & 10 - 15 & 15 - 20 & 20 - 25 & 25 - 30 & 30 - 35 & 35 - 40 & 40 - 45 & 45 - 50 & 50 - 60 & 60 - 70 \\
 & $\theta_{\pi}$ & 0 - & 5 - & 10 - & 15 - & 20 - & 25 - & 30 - & 35 - & 40 - & 45 - & 50 - & 60 - & 0 - & 5 - & 10 - & 15 - & 20 - & 25 - & 30 - & 35 - & 40 - & 45 - & 50 - & 60 - \\
 & (Degrees)      & 5   & 10  & 15   & 20   & 25   & 30   & 35   & 40   & 45   & 50   & 60   & 70   & 5   & 10  & 15   & 20   & 25	& 30   & 35   & 40   & 45	& 50   & 60   & 70   \\
\hline
\multirow{12}{*}{\numu}
 &   0 - 5 &  1.000 &  0.994 &  0.987 &  0.989 &  0.989 &  0.992 &  0.994 &  0.974 &  0.975 &  0.698 & -0.968 & -0.977 &  0.667 &  0.649 &  0.629 &  0.618 &  0.637 &  0.627 &  0.564 &  0.517 & -0.182 & -0.505 &  0.167 & -0.193 \\
 &  5 - 10 &  0.994 &  1.000 &  0.999 &  0.999 &  0.999 &  1.000 &  0.999 &  0.990 &  0.992 &  0.762 & -0.968 & -0.962 &  0.688 &  0.673 &  0.656 &  0.648 &  0.661 &  0.639 &  0.583 &  0.509 & -0.190 & -0.547 &  0.201 & -0.245 \\
 & 10 - 15 &  0.987 &  0.999 &  1.000 &  1.000 &  0.999 &  0.999 &  0.998 &  0.991 &  0.994 &  0.784 & -0.966 & -0.953 &  0.693 &  0.680 &  0.665 &  0.658 &  0.668 &  0.642 &  0.590 &  0.508 & -0.188 & -0.560 &  0.213 & -0.262 \\
 & 15 - 20 &  0.989 &  0.999 &  1.000 &  1.000 &  0.999 &  0.999 &  0.998 &  0.991 &  0.993 &  0.778 & -0.967 & -0.955 &  0.691 &  0.678 &  0.662 &  0.655 &  0.666 &  0.641 &  0.587 &  0.508 & -0.189 & -0.557 &  0.211 & -0.259 \\
 & 20 - 25 &  0.989 &  0.999 &  0.999 &  0.999 &  1.000 &  1.000 &  0.997 &  0.986 &  0.989 &  0.764 & -0.975 & -0.963 &  0.684 &  0.671 &  0.657 &  0.650 &  0.662 &  0.644 &  0.589 &  0.525 & -0.173 & -0.542 &  0.203 & -0.242 \\
\end{tabular}
\caption{ Correlation matrix for the $d\sigma/d\theta_{\pi}$ flux systematic uncertainty }
\end{table}
\end{turnpage}
\endgroup

\begingroup
\squeezetable
\begin{turnpage}
\begin{table}[h]
\tabcolsep=0.08cm
\begin{tabular}{cc|cccccccccccc|cccccccccccc}
 &  & \multicolumn{12}{c|}{\numu} & \multicolumn{12}{c}{\numubar} \\
%& $\theta_{\pi}$\ (Degrees) & 0 - 5 & 5 - 10 & 10 - 15 & 15 - 20 & 20 - 25 & 25 - 30 & 30 - 35 & 35 - 40 & 40 - 45 & 45 - 50 & 50 - 60 & 60 - 70 & 0 - 5 & 5 - 10 & 10 - 15 & 15 - 20 & 20 - 25 & 25 - 30 & 30 - 35 & 35 - 40 & 40 - 45 & 45 - 50 & 50 - 60 & 60 - 70 \\
 & $\theta_{\pi}$ & 0 - & 5 - & 10 - & 15 - & 20 - & 25 - & 30 - & 35 - & 40 - & 45 - & 50 - & 60 - & 0 - & 5 - & 10 - & 15 - & 20 - & 25 - & 30 - & 35 - & 40 - & 45 - & 50 - & 60 - \\
 & (Degrees)      & 5   & 10  & 15   & 20   & 25   & 30   & 35   & 40   & 45   & 50   & 60   & 70   & 5   & 10  & 15   & 20   & 25	& 30   & 35   & 40   & 45	& 50   & 60   & 70   \\
\hline
\multirow{12}{*}{\numu}
 &   0 - 5 &  1.000 &  0.935 &  0.438 &  0.379 &  0.172 &  0.246 &  0.697 &  0.583 & -0.187 & -0.388 & -0.135 & -0.037 &  0.799 &  0.881 &  0.827 &  0.824 &  0.337 &  0.299 &  0.356 &  0.445 &  0.345 &  0.221 & -0.135 &  0.228 \\
 &  5 - 10 &  0.935 &  1.000 &  0.360 &  0.329 &  0.183 &  0.275 &  0.767 &  0.618 & -0.267 & -0.489 & -0.137 & -0.037 &  0.812 &  0.897 &  0.900 &  0.890 &  0.302 &  0.261 &  0.264 &  0.320 &  0.249 &  0.186 & -0.303 &  0.282 \\
 & 10 - 15 &  0.438 &  0.360 &  1.000 &  0.919 &  0.725 &  0.567 &  0.185 &  0.029 & -0.038 & -0.070 & -0.238 & -0.301 &  0.293 &  0.367 &  0.372 &  0.360 &  0.692 &  0.603 &  0.474 &  0.536 &  0.324 &  0.448 &  0.138 &  0.460 \\
 & 15 - 20 &  0.379 &  0.329 &  0.919 &  1.000 &  0.828 &  0.634 &  0.192 &  0.004 & -0.071 & -0.111 & -0.337 & -0.417 &  0.261 &  0.312 &  0.345 &  0.362 &  0.733 &  0.609 &  0.497 &  0.466 &  0.380 &  0.458 & -0.015 &  0.571 \\
 & 20 - 25 &  0.172 &  0.183 &  0.725 &  0.828 &  1.000 &  0.878 &  0.321 &  0.164 &  0.180 &  0.069 & -0.144 & -0.247 &  0.160 &  0.083 &  0.231 &  0.247 &  0.516 &  0.514 &  0.435 &  0.281 &  0.338 &  0.396 & -0.237 &  0.439 \\
\end{tabular}
\caption{ Correlation matrix for the $d\sigma/d\theta_{\pi}$ non-flux systematic uncertainty }
\end{table}
\end{turnpage}
\endgroup

\begingroup
\squeezetable
\begin{turnpage}
\begin{table}[h]
\tabcolsep=0.11cm
\begin{tabular}{cc|cccccccccc|cccccccccc}
 &  & \multicolumn{10}{c|}{\numu} & \multicolumn{10}{c}{\numubar} \\
%& $Q^{2}$\ (GeV/c)$^{2}$ & 0 - 0.025 & 0.025 - 0.05 & 0.05 - 0.075 & 0.075 - 0.1 & 0.1 - 0.15 & 0.15 - 0.2 & 0.2 - 0.3 & 0.3 - 0.4 & 0.4 - 0.6 & 0.6 - 0.8 & 0 - 0.025 & 0.025 - 0.05 & 0.05 - 0.075 & 0.075 - 0.1 & 0.1 - 0.15 & 0.15 - 0.2 & 0.2 - 0.3 & 0.3 - 0.4 & 0.4 - 0.6 & 0.6 - 0.8 \\
 & $Q^{2}$\ (GeV/c)$^{2}$ & 0 -   & 0.025 - & 0.05 - & 0.075 - & 0.1 - & 0.15 - & 0.2 - & 0.3 - & 0.4 - & 0.6 - & 0  -	& 0.025 - & 0.05 - & 0.075 - & 0.1 - & 0.15 - & 0.2 - & 0.3 - & 0.4 - & 0.6 - \\
 &                        & 0.025 & 0.05    & 0.075  & 0.1     & 0.15  & 0.2    & 0.3   & 0.4   & 0.6   & 0.8   & 0.025 & 0.05    & 0.075  & 0.1	 & 0.15  & 0.2    & 0.3   & 0.4   & 0.6   & 0.8  \\
\hline
\multirow{10}{*}{\numu}
 &    0 - 0.025 &  1.000 &  0.102 & -0.115 & -0.076 & -0.034 & -0.016 & -0.006 & -0.002 & -0.001 & -0.000 &  0.000 &  0.000 &  0.000 &  0.000 &  0.000 &  0.000 &  0.000 &  0.000 &  0.000 &  0.000 \\
 & 0.025 - 0.05 &  0.102 &  1.000 &  0.383 & -0.025 & -0.086 & -0.052 & -0.026 & -0.007 & -0.002 & -0.001 &  0.000 &  0.000 &  0.000 &  0.000 &  0.000 &  0.000 &  0.000 &  0.000 &  0.000 &  0.000 \\
 & 0.05 - 0.075 & -0.115 &  0.383 &  1.000 &  0.570 &  0.073 & -0.082 & -0.059 & -0.021 & -0.008 & -0.002 &  0.000 &  0.000 &  0.000 &  0.000 &  0.000 &  0.000 &  0.000 &  0.000 &  0.000 &  0.000 \\
 &  0.075 - 0.1 & -0.076 & -0.025 &  0.570 &  1.000 &  0.614 & -0.010 & -0.103 & -0.047 & -0.015 & -0.004 &  0.000 &  0.000 &  0.000 &  0.000 &  0.000 &  0.000 &  0.000 &  0.000 &  0.000 &  0.000 \\
 &   0.1 - 0.15 & -0.034 & -0.086 &  0.073 &  0.614 &  1.000 &  0.509 & -0.008 & -0.097 & -0.047 & -0.009 &  0.000 &  0.000 &  0.000 &  0.000 &  0.000 &  0.000 &  0.000 &  0.000 &  0.000 &  0.000 \\
\end{tabular}
\caption{ Correlation matrix for the $d\sigma/dQ^{2}$ statistical uncertainty }
\end{table}
\end{turnpage}
\endgroup

\begingroup
\squeezetable
\begin{turnpage}
\begin{table}[h]
\tabcolsep=0.11cm
\begin{tabular}{cc|cccccccccc|cccccccccc}
 &  & \multicolumn{10}{c|}{\numu} & \multicolumn{10}{c}{\numubar} \\
%& $Q^{2}$\ (GeV/c)$^{2}$ & 0 - 0.025 & 0.025 - 0.05 & 0.05 - 0.075 & 0.075 - 0.1 & 0.1 - 0.15 & 0.15 - 0.2 & 0.2 - 0.3 & 0.3 - 0.4 & 0.4 - 0.6 & 0.6 - 0.8 & 0 - 0.025 & 0.025 - 0.05 & 0.05 - 0.075 & 0.075 - 0.1 & 0.1 - 0.15 & 0.15 - 0.2 & 0.2 - 0.3 & 0.3 - 0.4 & 0.4 - 0.6 & 0.6 - 0.8 \\
 & $Q^{2}$\ (GeV/c)$^{2}$ & 0 -   & 0.025 - & 0.05 - & 0.075 - & 0.1 - & 0.15 - & 0.2 - & 0.3 - & 0.4 - & 0.6 - & 0  -	& 0.025 - & 0.05 - & 0.075 - & 0.1 - & 0.15 - & 0.2 - & 0.3 - & 0.4 - & 0.6 - \\
 &                        & 0.025 & 0.05    & 0.075  & 0.1     & 0.15  & 0.2    & 0.3   & 0.4   & 0.6   & 0.8   & 0.025 & 0.05    & 0.075  & 0.1	 & 0.15  & 0.2    & 0.3   & 0.4   & 0.6   & 0.8  \\
\hline
\multirow{10}{*}{\numu}
 &    0 - 0.025 &  1.000 &  1.000 &  0.999 &  0.998 &  0.996 &  0.994 &  0.990 &  0.982 &  0.975 &  0.715 &  0.659 &  0.659 &  0.646 &  0.652 &  0.657 &  0.654 &  0.630 &  0.636 &  0.656 &  0.676 \\
 & 0.025 - 0.05 &  1.000 &  1.000 &  1.000 &  0.999 &  0.998 &  0.997 &  0.994 &  0.987 &  0.981 &  0.713 &  0.657 &  0.658 &  0.645 &  0.653 &  0.658 &  0.656 &  0.633 &  0.641 &  0.661 &  0.680 \\
 & 0.05 - 0.075 &  0.999 &  1.000 &  1.000 &  1.000 &  0.999 &  0.997 &  0.994 &  0.987 &  0.980 &  0.700 &  0.660 &  0.662 &  0.649 &  0.657 &  0.662 &  0.660 &  0.640 &  0.648 &  0.666 &  0.680 \\
 &  0.075 - 0.1 &  0.998 &  0.999 &  1.000 &  1.000 &  1.000 &  0.998 &  0.995 &  0.988 &  0.980 &  0.684 &  0.664 &  0.666 &  0.654 &  0.664 &  0.667 &  0.667 &  0.649 &  0.658 &  0.674 &  0.682 \\
 &   0.1 - 0.15 &  0.996 &  0.998 &  0.999 &  1.000 &  1.000 &  0.999 &  0.997 &  0.991 &  0.983 &  0.673 &  0.662 &  0.665 &  0.653 &  0.665 &  0.668 &  0.669 &  0.653 &  0.664 &  0.679 &  0.683 \\
\end{tabular}
\caption{ Correlation matrix for the $d\sigma/dQ^{2}$ flux systematic uncertainty }
\end{table}
\end{turnpage}
\endgroup

\begingroup
\squeezetable
\begin{turnpage}
\begin{table}[h]
\tabcolsep=0.11cm
\begin{tabular}{cc|cccccccccc|cccccccccc}
 &  & \multicolumn{10}{c|}{\numu} & \multicolumn{10}{c}{\numubar} \\
%& $Q^{2}$\ (GeV/c)$^{2}$ & 0 - 0.025 & 0.025 - 0.05 & 0.05 - 0.075 & 0.075 - 0.1 & 0.1 - 0.15 & 0.15 - 0.2 & 0.2 - 0.3 & 0.3 - 0.4 & 0.4 - 0.6 & 0.6 - 0.8 & 0 - 0.025 & 0.025 - 0.05 & 0.05 - 0.075 & 0.075 - 0.1 & 0.1 - 0.15 & 0.15 - 0.2 & 0.2 - 0.3 & 0.3 - 0.4 & 0.4 - 0.6 & 0.6 - 0.8 \\
 & $Q^{2}$\ (GeV/c)$^{2}$ & 0 -   & 0.025 - & 0.05 - & 0.075 - & 0.1 - & 0.15 - & 0.2 - & 0.3 - & 0.4 - & 0.6 - & 0  -	& 0.025 - & 0.05 - & 0.075 - & 0.1 - & 0.15 - & 0.2 - & 0.3 - & 0.4 - & 0.6 - \\
 &                        & 0.025 & 0.05    & 0.075  & 0.1     & 0.15  & 0.2    & 0.3   & 0.4   & 0.6   & 0.8   & 0.025 & 0.05    & 0.075  & 0.1	 & 0.15  & 0.2    & 0.3   & 0.4   & 0.6   & 0.8  \\
\hline
\multirow{10}{*}{\numu}
 &    0 - 0.025 &  1.000 &  0.975 &  0.904 &  0.839 &  0.810 &  0.644 &  0.496 &  0.097 &  0.059 & -0.166 &  0.550 &  0.530 &  0.511 &  0.405 &  0.325 &  0.345 &  0.313 &  0.189 &  0.140 &  0.100 \\
 & 0.025 - 0.05 &  0.975 &  1.000 &  0.943 &  0.864 &  0.827 &  0.689 &  0.570 &  0.169 &  0.083 & -0.157 &  0.579 &  0.559 &  0.534 &  0.464 &  0.408 &  0.431 &  0.381 &  0.276 &  0.233 &  0.124 \\
 & 0.05 - 0.075 &  0.904 &  0.943 &  1.000 &  0.962 &  0.896 &  0.755 &  0.672 &  0.349 &  0.176 & -0.108 &  0.418 &  0.377 &  0.345 &  0.319 &  0.341 &  0.412 &  0.370 &  0.266 &  0.257 &  0.249 \\
 &  0.075 - 0.1 &  0.839 &  0.864 &  0.962 &  1.000 &  0.956 &  0.824 &  0.720 &  0.437 &  0.229 & -0.047 &  0.313 &  0.298 &  0.278 &  0.277 &  0.321 &  0.423 &  0.415 &  0.327 &  0.289 &  0.255 \\
 &   0.1 - 0.15 &  0.810 &  0.827 &  0.896 &  0.956 &  1.000 &  0.940 &  0.826 &  0.443 &  0.284 & -0.014 &  0.379 &  0.375 &  0.412 &  0.392 &  0.410 &  0.507 &  0.568 &  0.495 &  0.328 &  0.195 \\
\end{tabular}
\caption{ Correlation matrix for the $d\sigma/dQ^{2}$ non-flux systematic uncertainty }
\end{table}
\end{turnpage}
\endgroup

\begingroup
\squeezetable
\begin{turnpage}
\begin{table}[h]
\tabcolsep=0.11cm
\begin{tabular}{cc|ccccccccc|ccccccccc}
 &  & \multicolumn{9}{c|}{\numu} & \multicolumn{9}{c}{\numubar} \\
%& $E_{\pi}$ \ (GeV) & 0 - 0.25 & 0.25 - 0.5 & 0.5 - 0.75 & 0.75 - 1.0 & 1.0 - 1.5 & 1.5 - 2.0 & 2.0 - 2.5 & 2.5 - 3.5 & 3.5 - 4.5 & 0 - 0.25 & 0.25 - 0.5 & 0.5 - 0.75 & 0.75 - 1.0 & 1.0 - 1.5 & 1.5 - 2.0 & 2.0 - 2.5 & 2.5 - 3.5 & 3.5 - 4.5 \\
 & $E_{\pi}$ \ (GeV) & 0 -  & 0.25 - & 0.5 - & 0.75 - & 1.0 - & 1.5 - & 2.0 - & 2.5 - & 3.5 - & 0 -  & 0.25 - & 0.5 - & 0.75 - & 1.0 - & 1.5 - & 2.0 - & 2.5 - & 3.5 -  \\
 &                   & 0.25 & 0.5    & 0.75  & 1.0    & 1.5   & 2.0   & 2.5   & 3.5   & 4.5   & 0.25 & 0.5    & 0.75  & 1.0    & 1.5   & 2.0   & 2.5   & 3.5   & 4.5    \\
\hline
\multirow{9}{*}{\numu}
 &   0 - 0.25 &  1.000 &  0.255 & -0.137 & -0.050 & -0.013 &  0.001 &  0.000 &  0.000 &  0.000 &  0.000 &  0.000 &  0.000 &  0.000 &  0.000 &  0.000 &  0.000 &  0.000 &  0.000 \\
 & 0.25 - 0.5 &  0.255 &  1.000 &  0.154 & -0.104 & -0.107 & -0.038 & -0.011 & -0.004 & -0.001 &  0.000 &  0.000 &  0.000 &  0.000 &  0.000 &  0.000 &  0.000 &  0.000 &  0.000 \\
 & 0.5 - 0.75 & -0.137 &  0.154 &  1.000 &  0.561 &  0.012 & -0.139 & -0.078 & -0.031 & -0.009 &  0.000 &  0.000 &  0.000 &  0.000 &  0.000 &  0.000 &  0.000 &  0.000 &  0.000 \\
 & 0.75 - 1.0 & -0.050 & -0.104 &  0.561 &  1.000 &  0.620 & -0.012 & -0.119 & -0.076 & -0.032 &  0.000 &  0.000 &  0.000 &  0.000 &  0.000 &  0.000 &  0.000 &  0.000 &  0.000 \\
 &  1.0 - 1.5 & -0.013 & -0.107 &  0.012 &  0.620 &  1.000 &  0.488 &  0.077 & -0.077 & -0.072 &  0.000 &  0.000 &  0.000 &  0.000 &  0.000 &  0.000 &  0.000 &  0.000 &  0.000 \\
\end{tabular}
\caption{ Correlation matrix for the $d\sigma/dE_{\pi}$ statistical uncertainty where the \numu cross section is weighted to the \numubar flux }
\end{table}
\end{turnpage}
\endgroup

\begingroup
\squeezetable
\begin{turnpage}
\begin{table}[h]
\tabcolsep=0.11cm
\begin{tabular}{cc|ccccccccc|ccccccccc}
 &  & \multicolumn{9}{c|}{\numu} & \multicolumn{9}{c}{\numubar} \\
%& $E_{\pi}$ \ (GeV) & 0 - 0.25 & 0.25 - 0.5 & 0.5 - 0.75 & 0.75 - 1.0 & 1.0 - 1.5 & 1.5 - 2.0 & 2.0 - 2.5 & 2.5 - 3.5 & 3.5 - 4.5 & 0 - 0.25 & 0.25 - 0.5 & 0.5 - 0.75 & 0.75 - 1.0 & 1.0 - 1.5 & 1.5 - 2.0 & 2.0 - 2.5 & 2.5 - 3.5 & 3.5 - 4.5 \\
 & $E_{\pi}$ \ (GeV) & 0 -  & 0.25 - & 0.5 - & 0.75 - & 1.0 - & 1.5 - & 2.0 - & 2.5 - & 3.5 - & 0 -  & 0.25 - & 0.5 - & 0.75 - & 1.0 - & 1.5 - & 2.0 - & 2.5 - & 3.5 -  \\
 &                   & 0.25 & 0.5    & 0.75  & 1.0    & 1.5   & 2.0   & 2.5   & 3.5   & 4.5   & 0.25 & 0.5    & 0.75  & 1.0    & 1.5   & 2.0   & 2.5   & 3.5   & 4.5    \\
\hline
\multirow{9}{*}{\numu}
 &   0 - 0.25 &  1.000 &  0.998 &  0.996 &  0.992 &  0.985 &  0.975 &  0.984 &  0.947 &  0.934 &  0.664 &  0.566 &  0.661 &  0.658 &  0.649 &  0.633 &  0.637 &  0.648 &  0.695 \\
 & 0.25 - 0.5 &  0.998 &  1.000 &  0.999 &  0.998 &  0.994 &  0.988 &  0.991 &  0.942 &  0.923 &  0.658 &  0.541 &  0.658 &  0.663 &  0.659 &  0.648 &  0.652 &  0.657 &  0.697 \\
 & 0.5 - 0.75 &  0.996 &  0.999 &  1.000 &  0.999 &  0.994 &  0.987 &  0.994 &  0.952 &  0.935 &  0.657 &  0.548 &  0.655 &  0.658 &  0.652 &  0.639 &  0.644 &  0.652 &  0.698 \\
 & 0.75 - 1.0 &  0.992 &  0.998 &  0.999 &  1.000 &  0.998 &  0.992 &  0.995 &  0.943 &  0.923 &  0.655 &  0.532 &  0.656 &  0.664 &  0.661 &  0.651 &  0.654 &  0.657 &  0.700 \\
 &  1.0 - 1.5 &  0.985 &  0.994 &  0.994 &  0.998 &  1.000 &  0.998 &  0.992 &  0.921 &  0.897 &  0.651 &  0.503 &  0.657 &  0.674 &  0.678 &  0.672 &  0.673 &  0.669 &  0.703 \\
\end{tabular}
\caption{ Correlation matrix for the $d\sigma/dE_{\pi}$ flux systematic uncertainty where the \numu cross section is weighted to the \numubar flux }
\end{table}
\end{turnpage}
\endgroup

\begingroup
\squeezetable
\begin{turnpage}
\begin{table}[h]
\tabcolsep=0.11cm
\begin{tabular}{cc|ccccccccc|ccccccccc}
 &  & \multicolumn{9}{c|}{\numu} & \multicolumn{9}{c}{\numubar} \\
%& $E_{\pi}$ \ (GeV) & 0 - 0.25 & 0.25 - 0.5 & 0.5 - 0.75 & 0.75 - 1.0 & 1.0 - 1.5 & 1.5 - 2.0 & 2.0 - 2.5 & 2.5 - 3.5 & 3.5 - 4.5 & 0 - 0.25 & 0.25 - 0.5 & 0.5 - 0.75 & 0.75 - 1.0 & 1.0 - 1.5 & 1.5 - 2.0 & 2.0 - 2.5 & 2.5 - 3.5 & 3.5 - 4.5 \\
 & $E_{\pi}$ \ (GeV) & 0 -  & 0.25 - & 0.5 - & 0.75 - & 1.0 - & 1.5 - & 2.0 - & 2.5 - & 3.5 - & 0 -  & 0.25 - & 0.5 - & 0.75 - & 1.0 - & 1.5 - & 2.0 - & 2.5 - & 3.5 -  \\
 &                   & 0.25 & 0.5    & 0.75  & 1.0    & 1.5   & 2.0   & 2.5   & 3.5   & 4.5   & 0.25 & 0.5    & 0.75  & 1.0    & 1.5   & 2.0   & 2.5   & 3.5   & 4.5    \\
\hline
\multirow{9}{*}{\numu}
 &   0 - 0.25 &  1.000 &  0.152 &  0.095 &  0.447 &  0.225 & -0.303 & -0.479 & -0.399 & -0.370 &  0.905 &  0.658 &  0.146 &  0.154 &  0.304 &  0.415 & -0.058 & -0.347 & -0.306 \\
 & 0.25 - 0.5 &  0.152 &  1.000 &  0.885 &  0.453 &  0.113 & -0.106 & -0.181 & -0.167 & -0.072 & -0.188 &  0.137 &  0.429 &  0.248 &  0.001 &  0.075 &  0.244 &  0.307 &  0.288 \\
 & 0.5 - 0.75 &  0.095 &  0.885 &  1.000 &  0.728 &  0.408 &  0.163 &  0.105 &  0.108 &  0.166 & -0.159 &  0.163 &  0.549 &  0.493 &  0.290 &  0.270 &  0.438 &  0.465 &  0.401 \\
 & 0.75 - 1.0 &  0.447 &  0.453 &  0.728 &  1.000 &  0.848 &  0.441 &  0.223 &  0.156 &  0.158 &  0.348 &  0.386 &  0.480 &  0.643 &  0.739 &  0.690 &  0.561 &  0.354 &  0.246 \\
 &  1.0 - 1.5 &  0.225 &  0.113 &  0.408 &  0.848 &  1.000 &  0.792 &  0.468 &  0.248 &  0.194 &  0.268 &  0.248 &  0.364 &  0.628 &  0.846 &  0.728 &  0.604 &  0.373 &  0.234 \\
\end{tabular}
\caption{ Correlation matrix for the $d\sigma/dE_{\pi}$ non-flux systematic uncertainty where the \numu cross section is weighted to the \numubar flux }
\end{table}
\end{turnpage}
\endgroup

\begingroup
\squeezetable
\begin{turnpage}
\begin{table}[h]
\tabcolsep=0.08cm
\begin{tabular}{cc|cccccccccccc|cccccccccccc}
 &  & \multicolumn{12}{c|}{\numu} & \multicolumn{12}{c}{\numubar} \\
% & $\theta_{\pi}$\ (Degrees) & 0 - 5 & 5 - 10 & 10 - 15 & 15 - 20 & 20 - 25 & 25 - 30 & 30 - 35 & 35 - 40 & 40 - 45 & 45 - 50 & 50 - 60 & 60 - 70 & 0 - 5 & 5 - 10 & 10 - 15 & 15 - 20 & 20 - 25 & 25 - 30 & 30 - 35 & 35 - 40 & 40 - 45 & 45 - 50 & 50 - 60 & 60 - 70 \\
 & $\theta_{\pi}$ & 0 - & 5 - & 10 - & 15 - & 20 - & 25 - & 30 - & 35 - & 40 - & 45 - & 50 - & 60 - & 0 - & 5 - & 10 - & 15 - & 20 - & 25 - & 30 - & 35 - & 40 - & 45 - & 50 - & 60 - \\
 & (Degrees)      & 5   & 10  & 15   & 20   & 25   & 30   & 35   & 40   & 45   & 50   & 60   & 70   & 5   & 10  & 15   & 20   & 25	& 30   & 35   & 40   & 45	& 50   & 60   & 70   \\
\hline
\multirow{12}{*}{\numu}
 &   0 - 5 &  1.000 &  0.018 & -0.058 & -0.023 & -0.007 & -0.005 & -0.004 & -0.002 & -0.002 & -0.003 &  0.003 &  0.002 &  0.000 &  0.000 &  0.000 &  0.000 &  0.000 &  0.000 &  0.000 &  0.000 &  0.000 &  0.000 &  0.000 &  0.000 \\
 &  5 - 10 &  0.018 &  1.000 &  0.056 & -0.065 & -0.026 & -0.012 & -0.006 & -0.004 & -0.003 &  0.003 & -0.003 &  0.000 &  0.000 &  0.000 &  0.000 &  0.000 &  0.000 &  0.000 &  0.000 &  0.000 &  0.000 &  0.000 &  0.000 &  0.000 \\
 & 10 - 15 & -0.058 &  0.056 &  1.000 &  0.090 & -0.063 & -0.027 & -0.014 & -0.007 & -0.005 & -0.002 &  0.003 &  0.001 &  0.000 &  0.000 &  0.000 &  0.000 &  0.000 &  0.000 &  0.000 &  0.000 &  0.000 &  0.000 &  0.000 &  0.000 \\
 & 15 - 20 & -0.023 & -0.065 &  0.090 &  1.000 &  0.113 & -0.067 & -0.033 & -0.013 & -0.008 &  0.002 & -0.000 &  0.007 &  0.000 &  0.000 &  0.000 &  0.000 &  0.000 &  0.000 &  0.000 &  0.000 &  0.000 &  0.000 &  0.000 &  0.000 \\
 & 20 - 25 & -0.007 & -0.026 & -0.063 &  0.113 &  1.000 &  0.135 & -0.065 & -0.024 & -0.014 &  0.012 & -0.007 &  0.004 &  0.000 &  0.000 &  0.000 &  0.000 &  0.000 &  0.000 &  0.000 &  0.000 &  0.000 &  0.000 &  0.000 &  0.000 \\
\end{tabular}
\caption{ Correlation matrix for the $d\sigma/d\theta_{\pi}$ statistical uncertainty where the \numu cross section is weighted to the \numubar flux }
\end{table}
\end{turnpage}
\endgroup

\begingroup
\squeezetable
\begin{turnpage}
\begin{table}[h]
\tabcolsep=0.08cm
\begin{tabular}{cc|cccccccccccc|cccccccccccc}
 &  & \multicolumn{12}{c|}{\numu} & \multicolumn{12}{c}{\numubar} \\
% & $\theta_{\pi}$\ (Degrees) & 0 - 5 & 5 - 10 & 10 - 15 & 15 - 20 & 20 - 25 & 25 - 30 & 30 - 35 & 35 - 40 & 40 - 45 & 45 - 50 & 50 - 60 & 60 - 70 & 0 - 5 & 5 - 10 & 10 - 15 & 15 - 20 & 20 - 25 & 25 - 30 & 30 - 35 & 35 - 40 & 40 - 45 & 45 - 50 & 50 - 60 & 60 - 70 \\
 & $\theta_{\pi}$ & 0 - & 5 - & 10 - & 15 - & 20 - & 25 - & 30 - & 35 - & 40 - & 45 - & 50 - & 60 - & 0 - & 5 - & 10 - & 15 - & 20 - & 25 - & 30 - & 35 - & 40 - & 45 - & 50 - & 60 - \\
 & (Degrees)      & 5   & 10  & 15   & 20   & 25   & 30   & 35   & 40   & 45   & 50   & 60   & 70   & 5   & 10  & 15   & 20   & 25	& 30   & 35   & 40   & 45	& 50   & 60   & 70   \\
\hline
\multirow{12}{*}{\numu}
 &   0 - 5 &  1.000 &  0.993 &  0.987 &  0.989 &  0.989 &  0.993 &  0.995 &  0.979 &  0.981 &  0.715 & -0.966 & -0.973 &  0.675 &  0.659 &  0.641 &  0.631 &  0.647 &  0.635 &  0.574 &  0.520 & -0.180 & -0.519 &  0.179 & -0.210 \\
 &  5 - 10 &  0.993 &  1.000 &  0.999 &  0.999 &  0.998 &  0.999 &  0.999 &  0.993 &  0.995 &  0.777 & -0.965 & -0.959 &  0.692 &  0.679 &  0.664 &  0.657 &  0.668 &  0.644 &  0.591 &  0.513 & -0.184 & -0.557 &  0.211 & -0.259 \\
 & 10 - 15 &  0.987 &  0.999 &  1.000 &  1.000 &  0.999 &  0.998 &  0.997 &  0.993 &  0.995 &  0.794 & -0.965 & -0.953 &  0.693 &  0.682 &  0.669 &  0.664 &  0.673 &  0.646 &  0.596 &  0.514 & -0.178 & -0.563 &  0.220 & -0.270 \\
 & 15 - 20 &  0.989 &  0.999 &  1.000 &  1.000 &  0.999 &  0.999 &  0.998 &  0.992 &  0.995 &  0.784 & -0.967 & -0.957 &  0.690 &  0.679 &  0.666 &  0.660 &  0.670 &  0.645 &  0.594 &  0.517 & -0.177 & -0.558 &  0.216 & -0.264 \\
 & 20 - 25 &  0.989 &  0.998 &  0.999 &  0.999 &  1.000 &  1.000 &  0.997 &  0.987 &  0.990 &  0.768 & -0.975 & -0.966 &  0.682 &  0.671 &  0.659 &  0.654 &  0.665 &  0.648 &  0.594 &  0.535 & -0.159 & -0.540 &  0.206 & -0.243 \\
\end{tabular}
\caption{ Correlation matrix for the $d\sigma/d\theta_{\pi}$ flux systematic uncertainty where the \numu cross section is weighted to the \numubar flux }
\end{table}
\end{turnpage}
\endgroup

\begingroup
\squeezetable
\begin{turnpage}
\begin{table}[h]
\tabcolsep=0.08cm
\begin{tabular}{cc|cccccccccccc|cccccccccccc}
 &  & \multicolumn{12}{c|}{\numu} & \multicolumn{12}{c}{\numubar} \\
%& $\theta_{\pi}$\ (Degrees) & 0 - 5 & 5 - 10 & 10 - 15 & 15 - 20 & 20 - 25 & 25 - 30 & 30 - 35 & 35 - 40 & 40 - 45 & 45 - 50 & 50 - 60 & 60 - 70 & 0 - 5 & 5 - 10 & 10 - 15 & 15 - 20 & 20 - 25 & 25 - 30 & 30 - 35 & 35 - 40 & 40 - 45 & 45 - 50 & 50 - 60 & 60 - 70 \\
 & $\theta_{\pi}$ & 0 - & 5 - & 10 - & 15 - & 20 - & 25 - & 30 - & 35 - & 40 - & 45 - & 50 - & 60 - & 0 - & 5 - & 10 - & 15 - & 20 - & 25 - & 30 - & 35 - & 40 - & 45 - & 50 - & 60 - \\
 & (Degrees)      & 5   & 10  & 15   & 20   & 25   & 30   & 35   & 40   & 45   & 50   & 60   & 70   & 5   & 10  & 15   & 20   & 25	& 30   & 35   & 40   & 45	& 50   & 60   & 70   \\
\hline
\multirow{12}{*}{\numu}
 &   0 - 5 &  1.000 &  0.937 &  0.450 &  0.381 &  0.178 &  0.250 &  0.695 &  0.577 & -0.162 & -0.378 & -0.172 & -0.058 &  0.797 &  0.879 &  0.821 &  0.819 &  0.336 &  0.293 &  0.352 &  0.438 &  0.342 &  0.212 & -0.136 &  0.222 \\
 &  5 - 10 &  0.937 &  1.000 &  0.379 &  0.336 &  0.193 &  0.284 &  0.768 &  0.615 & -0.237 & -0.475 & -0.177 & -0.055 &  0.807 &  0.895 &  0.894 &  0.885 &  0.302 &  0.256 &  0.262 &  0.318 &  0.246 &  0.178 & -0.300 &  0.272 \\
 & 10 - 15 &  0.450 &  0.379 &  1.000 &  0.926 &  0.734 &  0.571 &  0.189 &  0.028 & -0.046 & -0.088 & -0.260 & -0.319 &  0.289 &  0.360 &  0.367 &  0.359 &  0.693 &  0.588 &  0.463 &  0.511 &  0.313 &  0.428 &  0.118 &  0.455 \\
 & 15 - 20 &  0.381 &  0.336 &  0.926 &  1.000 &  0.832 &  0.636 &  0.190 & -0.004 & -0.081 & -0.133 & -0.353 & -0.427 &  0.248 &  0.303 &  0.338 &  0.357 &  0.734 &  0.595 &  0.480 &  0.443 &  0.362 &  0.438 & -0.031 &  0.566 \\
 & 20 - 25 &  0.178 &  0.193 &  0.734 &  0.832 &  1.000 &  0.876 &  0.319 &  0.160 &  0.177 &  0.061 & -0.148 & -0.253 &  0.155 &  0.083 &  0.229 &  0.246 &  0.519 &  0.508 &  0.427 &  0.274 &  0.329 &  0.387 & -0.240 &  0.437 \\
\end{tabular}
\caption{ Correlation matrix for the $d\sigma/d\theta_{\pi}$ non-flux systematic uncertainty where the \numu cross section is weighted to the \numubar flux }
\end{table}
\end{turnpage}
\endgroup

\begingroup
\squeezetable
\begin{turnpage}
\begin{table}[h]
\tabcolsep=0.11cm
\begin{tabular}{cc|cccccccccc|cccccccccc}
 &  & \multicolumn{10}{c|}{\numu} & \multicolumn{10}{c}{\numubar} \\
%& $Q^{2}$\ (GeV/c)$^{2}$ & 0 - 0.025 & 0.025 - 0.05 & 0.05 - 0.075 & 0.075 - 0.1 & 0.1 - 0.15 & 0.15 - 0.2 & 0.2 - 0.3 & 0.3 - 0.4 & 0.4 - 0.6 & 0.6 - 0.8 & 0 - 0.025 & 0.025 - 0.05 & 0.05 - 0.075 & 0.075 - 0.1 & 0.1 - 0.15 & 0.15 - 0.2 & 0.2 - 0.3 & 0.3 - 0.4 & 0.4 - 0.6 & 0.6 - 0.8 \\
 & $Q^{2}$\ (GeV/c)$^{2}$ & 0 -   & 0.025 - & 0.05 - & 0.075 - & 0.1 - & 0.15 - & 0.2 - & 0.3 - & 0.4 - & 0.6 - & 0 -	& 0.025 - & 0.05 - & 0.075 - & 0.1 - & 0.15 - & 0.2 - & 0.3 - & 0.4 - & 0.6 - \\
 &                        & 0.025 & 0.05    & 0.075  & 0.1     & 0.15  & 0.2    & 0.3   & 0.4   & 0.6   & 0.8   & 0.025 & 0.05    & 0.075  & 0.1	 & 0.15  & 0.2    & 0.3   & 0.4   & 0.6   & 0.8   \\
\hline
\multirow{10}{*}{\numu}
 &    0 - 0.025 &  1.000 &  0.079 & -0.122 & -0.066 & -0.025 & -0.010 & -0.004 & -0.001 & -0.001 & -0.000 &  0.000 &  0.000 &  0.000 &  0.000 &  0.000 &  0.000 &  0.000 &  0.000 &  0.000 &  0.000 \\
 & 0.025 - 0.05 &  0.079 &  1.000 &  0.348 & -0.052 & -0.089 & -0.042 & -0.016 & -0.004 & -0.001 & -0.001 &  0.000 &  0.000 &  0.000 &  0.000 &  0.000 &  0.000 &  0.000 &  0.000 &  0.000 &  0.000 \\
 & 0.05 - 0.075 & -0.122 &  0.348 &  1.000 &  0.549 &  0.037 & -0.088 & -0.050 & -0.014 & -0.005 & -0.001 &  0.000 &  0.000 &  0.000 &  0.000 &  0.000 &  0.000 &  0.000 &  0.000 &  0.000 &  0.000 \\
 &  0.075 - 0.1 & -0.066 & -0.052 &  0.549 &  1.000 &  0.597 & -0.041 & -0.104 & -0.037 & -0.009 & -0.002 &  0.000 &  0.000 &  0.000 &  0.000 &  0.000 &  0.000 &  0.000 &  0.000 &  0.000 &  0.000 \\
 &   0.1 - 0.15 & -0.025 & -0.089 &  0.037 &  0.597 &  1.000 &  0.482 & -0.040 & -0.092 & -0.035 & -0.005 &  0.000 &  0.000 &  0.000 &  0.000 &  0.000 &  0.000 &  0.000 &  0.000 &  0.000 &  0.000 \\
\end{tabular}
\caption{ Correlation matrix for the $d\sigma/dQ^{2}$ statistical uncertainty where the \numu cross section is weighted to the \numubar flux }
\end{table}
\end{turnpage}
\endgroup

\begingroup
\squeezetable
\begin{turnpage}
\begin{table}[h]
\tabcolsep=0.11cm
\begin{tabular}{cc|cccccccccc|cccccccccc}
&  & \multicolumn{10}{c|}{\numu} & \multicolumn{10}{c}{\numubar} \\
%& $Q^{2}$\ (GeV/c)$^{2}$ & 0 - 0.025 & 0.025 - 0.05 & 0.05 - 0.075 & 0.075 - 0.1 & 0.1 - 0.15 & 0.15 - 0.2 & 0.2 - 0.3 & 0.3 - 0.4 & 0.4 - 0.6 & 0.6 - 0.8 & 0 - 0.025 & 0.025 - 0.05 & 0.05 - 0.075 & 0.075 - 0.1 & 0.1 - 0.15 & 0.15 - 0.2 & 0.2 - 0.3 & 0.3 - 0.4 & 0.4 - 0.6 & 0.6 - 0.8 \\
 & $Q^{2}$\ (GeV/c)$^{2}$ & 0 -   & 0.025 - & 0.05 - & 0.075 - & 0.1 - & 0.15 - & 0.2 - & 0.3 - & 0.4 - & 0.6 - &  0 -   & 0.025 - & 0.05 - & 0.075 - & 0.1 - & 0.15 - & 0.2 - & 0.3 - & 0.4 - & 0.6 -\\
 &                        & 0.025 & 0.05    & 0.075  & 0.1     & 0.15  & 0.2    & 0.3   & 0.4   & 0.6   & 0.8   &  0.025 & 0.05    & 0.075  & 0.1	  & 0.15  & 0.2    & 0.3   & 0.4   & 0.6   & 0.8  \\
\hline
\multirow{10}{*}{\numu}
 &    0 - 0.025 &  1.000 &  0.999 &  0.999 &  0.998 &  0.996 &  0.993 &  0.987 &  0.978 &  0.972 &  0.731 &  0.661 &  0.661 &  0.649 &  0.652 &  0.657 &  0.654 &  0.628 &  0.632 &  0.648 &  0.667 \\
 & 0.025 - 0.05 &  0.999 &  1.000 &  1.000 &  0.999 &  0.998 &  0.996 &  0.992 &  0.984 &  0.978 &  0.725 &  0.661 &  0.661 &  0.650 &  0.655 &  0.660 &  0.657 &  0.634 &  0.639 &  0.655 &  0.671 \\
 & 0.05 - 0.075 &  0.999 &  1.000 &  1.000 &  1.000 &  0.999 &  0.997 &  0.993 &  0.985 &  0.978 &  0.713 &  0.664 &  0.665 &  0.654 &  0.660 &  0.664 &  0.662 &  0.641 &  0.646 &  0.660 &  0.671 \\
 &  0.075 - 0.1 &  0.998 &  0.999 &  1.000 &  1.000 &  1.000 &  0.998 &  0.995 &  0.987 &  0.980 &  0.698 &  0.668 &  0.669 &  0.658 &  0.666 &  0.670 &  0.669 &  0.650 &  0.656 &  0.667 &  0.674 \\
 &   0.1 - 0.15 &  0.996 &  0.998 &  0.999 &  1.000 &  1.000 &  0.999 &  0.997 &  0.991 &  0.983 &  0.682 &  0.667 &  0.669 &  0.659 &  0.669 &  0.671 &  0.672 &  0.657 &  0.664 &  0.674 &  0.674 \\
\end{tabular}
\caption{ Correlation matrix for the $d\sigma/dQ^{2}$ flux systematic uncertainty where the \numu cross section is weighted to the \numubar flux }
\end{table}
\end{turnpage}
\endgroup

\begingroup
\squeezetable
\begin{turnpage}
\begin{table}[h]
\tabcolsep=0.11cm
\begin{tabular}{cc|cccccccccc|cccccccccc}
 &  & \multicolumn{10}{c|}{\numu} & \multicolumn{10}{c}{\numubar} \\
%& $Q^{2}$\ (GeV/c)$^{2}$ & 0 - 0.025 & 0.025 - 0.05 & 0.05 - 0.075 & 0.075 - 0.1 & 0.1 - 0.15 & 0.15 - 0.2 & 0.2 - 0.3 & 0.3 - 0.4 & 0.4 - 0.6 & 0.6 - 0.8 & 0 - 0.025 & 0.025 - 0.05 & 0.05 - 0.075 & 0.075 - 0.1 & 0.1 - 0.15 & 0.15 - 0.2 & 0.2 - 0.3 & 0.3 - 0.4 & 0.4 - 0.6 & 0.6 - 0.8 \\
 & $Q^{2}$\ (GeV/c)$^{2}$ & 0 -   & 0.025 - & 0.05 - & 0.075 - & 0.1 - & 0.15 - & 0.2 - & 0.3 - & 0.4 - & 0.6 - & 0  -	& 0.025 - & 0.05 - & 0.075 - & 0.1 - & 0.15 - & 0.2 - & 0.3 - & 0.4 - & 0.6 - \\
 &                        & 0.025 & 0.05    & 0.075  & 0.1     & 0.15  & 0.2    & 0.3   & 0.4   & 0.6   & 0.8   & 0.025 & 0.05    & 0.075  & 0.1	 & 0.15  & 0.2    & 0.3   & 0.4   & 0.6   & 0.8  \\
\hline
\multirow{10}{*}{\numu}
 &    0 - 0.025 &  1.000 &  0.974 &  0.894 &  0.830 &  0.810 &  0.652 &  0.493 &  0.068 &  0.017 & -0.178 &  0.549 &  0.531 &  0.510 &  0.406 &  0.324 &  0.342 &  0.312 &  0.195 &  0.146 &  0.100 \\
 & 0.025 - 0.05 &  0.974 &  1.000 &  0.937 &  0.853 &  0.822 &  0.688 &  0.568 &  0.144 &  0.046 & -0.164 &  0.581 &  0.560 &  0.532 &  0.465 &  0.410 &  0.430 &  0.379 &  0.271 &  0.232 &  0.127 \\
 & 0.05 - 0.075 &  0.894 &  0.937 &  1.000 &  0.958 &  0.890 &  0.749 &  0.682 &  0.346 &  0.170 & -0.104 &  0.406 &  0.365 &  0.325 &  0.306 &  0.332 &  0.402 &  0.356 &  0.254 &  0.257 &  0.260 \\
 &  0.075 - 0.1 &  0.830 &  0.853 &  0.958 &  1.000 &  0.953 &  0.817 &  0.740 &  0.440 &  0.234 & -0.038 &  0.297 &  0.284 &  0.256 &  0.261 &  0.308 &  0.410 &  0.399 &  0.316 &  0.292 &  0.264 \\
 &   0.1 - 0.15 &  0.810 &  0.822 &  0.890 &  0.953 &  1.000 &  0.940 &  0.835 &  0.420 &  0.260 & -0.014 &  0.374 &  0.370 &  0.402 &  0.385 &  0.405 &  0.501 &  0.561 &  0.485 &  0.324 &  0.202 \\
\end{tabular}
\caption{ Correlation matrix for the $d\sigma/dQ^{2}$ non-flux systematic uncertainty where the \numu cross section is weighted to the \numubar flux }
\end{table}
\end{turnpage}
\endgroup

\fi

\end{document}